%% file: main_published.tex
\documentclass[aps,prd,showpacs,floatfix,nofootinbib,superscriptaddress,longbibliography,11pt]{revtex4-2}
\pdfoutput=1
\usepackage[T1]{fontenc}
\usepackage[utf8]{inputenc}
\usepackage[greek,english]{babel}
\usepackage{CJKutf8}
\usepackage{amsfonts}
\usepackage{tikz}
\usetikzlibrary{decorations}

\usepackage{amsmath}
\usepackage{amssymb}
\usepackage{verbatim,graphicx}
\usepackage{mathtools}
\usepackage{color}
\usepackage[all]{xy}
\usepackage{simplewick}
\usepackage{hyperref}
\usepackage{setspace}
\usepackage{array}
\usepackage{tkz-euclide}
\usepackage{amsthm}
\usepackage{enumitem}
\usepackage{bbm}
\usepackage{bbold}
\usepackage{subfigure}

\graphicspath{}

\newcounter{jvcc}

\newcounter{yjcc}

\newcommand{\be}{\begin{eqnarray}}
\newcommand{\ee}{\end{eqnarray}}

\newcommand{\bmat}{\left ( \begin{array}{cc} }
	\newcommand{\emat}{\end{array} \right ) }

\usetikzlibrary{shapes.misc}

\tikzset{cross/.style={cross out, draw=black, fill=none, minimum size=2*(#1-\pgflinewidth), inner sep=0pt, outer sep=0pt}, cross/.default={2pt}}

\usetikzlibrary{shapes.misc}

\newcommand{\beq}{\begin{equation}}
\newcommand{\beqs}{\begin{equation*}}
\newcommand{\eeq}{\end{equation}}
\newcommand{\eeqs}{\end{equation*}}

\allowdisplaybreaks

\input{victorsmacros}

\def\tu{\tilde{u}}
\def\tk{\tilde{k}}
\def\tb{\wt{\b}}

\newcommand{\subf}[2]{%
	{\begin{tabular}[t]{@{}c@{}}
			#1\\#2
		\end{tabular}}%
	}

\begin{document}

\begin{center}
\vspace{2cm}
\end{center}

\title{Euclidean-to-Lorentzian wormhole transition and\\ gravitational symmetry breaking in the Sachdev-Ye-Kitaev model}

\author{Antonio M. Garc\'\i a-Garc\'\i a}
\email{amgg@sjtu.edu.cn}
\affiliation{Shanghai Center for Complex Physics,
	School of Physics and Astronomy, Shanghai Jiao Tong
	University, Shanghai 200240, China}
\author{Victor Godet}
\email{victor.godet.h@gmail.com}
\affiliation{International Centre for Theoretical Sciences (ICTS-TIFR),\\ Tata Institute of Fundamental Research,
Shivakote, Hesaraghatta, Bangalore 560089, India}

\author{Can Yin\begin{CJK*}{UTF8}{gbsn}
		(殷灿)
\end{CJK*}}
\email{yin_can@sjtu.edu.cn}
\author{Jie Ping Zheng\begin{CJK*}{UTF8}{gbsn}
(郑杰平)
\end{CJK*}}
\email{jpzheng@sjtu.edu.cn}
\affiliation{Shanghai Center for Complex Physics,
	School of Physics and Astronomy, Shanghai Jiao Tong
	University, Shanghai 200240, China}
\begin{abstract}
\vspace{3cm}
We study a two-site Sachdev-Ye-Kitaev model with complex couplings and a weak inter-site interaction. At low temperatures, the system is dual to a Euclidean wormhole in Jackiw-Teitelboim gravity plus matter. Interestingly, the energy spectrum becomes real for sufficiently strong inter-site coupling despite the Hamiltonian being non-Hermitian. In gravity, this complex-to-real transition corresponds to a Euclidean-to-Lorentzian transition: a dynamical  restoration of the gravitational $\r{SL}(2,\R)$ symmetry of the Lorentzian wormhole, broken to $\r{U}(1)$ in the Euclidean wormhole.  We show this by  identifying an order parameter for the symmetry breaking and by matching the oscillating patterns of the Green's functions. Above the transition, the system can be continued to Lorentzian signature and  is dual to an eternal traversable wormhole. Additionally, we observe a thermal phase transition from the wormhole to two black holes and provide a detailed matching of the associated physical quantities. The analysis of level statistics reveals that in a broad range of parameters the dynamics is quantum chaotic in the universality class of systems with time reversal invariance.

\end{abstract}

\maketitle
\newpage

\makeatletter
\def\l@subsubsection#1#2{}
\makeatother

\tableofcontents

\section{Introduction}
The Sachdev-Ye-Kitaev (SYK) \cite{sachdev1993,kitaev2015} is a model  of fermions with infinite range interactions \cite{french1970,bohigas1971,bohigas1971a,french1971,mon1975,benet2001} in zero spatial dimension.  At large $N$ and low temperatures, it has a gravity  dual described by Jackiw-Teitelboim (JT) gravity \cite{jackiw1985, teitelboim1983}, a theory of two-dimensional gravity describing a universal sector of near-extremal black holes  \cite{Almheiri:2014cka, jensen2016, engels2016, maldacena2016a, Nayak:2018qej, Moitra:2019bub, Castro:2019crn, Iliesiu:2020qvm}. This toy version of AdS/CFT, known as nearly AdS$_2$ holography, has been used to investigate many questions in quantum gravity. The simplicity of this model of holography makes it an ideal playground to test ideas and learn new insights. For some pointers into the nearly AdS$_2$ literature, we can mention topics such as the quantum chaotic nature of gravity \cite{Cotler:2016fpe, garcia2016, garcia2017,Saad:2018bqo, Saad:2019lba, Saad:2019pqd}, traversable wormholes \cite{maldacena2017, maldacena2018, milekhin2019}, quantum cosmology \cite{Maldacena:2019cbz,Anous:2020lka, Chen:2020tes,Hartman:2020khs,Aalsma:2021bit, Kames-King:2021etp}, flat space holography \cite{Dubovsky:2017cnj,Afshar:2019axx, Gautason:2020tmk, Godet:2021cdl}, computational complexity \cite{Lin:2019kpf, Iliesiu:2021ari}
or the information paradox  \cite{Almheiri:2018ijj, Almheiri:2019psf, Almheiri:2019hni, almheiri2020, penington2020,almheiri2020a,Stanford:2020wkf, Gao:2021tzr}.

A theme that has emerged in recent years is that the semiclassical gravitational path integral appears to be dual to an ensemble average of theories \cite{Saad:2018bqo, Saad:2019lba}. The meaning of this average remains to be understood \cite{Witten:1999xp, maldacena2004}, see \cite{Saad:2021rcu, Saad:2021uzi, Mukhametzhanov:2021nea,Iliesiu:2021are, garcia2022, Blommaert:2021gha, Mukhametzhanov:2021hdi,Blommaert:2021fob,  Heckman:2021vzx, Schlenker:2022dyo, Collier:2022emf, Chandra:2022bqq} for a sample of recent discussions on this point. The inclusion of complex metrics appears necessary but it is unclear how to specify the contour of integration \cite{Halliwell:1989dy,Bousso:1998na, Sorkin:2009ka, Witten:2021nzp}. In this paper, we study the gravitational path integral in a setup where we have analytic control on the gravity side and a dual microscopic description in terms of SYK, in order to shed some light on the rules governing the gravity path integral.

A  notable example of nearly AdS$_2$ holography is the eternal traversable wormhole  \cite{maldacena2018} dual to a two-site SYK model with a weak inter-site coupling. It was shown that both the wormhole solution in JT gravity and the two-site SYK model were described by the same Schwarzian effective action.  The ground state was argued to be gapped and close to a thermofield double state. The gap induced by the weak inter-site coupling is enhanced by the strong interactions in each site, as can be observed in the real time probability of tunneling between the two sites \cite{plugge2020,milekhin2019} .
As the temperature is increased, the system experiences a first order phase transition to a phase with two black holes. This transition affects the quantum dynamics \cite{garcia2019} that is quantum chaotic only in the high temperature phase. Extensions of these results include replacing Majoranas with Dirac fermions \cite{pengfei2021,sahoo2020,zhang2020,garcia2021e} or the use of a sparse \cite{garcia2021c,swingle2020} two-site SYK \cite{caceres2021} while potential applications in condensed matter were addressed in \cite{pengfei2021a}.

A purely Euclidean version of this story was obtained in \cite{garcia2021} by studying a two-site non-Hermitian SYK model with complex couplings but without inter-site interaction. In this case, the Hamiltonian is non-Hermitian and does not define a Lorentzian system with unitary evolution. Nonetheless, the system can be studied as a purely Euclidean system, from the point of view of  statistical mechanics. The low temperature phase was shown to be dominated by replica symmetry breaking configurations \cite{garcia2021a,garcia2022a} corresponding to a Euclidean wormhole in JT gravity \cite{garcia2021}. Here, the imaginary part of the Hamiltonian gives imaginary sources in the wormhole. As there is no coupling between the two sites, the Euclidean wormhole has to be the result of the SYK average. This leads to a factorization puzzle which can be resolved by finding half-wormhole solutions \cite{garcia2022} which realize the proposal of \cite{Saad:2021rcu} in nearly AdS$_2$ holography. Recent studies on the non-Hermitian SYK model include a symmetry classification of quantum chaotic dynamics,
\cite{garcia2021d}, a generalization with additional $\r{U}(1)$ charge \cite{Rathi:2021mla}, measurement-induced transitions  \cite{zhang2021,pengfei2021c} and Lindbladian approach to dissipative quantum dynamics \cite{sa2022,kulkarni2022}.

In this paper, we study a two-site SYK model with both complex couplings and a weak inter-site interaction. This system combines the two effects  discussed above and is dual to a Euclidean wormhole in JT gravity. Our main result is the observation of a dynamical complex-to-real transition where the spectrum of the  Hamiltonian (obtained by exact diagonalization) becomes real for sufficiently strong inter-site coupling, despite the Hamiltonian being non-Hermitian. This shows that a purely Euclidean system can experience a transition above which the system has real energy spectrum and hence defines a Lorentzian system with unitary evolution.

  In JT gravity, we show that this complex-to-real transition corresponds to a Euclidean-to-Lorentzian transition: the dynamical restoration of the $\r{SL}(2,\R)$ symmetry of the Lorentzian wormhole, broken to $\r{U}(1)$ in the Euclidean wormhole.  To be more precise, there are two distinct $\r{SL}(2,\R)$ symmetries which come from the isometries of the Lorentzian wormhole (the  global AdS$_2$ geometry). There is an $\r{SL}(2,\R)$ gauge symmetry, which is part of the gravitational constraints of the theory, and an $\r{SL}(2,\R)$ global symmetry, corresponding to the physical action of the isometries on the asymptotic boundary \cite{Lin:2019qwu, Harlow:2021dfp}. These are gravitational symmetries because they are part of the diffeomorphism group of the theory. Both symmetries are broken to $\r{U}(1)$ in the Euclidean wormhole. The Euclidean-to-Lorentzian transition corresponds to a restoration of the gravitational $\r{SL}(2,\R)$ symmetry (both gauge and global) in the purely Euclidean system (defined by imposing only the $\r{U}(1)$ gauge constraint).

We study this complex-to-real transition by identifying, both in SYK and JT gravity, an order parameter for the gravitational symmetry breaking. We propose a mechanism for the  transition involving a change of contour in the gravity path integral. The transition can also be diagnosed in a phase shift measured by the Green's functions. Above the transition, the Hamiltonian is pseudo-Hermitian, \ie it has real eigenvalues despite being non-Hermitian, and defines a Lorentzian system with unitary evolution in the framework of pseudo-Hermitian quantum mechanics \cite{Mostafazadeh:2008pw}. This shows that a pseudo-Hermitian Hamiltonian can be holographic, and is dual here to an eternal traversable wormhole.

 In addition to the complex-to-real transition, we also observe a thermal phase transition from the wormhole phase to a phase with two black holes. This transition has been discussed in \cite{maldacena2018,garcia2021}  for limiting cases  of our system. Here, we study the parameters of this transition, such as the energy gap and critical temperature, as functions of the two parameters $\la$ and $\k$. We find an excellent match between the numerical SYK results and the analytical JT analysis. Note that this transition constitutes another example of gravitational symmetry breaking where the $\r{U}(1)\times \r{U}(1)$ symmetry of the two black holes is broken to the diagonal $\r{U}(1)$ of the Euclidean wormhole.

In this work we view JT gravity as an effective theory, an approximation of the (still elusive, see \eg \cite{Goel:2021wim}) exact holographic dual of SYK, in the same way that semi-classical gravity is viewed in higher-dimensional examples of AdS/CFT. This has to be contrasted with the exact path integral computations of \cite{Saad:2019lba} where JT gravity is viewed as an exact theory, dual to a random matrix model (RMT). The imaginary sources allow us to evaluate the path integral using saddle-points which is closer to what we can expect in higher dimensions. The gravitational symmetry breaking/restoration described in this paper, interpreted as a Euclidean-to-Lorentzian transition, might also be possible more generally. In higher dimensions, Euclidean wormholes with similar properties were constructed in \cite{marolf2021} and static traversable wormholes were obtained in \cite{Maldacena:2018gjk, Bintanja:2021xfs}.

Finally, we study the level statistics in the SYK model and find that the dynamics is quantum chaotic in a broad range of parameters as it is well described by the random matrix predictions for either real or complex eigenvalues.

\section{The two-site non-Hermitian SYK model}

The system we consider consists of two SYK models with complex couplings, labeled left (L) and right (R), and with a weak inter-site interaction. The Hamiltonian is
\begin{equation} \label{hami}
	H=H_L+H_R+H_I
	\end{equation}
	 with
\begin{equation}\begin{aligned}\label{hamie}
H_L &=-\sum_{i<j<k<l}(J_{ijkl}+i\kappa M_{ijkl})\psi^L_i \psi^L_j \psi^L_k \psi^L_l \\
H_R &=-\sum_{i<j<k<l}(J_{ijkl}-i\kappa M_{ijkl})\psi^R_i \psi^R_j \psi^R_k \psi^R_l \\
H_I &=i\lambda\sum_i \psi^L_i\psi^R_i,
\end{aligned}\end{equation}
where $\psi_i^L,\psi_i^R,i=1,\dots,N$ are Majorana fermions. After ensemble average, and using the standard decoupling procedure with
\begin{equation}\begin{aligned}
\langle J_{ijkl}\rangle=\langle M_{ijkl}\rangle=0, \qquad \langle J_{ijkl}^2\rangle=\langle M_{ijkl}^2\rangle=\frac{3!J^2}{N^3}
\end{aligned}\end{equation}
we obtain the following action,
\begin{equation}\begin{aligned}
I_{eff}= -\frac{1}{2}\log\det(\delta_{ab}\partial-\Sigma_{ab}) +\frac{1}{2}\sum_{ab}\int\!\!\!\int\left(\Sigma_{ab}G_{ab}-\frac{(1-t_{ab}\kappa^2)J^2}{4}G_{ab}^4 \right) +\frac{i\lambda}{2}\int G_{LR}(\tau,\tau)-G_{RL}(\tau,\tau).
\label{eq:effective_action}
\end{aligned}\end{equation}
where $a = L, R, b = L, R$ and $t_{LL} = t_{RR} = 1, t_{LR} = t_{RL} = -1$.

In the large $N$ limit, the saddle-point Schwinger-Dyson (SD) equations are given by,
\begin{equation}\left\{\begin{aligned}\label{sde}
& -i\omega G_{LL}-\Sigma_{LL}G_{LL}-\Sigma_{LR}G_{RL}=1, \quad -i\omega G_{LR}-\Sigma_{LL}G_{LR}-\Sigma_{LR}G_{RR}=0 \\
& \Sigma_{LL}(\tau)=(1-\kappa^2)J^2 G_{LL}^3(\tau), \qquad \Sigma_{LR}(\tau)=(1+\kappa^2)J^2 G_{LR}^3(\tau) -i\lambda\delta(\tau)
\end{aligned}\right.\end{equation}
where the first two equations are expressed in the frequency domain, while the last two are in the imaginary time one.

\subsection{Thermodynamic properties}

In our system, there are two parameters $\lambda$ and $\kappa$. The parameter $\lambda$ controls the strength of direct hopping between the two systems. As mentioned earlier, the wormhole phase is characterized \cite{maldacena2018} by a non-trivial dependence of the gap on $\lambda$. By contrast, $\kappa$ controls the strength of the imaginary part of each separate SYK Hamiltonian \cite{garcia2021a,garcia2021} and it is not directly related to the coupling between the two systems.

In this section, we study the combined effect of the two couplings in the thermodynamic properties of the system. More specifically, we compute the free energy and the spectral gap $E_g$ as a function of temperature and the parameters $\lambda$ and $\kappa$.

The free energy is obtained from the on-shell action where the Green's functions in the action are given by the solutions of the saddle-point SD equations (\ref{eq:effective_action}),
\begin{equation}\begin{aligned}
		-\beta F =& \log 2 + \frac{1}{2}Tr\log\frac{(-i\omega-\Sigma_{LL})(-i\omega-\Sigma_{RR})-\Sigma_{LR}\Sigma_{RL}}{(-i\omega)(-i\omega)} \\
		& -\frac{1}{2}\left(1-\frac{1}{q}\right)J^2 \int\!\!\!\int (1-\kappa^2)(G_{LL}^4 +G_{RR}^4) +(1+\kappa^2)(G_{LR}^4 +G_{RL}^4). \\
\end{aligned}\end{equation}

We know that for $\lambda=0$ and finite $\kappa$ \cite{garcia2021a,garcia2021}, the system will stay in the wormhole phase, whose low-temperature free energy is lower than that of black hole phase. In the wormhole phase, the free energy is independent of the temperature.

The reason of the existence of the wormhole phase even without an explicit coupling term ($\lambda=0$) can be understood from the SD equations by rewriting them as
\begin{equation}\begin{aligned}
		\Sigma_{LL}(\tau)=J'^2 G_{LL}^3(\tau), \qquad \Sigma_{LR}(\tau)=J'^2 G_{LR}^3(\tau) +\frac{2\kappa^2}{1-\kappa^2}J'^2 G_{LR}^3(\tau)
\end{aligned}\end{equation}
where $J'^2 = (1-\kappa^2)J^2$. Solutions of these equations are Green's functions with an exponential $\tau$ dependence. Specifically, if we assume $G_{LR}$ is purely imaginary and exponential around $\tau=0$, $G_{LR}$ will be non-zero only in the neighborhood of $\tau=0$. Therefore, we can make the approximation $\frac{2\kappa^2}{1-\kappa^2}G_{LR}^3(\tau)\sim i\lambda_{eff}\delta(\tau)$ where $\lambda_{eff}$ can be identified as the effective coupling constant, and its sign depends on that of $\text{Im}[G_{LR}]$. Comparing with the SD equations in \cite{maldacena2018}, such approximation will lead to the exponential solutions of $G_{LL},iG_{LR}\in\mathbb{R}$ characteristic of the wormhole phase.
Indeed, the gap $E_g$ that separates the wormhole ground state from excited states can be extracted from the exponential decay of $G_{LR}$, $E_g=-\lim_{\tau \to \infty}\log|G_{LR}|/\tau$.

\begin{figure}[h]
	\centering
	\subfigure[]{\includegraphics[scale=.5]{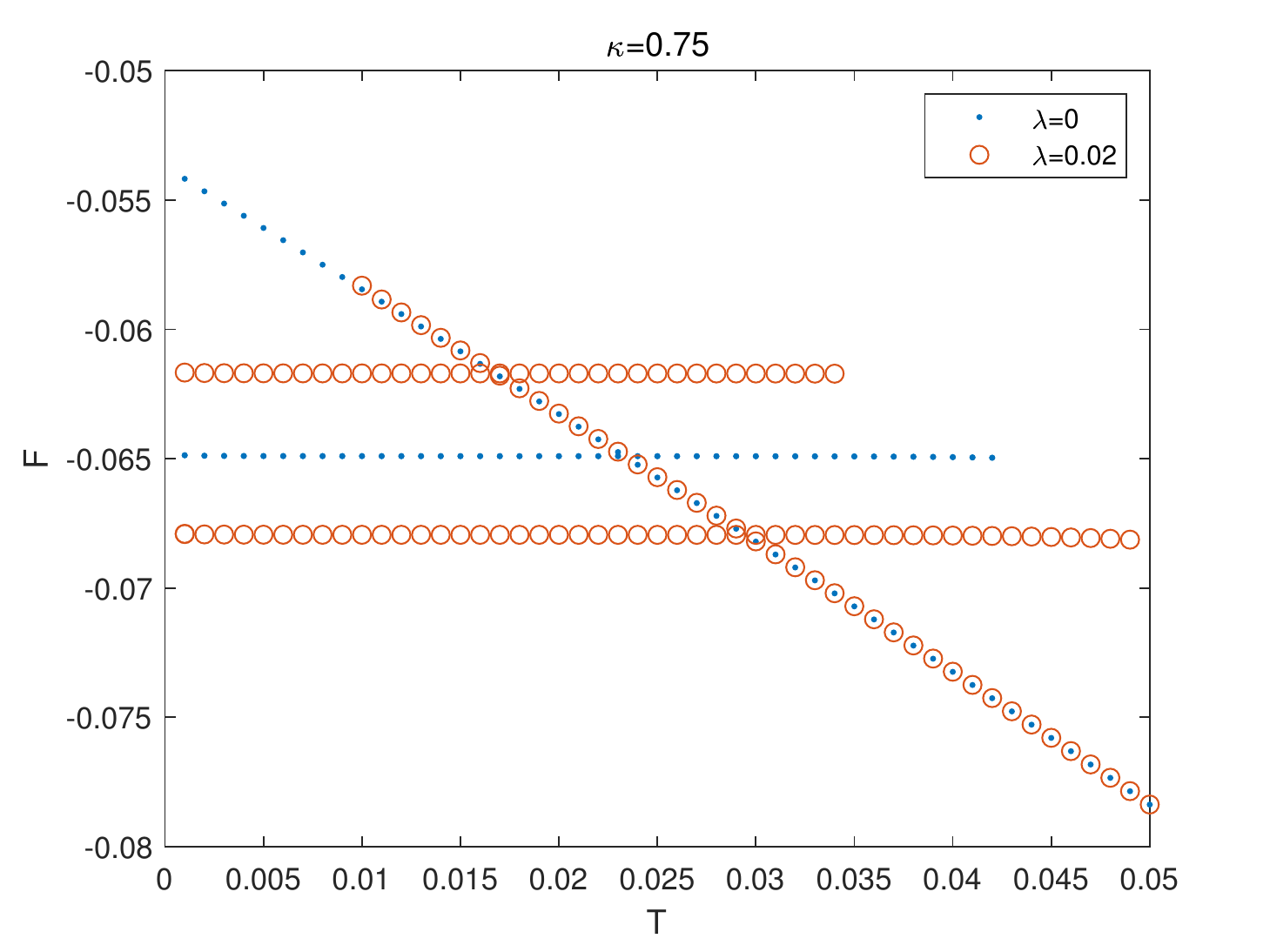}}
	\subfigure[]{\includegraphics[scale=.5]{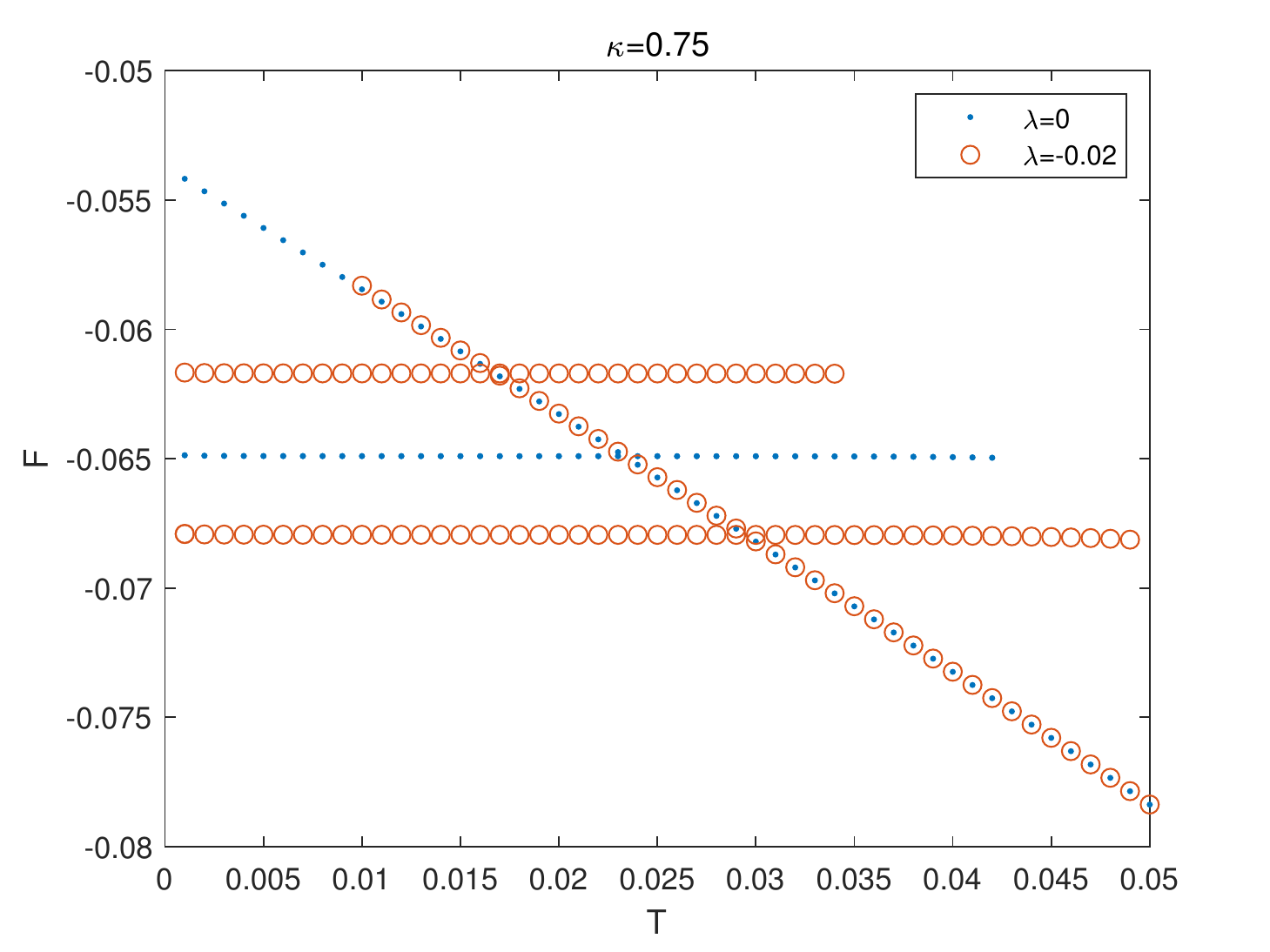}}
	\caption{$F$ versus $T$ for $\kappa = 0.75$. Left: $\lambda=0.02$. Right: $\lambda=-0.02$. For comparison, we also show the result for $\lambda = 0$ (blue dots). As is observed, $F$ does not depend on the sign of $\lambda$ and a finite $\lambda$ lowers the free energy. }\label{fig:F_ka_p75_ld_02&m02}
\end{figure}

\begin{figure}[h]
	\centering
	\subfigure[]{\includegraphics[scale=.5]{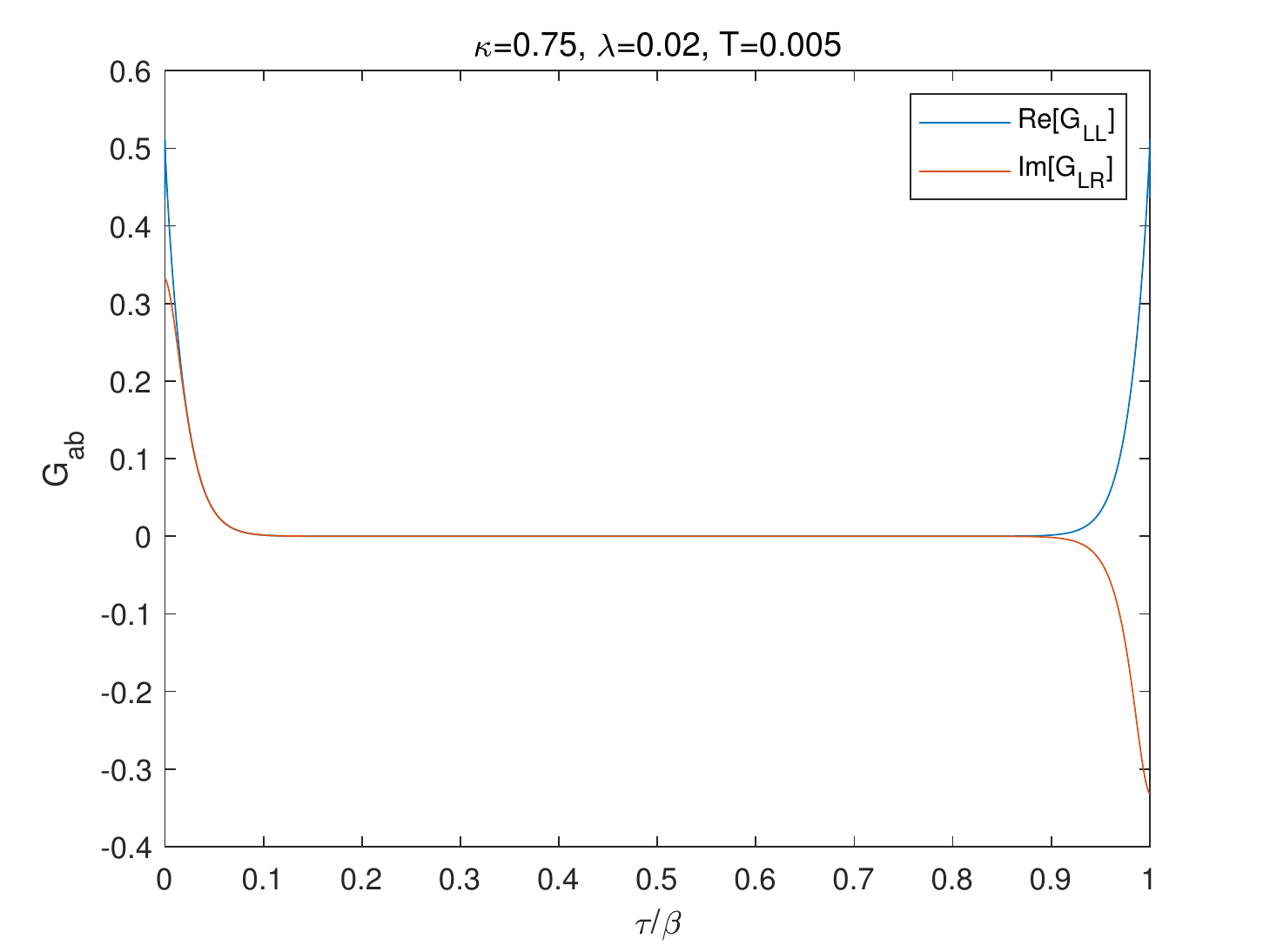}}
	\subfigure[]{\includegraphics[scale=.5]{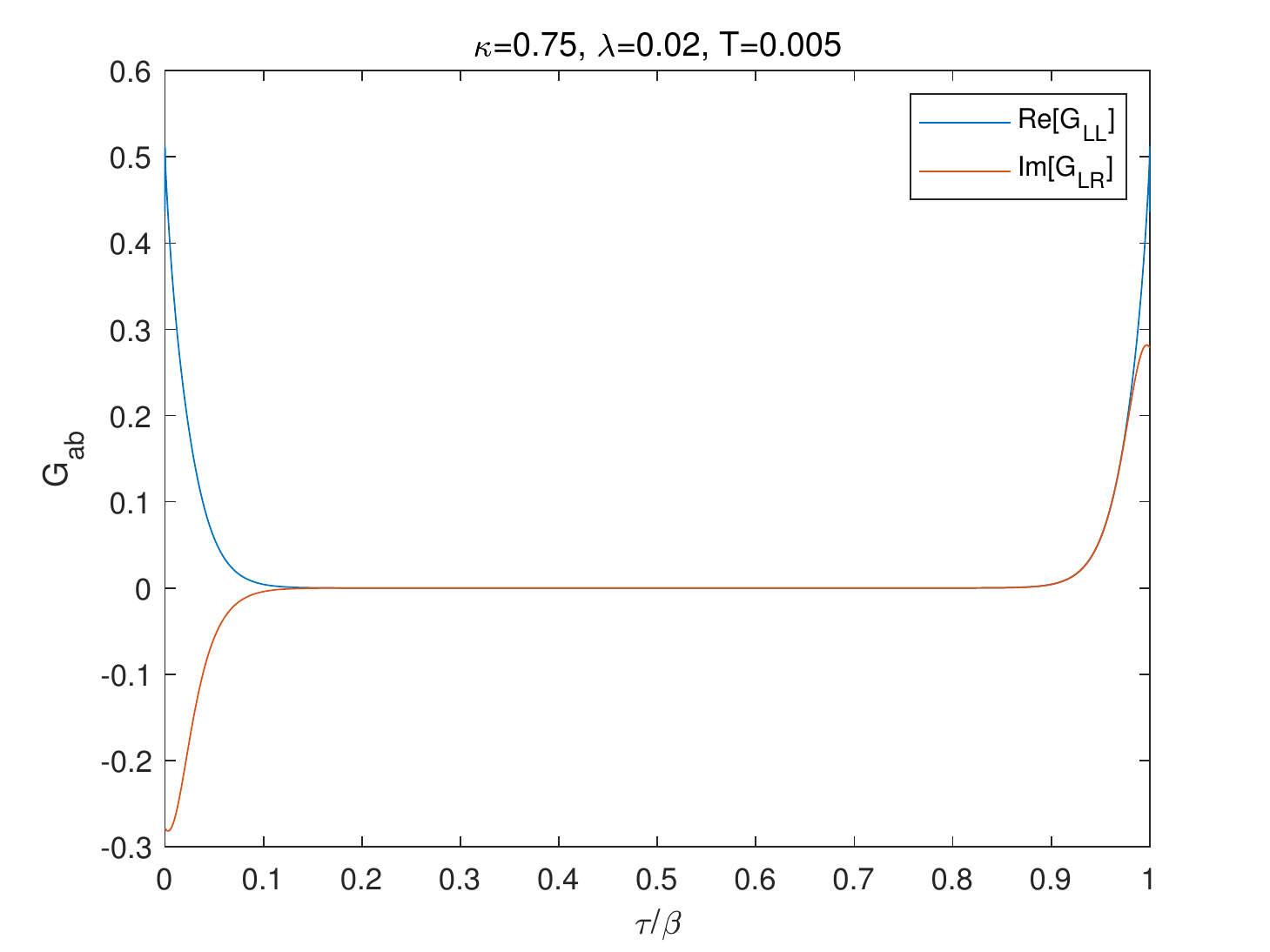}}
	\caption{$G_{ab}$ for $\kappa = 0.75$, $\lambda = 0.02$ and $T=0.005$ for the two branches shown in Fig.~\ref{fig:F_ka_p75_ld_02&m02}. Different sign of $G_{LR}$ will enhance/reduce the effective coupling, leading to the same behavior for $F$ for different signs of $\lambda$. When $\lambda$ is positive, the left solution is chosen, and vice versa.  }\label{fig:G_ka_p75_ld_02_T_005}
\end{figure}

\begin{figure}[h]
	\centering
	\subfigure[]{\includegraphics[scale=.5]{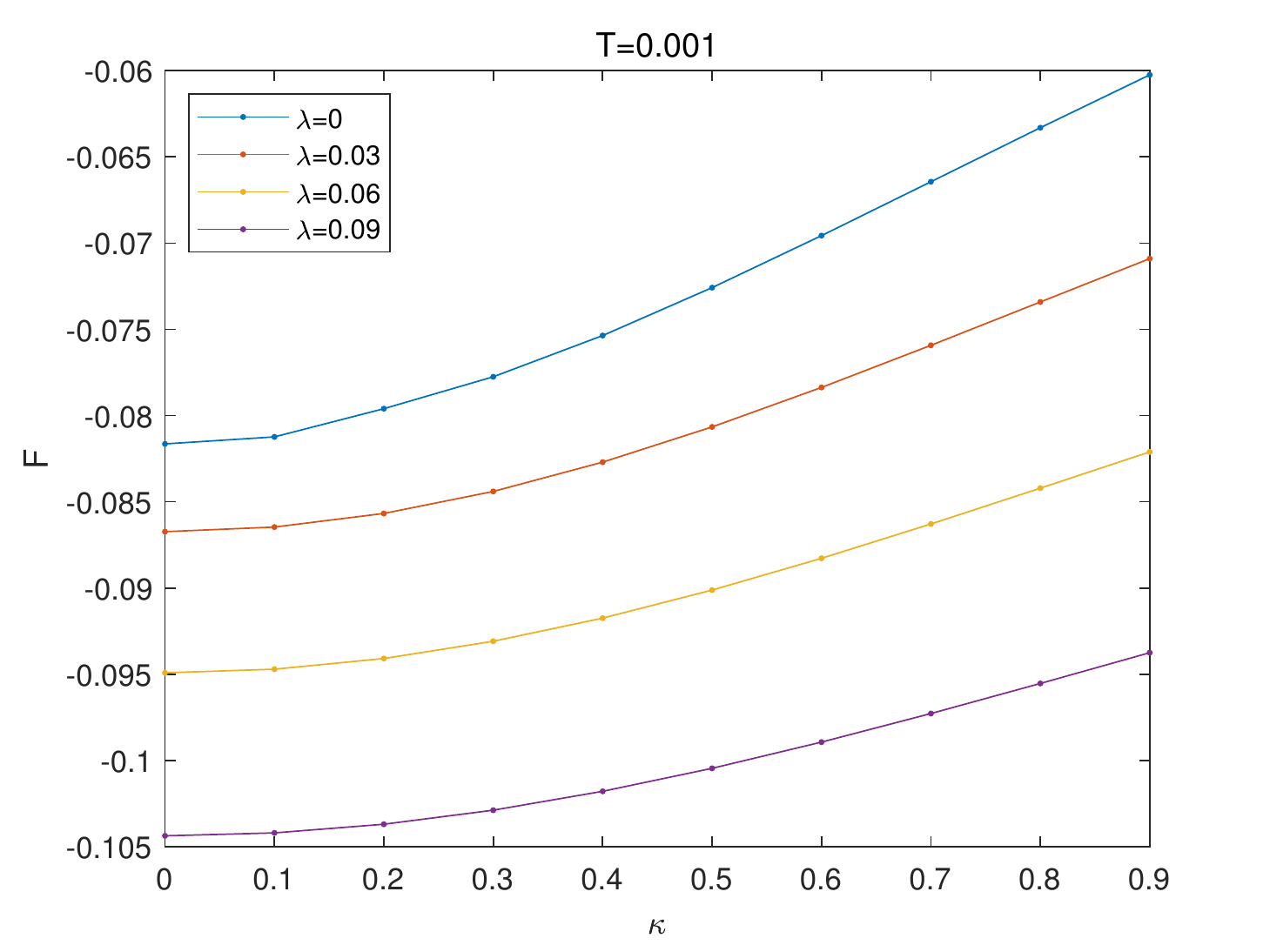}}
	\subfigure[]{\includegraphics[scale=.5]{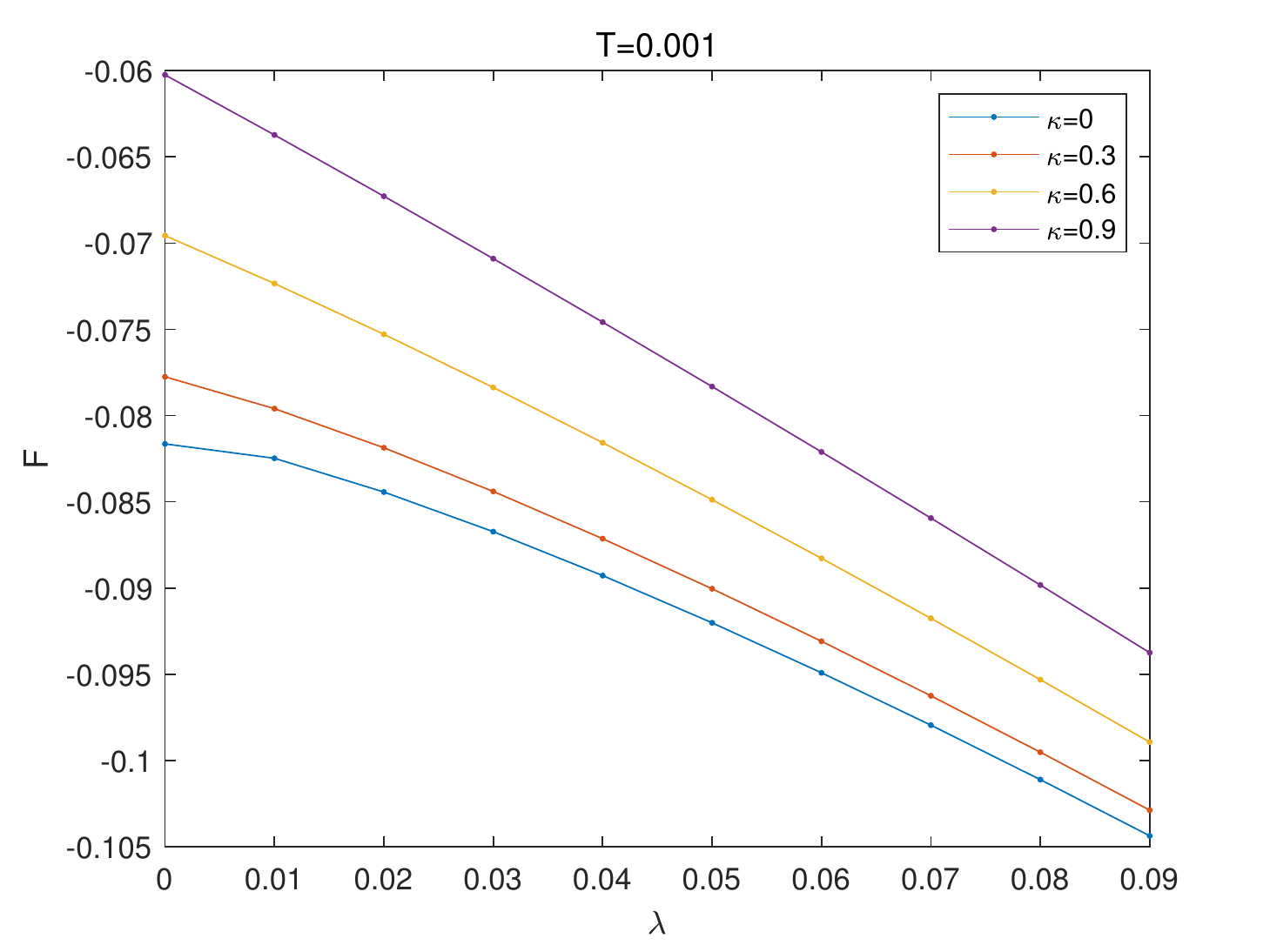}}
	\caption{(a) $F$ versus $\kappa$ for various $\lambda$ and $T=0.001$; (b) $F$ versus $\lambda$ for various $\kappa$ and $T=0.001$. For a given temperature, the free energy $F$ increases with $\kappa$, while it decreases with $\lambda$.}\label{fig:F0}
\end{figure}

\begin{figure}[h]
	\centering
	\subfigure[]{\includegraphics[scale=.5]{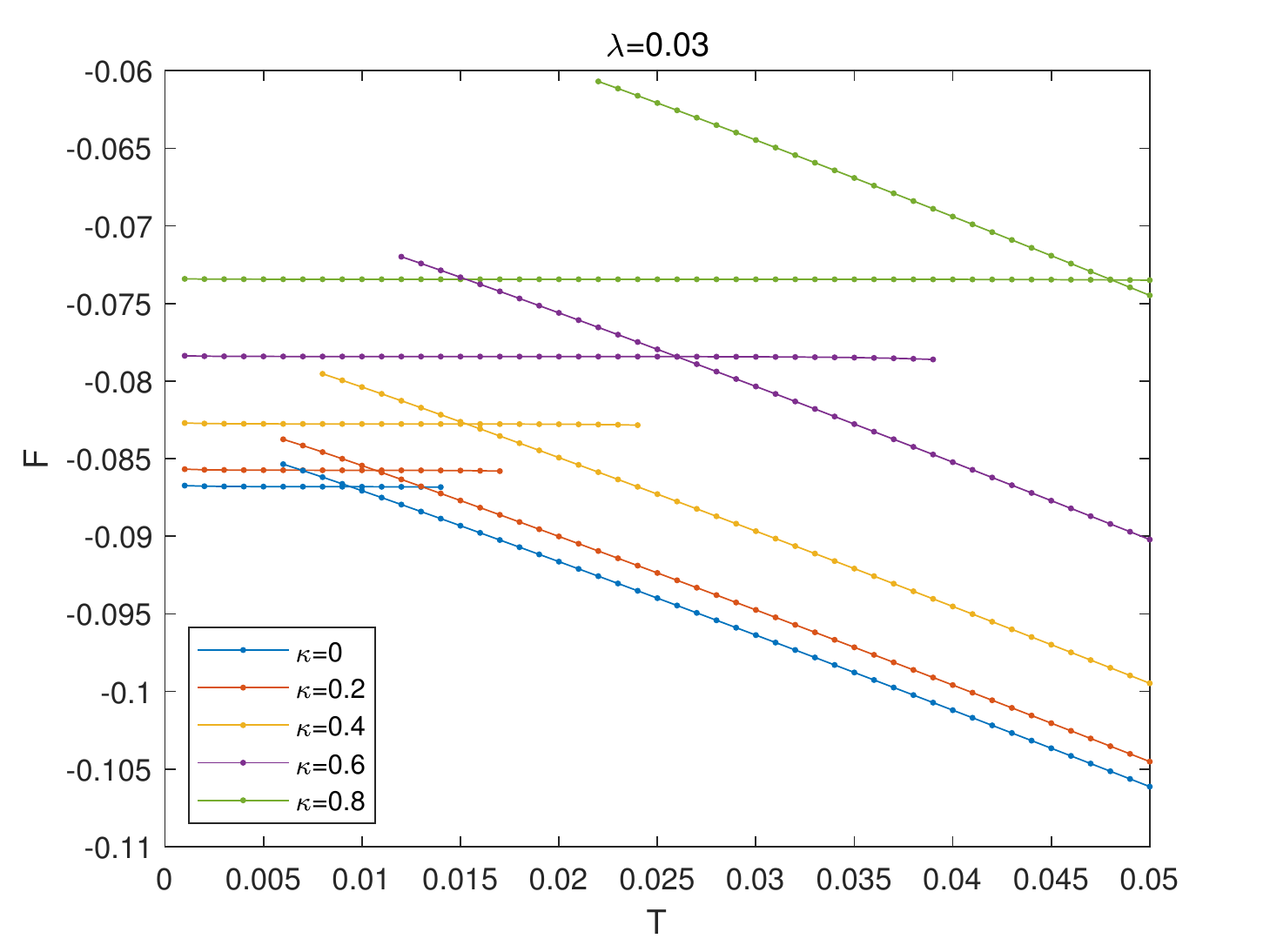}}
	\subfigure[]{\includegraphics[scale=.5]{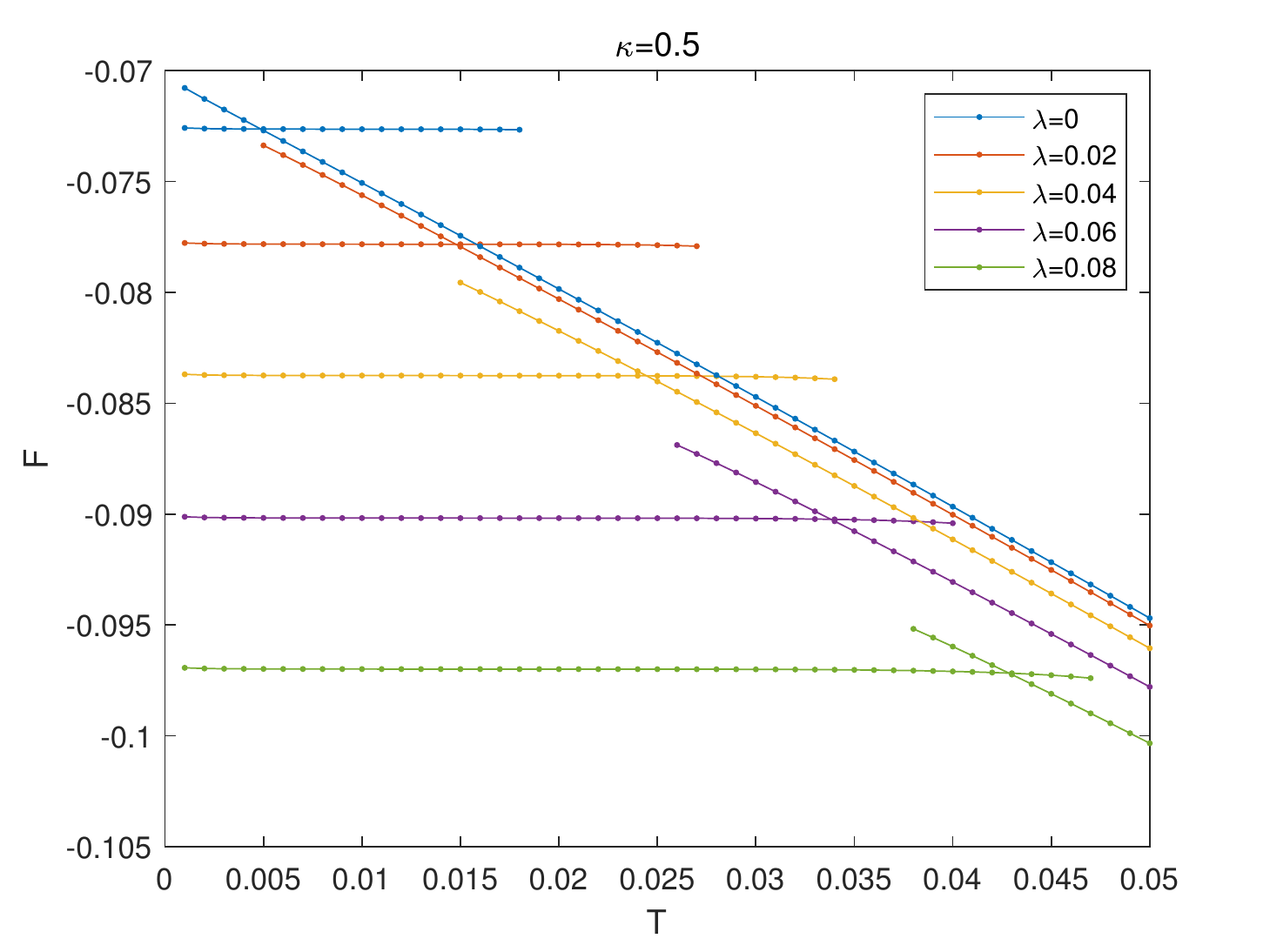}}
	\caption{(a) Free energy $F$ as a function of temperature $T$ for various $\kappa$'s and $\lambda=0.03$; (b) Free energy $F$ as a function of temperature $T$ for various $\lambda$'s and $\kappa=0.5$. In all cases, we observe a first order phase transition with a critical temperature that is an increasing function of $\lambda$ and $\kappa$. }\label{fig:F_T}
\end{figure}

We now proceed to investigate the case when $\lambda$ is also turned on.
In principle, from the form of the SD equations, one may think that the effect of a finite $\lambda$ on the free energy may depend on its sign and it may not always enhance the gap that characterizes the the wormhole phase.
However, our results clearly indicate that this is not the case. Finite $\lambda$ will always lower the free energy for any given $\kappa$. The reason is that, for given $\lambda$, $\kappa$, $T$, there are three solutions and only the one with the lowest free energy will be chosen since it is the dominant saddle point solution when evaluating partition function. Two among these three solutions, denoted by $G_{LR}^{(1)}$ and $G_{LR}^{(2)}$, give approximately constant free energy and represent the wormhole phase, the third one corresponding to the black hole phase.  When $\lambda=0$, two wormhole solutions satisfy $G_{LR}^{(1)}=-G_{LR}^{(2)}$ and they give rise to the same free energy. However, if $\lambda\neq 0$, this equality becomes approximately correct, \ie $G_{LR}^{(1)}\approx -G_{LR}^{(2)}$, and the free energy will split, so one of the solutions for $G_{LR}$ increases the free energy while the other decreases it for any temperature or $\kappa$ (see results depicted in Fig.~\ref{fig:F_ka_p75_ld_02&m02}). Specifically, when $\lambda>0$, the solution with positive imaginary part $\text{Im}[G_{LR}(\tau)]>0$ in $\tau\in(0,\epsilon)$, as illustrated in the left diagram of Fig.~\ref{fig:G_ka_p75_ld_02_T_005}, is preferred for lower free energy, while $\lambda<0$ we have opposite selection for the same reason. Therefore, finite $\lambda$ can always make the free energy lower by choosing proper solutions. We will see in the next section that the same mechanism takes place in JT gravity: two wormhole saddle-points are exchanged under a change of sign of $\la$, ensuring that thermodynamical quantities are even functions of $\la$.

In the large $N$ limit, the path integral, and therefore the free energy, is dominated by the saddle-point configurations given by the solutions of the SD equations. The physical solution is the one with the lowest free energy. Therefore, as illustrated in Fig. \ref{fig:F0}, the free energy will always decrease as $\lambda$ is increased from zero, so the wormhole phase becomes more thermodynamically stable by combining the effect of  $\lambda$ and $\kappa$.

Having provided a qualitative description of the impact of a finite $\lambda$, we now proceed to a more systematic analysis of the free energy $F(T)$ for various $\kappa$ and $\lambda$. In all cases, see Fig.~\ref{fig:F_T}, there always exists a first order phase transition separating a nearly flat free energy in the low temperature limit from a high temperature phase which is approximately linear. The critical temperature is a increasing function of $\kappa$ and $\lambda$ reinforcing the idea that the two effects are additive regarding the stability of the wormhole low temperature phase.

\begin{figure}[h]
	\centering
	\includegraphics[scale=.7]{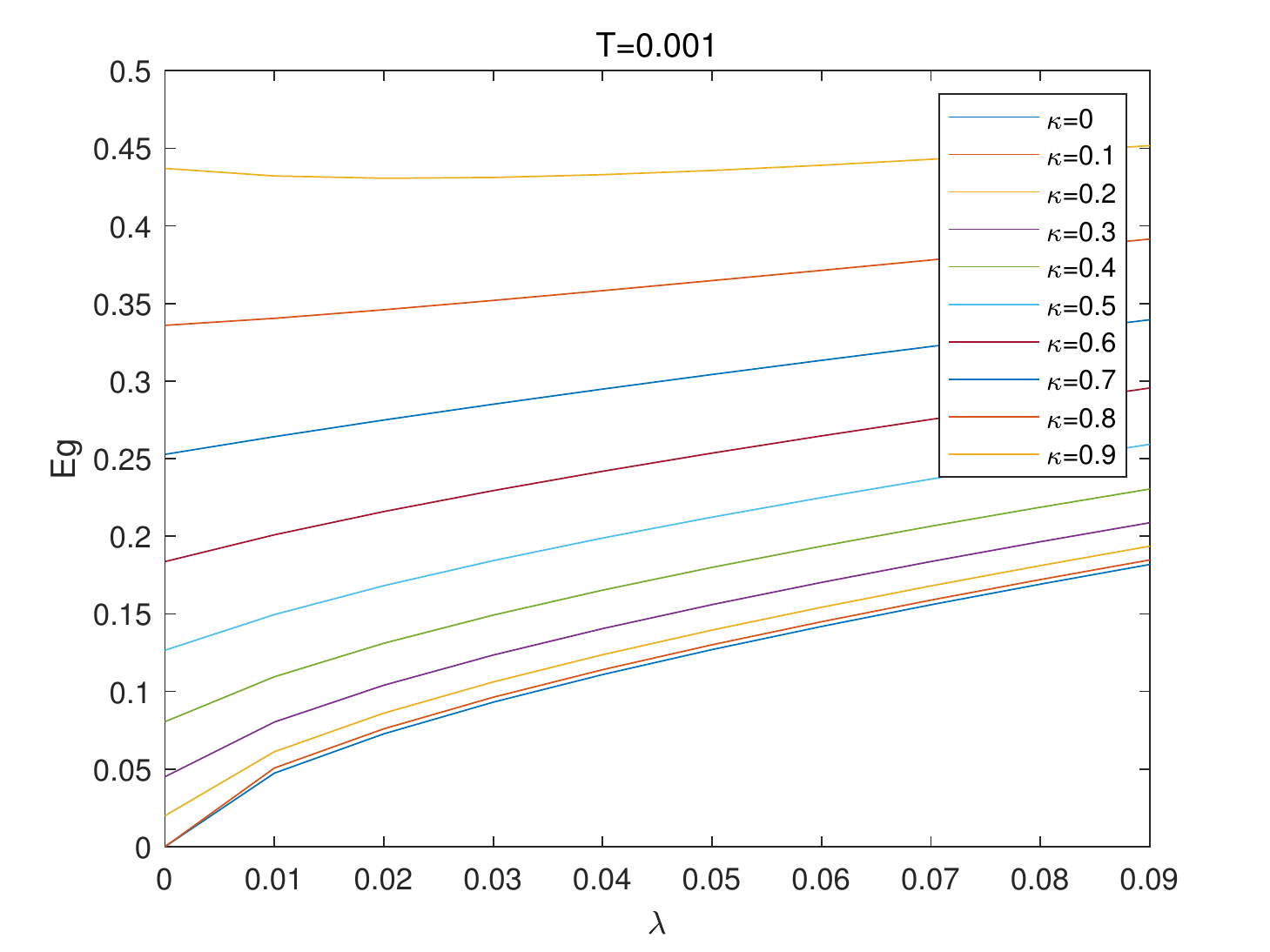}
	\caption{$E_g$ versus $\lambda$ for $T=0.001$. For $\kappa\leq 0.6$, and $\lambda \geq 0.02$, $E_g$ depends weakly on $\lambda$ and the curves for different $\kappa$ are near parallel which indicates $E_g$'s dependence on $\kappa$ and $\lambda$ is approximately separable. }\label{fig:Eg_lambda}
\end{figure}

\begin{figure}[h]
	\centering
	\includegraphics[scale=.7]{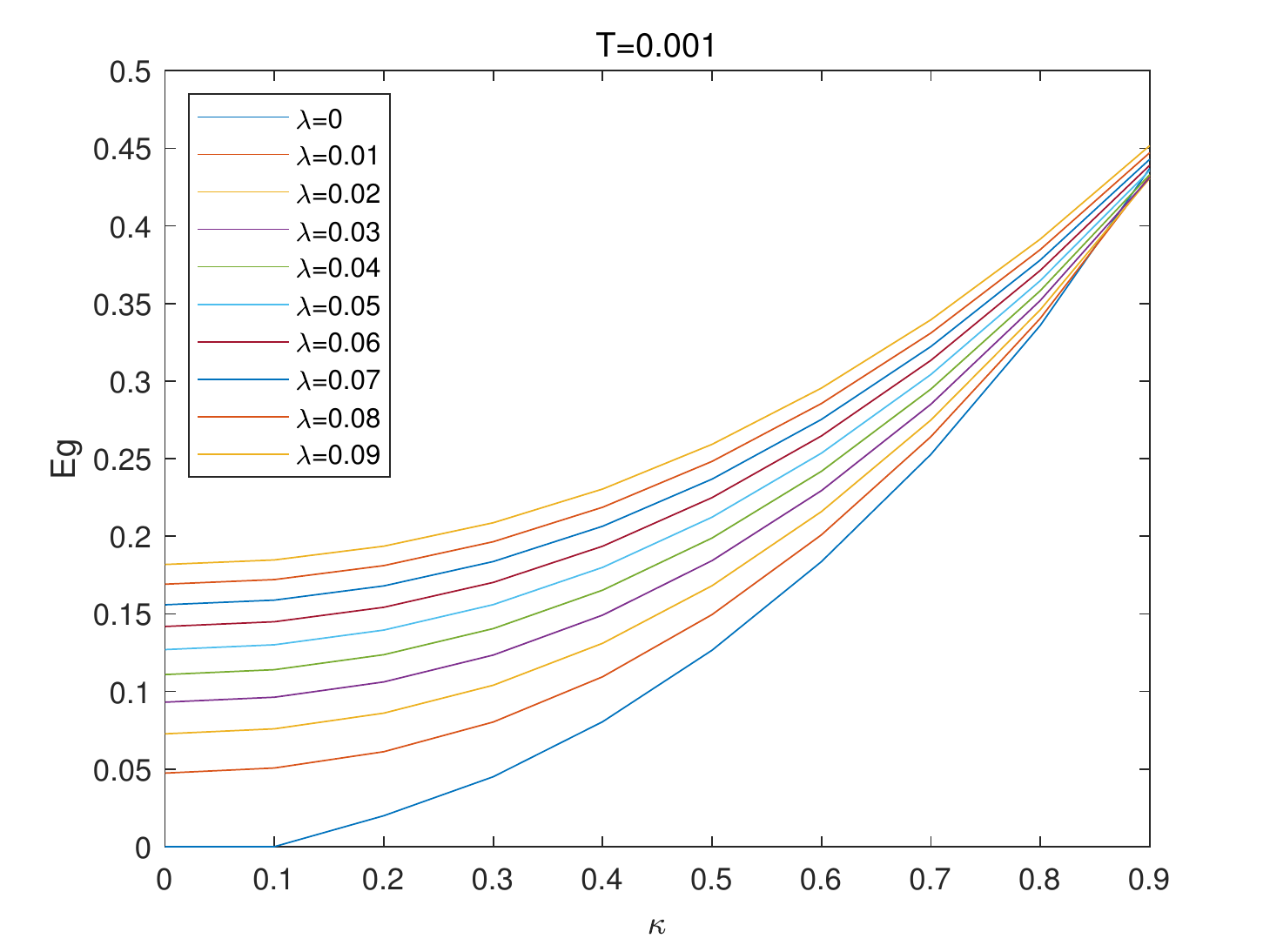}
	\caption{$E_g$ versus $\kappa$ for $T=0.001$ and different $\lambda $.
		As in the previous figure, we observe an almost flat dependence on $\kappa$ for sufficiently small $\lambda$. This quasi-flat region increases with $\lambda$.
	}\label{fig:Eg_kappa}
\end{figure}

\begin{figure}[h]
	\centering
	\subfigure[]{\includegraphics[scale=.7]{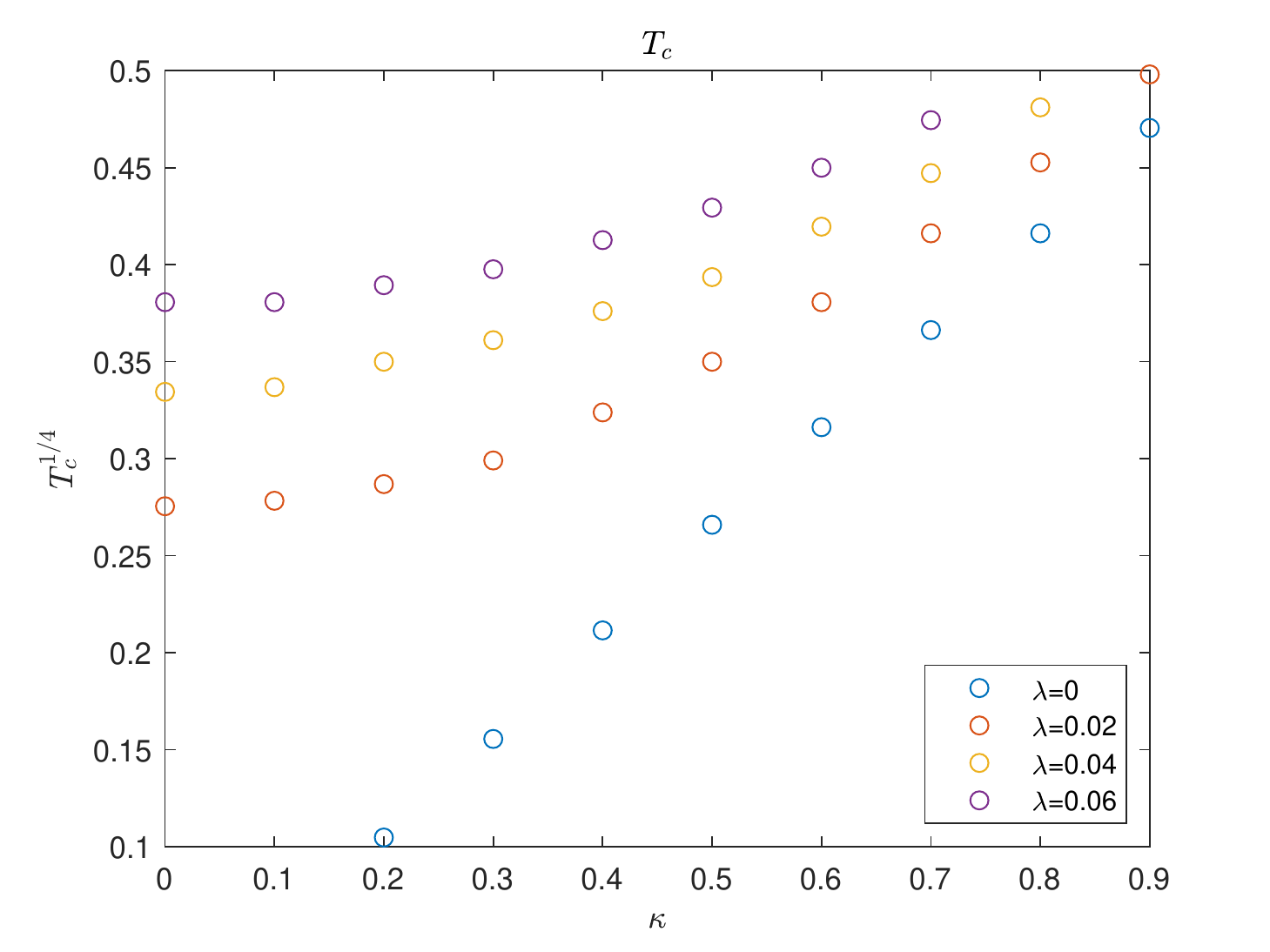}}
	\caption{$T_c^{1/4}$ versus $\kappa$ for various  $\lambda$. $T_c$ is critical temperature of phase transition. For $\lambda=0$, our results indicate $T_c\sim \kappa^4$, which is consistent with the analytic result of Ref.~\cite{garcia2021a}. For finite $\lambda$, we observe strong deviations from this behavior. The analytical prediction \eqref{eq:Tc_gravity} from the JT gravity (the effective Schwarzian action)  matches well with the SYK results, see Fig.~\ref{fig:Tc_ew_kappa_lambda}. }\label{fig:Tc_lambda_kappa}
\end{figure}

Likewise, the dependence of $E_g$ on $\kappa$ and $\lambda$ in the low temperature limit is summarized in Fig.~\ref{fig:Eg_lambda} and Fig~ .\ref{fig:Eg_kappa}. Inspired by the analytical dependence of $E_g$ on $\lambda$ and $\kappa$ in the $\kappa = 0$ \cite{maldacena2018} and $\lambda = 0$ \cite{garcia2021a,garcia2021} limit, we first test whether the combined effect of both couplings is $E_g=A\lambda^{\frac{2}{3}}+B\kappa^2$. However, this relation seems to work only in the small $\kappa$ region.
From the study of the gravity dual, we shall derive an analytical expression for $E_g$ which is in agreement with the ansatz above for small $\kappa$ and also describe well $E_g$ for other values of $\lambda$ and $\kappa$, as shown in Fig~.\ref{fig:Eg_ew_kappa_lambda}.

We also study the critical temperature $T_c(\kappa,\lambda)$ of the first order phase transition. In Fig.~\ref{fig:Tc_lambda_kappa}, we observe that for sufficiently small $\lambda$ and large $\kappa$ we have $T_c\sim\kappa^4$. This is consistent with a recent analytical prediction \cite{garcia2021a} for $\lambda = 0$. In the gravity section, we will derive an analytic expression of $T_c$ given in (\ref{eq:Tc_gravity})  which is in agreement with these numerical results even at finite $\lambda$.

\subsection{Spectrum, spin symmetry and complex-to-real transition}
We now study qualitative features, and the impact of symmetries of the spectrum of the Hamiltonian \eqref{hami}. In the $\lambda = 0$ limit, the Hamiltonian has an approximate spin-like symmetry represented by $\hat S =\sum_{i}^{N}\psi_i^L\psi_i^R$, which is exact for $\kappa =0$. For a certain $\kappa \neq 0$, eigenvalues of the Hamiltonian tend to cluster around those of the operator $\hat S$ for sufficiently large $\lambda$. This complicates the calculation of spectral correlations, for if $\hat S$ strictly commutes with the Hamiltonian, the spectral analysis must be restricted to eigenvalues within each cluster.  In Ref.~\cite{garcia2019}, this problem was solved by breaking this spin symmetry completely by considering couplings in each of the SYK's that differ by an overall constant $\alpha \neq 1$. If $\alpha \approxeq 1$, the wormhole phase still survives \cite{maldacena2018} but the gap becomes smaller.

Another solution is to choose a basis in which the Hamiltonian is
block-diagonalized where each block corresponds to an eigenvalue of $\hat S$. Even if the symmetry is broken for $\kappa \neq 0$, there still exists the parity symmetry which corresponds to a spin-like operator $e^{i\pi S}$ ~\cite{garcia2019}. For numerical convenience, we choose the following Hamiltonian, which is  equivalent to \eqref{hamie}:
\begin{equation}
	\begin{aligned}
        H_L &=-\sum_{i<j<k<l}(J_{ijkl}+i\kappa M_{ijkl})\psi_i \psi_j \psi_k \psi_l \qquad H_R =H_L^*\\
		H &=E\otimes H_L +H_R \otimes E + i\lambda \hat S\\
		\psi^L_i &= E \otimes \psi_i \qquad \psi^R_i = \psi_i \otimes \psi_c\\
		\psi_c &=(-i)^{N/2}\prod_{i=1}^N \psi_i
	\end{aligned}
\end{equation}
Here, $E$ is the identity matrix  with the size $2^{\frac{N}{2}}\times 2^{\frac{N}{2}}$. The parity operator $\psi_c$ is conserved($[\psi_c,H]=0$) with eigenvalues $\pm 1$. So we only need to reorder diagonal elements in descending order, then use exactly the same ordering to reorganize the Hamiltonian into block-matrix form of two parity sectors. Thus we can carry out the level statistic analysis on one of the two blocks separately. In case we are interested to study low temperature properties, we must choose the block that includes the ground state.

\begin{figure}[h]
	\centering
	\includegraphics[scale=0.65]{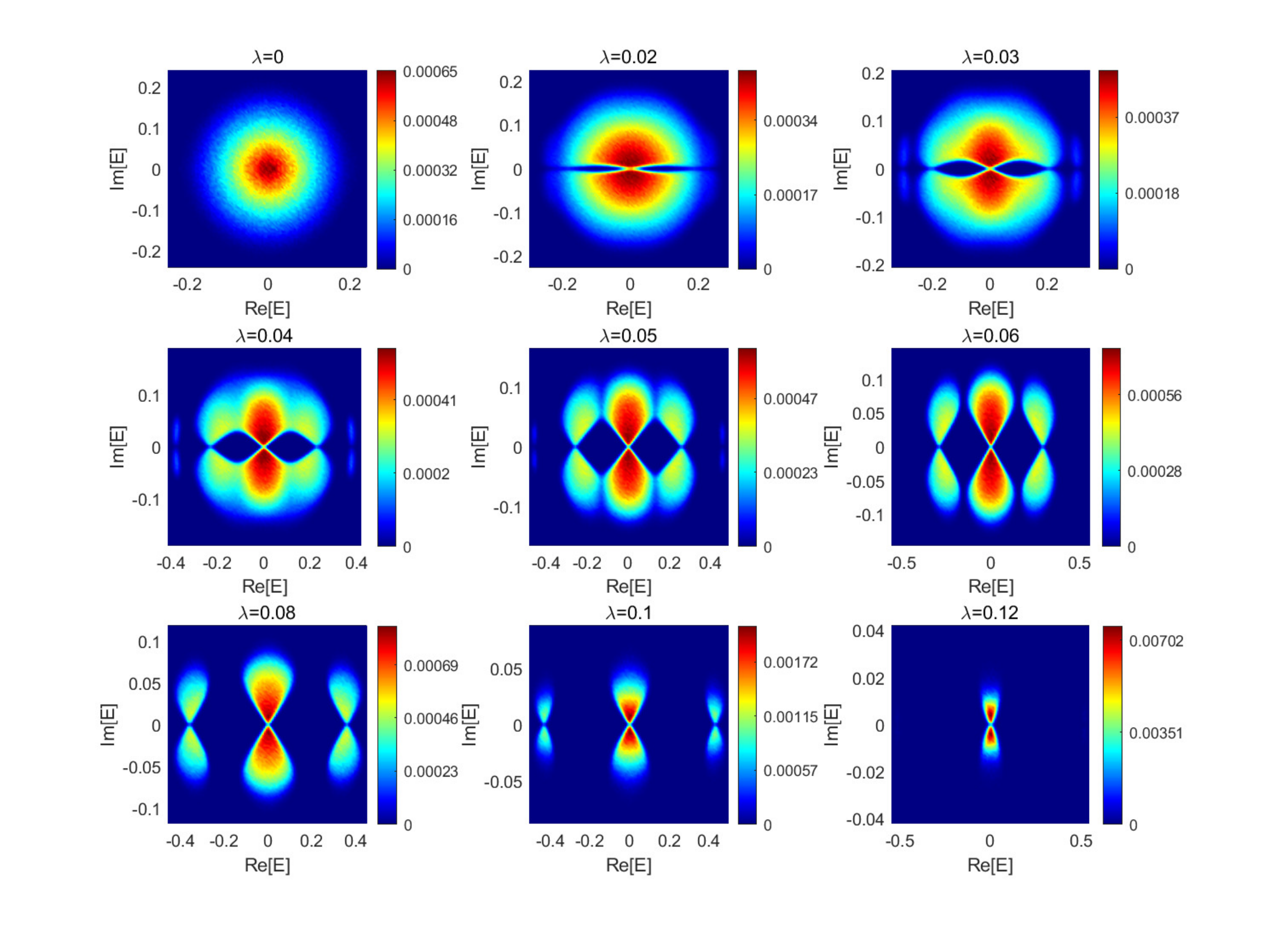}
	\caption{Complex spectral density for $\kappa = 1$ and different values of
$\lambda < \lambda_c$, we do not plot full real eigenvalues. The spectrum becomes increasingly organized in separated blobs in the complex plane. For sufficiently large $\lambda > 0.145$ it becomes real.}\label{fig:spectrum}
\end{figure}

In order to assess the importance of the spin-symmetry mentioned above in our Hamiltonian (\ref{hami}), we represent in Fig.~\ref{fig:spectrum} the spectral density for $\kappa =1$ and different values of the explicit coupling $\lambda$. We observe a rather symmetric distribution for $\lambda = 0$ which is in agreement with the results of Ref.~\cite{garcia2021,garcia2021a,garcia2021d}. As $\lambda$ increases, gaps around the real axis start to form. The spectral density for larger values of $\lambda$ shows that a growing part of the spectrum becomes real but we still observe complex eigenvalues in certain regions. The existence of the latter is directly related to the spin operator $\hat S$ which is an approximate symmetry of the Hamiltonian for sufficiently large $\lambda$.  Complex eigenvalues are restricted to the area between nearby eigenvalues of $\hat S$ on the real axis, and hence their number depends on $N$. Real eigenvalues starts to cluster around the eigenvalues of the spin operator (which we do not plot explicitly). By contrast, the maximum density of complex spectrum is located around double cones with tips on the real axis between the nearby eigenvalues of $\hat S$.

\begin{figure}[h]
		\centering
		\includegraphics[scale=0.4]{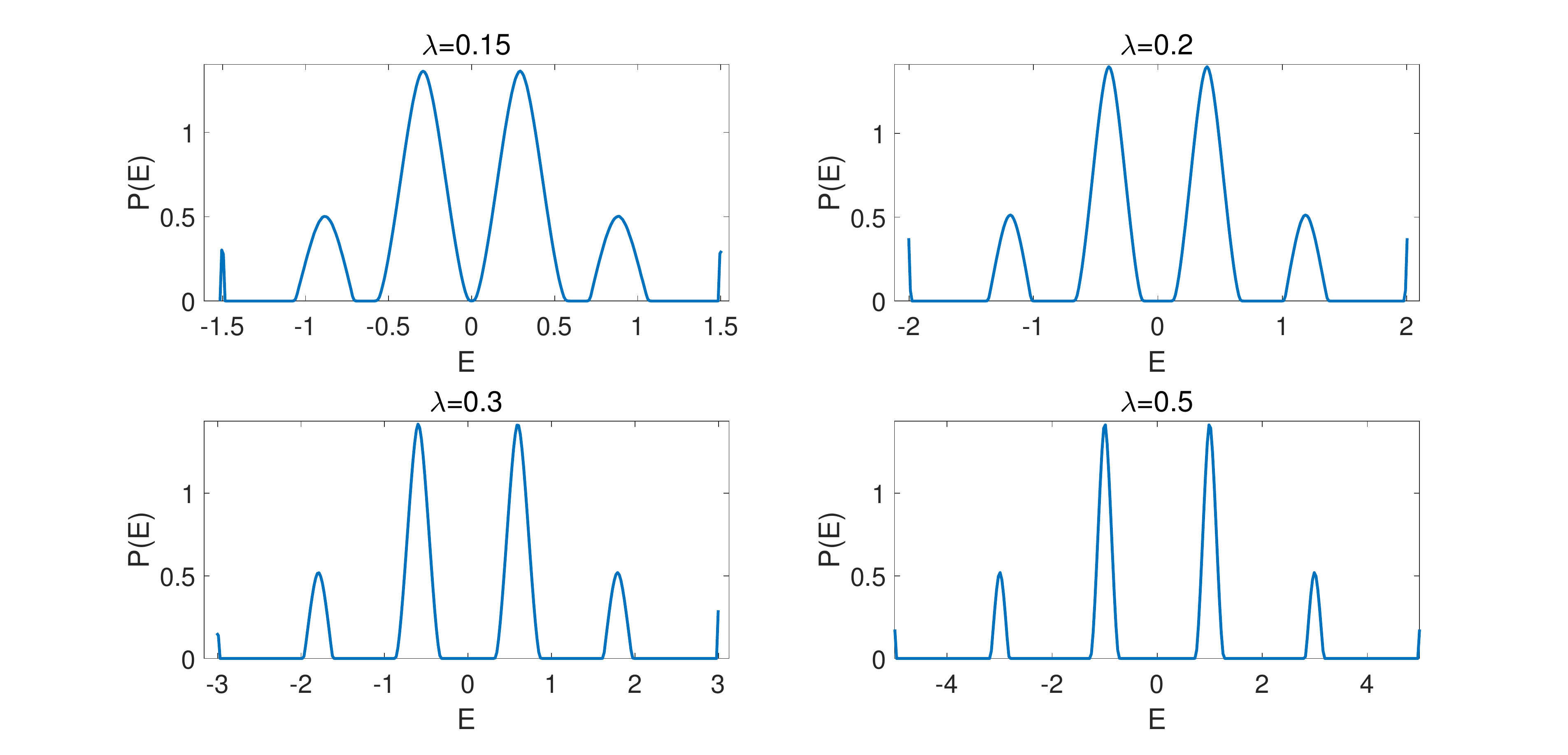}
		\caption{Spectral density for $\kappa = 1$ and different values of $\lambda$ ($\lambda > \lambda_c \approx 0.145$). The spectrum becomes real even though the Hamiltonian is non-Hermitian. As $\lambda$ increases, see especially $\lambda =0.5$, the spectrum is clustered around the  eigenvalues ($\pm (2n+1), n= 0,1,2$) of the spin operator for odd parity sector.}\label{fig:denreal}
	\end{figure}

Interestingly, a further increase in $\lambda$ leads to an unexpected result. Even though the Hamiltonian is non-Hermitian, the whole spectrum becomes real for $\lambda > \lambda_c(\kappa)$, see Fig.~\ref{fig:denreal}. Upon a further increase in $\lambda$, the spectral support of the already real spectrum is split in separate intervals. For $\lambda$ large enough, we observe that these intervals are centered around the eigenvalues of the spin operator. As was shown in Ref. \cite{garcia2019}, this clustering can be shown analytically by taking $\lambda \gg 1$ so that the other terms in the Hamiltonian are a small perturbation.

\begin{figure}[h]
	\centering
	\subfigure[]{\includegraphics[width=8.2cm]{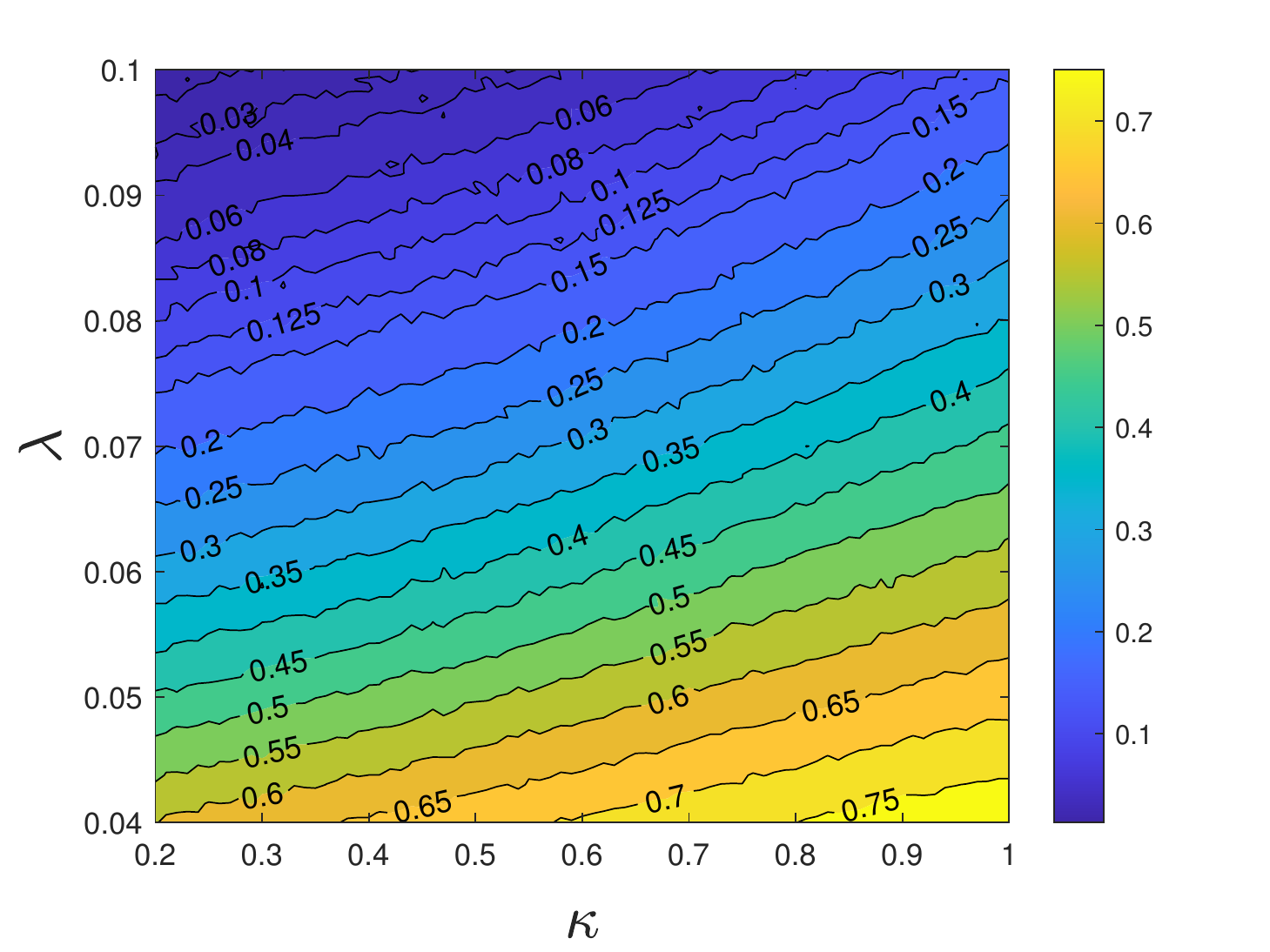}}
	\subfigure[]{\includegraphics[width=7.8cm]{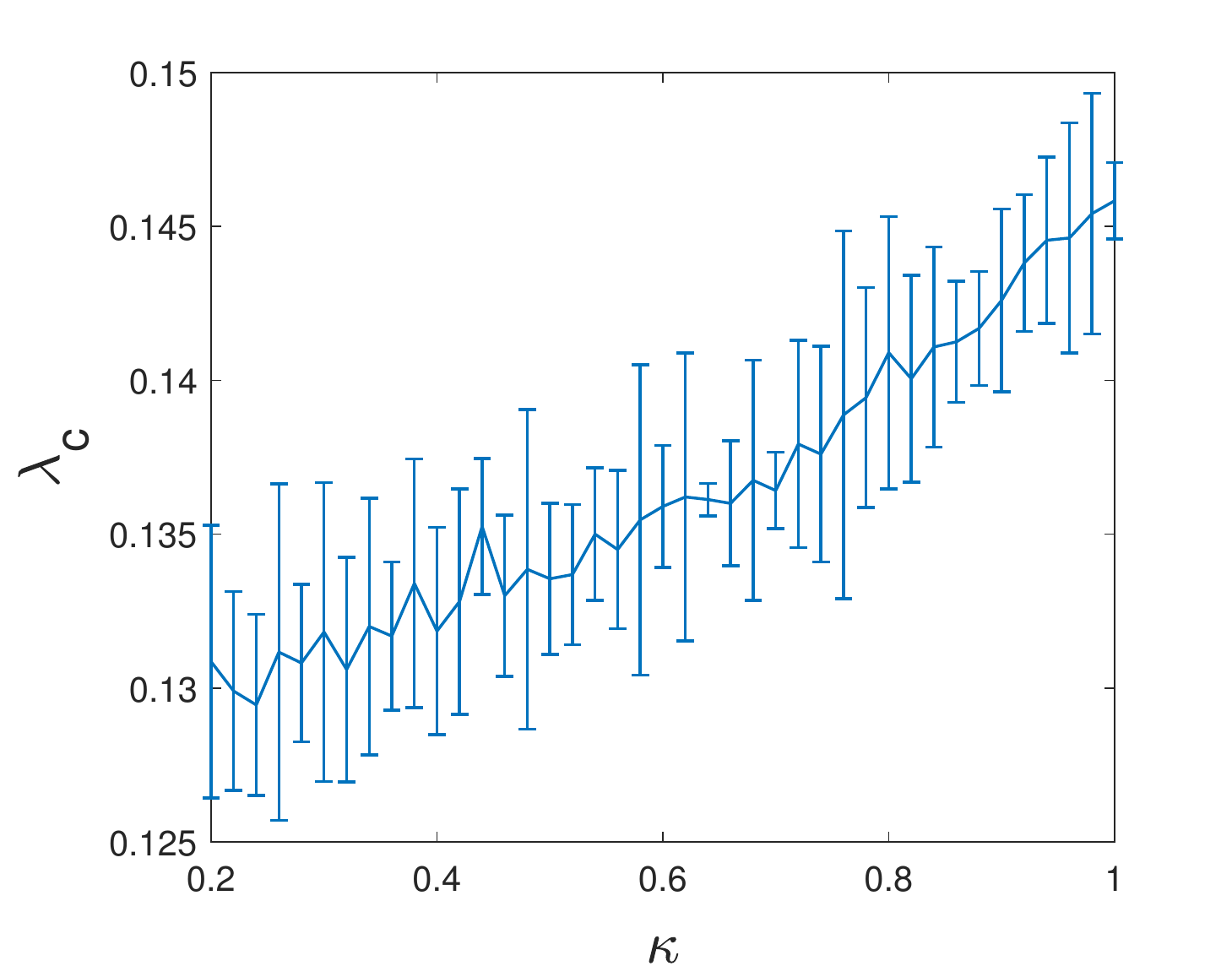}}
	\caption{Proportion of complex spectrum respect to the full spectrum. (a): percentage of eigenvalues with a non-zero complex part for different values of $\lambda$ and $\kappa$. For a given $\kappa$ and $\lambda$, we have employed $81\times 81=6561$ sets each having $100$ disorder realizations. (b): The critical $\lambda = \lambda_c$ for which the full spectrum becomes real for different values of $\kappa$. The blue line stands for the minimum $\lambda = \lambda_c$ for which all eigenvalues of $100$ disorder realizations are real. The error bars are the variance of this distribution. }\label{fig:null_kl}
\end{figure}

	In Fig.~\ref{fig:null_kl}(a), we present the ratio of complex and real eigenvalues for different $\kappa$ and $\lambda$.

	 We also find that though the percentage of the real spectrum increases with $\lambda$ monotonically, this increase slows down when the percentage of complex eigenvalues is small ($<10\%$). In Fig.~\ref{fig:null_kl}(b), we depict $\lambda_c(\kappa)$, the minimum $\lambda$ for which all eigenvalues are real for a given $\kappa$. As was expected, it shares similarities with the contours in Fig.~\ref{fig:null_kl}(a).

We also note that the existence of a critical $\lambda = \lambda_c$ is not an approximate result: the imaginary parts of the eigenvalues are strictly zero within the numerical precision $10^{-15}$.  We shall see in the following section that this transition has an observable impact on the oscillation patterns of real time Green's functions, which are related to quantum tunneling for $\kappa = 0$. Moreover, the transition does not require large $N$ or disorder average, although $\lambda_c$ is sensitive to the disorder realization.

It is instructive to consider the case $N=4$ for which the complex-to-real transition can be seen explicitly. In this case we have simply $H_L=  -(J+iK)\psi_1 \psi_2 \psi_3 \psi_4$ where $J$ and $K$ are arbitrary real numbers. The Hamiltonian is easily diagonalized and the eigenvalues are
\bea
\pm 2J ,\qq   \pm 2\sqrt{J^2 +4\lambda^2},\qq \pm 2\sqrt{-K^2 +\lambda^2}~,\qq (N=4)
\eea
where the first eigenvalue is threefold degenerate and the third one is fourfold degenerate.
The complex-to-real transition in this case corresponds to the fact that the spectrum becomes real for $\la>\la_c = K$.



\newpage

An order parameter for the complex-to-real transition is the thermal expectation value of the imaginary part of the Hamiltonian, which is defined as
\be
\r{Im}\,H = \k \sum_{i<j<k<\l} M_{ijk\l} ( \psi_i^R\psi_j^R\psi_k^R\psi_\l^R-\psi_i^L\psi_j^L\psi_k^L\psi_\l^L )~.
\ee
and its expectation value  can be computed as the derivative of the partition function with respect to $\k$:
\be\label{ImHexpct}
\ln \r{Im}\,H \rn_\b = {i\k\/\b} {\p \/\p\k}\log Z,
\ee
where $Z=\r{Tr}\,e^{-\b H}$. This is an order parameter for the transition since $\ln \r{Im}\,H \rn_\b $ vanishes at the transition point, because the spectrum of the Hamiltonian becomes real. This quantity can be derived from the partition function so it will be possible to compute it in the next section using the  gravity path integral.

\begin{figure}
	\centering
	\includegraphics[scale=.7]{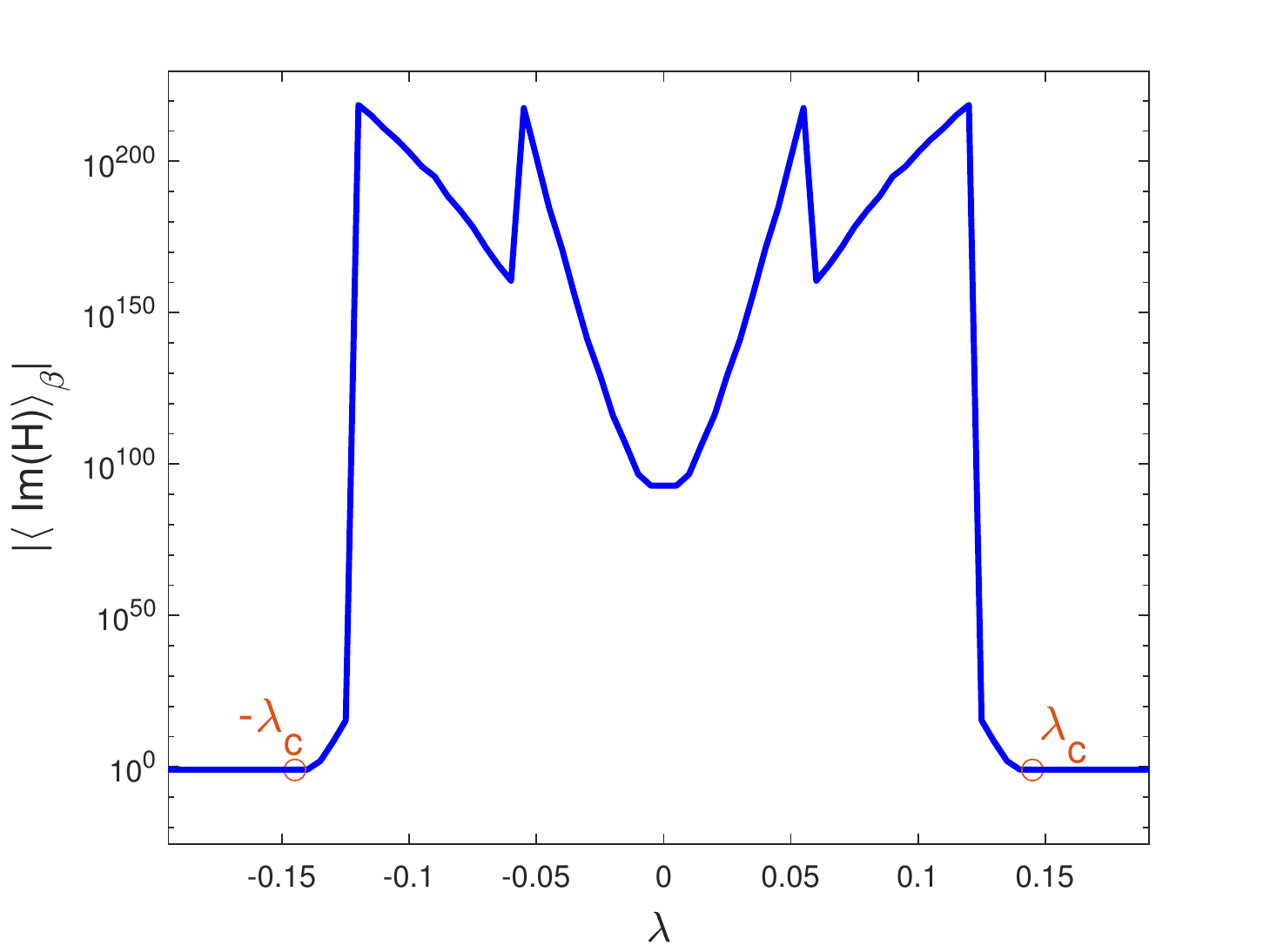}
	\caption{Order parameter $\langle O_{LR} \rangle = \frac{ \partial \log Z}{\partial k}$ of the complex-to-real transition in the SYK model. We use $N=12$ and $T\sim 10^{-3}$ to suppress statistical fluctuations. The Euclidean-to-Lorentzian in JT gravity gives a similar plot Fig.~ \ref{Fig:orderparam}. The extra peaks in the SYK plot are due to the non-universal spin symmetry of the SYK model. }\label{fig:orderpOLR}
\end{figure}

We plot the order parameter in Fig.~\ref{fig:orderpOLR}. It initially increases, then exhibits a rather sharp decrease for some small $\lambda$ from the effect of the spin-operator. It increases again within the range of $\lambda \sim (0.06,0.12)$, then abruptly decreases to nearly zero and finally vanishes at $\lambda = \lambda_c$. The value of $\lambda_c$ is consistent with the one obtained from the level statistics or the analysis of the real time Green's functions. It is also possible to study the higher-moments of $\r{Im}\,H$ by taking additional derivatives with respect to $\k$. The fact that all the moments vanish at $\la=\la_c$ then implies that the spectrum becomes real.

\subsection{Tunneling amplitude and real time evolution}

In this section we study the evolution in real time of the SYK model (\ref{hami}) by solving the SD equation in real time. In this context, a similar study was first carried out in Ref. \cite{sahoo2020} for the case of a two-site Hermitian SYK model dual to traversable wormholes. We first present results for $G^>_{ab}(t)$ when $\kappa=0$ or $\lambda=0$ respectively. We then provide a heuristic description of its main features. Finally, we carry out a detailed numerical computation of the combined effect of a finite $\lambda$ and $\kappa$ in real time Green's functions. Note that although the system only makes sense in Euclidean signature, it is a well-defined procedure to analytically continue the Green's function to real time.

\begin{figure}
	\centering
	\subfigure[]{\includegraphics[scale=.4]{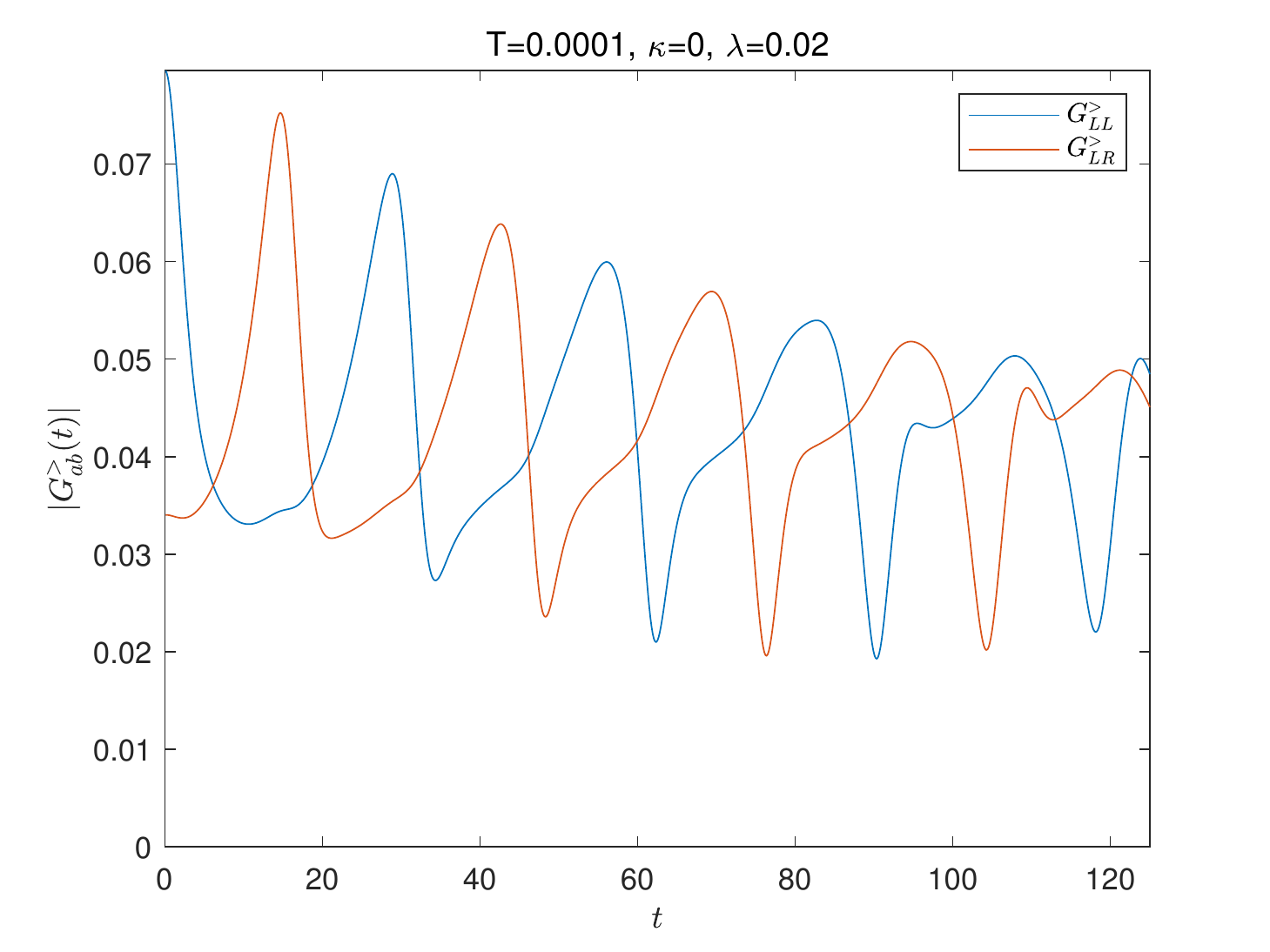}}
	\subfigure[]{\includegraphics[scale=.4]{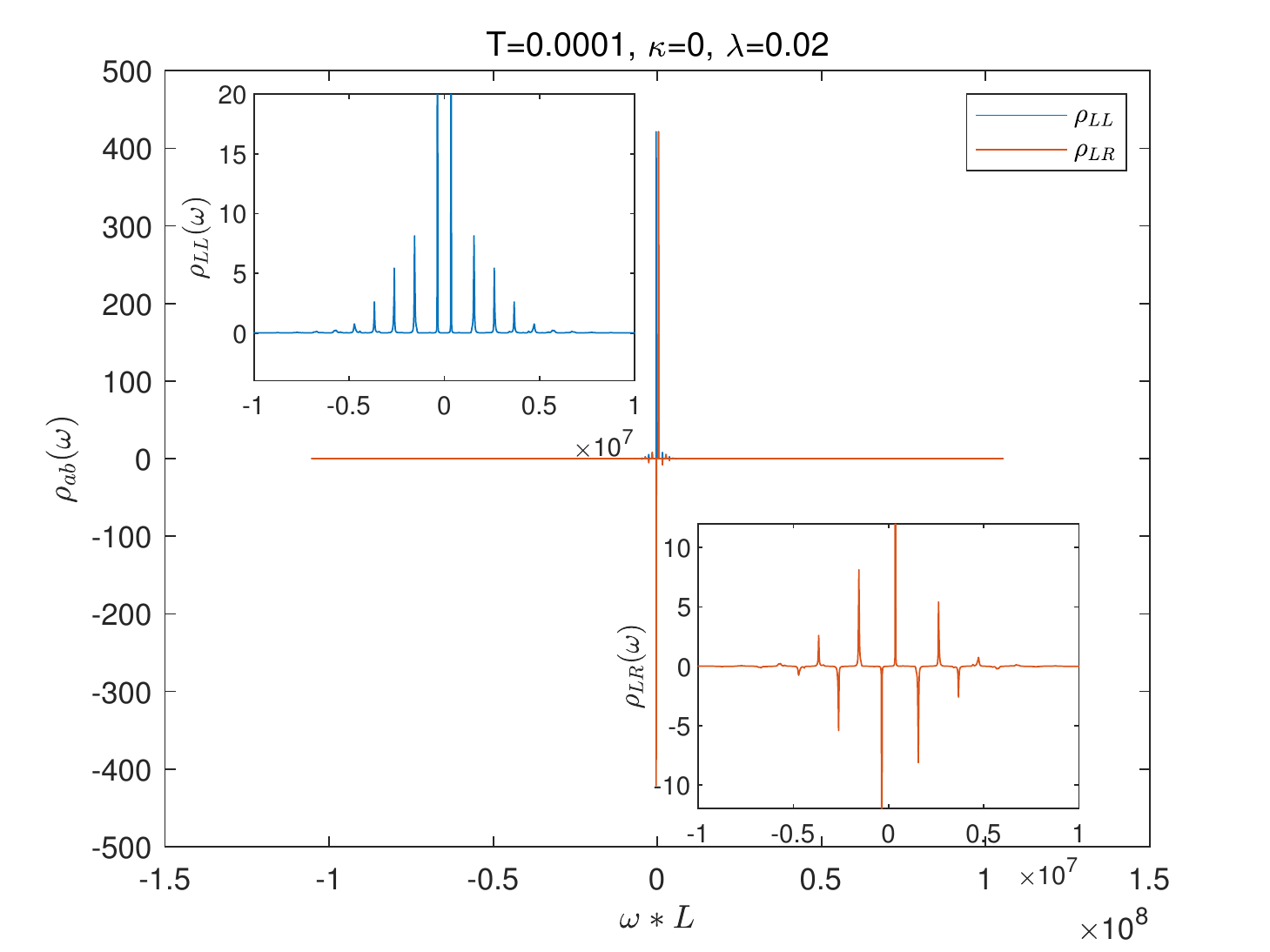}}
	\caption{Left: $|G^{>}_{ab}(t)|$(left) and $\rho_{LL}(\omega)$(right) for $\kappa=0$, $\lambda=0.01$, $\beta=10^4$. $\epsilon =  2\times10^{-5}$ and the time cutoff  $L=5\times10^5$. Right: $\rho_{LL}$ and $\rho_{LR}$ versus $\omega$ in the small $\omega$ region. }\label{fig:Gr_rholl_ka_0_ld_02_beta_1e4_lx_5e6_eta_2e-4_dm_25}
\end{figure}

Before proceeding, we firstly review the results for $\kappa = 0$ and finite $\lambda$ as discussed in Ref. \cite{sahoo2020}. The quantum dynamics in this case is controlled by the SD equations in real time
resulting from the analytical continuation of (\ref{sde}),
\begin{equation}\begin{aligned}
		\rho_+(\omega) &= -\frac{1}{\pi}\text{Im}G^r_+(\omega) \\
		\rho_{LL/LR}(\omega) &= \frac{1}{2}(\rho_+(\omega) \pm \rho_+(-\omega)) \\
		n_{LL/LR}(t) &= \int_{-\infty}^{\infty}d\omega \rho_{LL/LR}(\omega)n_F(\omega)e^{-i\omega t}, \qquad n_F(\omega)=\frac{1}{e^{\beta\omega}+1} \\
		\Sigma^r_+(\omega) &= -2iJ^2\int_0^{\infty}e^{i(\omega+i\epsilon)t}[\text{Re}[n_{LL}^3(t)]-i\text{Im}[n_{LR}^3]]\\
		G^r_{+}(\omega) &= \frac{1}{\omega + i\epsilon -\Sigma^{r}_{+} -\lambda}  \\
		\label{eq:iterative_formula_coupling}
\end{aligned}\end{equation}
We define
\be
G^>_{ab}(t) = -i\frac{1}{N}\sum_i\langle\psi_{i,a}(t)\psi_{i,b}(0)\rangle
\ee
whose Fourier trasnform is
\be
 G^>_{ab}(\omega) = -i(1-n_F(\omega))\rho_{ab}(\omega)
\label{eq:Gfwd_rho}
\ee
with $a,b = L,R$.
Results for $\kappa = 0$, the case already studied in Ref.~\cite{sahoo2020},  are depicted in Fig.~\ref{fig:Gr_rholl_ka_0_ld_02_beta_1e4_lx_5e6_eta_2e-4_dm_25}. As is observed,  $|G^>_{LL}|$ and $|G^>_{LR}|$ are out of phase, with maxima (minima) of one the functions corresponding to minima (maxima) of the other. The maxima of $G^>_{ab}$ appear in different $t$, which can be understood, when the system is dual to an eternal traversable wormhole, as the propagating time through the bulk. These real time results in Fig. \ref{fig:Gr_rholl_ka_0_ld_02_beta_1e4_lx_5e6_eta_2e-4_dm_25} are directly related to the imaginary solutions in Ref. \cite{maldacena2018} by applying a Wick rotation.

\begin{figure}
	\centering
	\subfigure[]{\includegraphics[scale=.6]{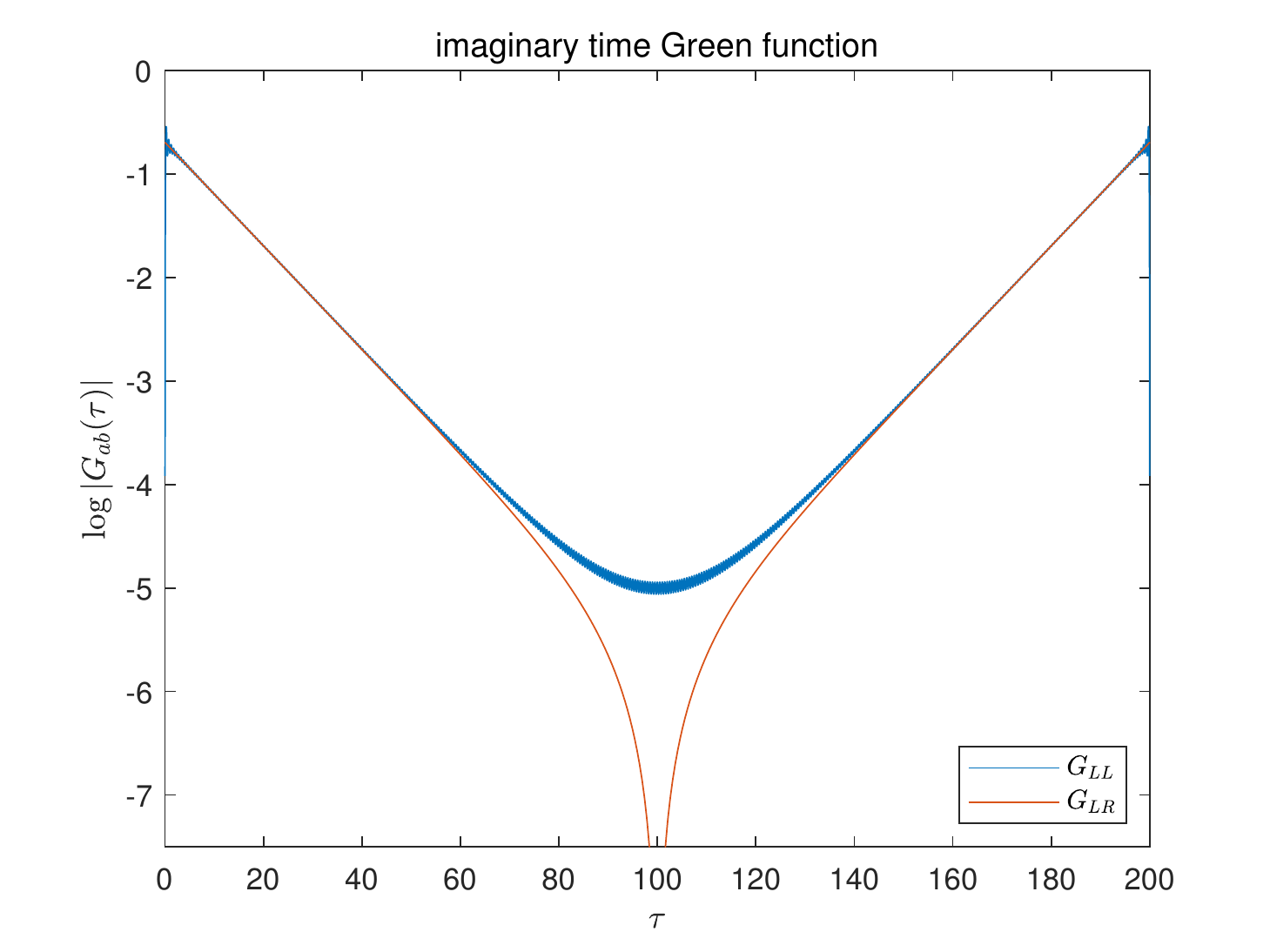}}
	\caption{$G_{ab}(\tau)$ after a Wick rotation of $G^R_{ab}(\omega)$ and $\rho_{LL}\approx 0.5\delta(\omega-\omega_0)+0.5\delta(\omega+\omega_0)$, $\rho_{LR} \approx 0.5 i\delta(\omega-\omega_0)-0.5 i\delta(\omega+\omega_0)$ }\label{fig:gf_wick_rotation}
\end{figure}

We have found out, see Fig. \ref{fig:Gr_rholl_ka_0_ld_02_beta_1e4_lx_5e6_eta_2e-4_dm_25}(b), that the largest two peaks of $\rho_{ab}(\omega)$, have the lowest frequency  and are symmetric with respect to $\omega=0$. Therefore, $\rho_{ab}$ is well approximated by
\begin{equation}\begin{aligned}
\rho_{LL}(\omega') &\approx A\delta(\omega'-E_0)+A\delta(\omega'+E_0) ~,\\
\rho_{LR}(\omega') &\approx iA\delta(\omega'-E_0) -iA\delta(\omega'+E_0)~,
\end{aligned}\end{equation}
with $A = 0.5$.  By employing the definition
\begin{equation}\begin{aligned}
G^R_{ab}(\omega)=\int_{-\infty}^{\infty} d\omega' \frac{\rho_{ab}(\omega')}{\omega-\omega'+i\eta}
\end{aligned}\end{equation}
for the retarded Green's function, we obtain
\begin{equation}\begin{aligned}
G^R_{LL}(\omega) &=\frac{A}{\omega -E_0 +i\eta} +\frac{A}{\omega +E_0 +i\eta} ~,\\
-i G^R_{LR}(\omega) &=\frac{A}{\omega -E_0 +i\eta} -\frac{A}{\omega +E_0 +i\eta}~.
\end{aligned}\end{equation}
Therefore, the imaginary time Green's functions are given by,
\begin{equation}\begin{aligned}
G_{LL}(\omega_n) &=\frac{A}{i\omega_n -E_0} +\frac{A}{i\omega_n +E_0}  \\
-i G_{LR}(\omega_n) &=\frac{A}{i\omega_n -E_0} -\frac{A}{i\omega_n +E_0}
\end{aligned}\end{equation}
and
\begin{equation}\begin{aligned}
G_{LL}(\tau) &=\frac{1}{\beta}\sum_{n=-\infty}^{\infty} \left( \frac{A}{i\omega_n -E_0} +\frac{A}{i\omega_n +E_0}\right) e^{i\omega_n\tau} =-\frac{A}{\beta}\sum_{n=-\infty}^{\infty} \frac{2i\omega_n}{\omega_n^2 +E_0^2} e^{i\omega_n\tau} \\
&= -\frac{A}{\beta}\sum_{n=-\infty}^{\infty} \frac{2i\pi(2n+1)/\beta}{\pi^2(2n+1)^2/\beta^2 +E_0^2} e^{i\pi(2n+1)\tau/\beta} = A\sum_{n=0}^{\infty} \frac{4\pi(2n+1)\,\sin\frac{\pi(2n+1)\tau}{\beta}}{\pi^2(2n+1)^2 +\beta^2 E_0^2}    \\
-i G_{LR}(\tau) &=\frac{1}{\beta}\sum_{n=-\infty}^{\infty} \left( \frac{A}{i\omega_n -E_0} -\frac{A}{i\omega_n +E_0}\right) e^{i\omega_n\tau} =-\frac{A}{\beta}\sum_{n=-\infty}^{\infty} \frac{2 E_0}{\omega_n^2 +E_0^2} e^{i\omega_n\tau} \\
&= -\frac{A}{\beta}\sum_{n=-\infty}^{\infty} \frac{2 E_0}{\pi^2(2n+1)^2/\beta^2 +E_0^2} e^{i\pi(2n+1)\tau/\beta} = -A\sum_{n=0}^{\infty} \frac{4 E_0\beta \,\cos\frac{\pi(2n+1)\tau}{\beta}}{\pi^2(2n+1)^2 +\beta^2 E_0^2}
\label{eq:Gab_wick_rotation}
\end{aligned}\end{equation}
As is illustrated in Fig.~\ref{fig:gf_wick_rotation}, $G_{ab}(\tau)$, calculated from (\ref{eq:Gab_wick_rotation}), is real for $a = b = L$ or $a = b = R$ and purely imaginary for $a = L, b = R$. Moreover, $|G_{ab}|$ decays exponentially in the small $\tau$ region, which is consistent with previous numerical results \cite{maldacena2018}. For finite $\kappa$, we expect \cite{garcia2021a} that $G_{ab}(\tau)$, and therefore $\rho_{ab}(\omega)$, has also similar properties. The real time Schwinger-Dyson equations are  derived by extending the method of Ref.~\cite{sahoo2020} to a finite $\kappa$,

\begin{equation}\begin{aligned}
		\rho_+(\omega) &= -\frac{1}{\pi}\text{Im}G^r_+(\omega) \\
		\rho_{LL/LR}(\omega) &= \frac{1}{2}(\rho_+(\omega) \pm \rho_+(-\omega)) \\
		n_{LL/LR}(t) &= \int_{-\infty}^{\infty}d\omega \rho_{LL/LR}(\omega)n_F(\omega)e^{-i\omega t}, \qquad n_F(\omega)=\frac{1}{e^{\beta\omega}+1} \\
		\Sigma^r_+(\omega) &= -2iJ^2\int_0^{\infty}e^{i(\omega+i\epsilon)t}[(1-\kappa^2)\text{Re}[n_{LL}^3(t)]-i(1+\kappa^2)\text{Im}[n_{LR}^3]]\\
		G^r_{+}(\omega) &= \frac{1}{\omega + i\epsilon -\Sigma^{r}_{+} -\lambda}  \\
		\label{eq:iterative_formula_nonHermitian}
\end{aligned}\end{equation}

\begin{figure}
	\centering
	\subfigure[]{\includegraphics[scale=.4]{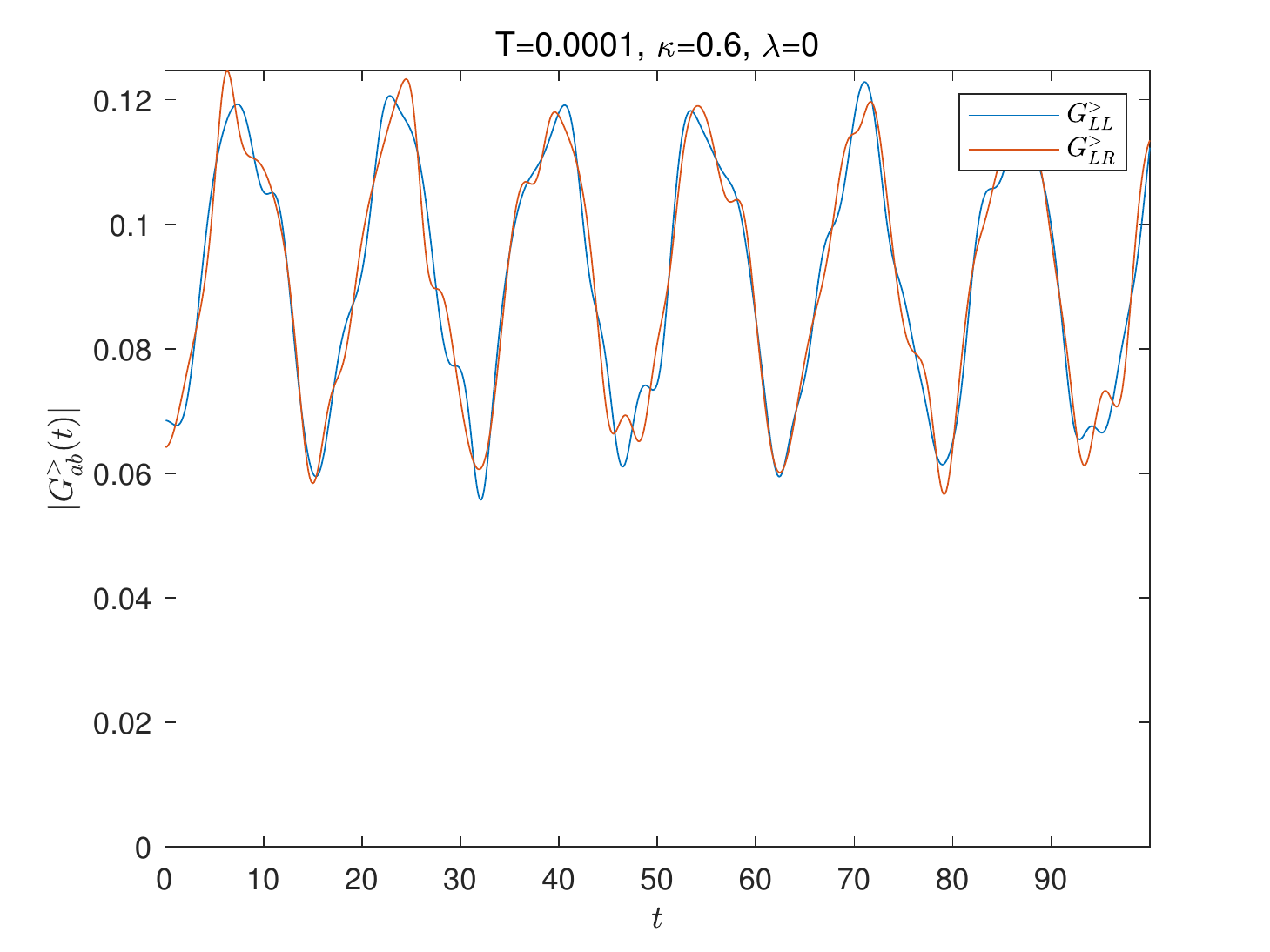}}
	\subfigure[]{\includegraphics[scale=.4]{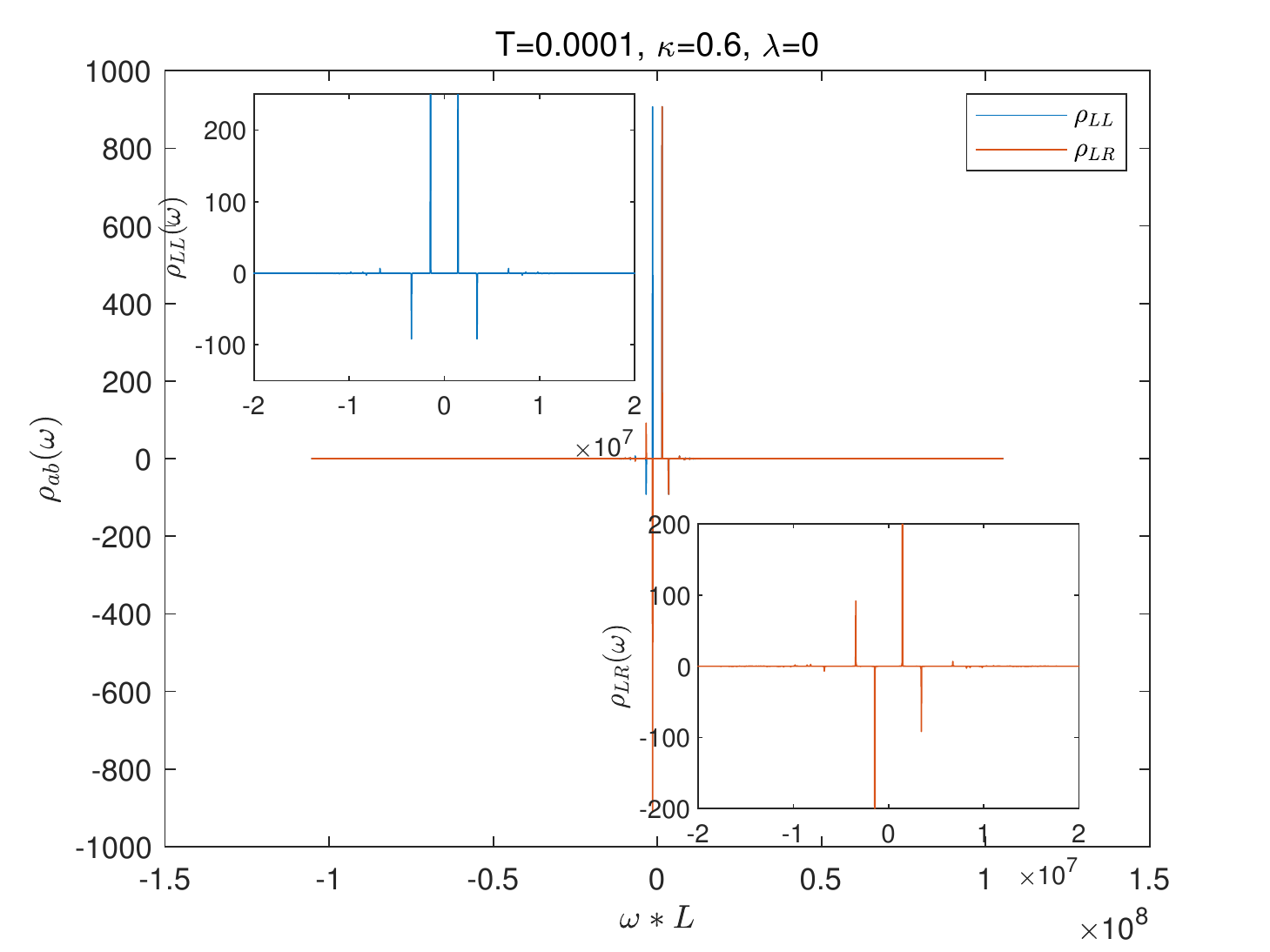}}
	\caption{$|G^{>}_{ab}(t)|$ (left) and $\rho_{LL}(\omega)$ (right) for $\kappa=0.6$, $\lambda=0$, $\beta=10^4$, $\epsilon=2\times10^{-4}$ and time cutoff $L=5\times10^6$. Insets on the right panel shows $\rho_{LL}(\omega)$ and $\rho_{LL}(\omega)$ in the small $\omega$ region. }\label{fig:Gr-rholl_ld_0_ka_p6_beta_1e4_lx_5e5_eta_2e-5_dm_26}
\end{figure}

The results for the case $\lambda=0$, $\kappa\neq 0$, depicted  in Fig.~\ref{fig:Gr-rholl_ld_0_ka_p6_beta_1e4_lx_5e5_eta_2e-5_dm_26}, also display an oscillatory behavior but with a crucial difference: both functions are now in-phase. Physically, it is an indication that there is no real tunneling between left and time particles but rather a synchronization of the dynamics of both sites despite the fact that they are not directly coupled. The only coupling arise after ensemble average. This in-phase pattern will be explained in the next section as a localization in the gravity path integral.

Since $|G^>_{ab}(t)|\sim|\langle\psi_a(t)\psi_b(0)\rangle|$ is the probability amplitude of observing a particle again at time $t$ if this particle appears at $t=0$ once, the overlap of $|G^>_{ab}(t)|$ with $a, b = L,R$ indicates the synchronization of the dynamics of the left and right sites even though they are not physically coupled.

Another interesting feature in the $\lambda = 0$ limit, see Fig.~\ref{fig:Gr-rholl_ld_0_ka_p6_beta_1e4_lx_5e5_eta_2e-5_dm_26}, is that $\rho_{LL}(\omega)$ has negative peaks. Superficially, this result looks surprising since for $\epsilon > 0$, all the peaks by definition must be positive from the very definition of $\rho_{ab}(\omega)$.  However, this applies only to $\kappa = 0$ where the Hamiltonian is Hermitian and the eigenvalues are real. For $\kappa \neq 0$, the eigenvalues are in general complex and therefore, peaks can be either positive or negative.

\begin{figure}
\centering
\subfigure[]{\includegraphics[scale=.3]{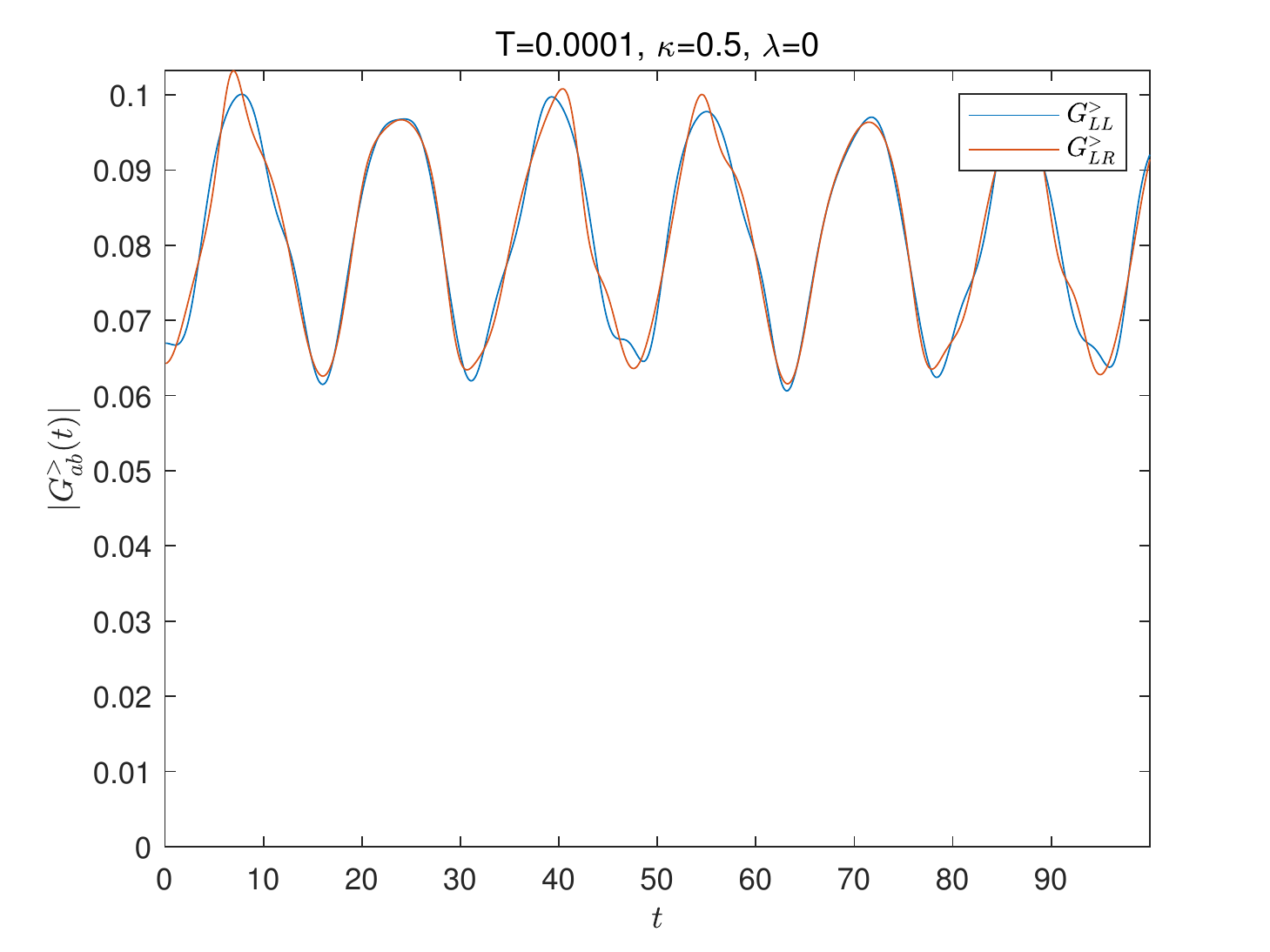}}
\subfigure[]{\includegraphics[scale=.3]{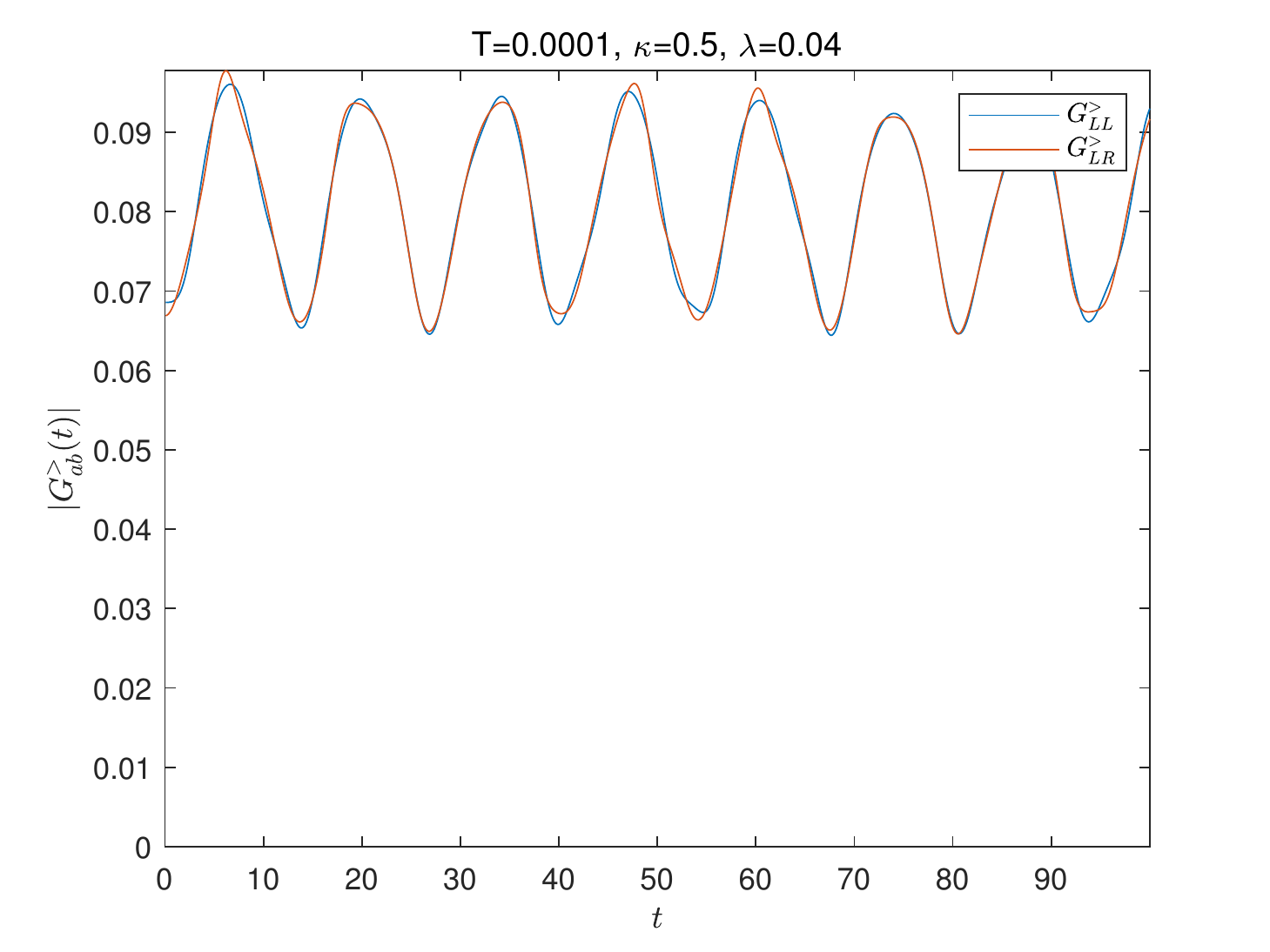}}
\subfigure[]{\includegraphics[scale=.3]{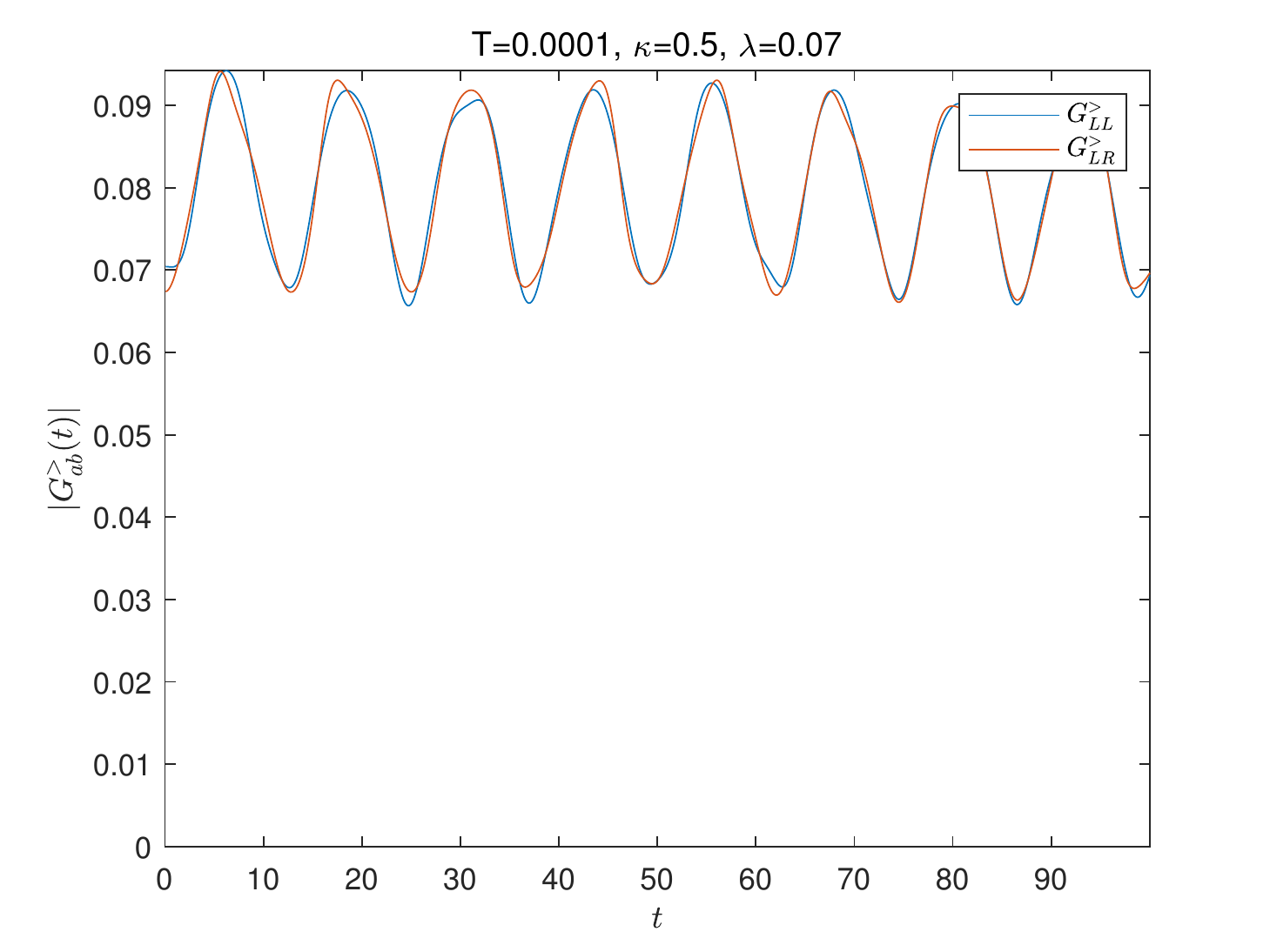}}
\subfigure[]{\includegraphics[scale=.3]{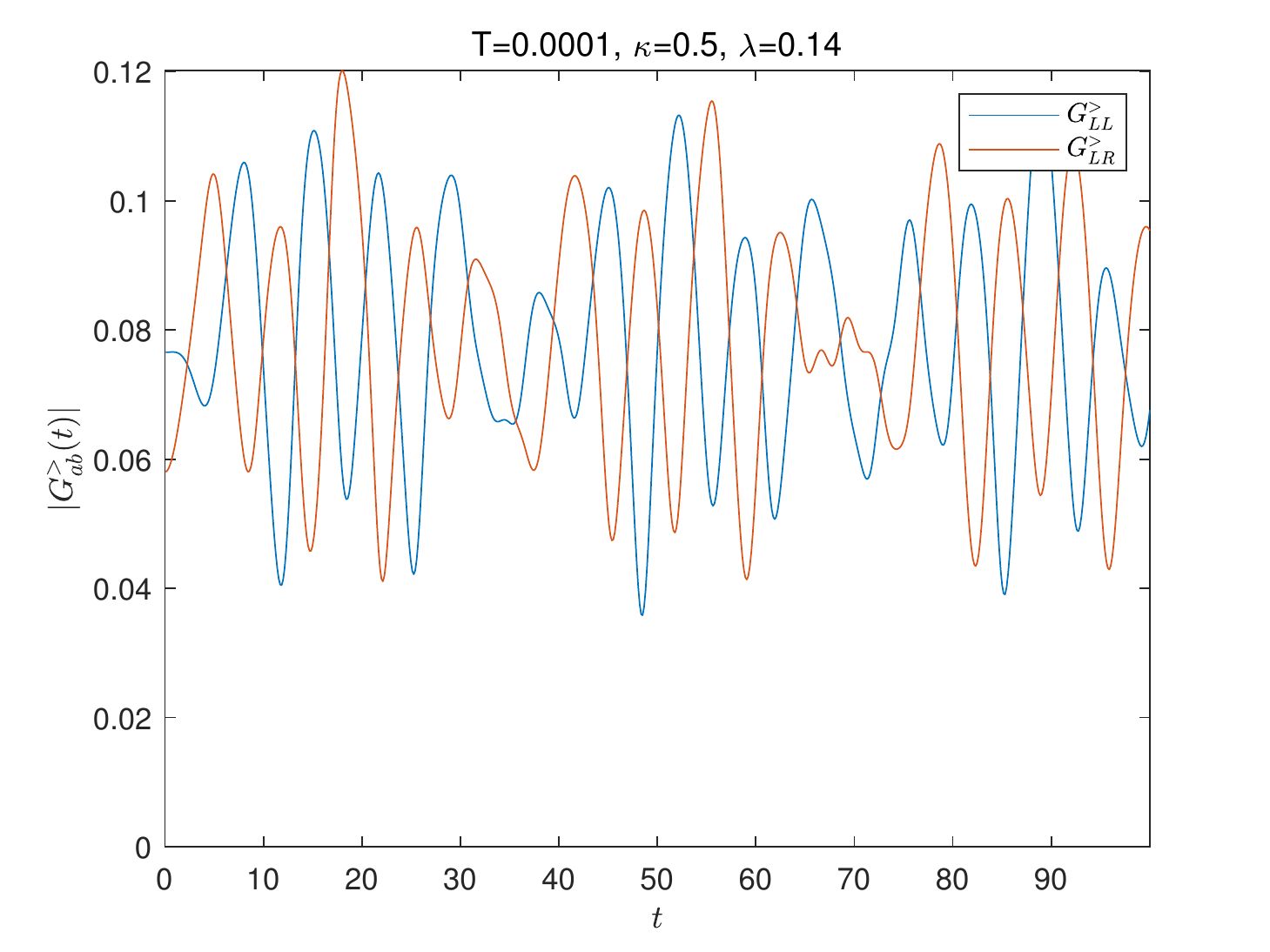}}
\subfigure[]{\includegraphics[scale=.3]{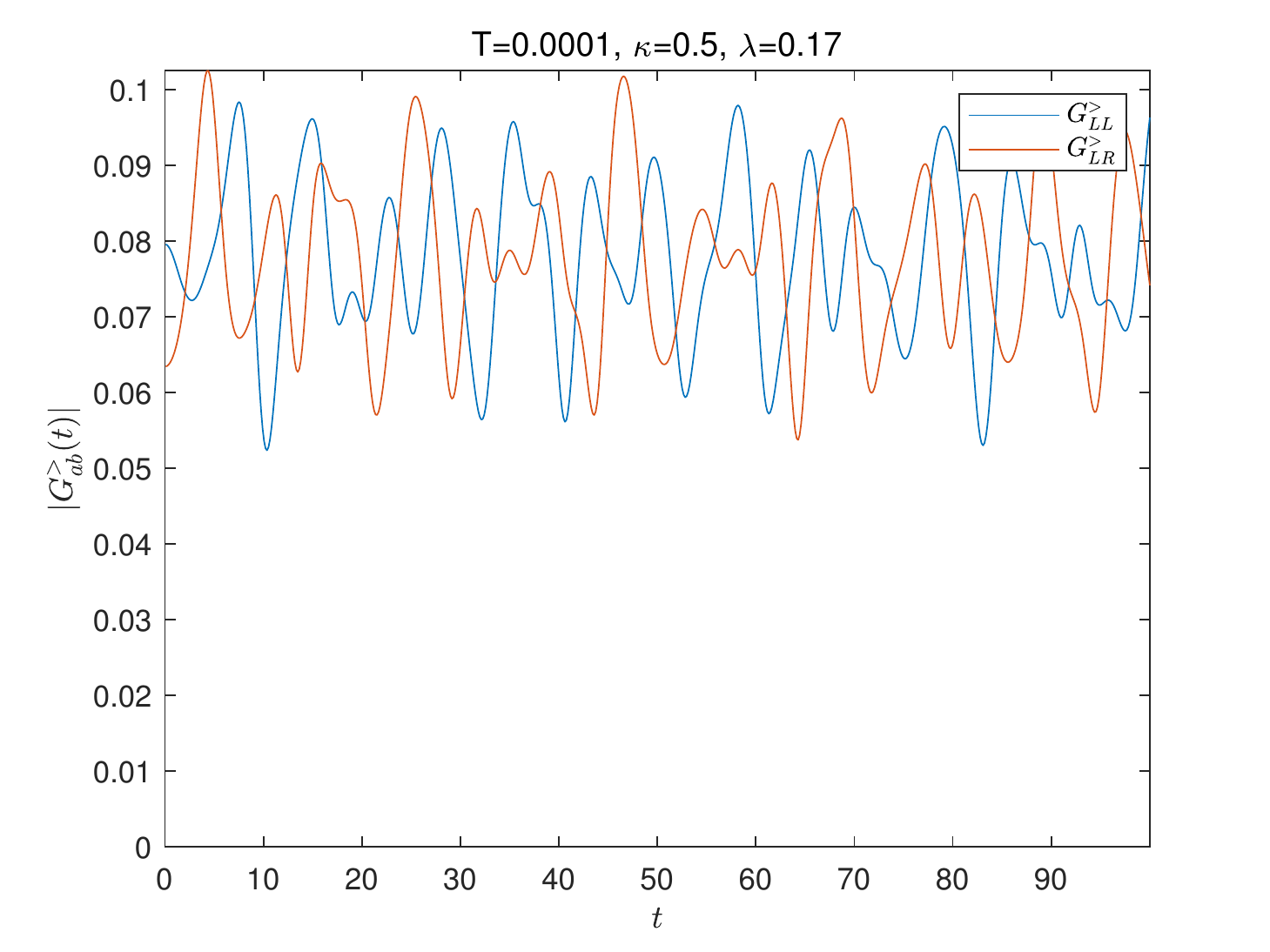}}
\subfigure[]{\includegraphics[scale=.3]{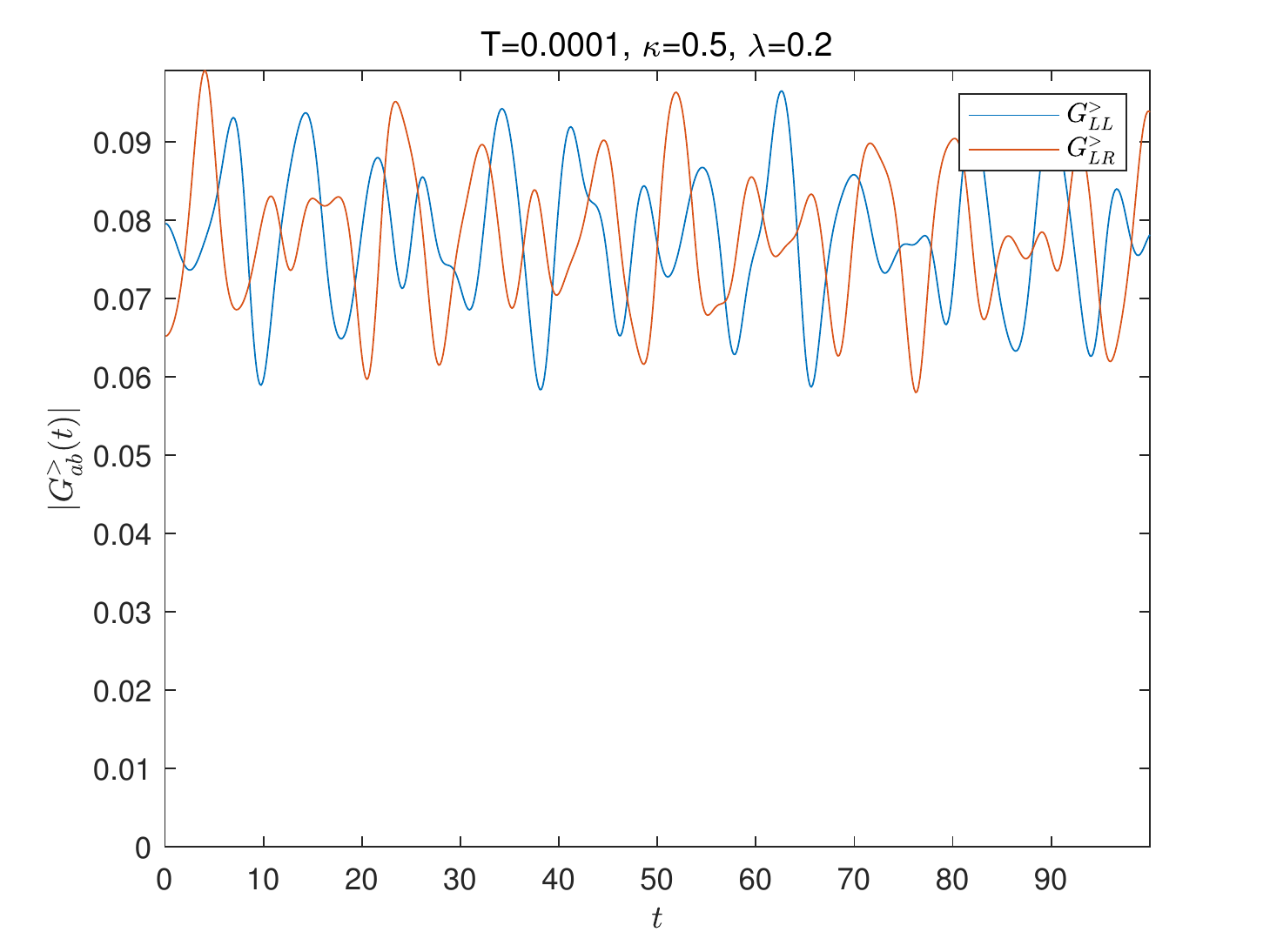}}
--------------------------------------------------------------------------------------------------------
\subfigure[]{\includegraphics[scale=.3]{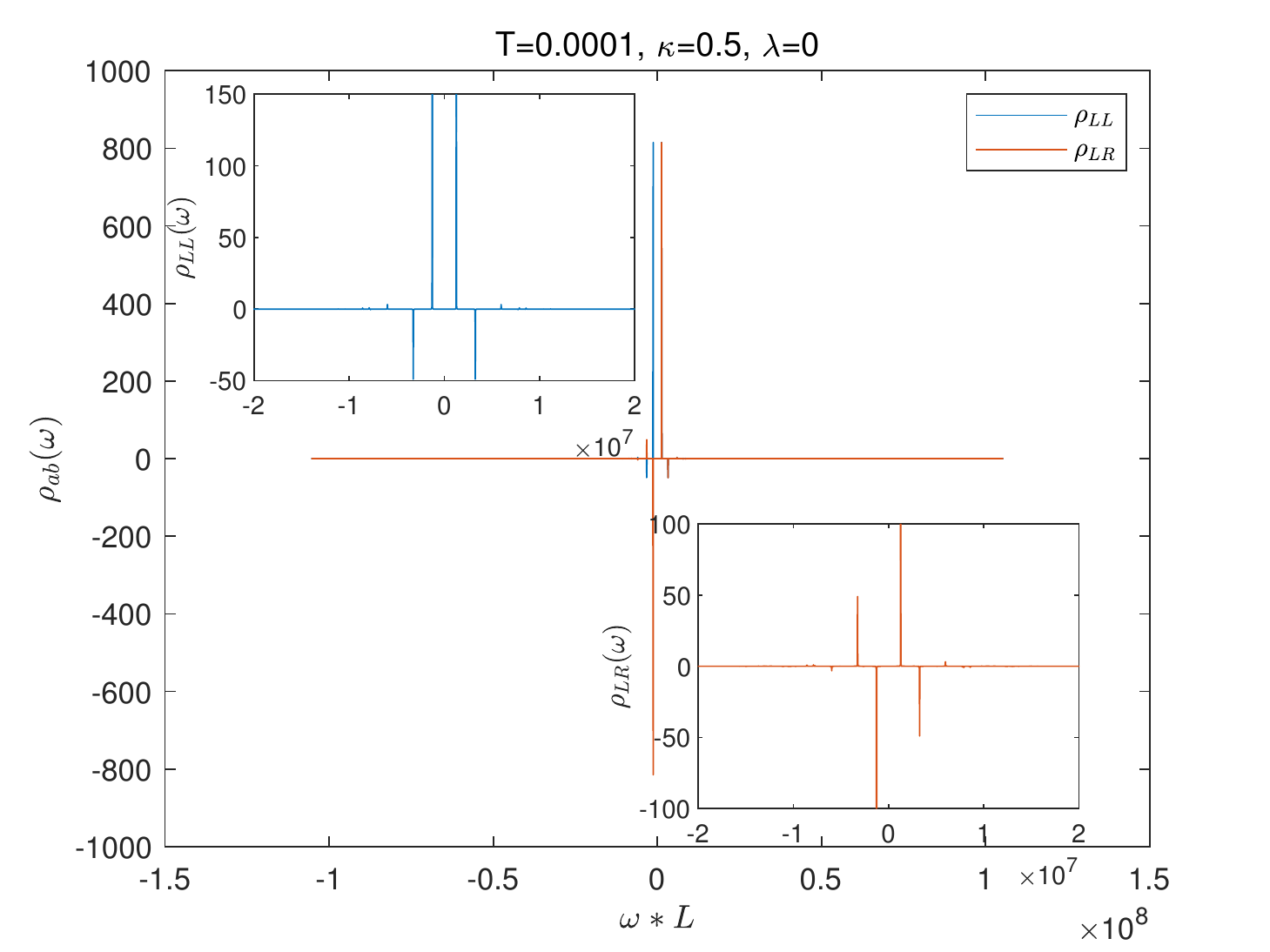}}
\subfigure[]{\includegraphics[scale=.3]{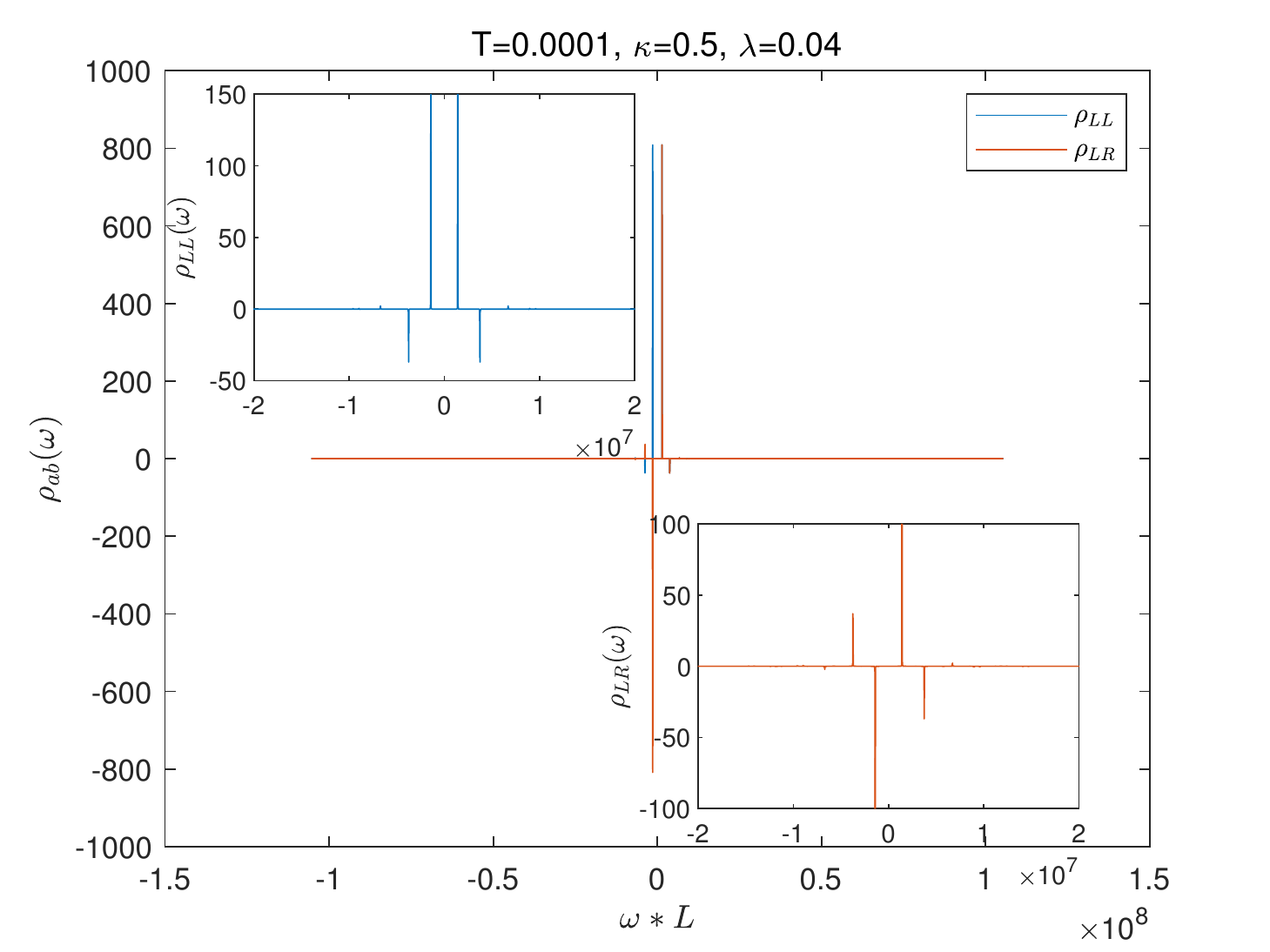}}
\subfigure[]{\includegraphics[scale=.3]{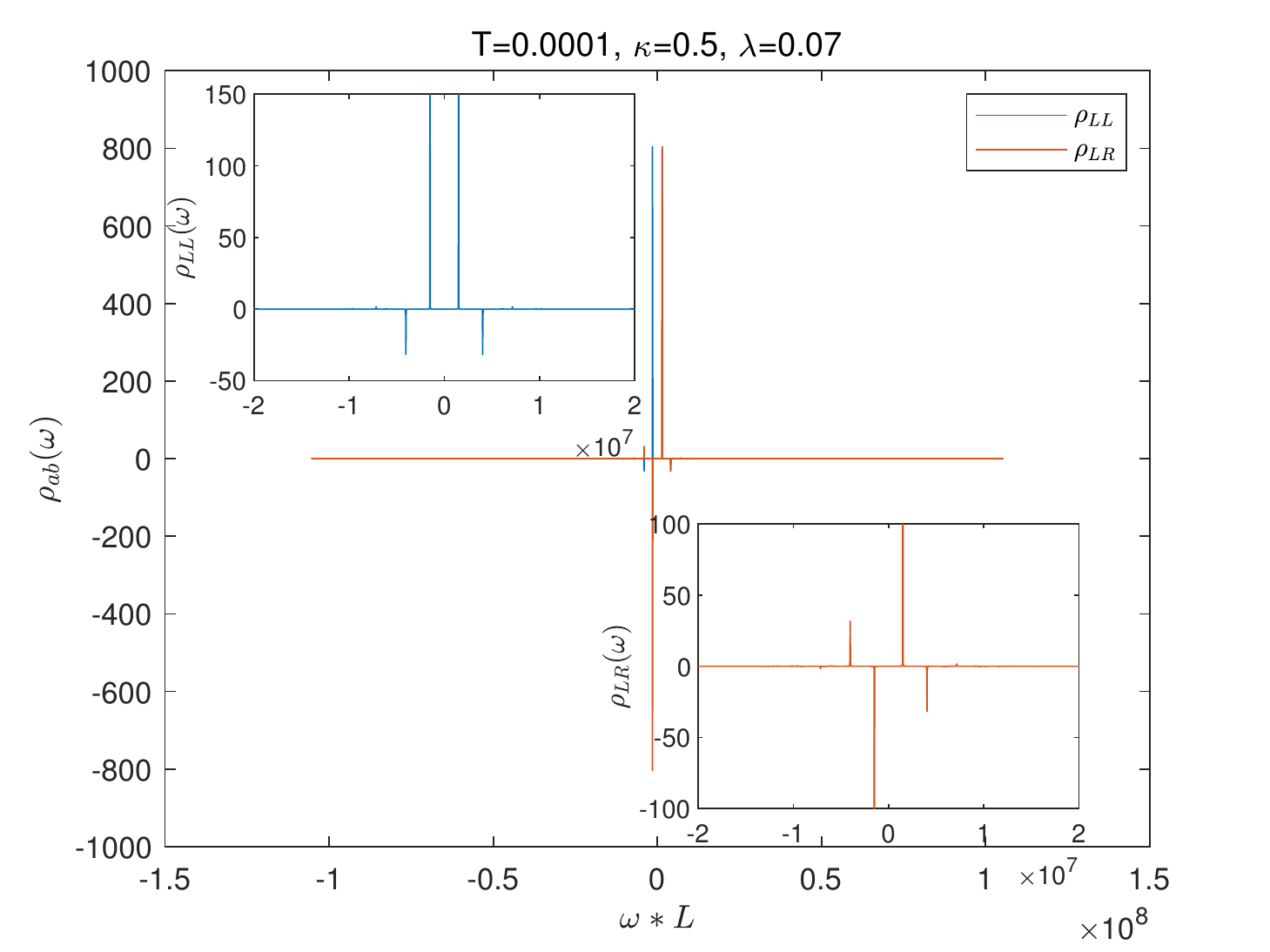}}
\subfigure[]{\includegraphics[scale=.3]{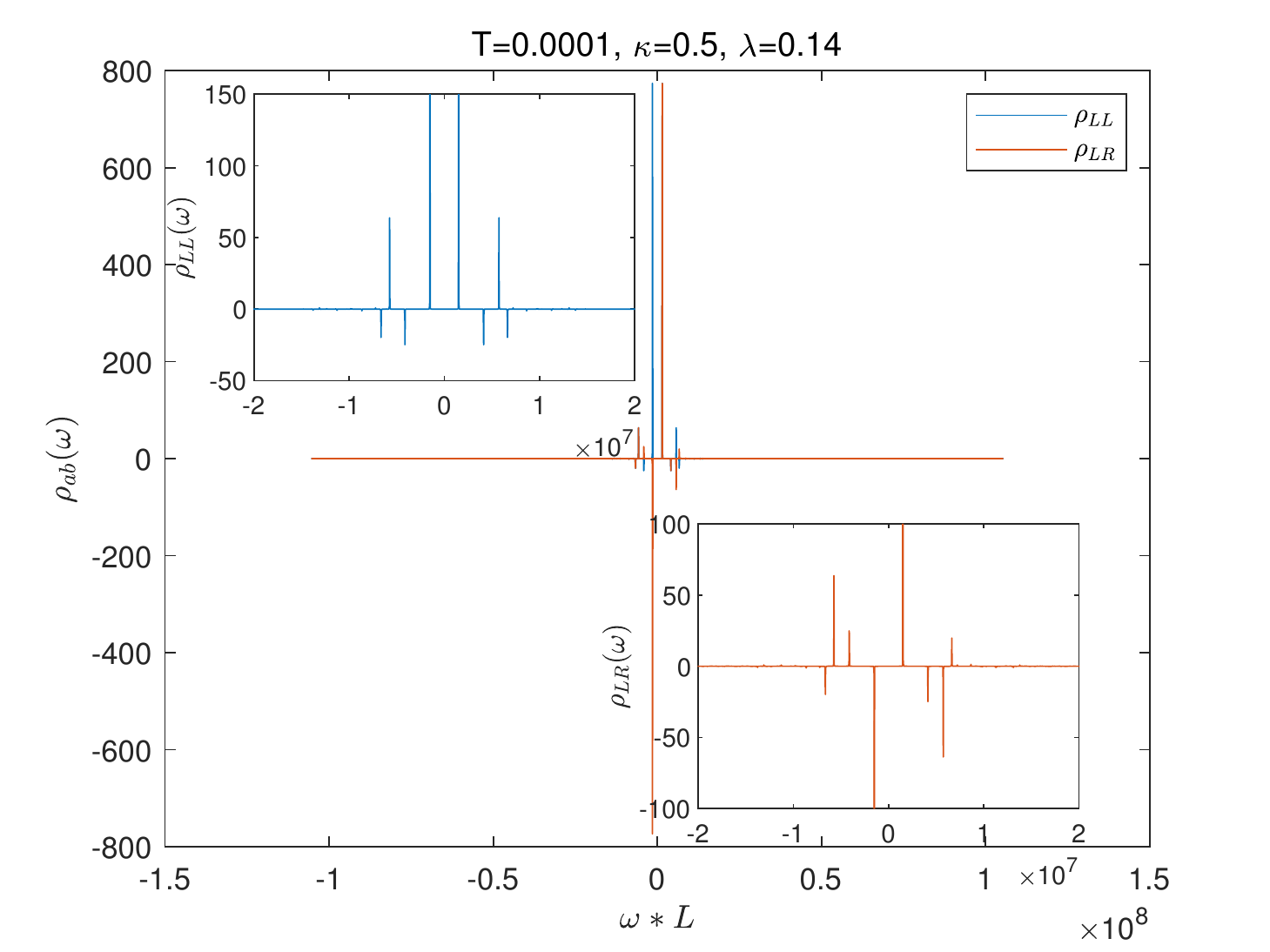}}
\subfigure[]{\includegraphics[scale=.3]{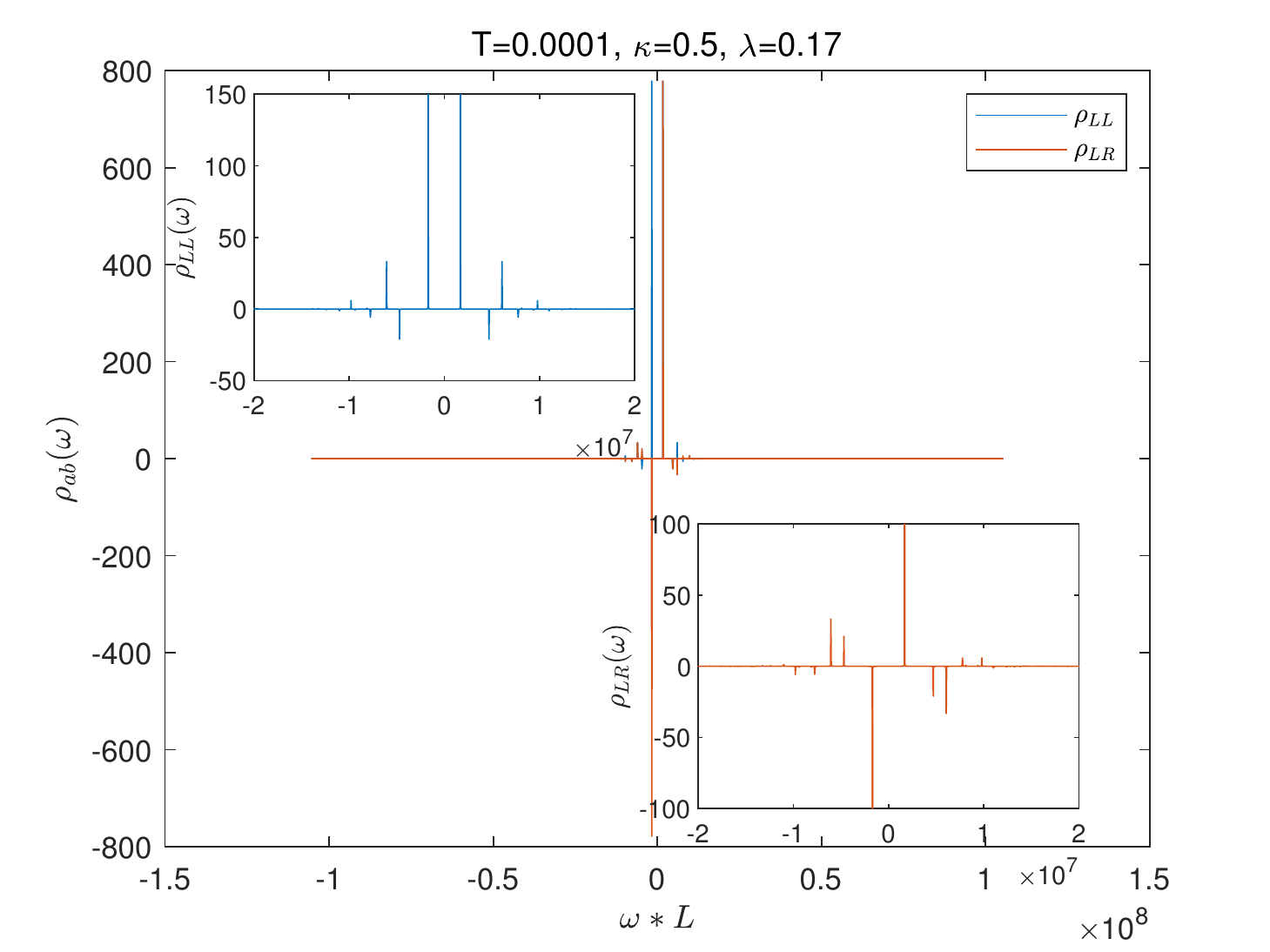}}
\subfigure[]{\includegraphics[scale=.3]{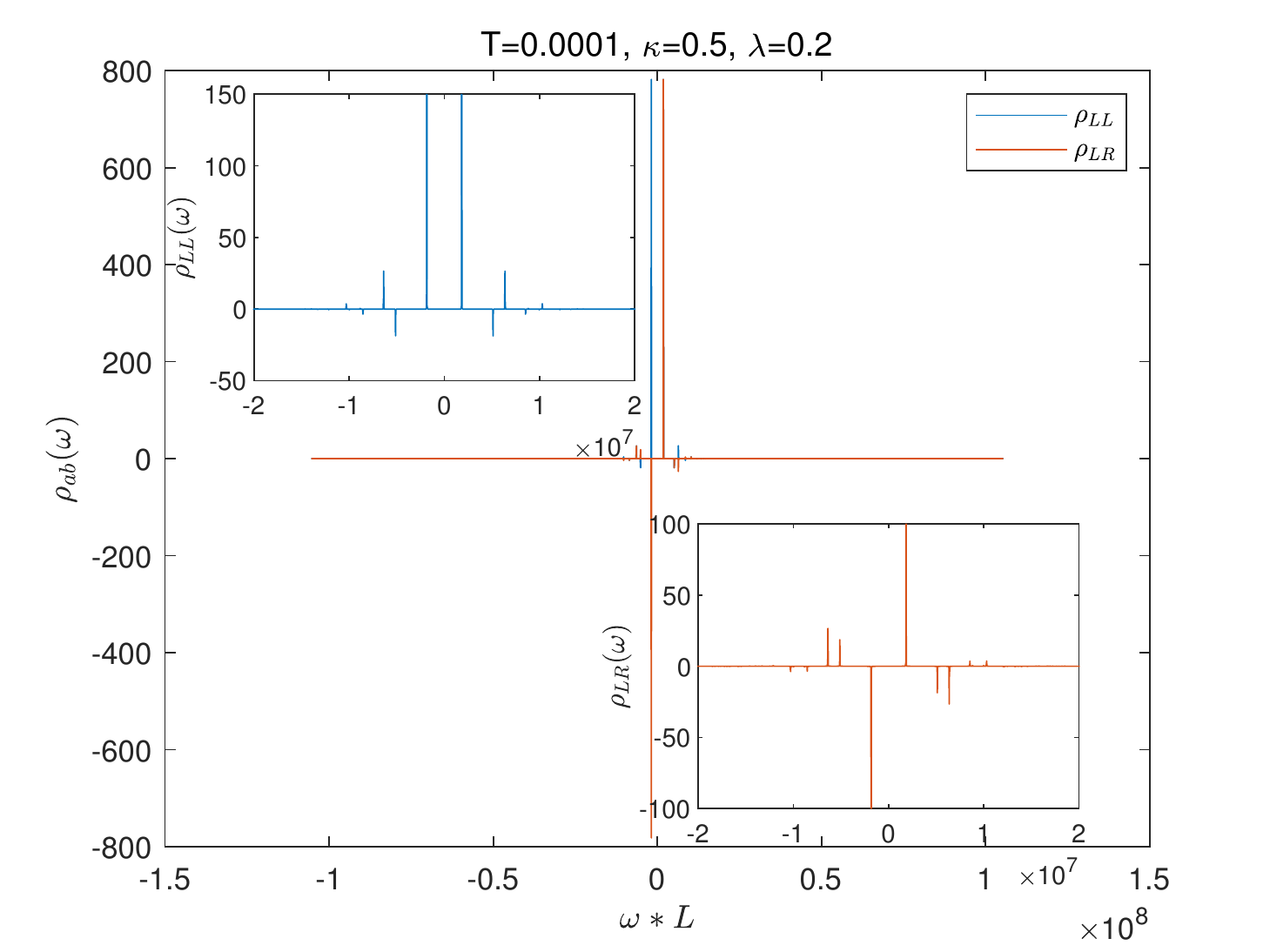}}
\caption{$|G^{>}_{ab}(t)|$ and $\rho_{ab}(\omega)$ for $\kappa=0.5$, $\lambda=0,~0.4,~0.7,~0.14,~0.17,~0.2$, $N=2^{25}$, $\epsilon=2\times10^{-4}$, $L=5\times10^6$, $\beta=10^4$. In the diagrams, we can observe a transition when we increase the value of $\lambda$, which corresponds to more peaks appear in $\rho_{ab}$.  }\label{fig:Gr_rholl_ka_p5_beta_1e4_lx_5e6_eta_2e-4_dm_25}
\end{figure}

We now turn to the study of the combined effect of a finite $\kappa$ and $\lambda$. In Fig.~\ref{fig:Gr_rholl_ka_p5_beta_1e4_lx_5e6_eta_2e-4_dm_25}, we can clearly observe that, for a fixed $\kappa$, the behavior of $|G^>_{ab}(t)|$ are different for small and large $\lambda$. For small $\lambda$, $|G^>_{ab}(t)|$ are similar to that in the case of $\lambda = 0$, namely, both Green's functions are in-phase. Likewise, $\rho_{ab}(\omega)$ have a simple structure: two leading peaks and two subleading peaks whose sign depends one whether $a = b$ or $a \neq b$ are dominant.
However, when $\lambda$ is sufficiently large, the oscillations of $|G^>_{ab}(t)|$ become qualitatively different, as for $\kappa = 0$, $|G^>_{LL}(t)|$, $|G^>_{LR}(t)|$ are out-of-phase
though the oscillating pattern becomes more intricate. This is fully consistent with
 $\rho_{ab}(\omega)$ where more subleading peaks are observed. More specifically, a finite $\lambda$ ($\kappa$) is responsible for positive (negative) subleading peaks in $\rho_{LL}(\omega)$. These peaks are suppressed when $\lambda$ gets larger. We could not study the nature of the transition between these two regimes because, for a fixed $\kappa$, we could not find solutions in this critical region $\lambda \sim 0.1$. A possible reason is that enhanced oscillations around the transition makes difficult to find $|G^>_{ab}(t)|$ numerically.

\begin{figure}[H]
\centering
\subfigure[]{\includegraphics[scale=.3]{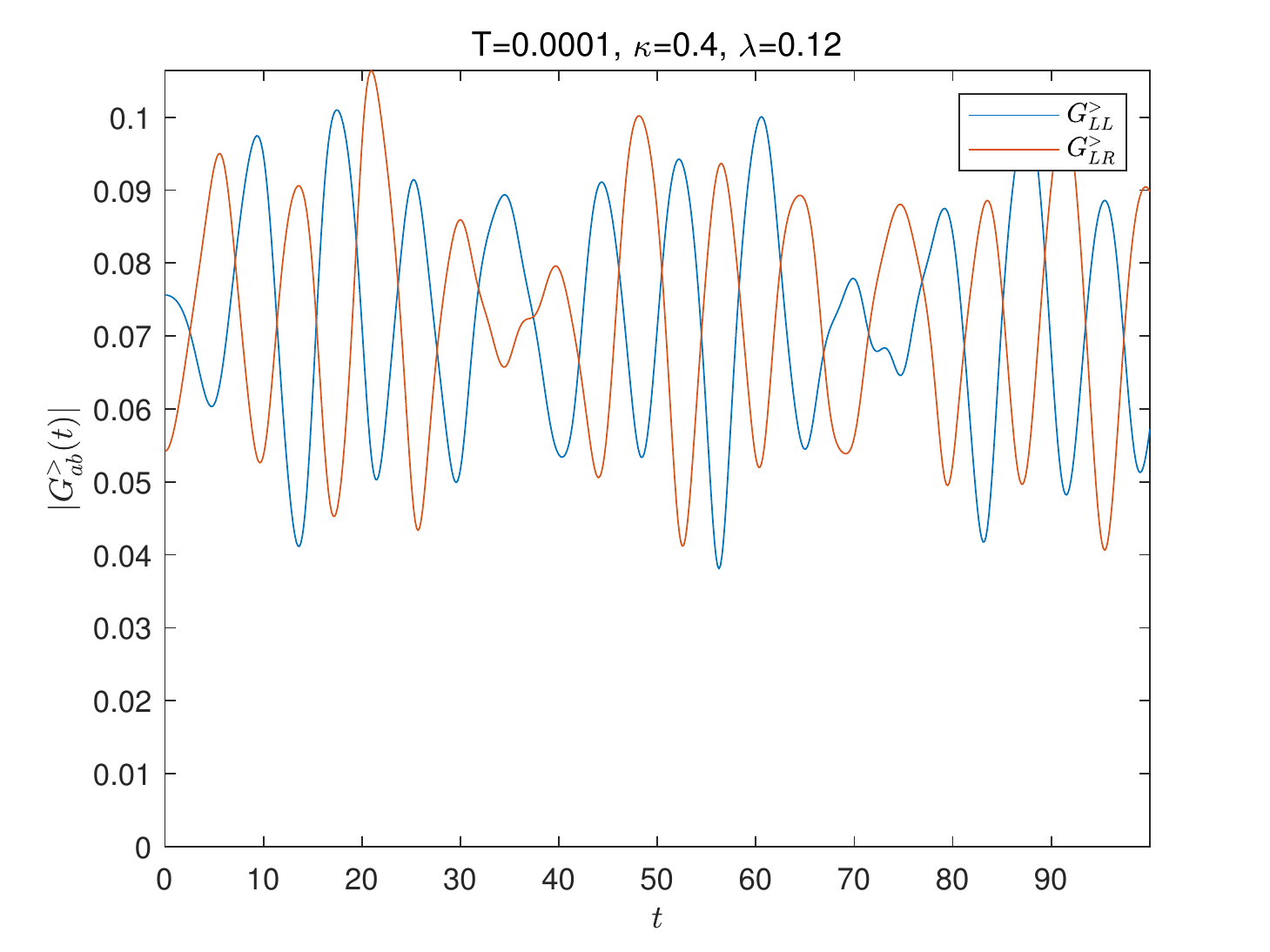}}
\subfigure[]{\includegraphics[scale=.3]{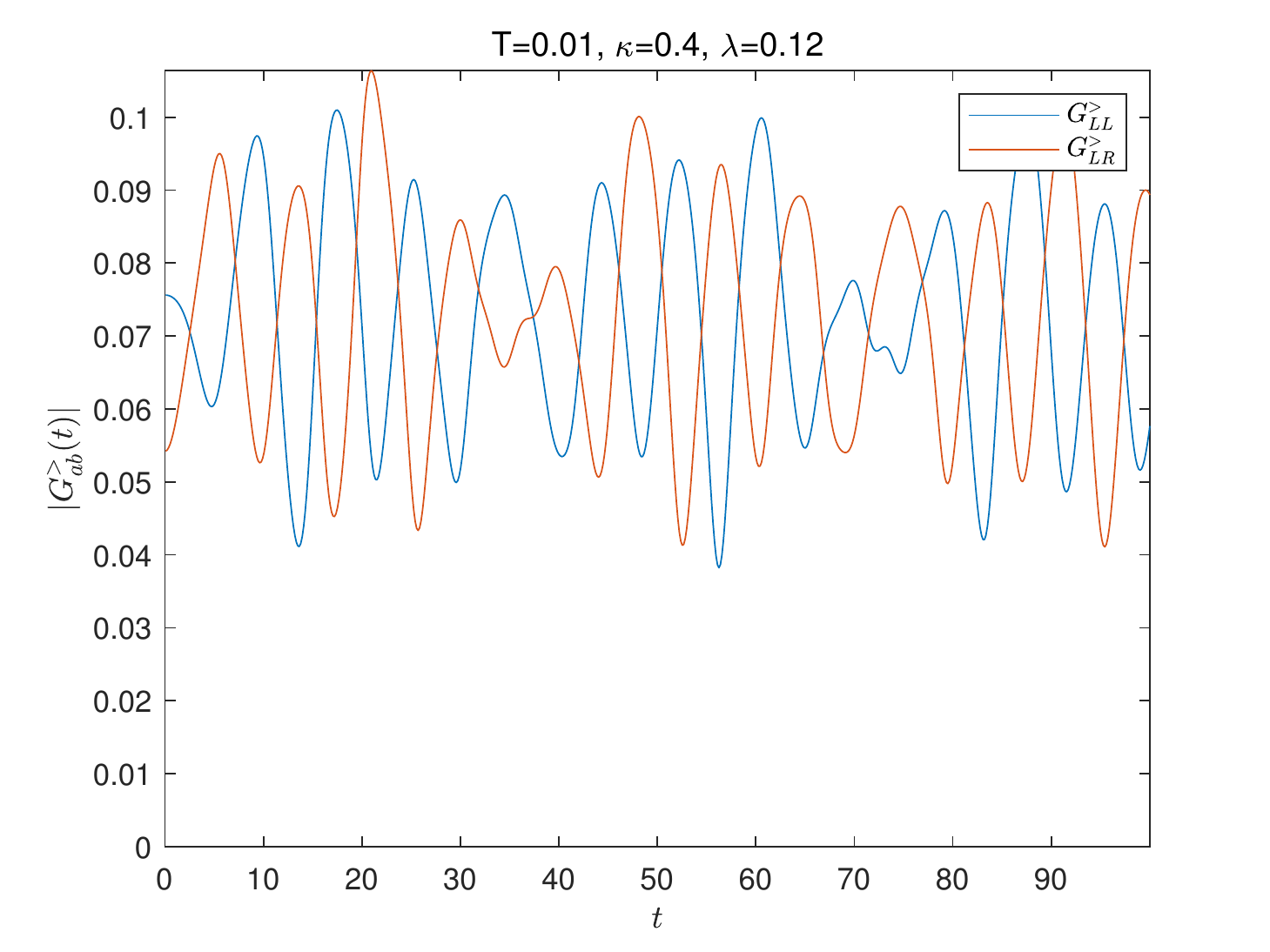}}
\subfigure[]{\includegraphics[scale=.3]{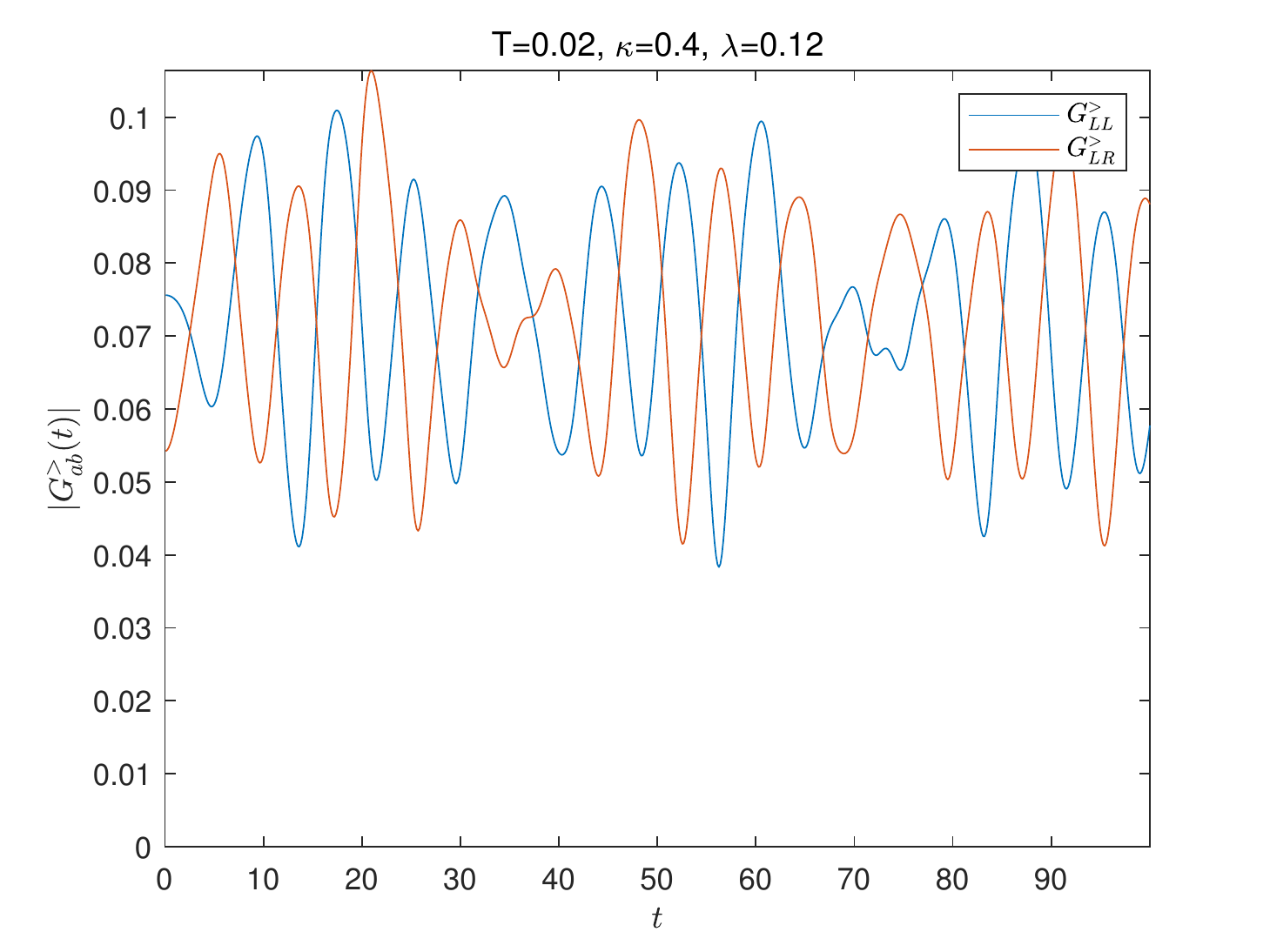}}
\subfigure[]{\includegraphics[scale=.3]{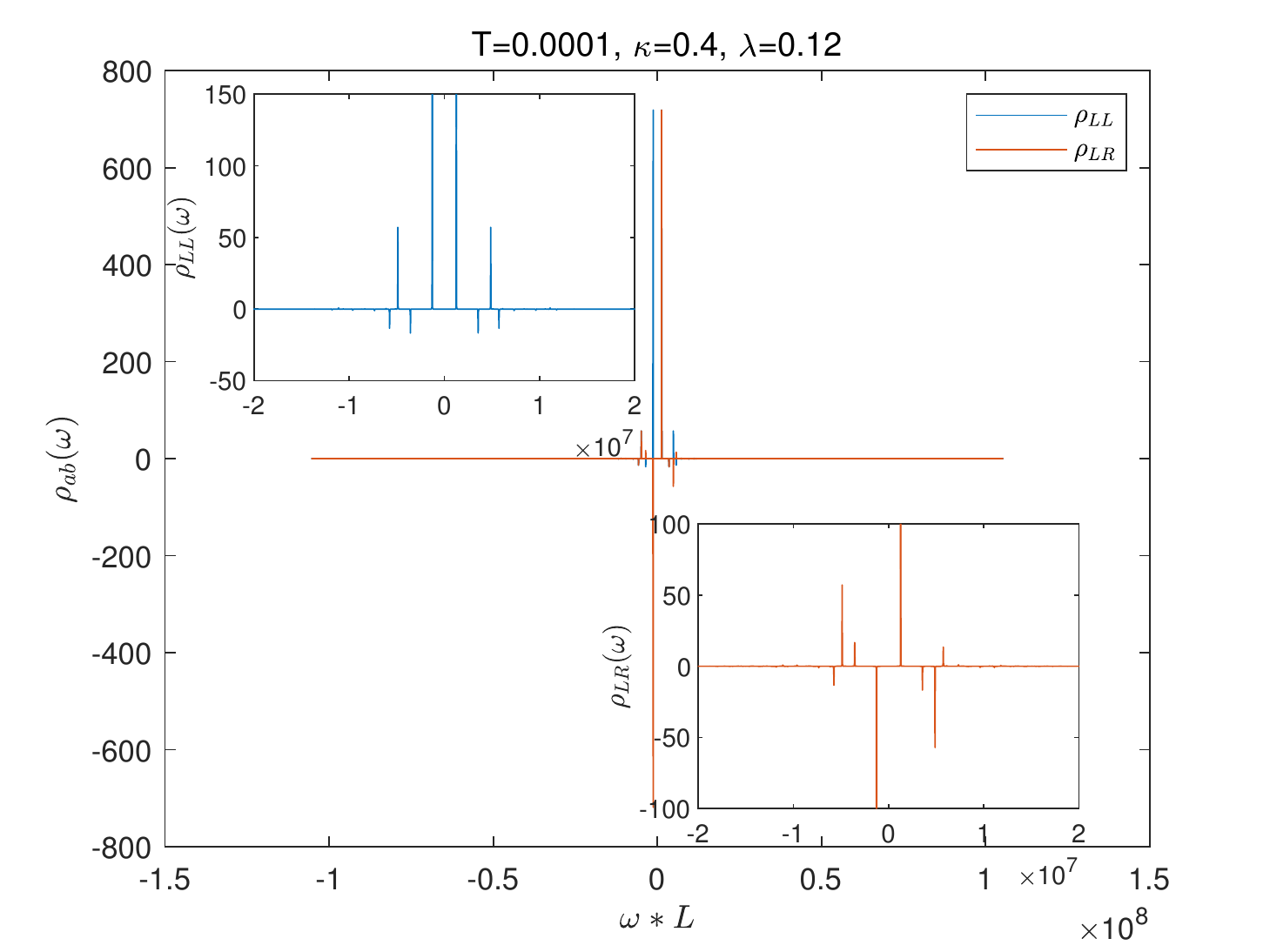}}
\subfigure[]{\includegraphics[scale=.3]{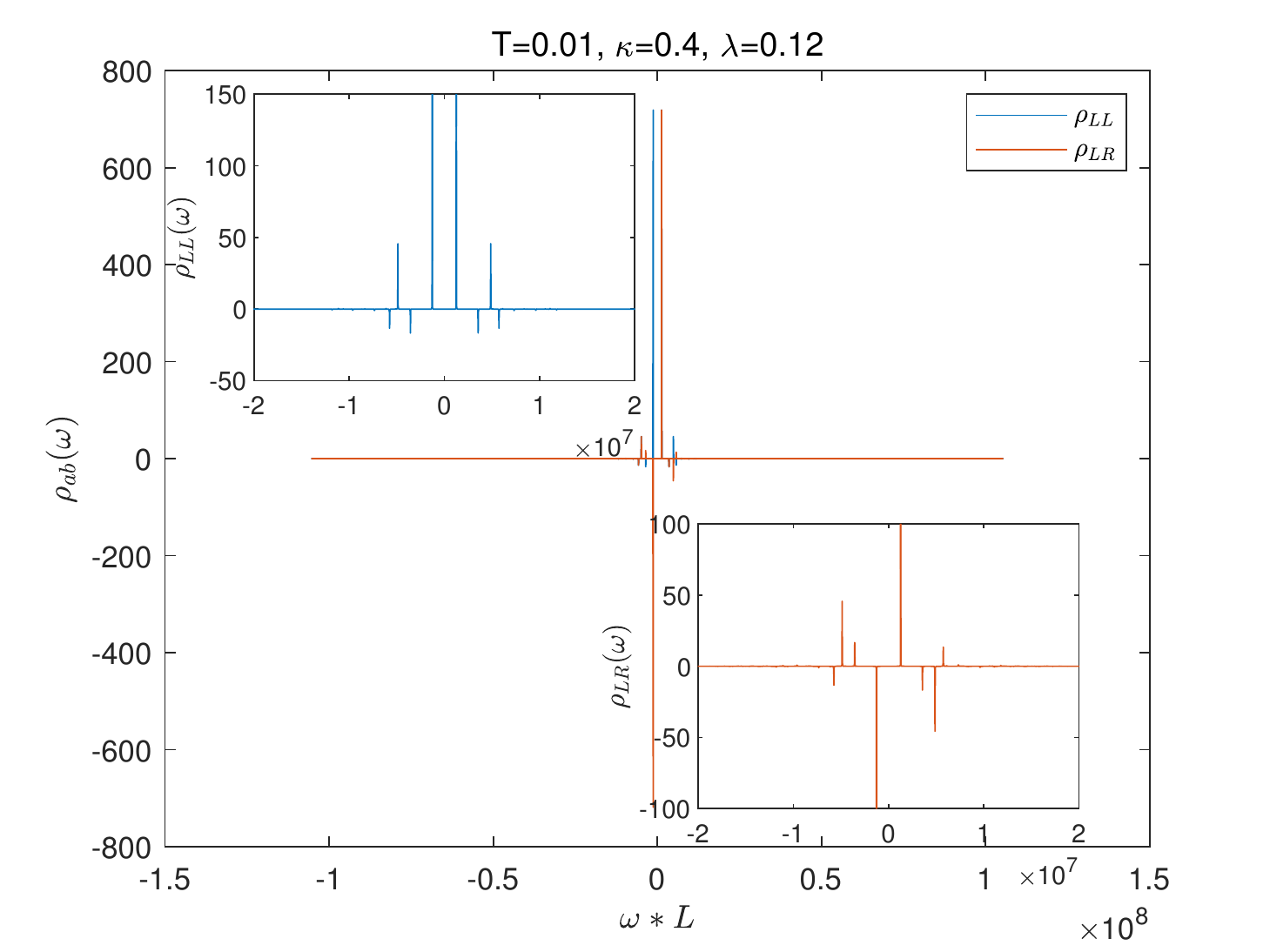}}
\subfigure[]{\includegraphics[scale=.3]{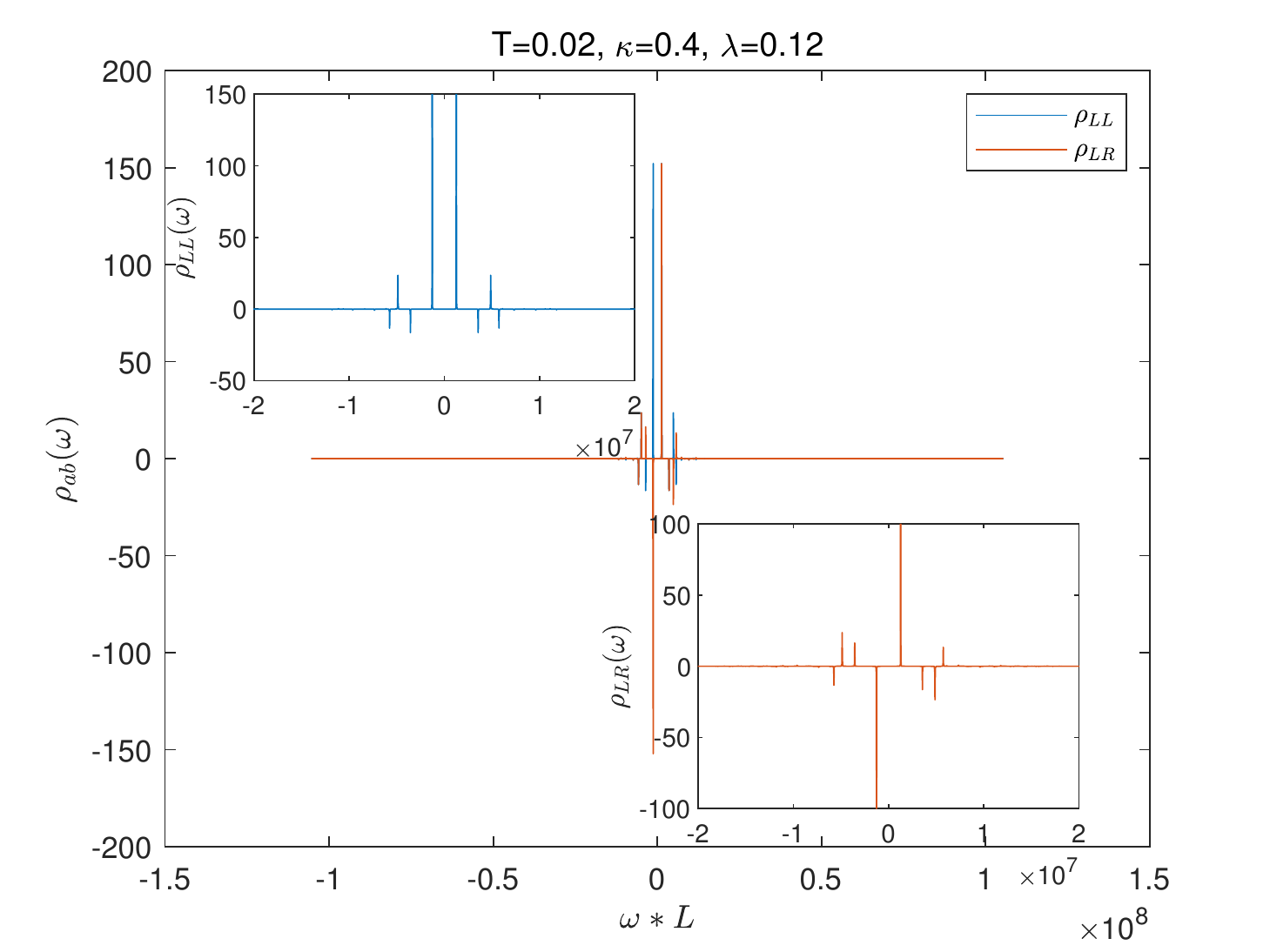}}
\subfigure[]{\includegraphics[scale=.3]{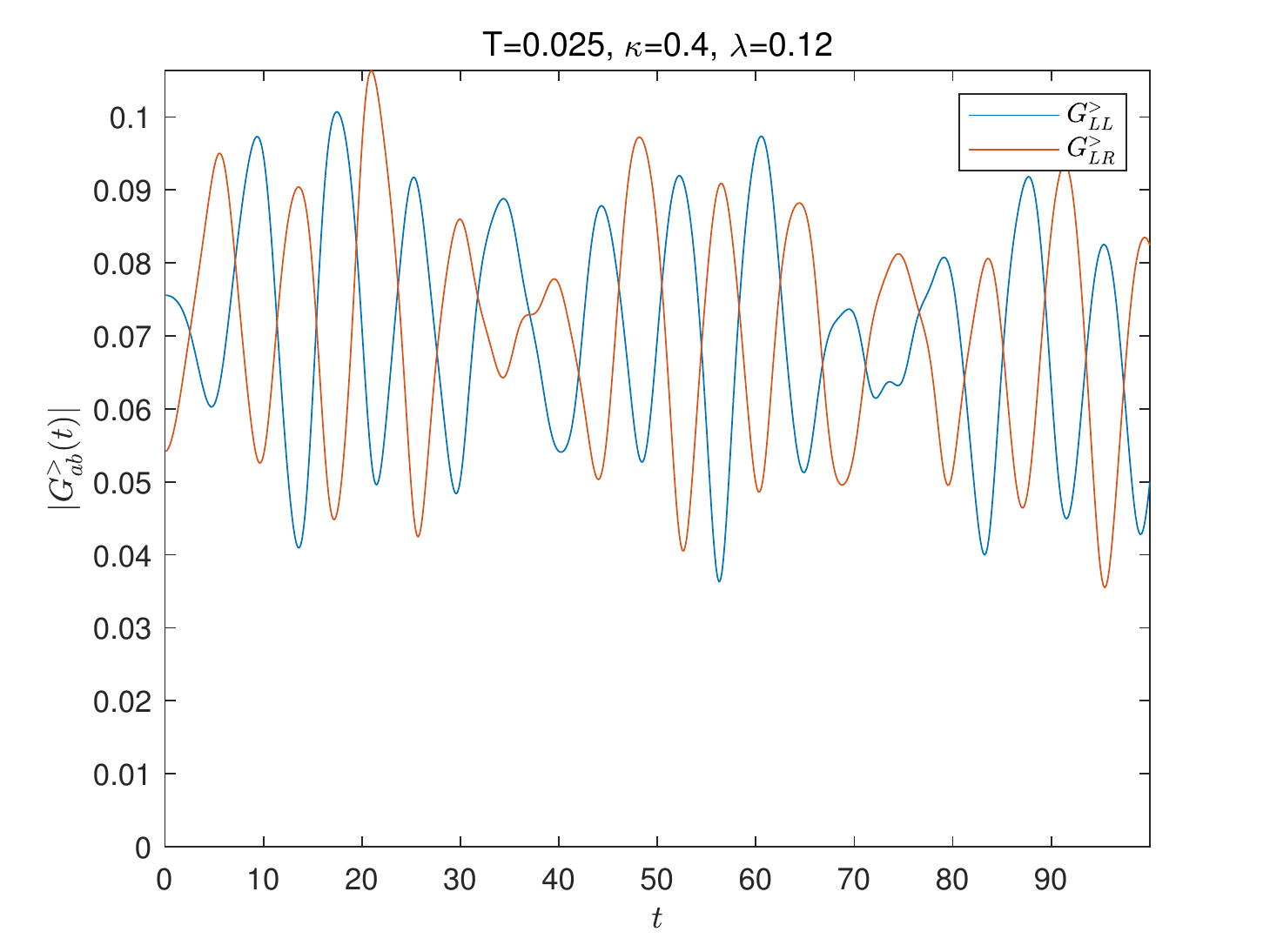}}
\subfigure[]{\includegraphics[scale=.3]{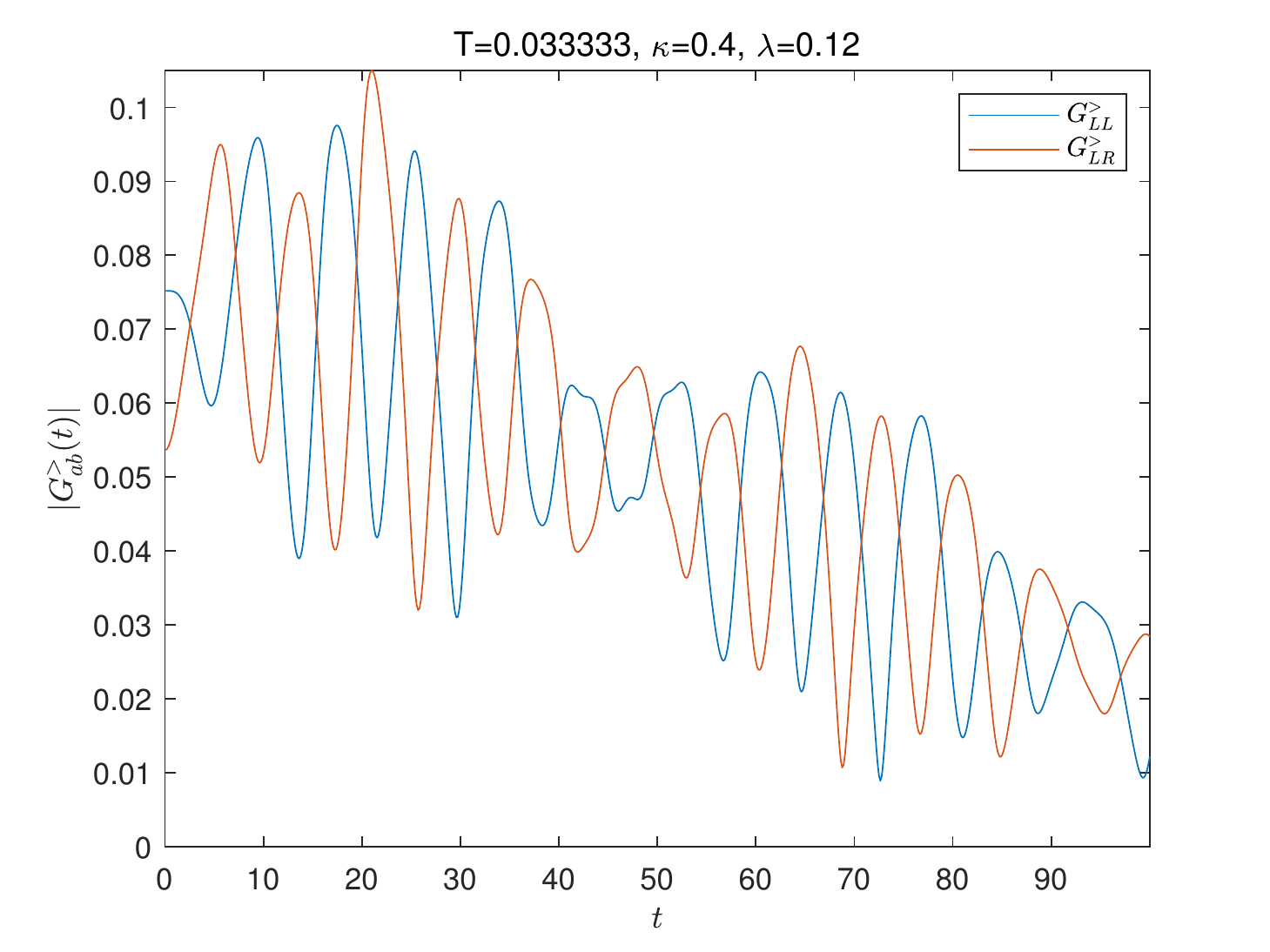}}
\subfigure[]{\includegraphics[scale=.3]{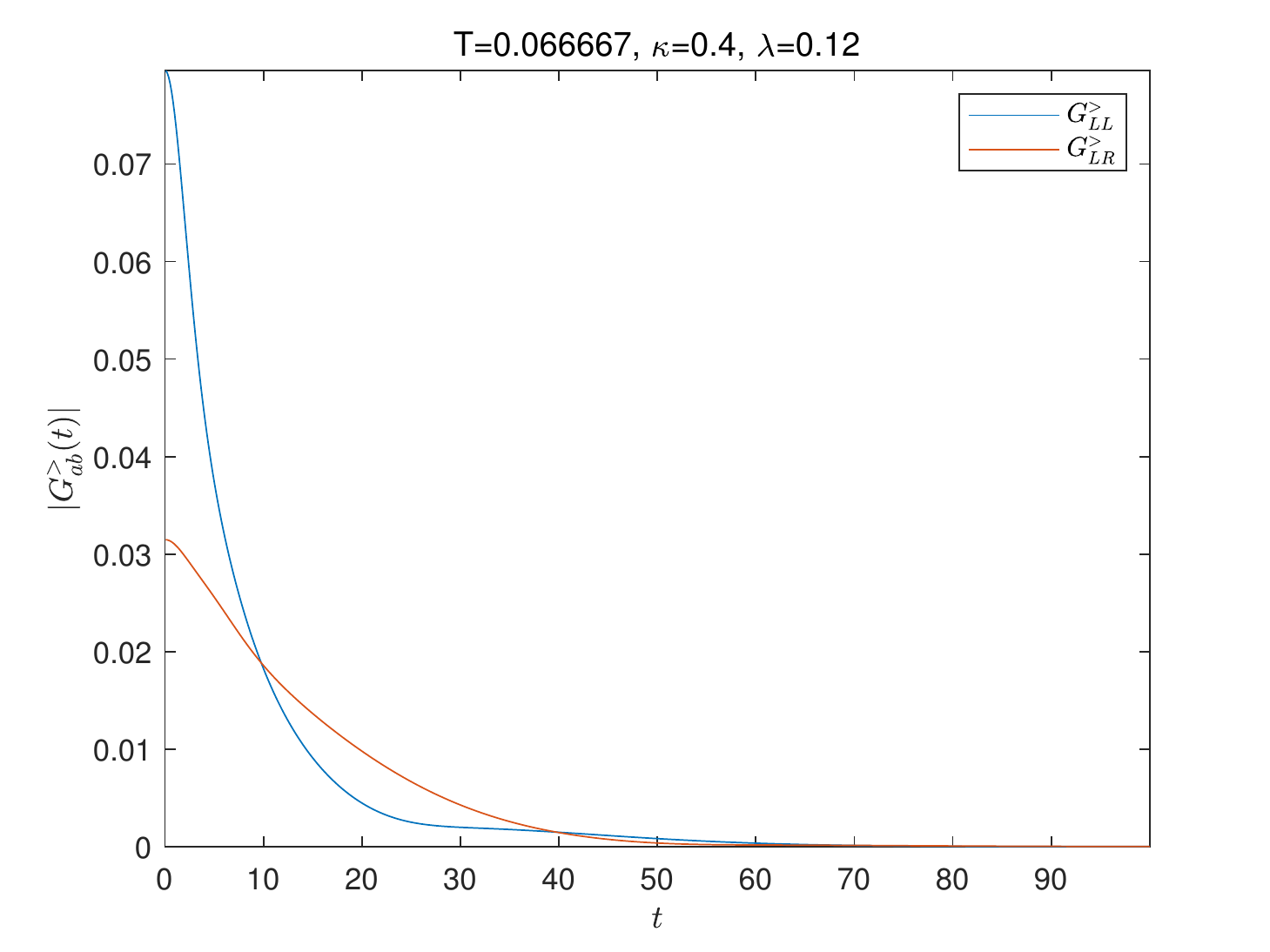}}
\subfigure[]{\includegraphics[scale=.3]{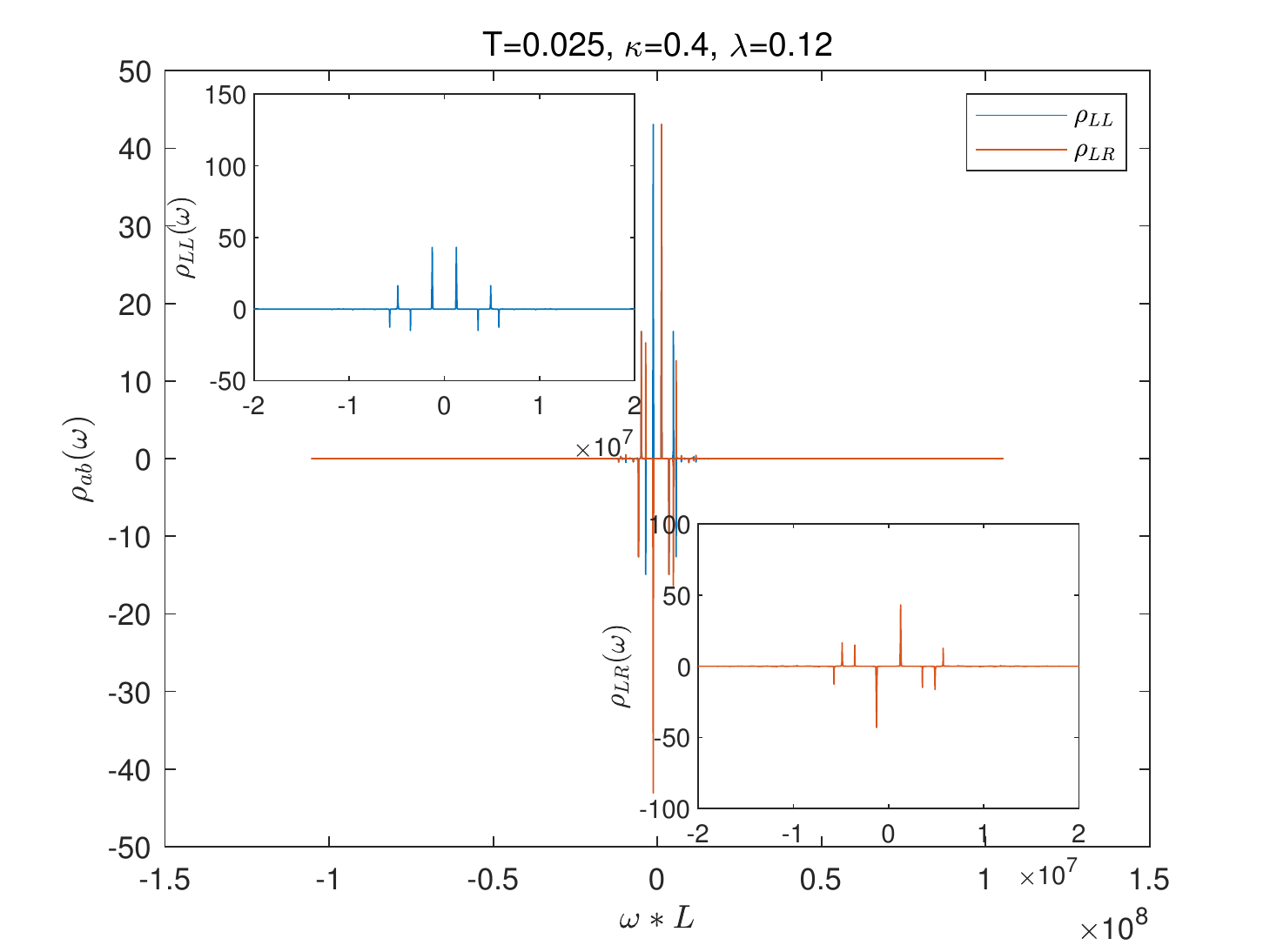}}
\subfigure[]{\includegraphics[scale=.3]{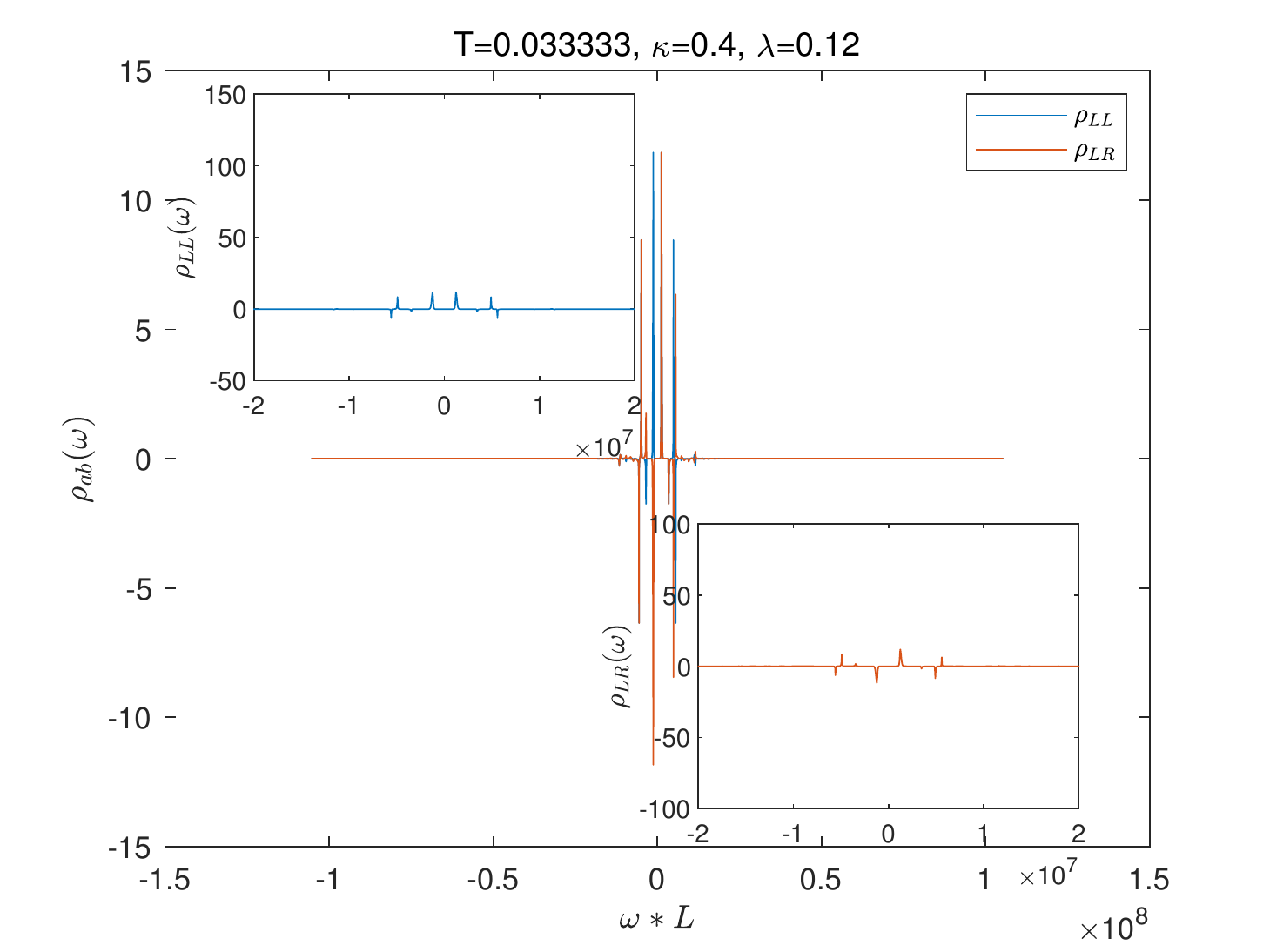}}
\subfigure[]{\includegraphics[scale=.3]{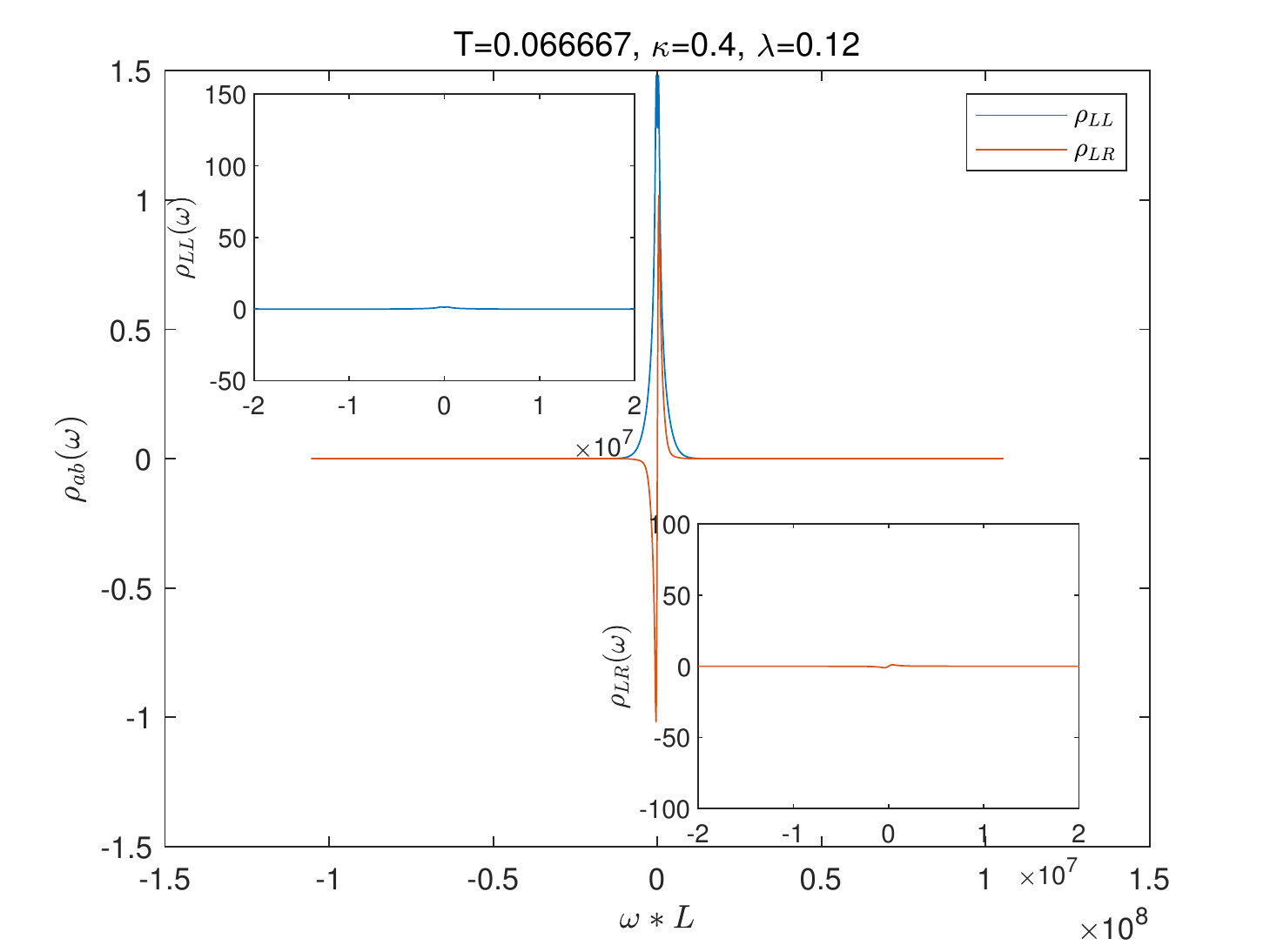}}
\subfigure[]{\includegraphics[scale=.3]{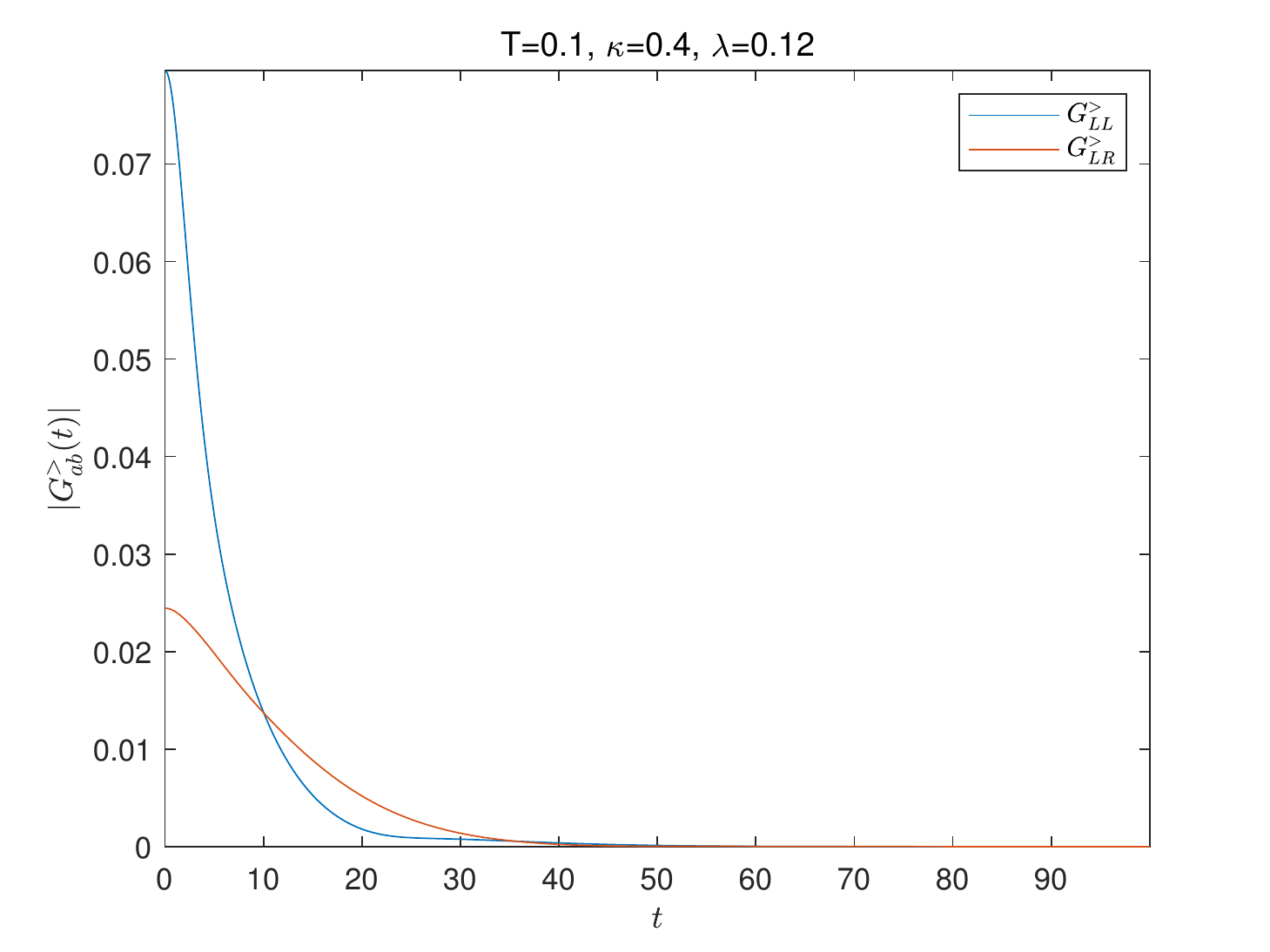}}
\subfigure[]{\includegraphics[scale=.3]{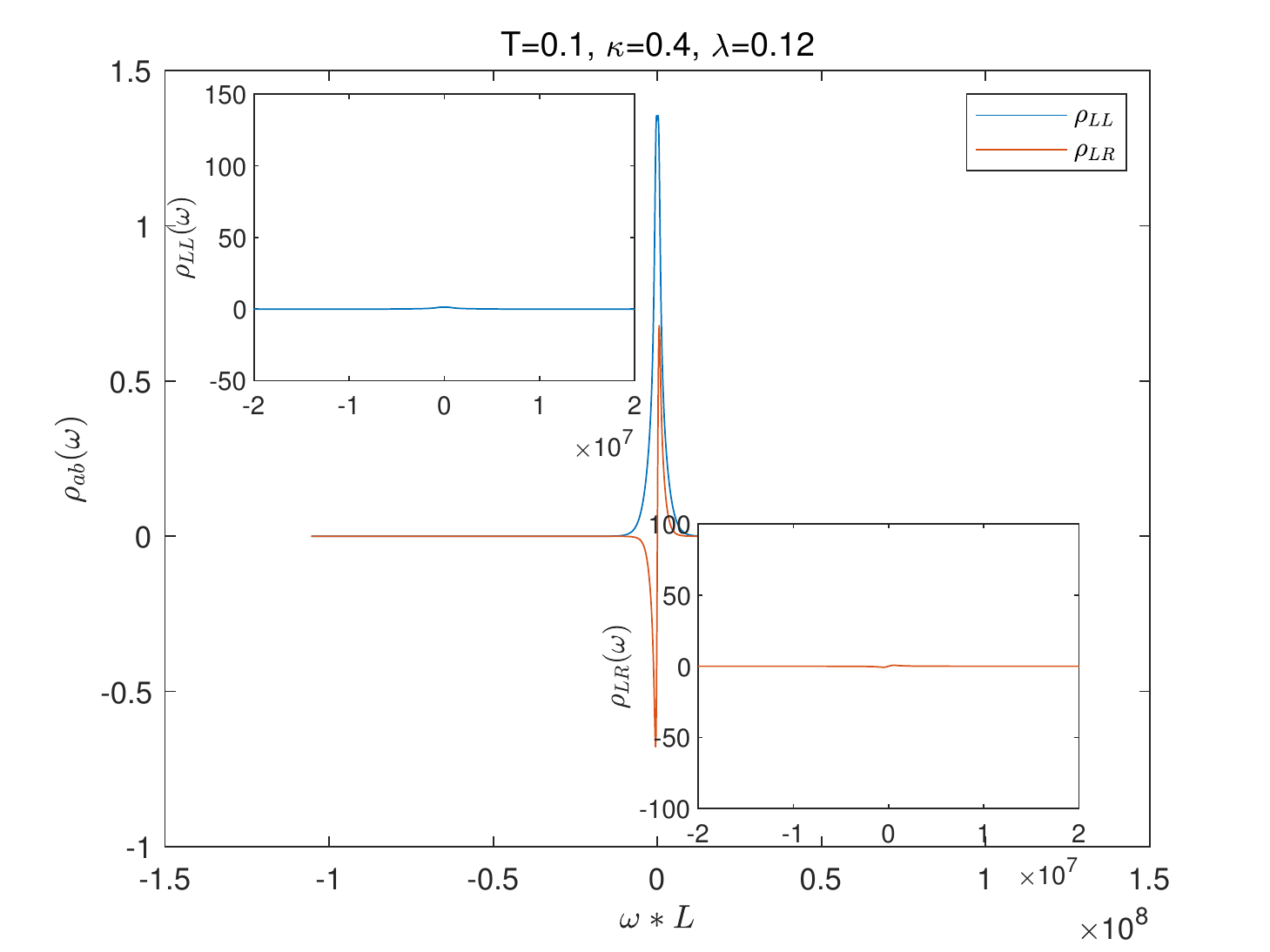}}
\subfigure[]{\includegraphics[scale=.3]{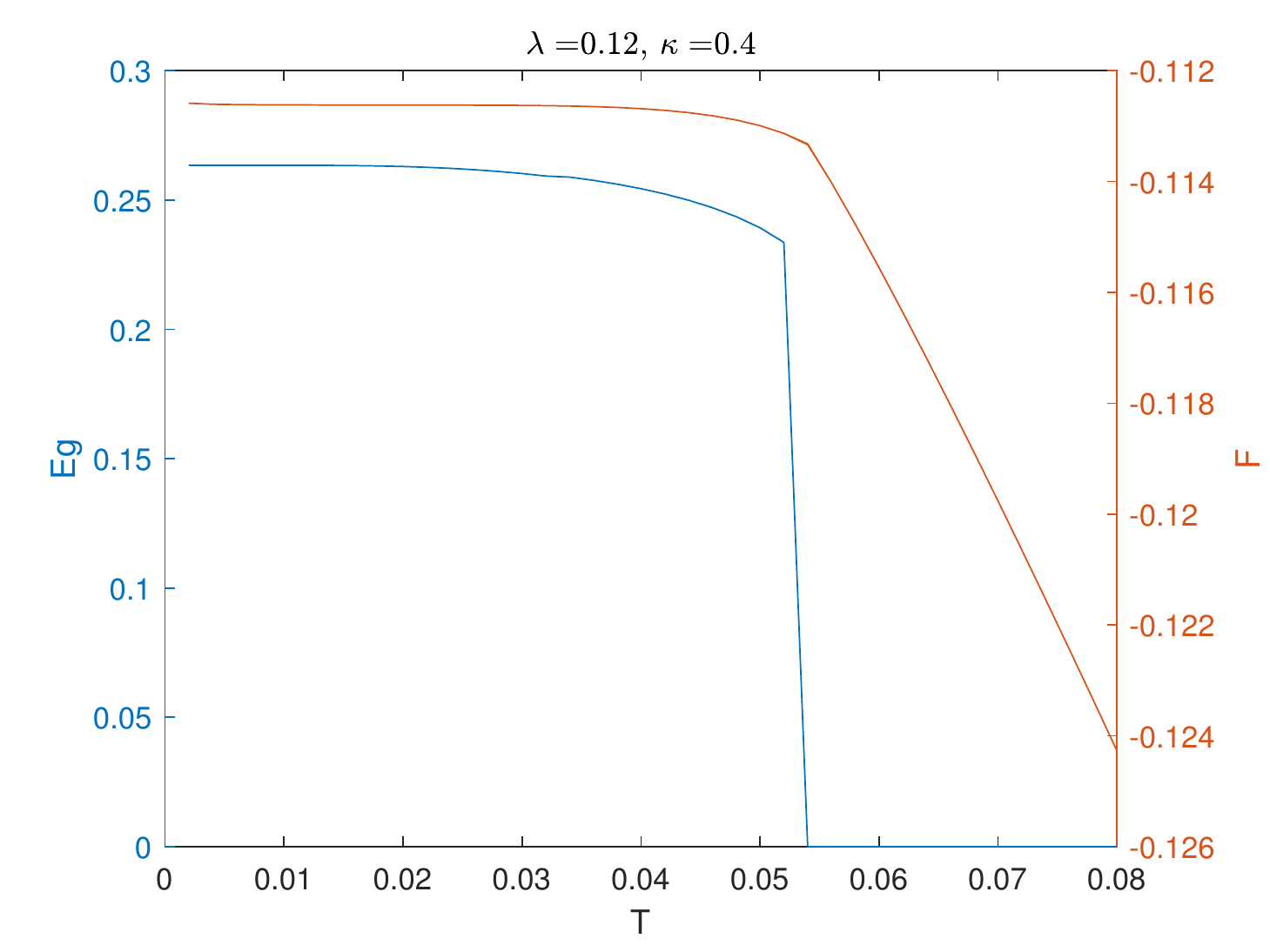}}
\caption{$|G^{>}_{ab}(t)|$ and $\rho_{ab}(\omega)$ for $\kappa=0.4$, $\lambda=0.12$, $N=2^{25}$, $\epsilon=2\times 10^{-4}$, $L=5\times 10^6$, when $\beta=10^4,~10^2,~50,~40,~30,~15,~10$, as well as $F$ and $E_g$ in the last figure. From the real time and imaginary time calculations, we can see when $T$ increases the system suffers a phase transition, and their transition temperatures are close, so both results are basically consistent.  }\label{fig:Gr_rholl_Eg_T_ka_p4_ld_p12}
\end{figure}

Finally, we check the consistency of our results by comparing the temperature dependence of the real time Green's function with the thermodynamic properties investigated previously. Results for $F(T)$ and $E_g(T)$, depicted again in Fig.~\ref{fig:Gr_rholl_Eg_T_ka_p4_ld_p12}(e) indicate that the thermodynamic phase transition temperature occurs at $ T_c \sim 0.055$. Previous real time Green's functions were considered in the low temperature region $T \to 0$ where the system is in the wormhole phase and the pattern of oscillations of $|G^>_{ab}(t)|$ is quite rich. As temperature is increased ($T > 0.02$), we observe a gradual suppression of oscillations.  This suppression is directly related to a broadening of the peaks in $\rho_{ab}(\omega)$. This behavior has already been predicted in Ref.~\cite{qi2020} for the $\kappa = 0$ case. Oscillations disappears already for $T=0.0666$
which suggests a transition to the black hole phase.

\begin{figure}
\centering
\subfigure[]{\includegraphics[scale=.4]{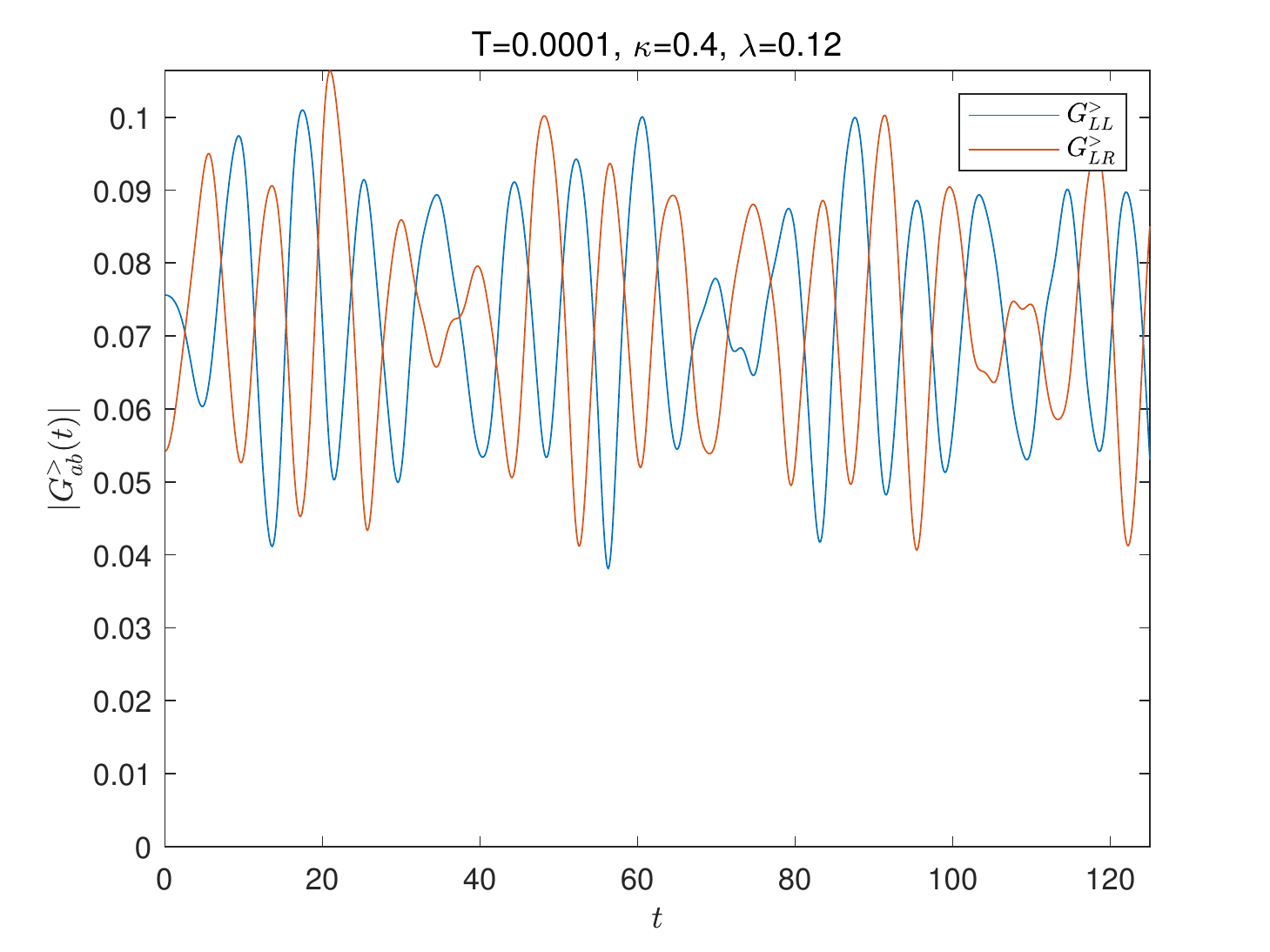}}
\subfigure[]{\includegraphics[scale=.4]{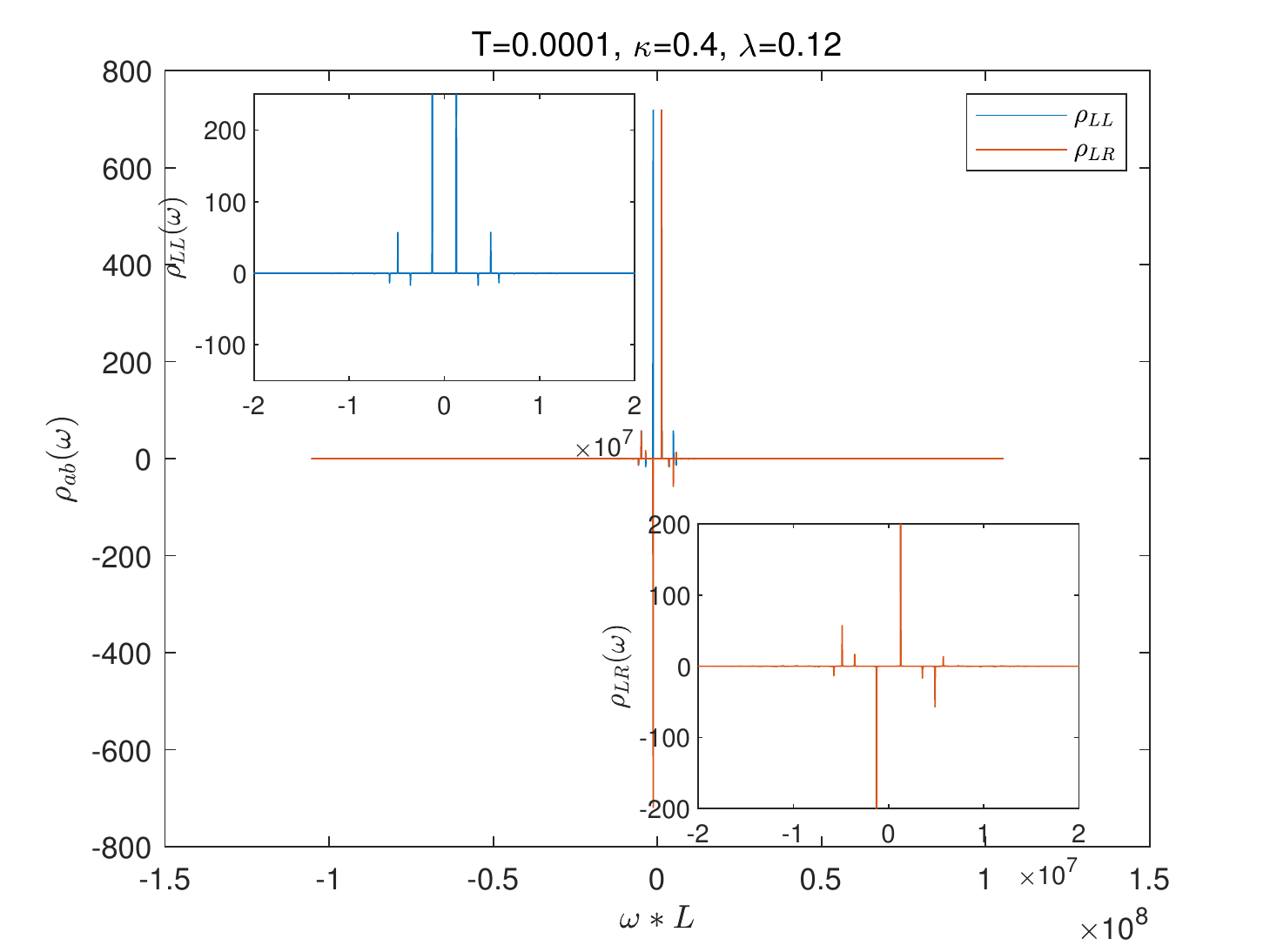}}
\subfigure[]{\includegraphics[scale=.4]{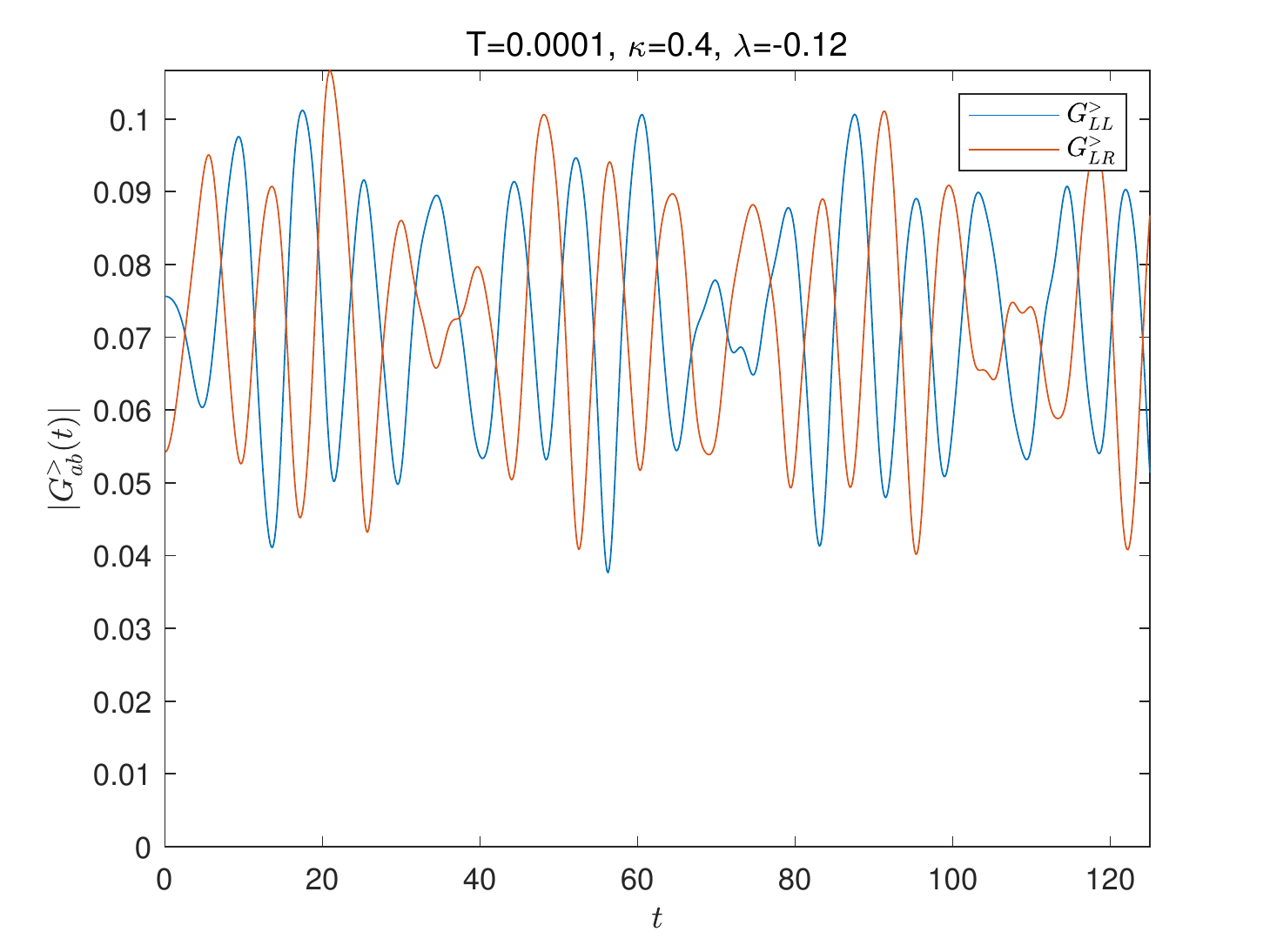}}
\subfigure[]{\includegraphics[scale=.4]{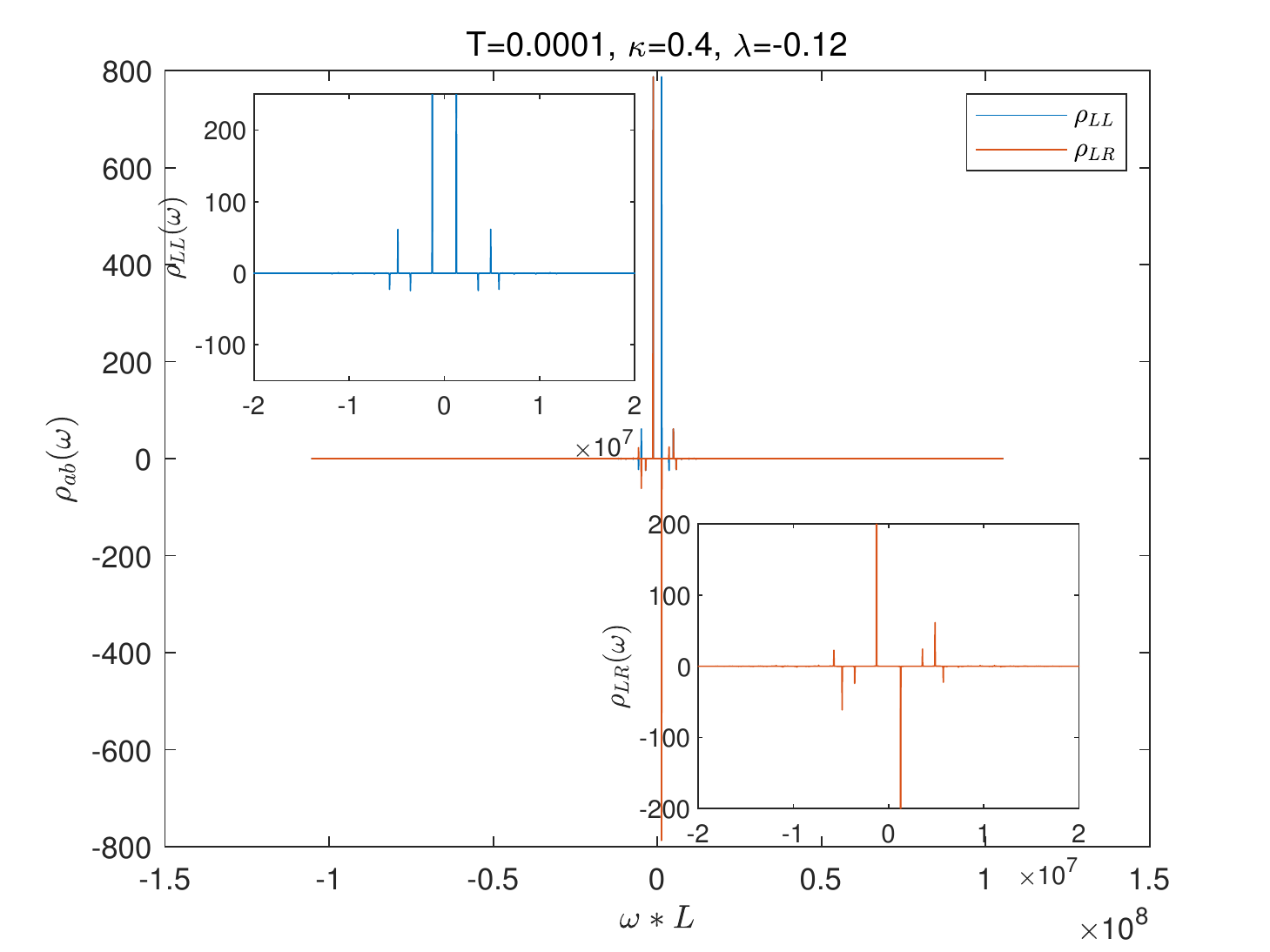}}
\caption{$|G^{>}_{ab}(t)|$ and $\rho_{ab}(\omega)$ for $\kappa=0.4$, $\lambda=0.12$ (top) and $\lambda=-0.12$ (bottom), with $N=2^{25}$, $\epsilon=2\times10^{-4}$, $L=5\times10^6$, $\beta=10^4$.  }\label{fig:Gr_rholl_ld_p12_ka_p4}
\end{figure}

Finally, we study the effect of sign switch $\lambda \to -\lambda$ in the real time evolution. In  Fig. \ref{fig:Gr_rholl_ld_p12_ka_p4}, we depict $|G^>_{ab}(t)|$ and $\rho_{ab}(\omega)$ for positive and negative values of $\lambda$. As is observed, the sign flip in $\lambda$ just changes the sign of $\rho_{LR}$ and leaves $|G^>_{ab}(t)|$ invariant.
This is reasonable since we expect that $\lambda \to -\lambda$ induces and overall sign difference in $G^R_{LR}(t) \to -G^R_{LR}(t)$.

 One interesting question is whether the observed transition between different oscillation patterns is accompanied with a change in the free energy $F$. The answer to this question is negative. The reason is that the exponential behavior of the imaginary time Green's function $G_{ab}(\tau)$ is related to the leading peak of the corresponding $\rho_{ab}(\omega)$.  However the oscillating patterns depend instead on the superposition regarding the leading and all the  subleading peaks.
  Since $F$ is a function of $G_{ab}(\tau)$, we do not expect $F$ to experience any significant change during the complex-to-real transition.  Indeed, the thermal phase transition is different from the complex-to-real transition. As we will see next in gravity, the former is a transition between the  wormhole and two black holes while the latter can be interpreted as a  Euclidean-to-Lorentzian transition in the wormhole phase.

\newpage

\section{Gravity dual}

In this section, we study the gravity dual of the SYK system described by the Hamiltonian \eqref{hami}.  At low temperatures, it is dual to a Euclidean wormhole in JT gravity with two parameters $\eta$ and $k$, respectively dual to $\la$ and $\k$.  This is a generalization of the eternal traversable wormhole \cite{maldacena2018} (corresponding to $k=0$) and the Euclidean wormhole without interaction \cite{garcia2021} (corresponding to $\eta=0$). The combined effect of $\eta$ and $k$ studied in this paper leads to a wormhole with similar thermodynamical properties and we will find an excellent match between the SYK and JT results.

Our system is a purely Euclidean system studied from the point of view of statistical mechanics. It doesn't have a Lorentzian interpretation (with unitary evolution) as the energy spectrum  is generally complex. Note that the Euclidean quantities can still be viewed as suitably analytically-continued versions of Lorentzian observables, which is akin to studying a partition function at imaginary value of the chemical potential, see \eg \cite{Roberge:1986mm, deForcrand:2002hgr}.

The main result of the previous section is the observation of a complex-to-real transition, where the energy spectrum becomes real for sufficiently strong inter-site coupling, despite the Hamiltonian being non-Hermitian. The Euclidean wormhole has a gravitational $\r{U}(1)$ symmetry while the Lorentzian wormhole has a gravitational $\r{SL}(2,\R)$ symmetry \cite{Lin:2019qwu, Harlow:2021dfp}. We will show that the complex-to-real transition corresponds to the dynamical restoration of the $\r{SL}(2,\R)$ symmetry of the Lorentzian wormhole, and thus can be interpreted as a Euclidean-to-Lorentzian transition.

\subsection{Wormhole solutions}
The theory we consider is Jackiw-Teitelboim gravity with a massless scalar field $\chi$ and a static Gao-Jafferis-Wall interaction \cite{gao2016, maldacena2018} involving  $N$ fields of dimension $\D$. To compare with SYK, we should take $\D={1/4}$. The theory is described by the action
\be\label{gravityS}
S = S_\r{JT} +S_\chi + S_\r{int}~,
\ee
where
\bea\nt
S_\r{JT} \= -S_0\,\chi_\r{Euler} -{1\/2} \int_M d^2 x\sqrt{g}\, \Phi(R+2) - \int_{\p M} d\tau\sqrt{h} \, \Phi (K-1)~,
\\\label{actionJT}
S_\chi \= {1\/2}\int_M d^2 x\sqrt{g} \,(\p\chi)^2~,\-
S_\r{int} \= g \sum_{i=1}^N \int du \,O_L^{(i)}(u)O_R^{(i)}(u)~.
\eea
The solution we consider is the Euclidean wormhole
\be
ds^2 = {d\tau^2 + d\rho^2\/\r{cos}^2\rho},\qq -{\pi\/2}< \rho <{\pi\/2},\qq \tau\sim \tau+b~.
\ee
Following \cite{garcia2021}, we deform the theory with boundary sources taken to be imaginary:
\be
\lim_{\rho\to-{\pi/2}}\chi = i \tk,\qq
\lim_{\rho\to{\pi/2}}\chi = -i \tk~.
\ee
The imaginary sources model the imaginary part of the SYK couplings as they correspond to a deformation of the Hamiltonian
\be
\d H = i \tk\, \cO_L - i \tk\, \cO_R
\ee
where $\cO_{L}, \cO_R$ are the marginal operators dual to $\chi$ on each boundary. We see that this is a good model of the imaginary part of the SYK Hamiltonian \eqref{hami} as we can identify the marginal operators with the SYK operators
\be
\cO_L\sim M_{ijk\l}\psi_i^\r{L}\psi_j^\r{L}\psi_k^\r{L}\psi_\l^\r{L},\qq \cO_R \sim  M_{ijk\l}\psi_i^\r{R}\psi_j^\r{R}\psi_k^\r{R}\psi_\l^\r{R}~.
\ee
The boundary conditions for JT gravity are
\be
ds^2 = {d\tu^2\/\e^2},\qq \Phi = {\phi_r\/\e}~,
\ee
and we study the theory in Euclidean signature with the periodicity condition
\be
\tu \sim \tu + \tb~.
\ee
\subsubsection{Schwarzian effective action}
Nearly AdS$_2$ holography is a theory of a boundary graviton, or reparametrization mode, which can be described by an effective Schwarzian action \cite{jensen2016, engels2016, maldacena2016a}. The wormhole has two boundaries so it is described by an action for  two reparametrization modes $\tau_L(u)$ and $\tau_R(u)$ after integrating out the matter degrees of freedom.

Without boundary sources ($\tilde{k}=0$), the system is  the eternal traversable wormhole whose action was derived in \cite{maldacena2018} in both SYK and JT gravity.  The boundary sources give an additional  contribution that can be computed by evaluating the action for a general solution for $\chi$ in the wormhole as a function of boundary sources $\chi_L$ and $\chi_R$:
\be
\chi(\tau,\rho) = \int_\R d\tau_L \,K_L(\tau,\rho;\tau_L)\chi_L(\tau_L)+ \int_\R d\tau_R \,K_R(\tau,\rho;\tau_R)\chi_R(\tau_R)
\ee
where the bulk-to-boundary propagators in AdS$_2$ are
\be
K_L(\tau,\rho;\tau_L) = {1\/2\pi} \le({\r{cos}\,\rho\/\r{cosh}(\tau-\tau_L)+\r{sin}\,\rho} \ri),\quad K_R(\tau,\rho;\tau_R) = {1\/2\pi} \le({\r{cos}\,\rho\/\r{cosh}(\tau-\tau_R)-\r{sin}\,\rho} \ri)~.
\ee
The value of the sources chosen here are
\be
\chi_L = i \tk,\qq \chi_R = -i\tk~.
\ee
The contribution of the scalar field is then obtained by evaluating the on-shell action after acting with the diffeomorphisms corresponding to the two Schwarzian modes \cite{maldacena2016a}, see Appendix A of \cite{garcia2022} for additional details.

At the end, the effective Schwarzian action of the system takes the form
\bea\nt
S \= - N \int_0^\b d u \le[ \le\{ \r{tanh}(\tfrac12 \tau_L(u)),u\ri\}+ \le\{ \r{tanh}(\tfrac12 \tau_R(u)),u\ri\}+\eta \le({\tau_L'(u)\tau_R'(u)\/\r{cosh}^2(\tfrac12(\tau_L(u)-\tau_R(u)))} \ri)^\D+{3\/2} k^2{b\/\b}\ri]\\\label{SchwAction}
\eea
which we study in Euclidean signature.  We see the effect of the boundary sources is to add a constant term in the action proportional to the wormhole size $b$. We use the physical time $u$ which is related to the coordinate $\tilde{u}$ via
\be
u = {\cJ\/\a_S} \tu = {N\/\phi_r}\tu,{}
\ee
and which is periodically identified $
u\sim u+\b$ where $\b = \tilde\b N$ is the physical temperature. The coupling constants are
\be\label{couplingsScaling}
\eta = {g\/2^{2\D}}\le({N\/\phi_r} \ri)^{2\D-1},\qq k^2= {2\/\pi N}\tk^2
\ee
which are taken fixed in the large $N$ limit. As in \cite{maldacena2018}, the validity of the action requires that $\eta \ll 1$ and that the system develops an approximate conformal symmetry close to the ground state, which will be assumed here.  The gravitational regime corresponds to large $N$. The overall factor of $N$ ensures that the path integral localizes on its saddle-points in the large $N$ limit.

 We have not been able to derive the Schwarzian action directly from the SYK model because, unlike in JT gravity, it is harder to split the real and imaginary couplings which enter in a rather symmetric way. Nonetheless, we expect, due to the strikingly similar properties of both systems, that the Schwarzian effective action will be the same in SYK, with $k$ proportional to $\k$. This is also expected from universality if we view the Schwarzian action as a type of effective hydrodynamics, the contribution from the imaginary sources corresponding to a marginal deformation.

\subsubsection{Wormhole solutions}

The Euclidean wormhole corresponds to the solution
\be
\tau_L(u)=\tau_R(u) = {b\/\b}\,u~,
\ee
which gives the Euclidean action
\be
S = N\le( {b^2\/\b} - {3\/2}k^2 b - \eta \b^{1-2\D} b^{2\D}\ri)~.
\ee
This action needs to be minimized with respect to the wormhole size $b$. For this purpose, it is useful to introduce the variable $X$ defined from the relation
\be
b=\b X^2,
\ee
so that the action becomes
\be
S = \b N\le(X^4 -{3\/2}k^2 X^2 -\eta X \ri)
\ee
where we have set $\D={1\/4}$. The action is a quartic polynomial in $X$.  The saddle-point in $X$ gives a cubic equation
\be\label{cubicEq}
X^3 -{3k^2\/4} X- {\eta\/4}=0.
\ee
This equation is also equivalent to the vanishing of $\r{U}(1)$ charge
\be\label{Q00}
Q_0=0~,
\ee
which is required as the $\r{U}(1)$ symmetry is a gauge symmetry. In fact, \eqref{Q00} is the integrated Hamiltonian constraint of JT gravity and implies the gravitational equations of motion. For JT gravity with matter, it takes the form \cite{Lin:2019qwu, Harlow:2021dfp}
\be
0 = Q_0 =- N (E[\tau_L(u)] +E[\tau_R(u)])+ \int_\S dx\, T_{00}^\r{matter}
\ee
on a Cauchy slice $\S$. Here, the functional
\be\label{energyFunctional}
E[\tau(u)] = {\tau^{(3)}(u)\/\tau'(u)^2} - {\tau''(u)^2 \/\tau'(u)^3} - \tau'(u)
\ee
measures the energy at each boundary in terms of the boundary graviton.  As an aside, we note that this form is similar to the integrated Hamiltonian constraint in higher-dimensional AdS. In \cite{Chowdhury:2021nxw}, this was used to prove a perturbative version of the holography of information \cite{Raju:2020smc}. This suggests that a similar statement should be possible in JT gravity with matter for excitations of the eternal traversable wormhole, \ie on the solution corresponding to the global AdS$_2$ geometry.

The cubic equation can be solved analytically using Cardano's method \cite{cardano1968ars}. The three roots can be written as
\be\label{threeX}
X_1= C_1 +C_2,\qq
X_2= C_1 j +C_2 j^2,\qq
X_3= C_1 j^2 +C_2 j
\ee
where $j=e^{2i\pi /3}$ and
\be\label{C1C2}
C_1 = {1\/2} (\eta+\sqrt{\eta^2-k^6})^{1/3},\qq
C_2 = {k^2\/2(\eta+\sqrt{\eta^2-k^6})^{1/3}}~.
\ee
The discriminant of the equation vanishes when $\eta=\eta_c$ with
\be
\eta_c=k^3~.
\ee
We will argue that $\eta_c$ is the counterpart of $\la_c$ in SYK, the critical value for the complex-to-real transition.

\begin{figure}
\begin{center}
	\hspace{-0.5cm}\begin{tabular}{cc}
		\subf{\includegraphics[width=7cm]{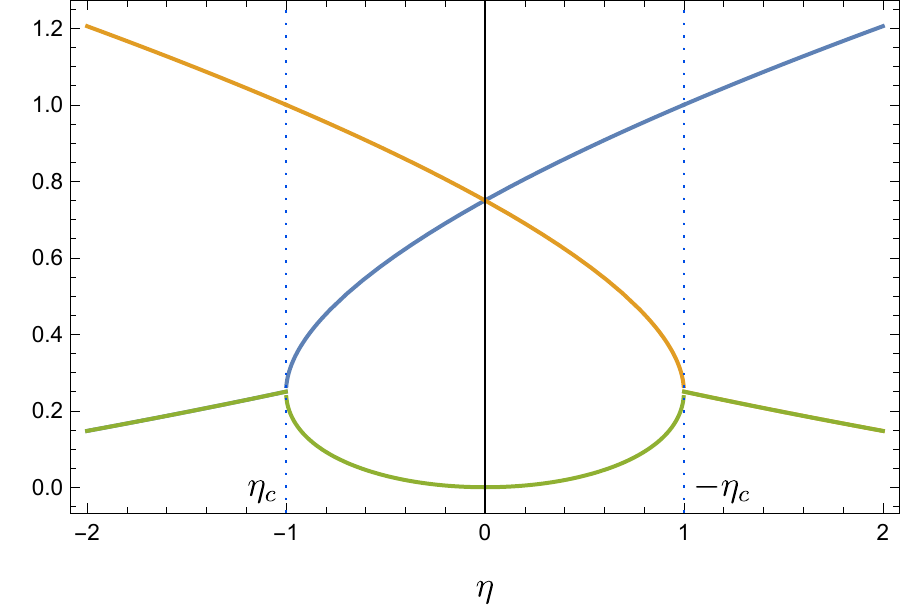}}{a. $\r{Re}\,X^2$} &
		\subf{\includegraphics[width=8.4cm]{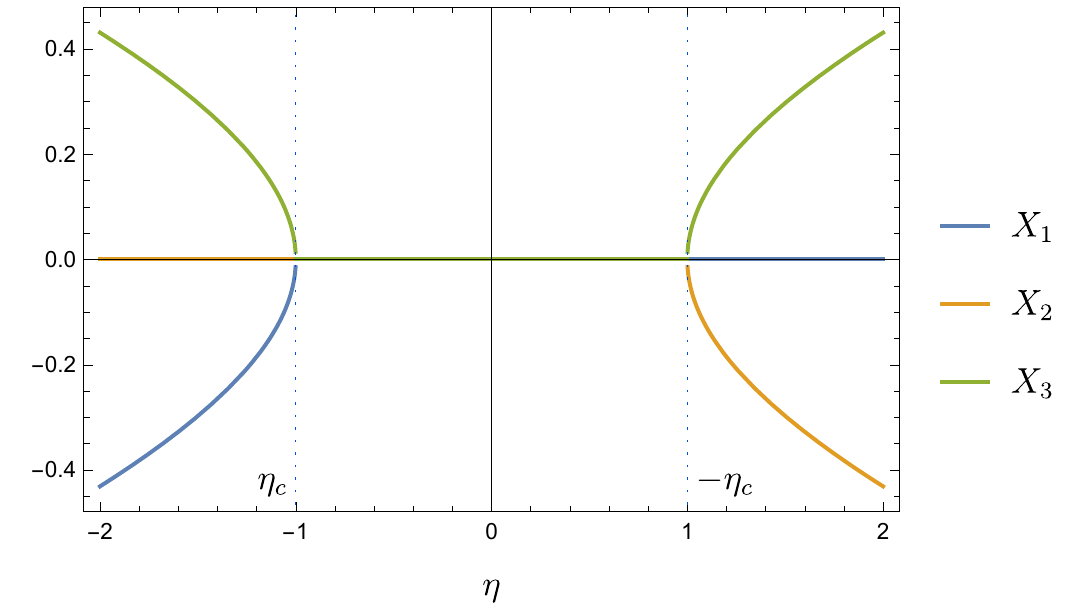}}{b. $\r{Im} \,X^2$}
	\end{tabular}
\end{center}
\caption{Real and imaginary parts of the wormhole size $b = \b X^2$.}\label{FigSize}
	\end{figure}

These three solutions give rise to three wormhole solutions which, for lack of a better terminology, we will refer to as the first, second and third saddle-points. We can see from Fig.~\ref{FigSize} that the size $b$ remains real for $|\eta|<\eta_c$ but can become complex above the transition, even though the dominant solution (in the canonical ensemble) always have real $b$. The appearance of similar complex saddle-points was observed in \cite{garcia2022} and will be related here to the complex-to-real transition.

	\subsubsection{Free energy}

\begin{figure}[H]
\begin{center}
		\hspace{-0.5cm}\begin{tabular}{cc}
		\subf{\includegraphics[width=7cm]{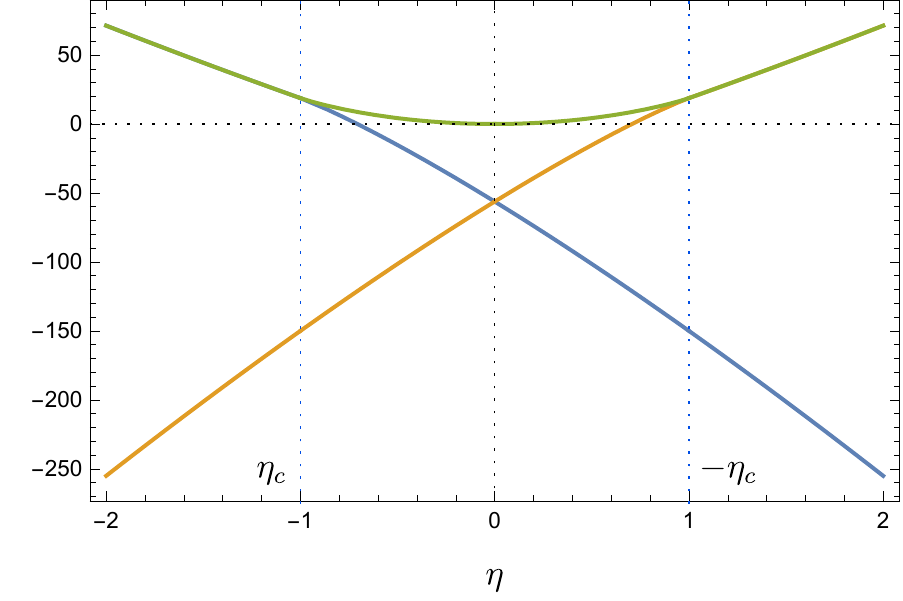}}{a. $\r{Re}\,F$} &
		\subf{\includegraphics[width=8.4cm]{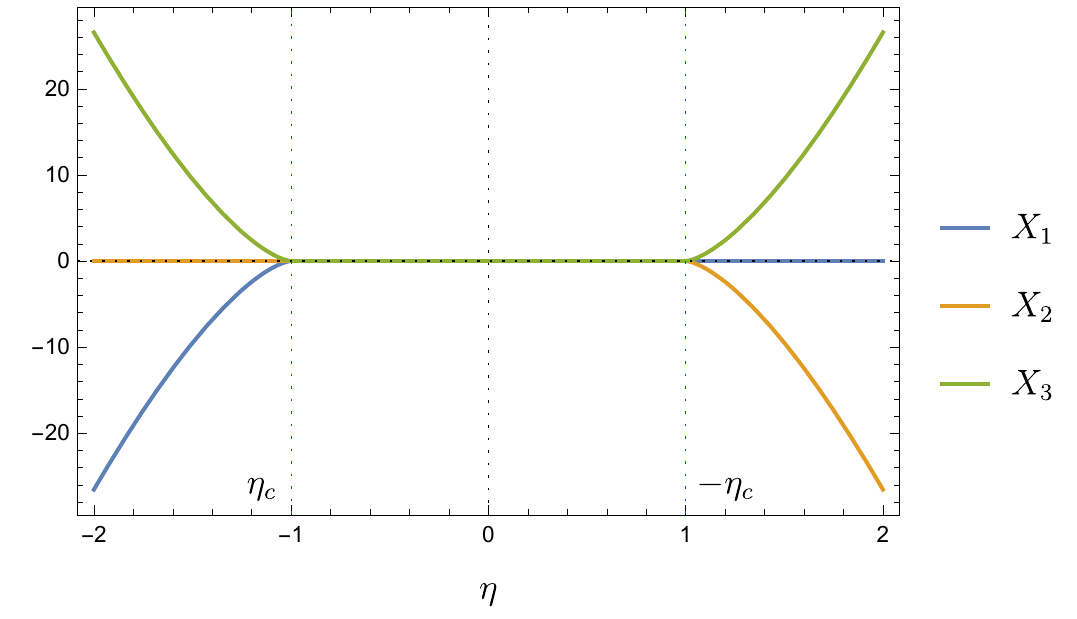}}{b. $\r{Im} \,F$}
	\end{tabular}
\end{center}
\caption{Real and imaginary parts of the free energy for the three wormhole solutions.}\label{Fig:WHfreeenergies}
	\end{figure}

The free energy of the wormhole is
\be
F = T S = N\le(X^4 -{3\/2}k^2 X^2 -\eta X \ri)
\ee
We see that the free energy is independent of the temperature which reflects that the phase is gapped. The real and imaginary parts of the free energies for the three wormhole saddle-points are plotted in Fig.~\ref{Fig:WHfreeenergies}. For $|\eta|>\eta_c$, the free energy of the two subleading wormholes become complex, although the total free energy remains real. This reflects the fact that these subleading wormholes become complex geometries as $b$ acquires an imaginary part.

\begin{figure}
	\begin{center}\begin{tabular}{cc}
\hspace{-0.5cm}		\subf{\includegraphics[width=8cm]{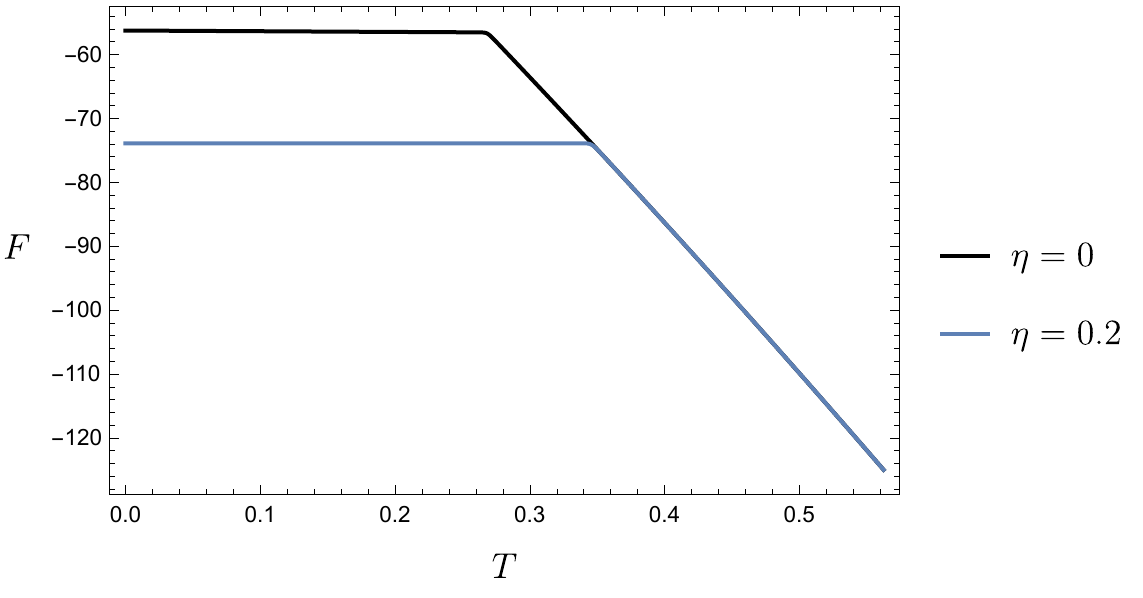}}{} &
		\subf{\includegraphics[width=8cm]{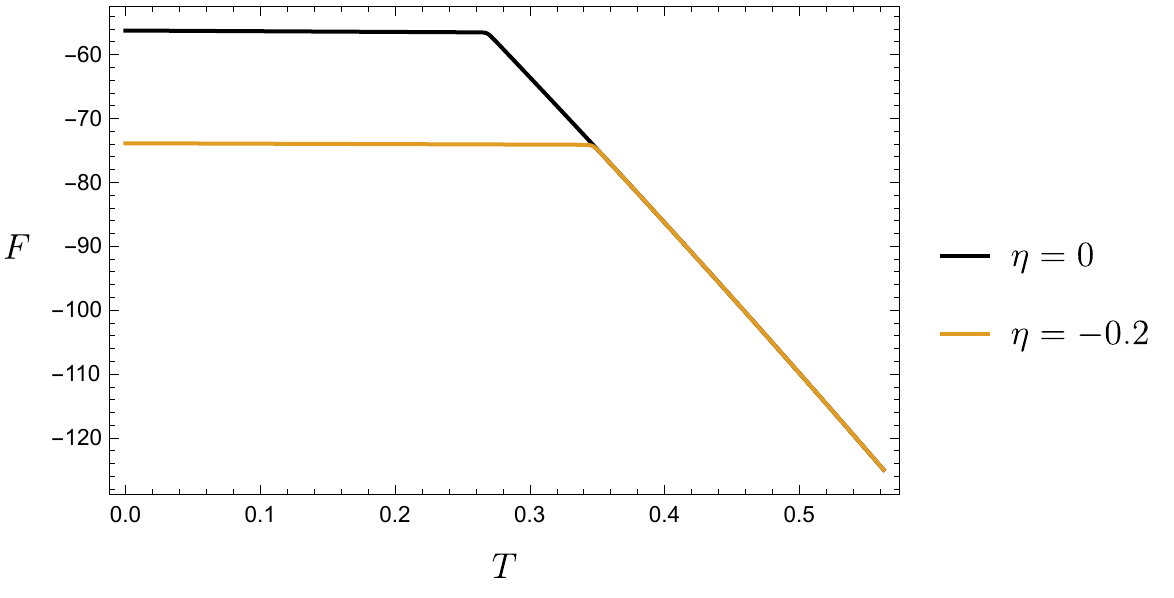}}{}
	\end{tabular}
\end{center}
\caption{Transition between two black holes and the wormhole. For $\eta>0$, the dominant saddle-point is the blue wormhole. Upon changing the sign of $\eta$, the blue and orange wormhole are exchanged so that the system behaves in a symmetric way under $\eta\ra -\eta$. This free energy is in excellent agreement with the SYK result depicted in Fig.~\ref{fig:F_T}.}\label{Fig:transitionF}
	\end{figure}

For $\eta>0$, we use here the dominant solution $X=X_1$. For $\eta<0$, we should use $X=X_2$ as the two saddle-points get exchanged. This mechanism was also observed in SYK and implies that the thermodynamic quantities will be symmetric under $\eta\ra-\eta$ as illustrated in Fig.~\ref{Fig:transitionF}. For this reason, it is enough to  focus on the region $\eta>0$.

\subsubsection{Energy gap}

The energy gap is given, for $\eta>0$, by \cite{maldacena2018}
\be\label{eq:EgX}
E_\r{gap} = \D {b\/\b}= \D X^2_1 ~,
\ee
where we take $\D={1/4}$.

\begin{figure}
\begin{center}
		\hspace{-0.5cm}\begin{tabular}{cc}
		\subf{\includegraphics[width=8.2cm]{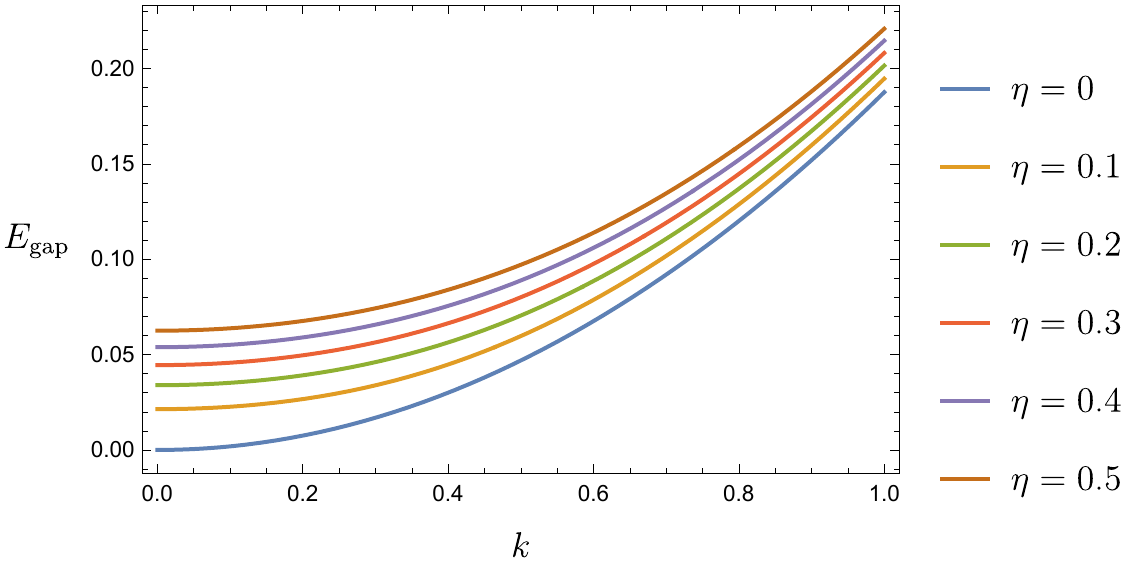}}{a.} &
		\subf{\includegraphics[width=8.1cm]{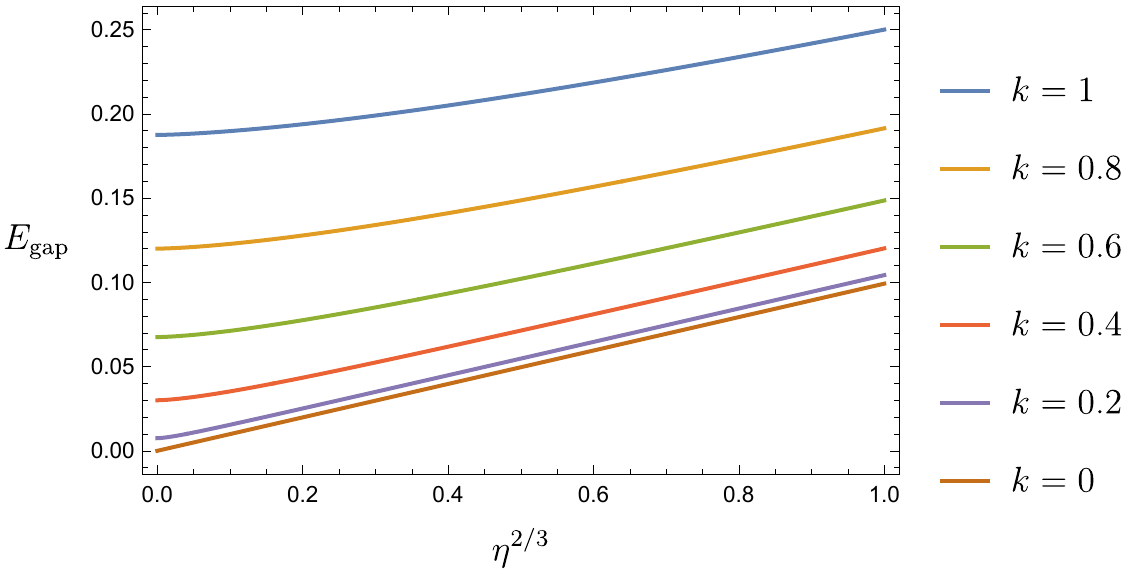}}{b.}
	\end{tabular}
\end{center}\caption{$E_\r{gap}$ as a function of $\eta$ and $k$.} \label{EgapPlot}
\end{figure}

\begin{figure}
	\centering
	\includegraphics[scale=.7]{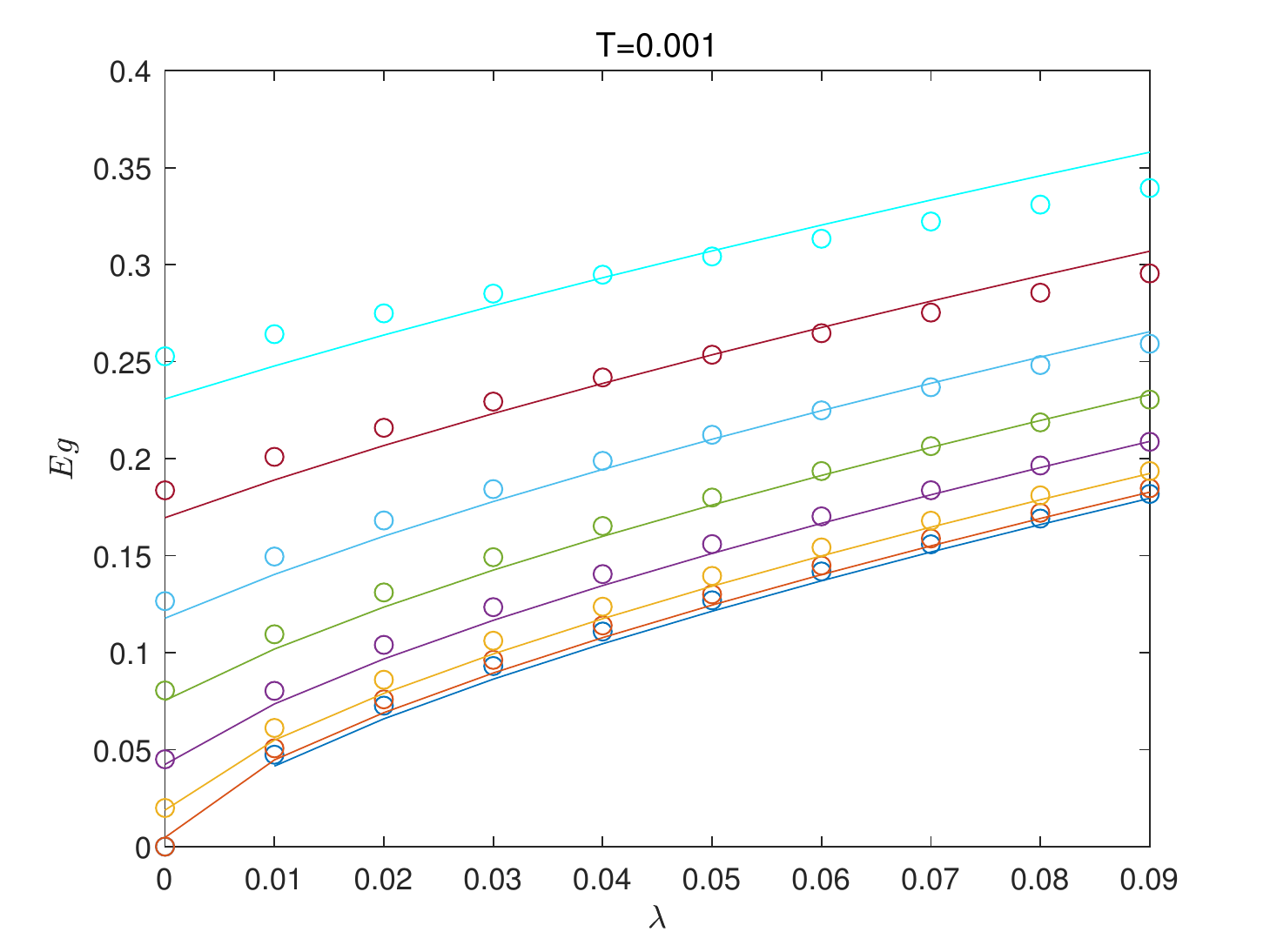}
	\caption{$E_g$ versus $\lambda$. Dots and circles from bottom to top (i.e. blue, red, yellow, ..., azure) correspond to results for $\kappa=0,0.1,\cdots,0.7$. Dots are numerical SYK results and circles correspond with the fitting to the gravity prediction (\ref{eq:gapgravity}), with $k=A\kappa$, $\eta=B\lambda$ and $A=1.58497$ and $B=27.05755$ best fitting parameters in all cases.}\label{fig:Eg_ew_kappa_lambda}
\end{figure}

Explicitly, the energy gap takes the form
\be
E_\r{gap}  ={ \Big(k^2 + \le(\eta+\sqrt{\eta^2-k^6}\ri)^{2/3}\Big)^2 \/16\le(\eta+\sqrt{\eta^2-k^6}\ri)^{2/3}} \label{eq:gapgravity}~.
\ee
The energy gap for various values of $\eta$ and $k$ is plotted in Fig.~\ref{EgapPlot}. It is similar to previous SYK results. A more quantitative comparison by using $k=A\kappa$, $\eta=B\lambda$ between the JT and SYK parameters where the coefficients can be fixed in various ways. In Fig.~\ref{fig:Eg_ew_kappa_lambda}, we compare $E_g$ in gravity (\ref{eq:gapgravity}) with the numerical SYK predictions for $E_g$ using $A, B$ as fitting parameters for $\lambda \in ]0, 0.09]$ and different $\lambda$. For $\kappa$ sufficiently small, we find an excellent agreement between the gravity and the SYK predictions for $A \approx 1.585$ and $B \approx 27.056$. Details of the fitting are given in Appendix~\ref{app:Eg}. The fact that $k = A\k$ should be taken small is a consequence of the scaling regime \eqref{couplingsScaling}. As $\tk$ should be fixed in the gravitational theory, $k=A\k$ scales as $N^{-1/2}$ and must be small in the large $N$ limit. For larger $\k$, we observe deviations in the SYK model, for example due to the fact that the Schwarzian terms are renormalized by a factor $1 - \kappa^2$. This effect is subleading in the large $N$ limit. It should be possible to interpret it as a subleading (\eg one-loop) effect in gravity but in this work we focus on the leading large $N$ behavior.

\subsubsection{Transition to two black holes}
The system with the chosen boundary conditions has another saddle-point corresponding to two black holes with free energy
\be
F_\r{BH} = -2 S_0 T-4\pi^2 T^2~.
\ee
In the canonical ensemble, the dominant solution is the one with the smallest free energy. We observe a phase transition at low temperatures where the two black holes become a wormhole. This transition was already studied in limiting cases in \cite{maldacena2018, garcia2021}.

The critical temperature is the value $T_c$ for which the free energy of the dominant wormhole is equal to the free energy of the two black holes
\be
T_c = -{\r{Re}\,F_\r{WH}\/2S_0}~,
\ee
where we assume $S_0\gg 1$. For $\eta>0$, the dominant saddle-point is $X_1$ in \eqref{threeX}. Explicitly, the critical temperature takes the form
\be \label{eq:Tc_gravity}
T_c= {3N\/8 S_0}  (\eta X_1 +k^2 X_1^2)~.
\ee
For negative $\eta$, the dominant saddle-point is the second one, so $X_1$ should be replaced by $X_2$ in the expression of $T_c$.

\begin{figure}
	\centering
	\includegraphics[width=12cm]{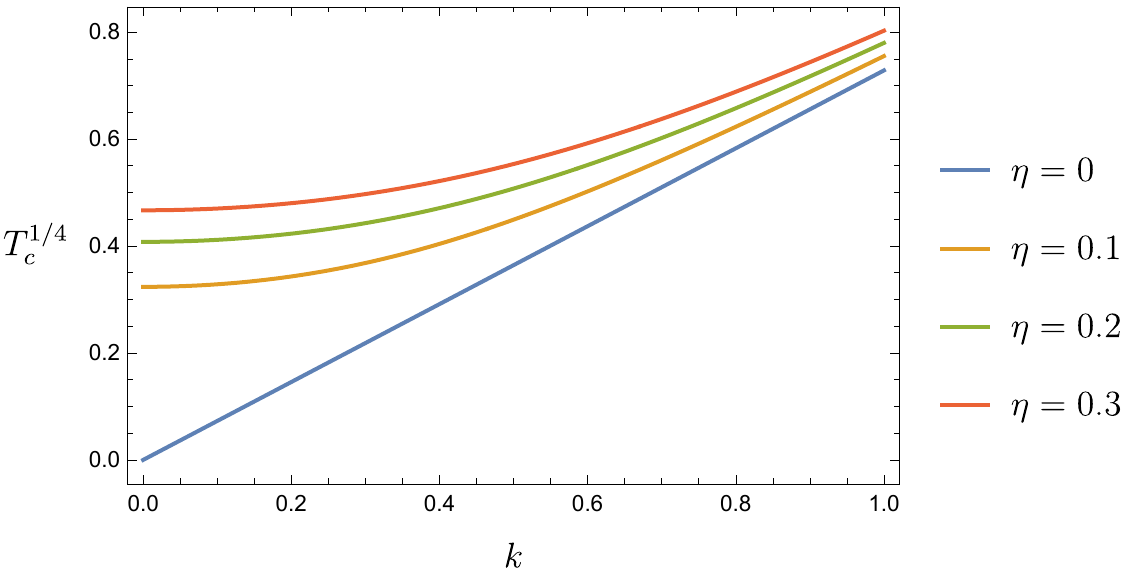}
	\caption{Critical temperature as a function of $k$ for various values of $\eta$.}\label{fig:TcJTa}
\end{figure}

\begin{figure}
	\centering
	\includegraphics[scale=.7]{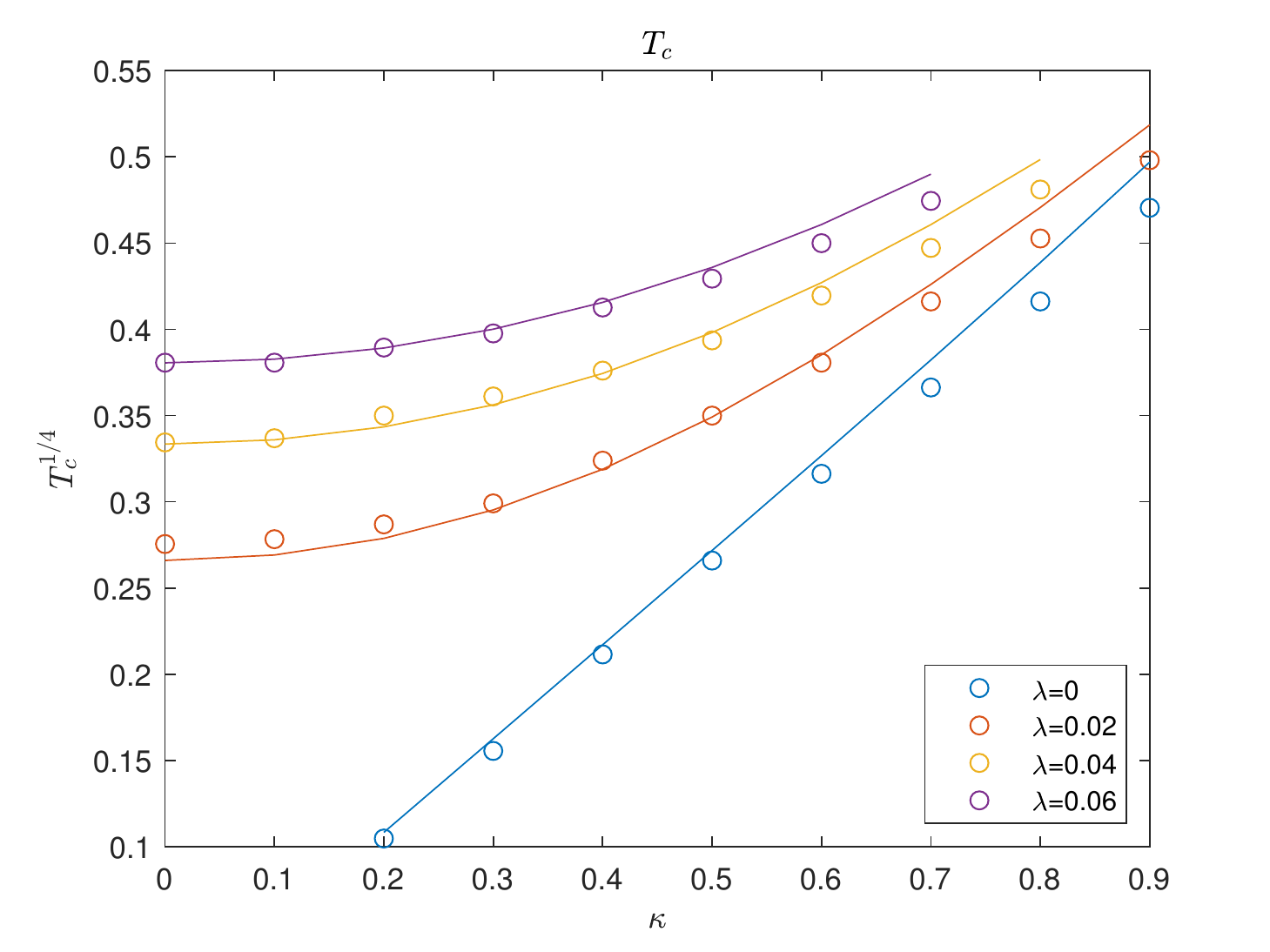}
	\caption{$T_c$ versus $\k$ for different $\la$. Lines are the numerical SYK results and circles represent the gravity prediction (\ref{eq:Tc_gravity}), with $k=A\kappa$, $\eta=B\lambda$ and $A=1.58497$ and $B=27.05755$ obtained from the fitting of $E_g$, see Appendix~\ref{app:Eg}.  }\label{fig:Tc_ew_kappa_lambda}
\end{figure}

Results for $T_c(k)$, for different values of $\eta$ are depicted in Fig.~\ref{fig:TcJTa}. The critical temperature obtained from  SYK in Fig.~\ref{fig:Tc_lambda_kappa} has a similar behavior.
A quantitive comparison is performed in Fig.~\ref{fig:Tc_ew_kappa_lambda}, where we compare the analytic expression (\ref{eq:Tc_gravity}) for $T_c$ with the numerical results from SYK. We plot the expression \eqref{eq:Tc_gravity} using $k = A\k, \eta = B\lambda$, where $A$ and $B$ are the values determined from $E_\r{gap}$, and where the overall factor is fixed  by matching the numerical results at a particular point (in this case $\kappa=0$, $\lambda=0.06$). We also find a good  agreement. A similar agreement is observed if other parameters are chosen to fix the overall factor.

\subsection{Gravitational symmetry breaking}

\subsubsection{Lorentzian phase}
In JT gravity, the $\r{SL}(2,\R)$ symmetry corresponding to the isometries of AdS$_2$ has to viewed as a gauge symmetry. In the path integral formulation, this is because configurations related by an $\r{SL}(2,\R)$ symmetry should be viewed as equivalent \cite{maldacena2016a}. From a Lorentzian point of view, this $\r{SL}(2,\R)$ symmetry is part of the diffeomorphism group. As a result, the associated charges must vanish
\be
Q_0 = 0,\qq Q_+ =0,\qq  Q_- = 0~.
\ee
This is part of the gauge constraints of the theory and can be derived, for example, by integrating the gravitational constraints (or equations of motion) on a Cauchy slice.

Imposing $Q_\pm = 0$ implies that $\tau_L(u)=\tau_R(u)\equiv \tau(u)$ so the left and right boundary modes get identified \cite{maldacena2018}. The last constraint gives
\be
Q_0 = e^{\vphi}N \le( -2\vphi'' + \p_\vphi V\ri) =0
\ee
in terms of the Liouville variable $\vphi(u)= \log\tau'(u)$ and where the potential is
\be
V(\vphi) = e^{2\vphi} - {3\/2}k^2 e^{\vphi}-\eta e^{2\D\vphi}~.
\ee
In this case, the Euclidean action can be written as a Liouville action
\be
S = N \int du \le(\vphi'(u)^2 + V(\vphi(u))\ri)
\ee
so we see that indeed the vanishing of $Q_0$ is equivalent to the equations of motion.

\subsubsection{Euclidean broken phase}

In the Euclidean wormhole, we don't have to impose the constraints $Q_\pm=0$ since these generators are not isometries of the geometry. This is the identification $\tau\sim \tau+b$ breaks the $\r{SL}(2,\R)$ symmetry to $\r{U}(1)$. We should still impose $Q_0=0$ which gives explicitly
\bea
0=- N^{-1} Q_0 \= E[\tau_L] + E[\tau_R] +\eta \D \le({1\/\tau_L'(u)}+{1\/\tau_R'(u)}\ri) \le( {\tau_L'(u)\tau_R'(u)\/ \r{cosh}^2({1\/2} (\tau_L(u)-\tau_R(u))}\ri)^\D + {3k^2\/2}\hspace{1cm}
\eea
where $E$ is the energy functional defined in \eqref{energyFunctional}. We can see that this corresponds to a system of two interacting boundary modes $\tau_L(u)$ and $\tau_R(u)$.

For $\eta=0$, the system has a simpler description in terms of Liouville variables $\vphi = \log\tau'$ given by the equation
\be
-N^{-1}Q_0 = (e^{-\vphi_L(u)}\vphi_L''(u) -e^{\vphi_L(u)})+(e^{-\vphi_R(u)}\vphi_R''(u) -e^{\vphi_R(u)}) + {3k^2\/2},\qq (\eta=0)~.
\ee
which corresponds to two correlated (but non-interacting) Liouville particles.

\subsubsection{Phase shift}
In the broken phase, additional classical solutions are possible because we don't impose $\tau_L=\tau_R$. In particular, we can have a shift by an arbitrary constant $\a$
\be\label{tauEqShift}
\tau_L(u) = {b\/\b} u +\a,\qq
\tau_R(u) = {b\/\b} u
\ee
This corresponds to a global $\r{U}(1)$ symmetry, which can be interpreted as relative shifts between the two sides. This symmetry is present in the Euclidean wormhole at $\eta=0$ but is explicitly broken by the inter-site coupling.

In the path integral, $\a$ enters as another moduli on which we should integrate. After evaluating the action on the classical solutions \eqref{tauEqShift}, we should perform the path integral over $b$ and $\a$. We can consider doing first the integral over $\a$. This is
\be
\int_\R d\a\, \r{exp}\le( - \b N \,\eta \le({b\/\b\,\r{cosh}(\a/2)}\ri)^{2\D}+\dots\ri)
\ee
where we highlighted the $\a$-dependence. This integral has a saddle-point at $\a=0$ and this will be the dominant solution in the Euclidean path integral. As a result, the effect of non-zero $\a$ is subleading in the thermodynamics and cannot be easily measured in Euclidean signature.

Rather, we will see that the variable $\a$ can be measured in the Wick-rotated two-point functions. We first define
\be
G_{LL}(u_1,u_2) = \ln O_L(u_1) O_L(u_2)\rn,\qq G_{LR}(u_1,u_2) = \ln \cO_L(u_1) \cO_R(u_2)\rn~,
\ee
and the classical solution \eqref{tauEqShift} gives the contribution
\bea
G_{LL}(u_1,u_2) \= \le( {\tau_L'(u_1) \tau_L'(u_2)\/\r{sinh}^2 ({1\/2}(\tau_L(u_1)-\tau_L(u_2) )} \ri)^\D= \le( {b^2 \/\b^2\,\r{sinh}^2 ({b\/2\b}(u_1-u_2 )} \ri)^\D~,\\
G_{LR}(u_1,u_2) \= \le( {\tau_L'(u_1) \tau_R'(u_2)\/\r{cosh}^2 ({1\/2}(\tau_L(u_1)-\tau_R(u_2) )} \ri)^\D = \le( {b^2 \/\b^2 \,\r{cosh}^2 ({b\/2\b}(u_1 -u_2)  +{\a\/2} ))} \ri)^\D~.
\eea
The effect of $\a$ can be studied by continuing to real time. This is a well-defined procedure, even though we don't expect the theory to always have a Lorentzian interpretation. The Wick rotation can be achieved by using $u_1 - u_2 = i v$ and we obtain
\be
G_{LL}(v)  = { \sqrt{2\w}\/ \r{sin}^2 \le( \w v - i\e \ri) },\qq G_{LR}(v)  = { \sqrt{2\w}\/ \r{cos}^2 \le( \w v + {1\/2}\tilde\a- i\e \ri) }
\ee
where we used $\D={1\/4}$ and defined $\a= i \tilde\a $. This shows that $\a$ measures the phase-shift between $G_\r{LL}$ and $G_\r{LR}$ in real time. Here, the frequency of the oscillations is given by
\be
\w = {b\/2\b } = 2 E_\r{gap}~.
\ee
The Wick rotation also acts on $\a$ since it is defined as the difference of two times. In other words, the real time Green's functions should be computed by integrating over the contour $\a\in i\R$. For $\tilde\a=0$, the Green's function are in-phase while for $\tilde\a=\pi$ they are out-of-phase.

 In fact, since the integrand is periodic, we should only integrate on the circle $\a\sim\a + 2i\pi$. In terms of the variable $\tilde{\a} = - i \a$, the integral becomes
\be\label{factorPI}
G_{LR}(v) = {1\/Z}\int_0^{2\pi} d\tilde\a\le( e^{-S_1} G_{LR}^{(1)}+e^{-S_2} G_{LR}^{(2)}+e^{-S_3} G_{LR}^{(3)}\ri)
\ee
where the three saddle-points in the $b$-integral corresponding to the three roots \eqref{threeX}.

Note that for $\eta=0$, the three saddle-points reduce to a single one. In this case, the different choices of $\a$ are exactly degenerate which corresponds to a global $\r{U}(1)_\r{axial}$ symmetry. To explain this, note that translations on $\tau_L$ and $\tau_R$ give a $\r{U}(1)\times \r{U}(1)$ symmetry. The inter-site interaction only preserves the diagonal $\r{U}(1)_\r{diag}$ symmetry (which correspond to the $\r{U}(1)$ isometry of the Euclidean wormhole) and explicitly breaks the $\r{U}(1)_\r{axial}$ symmetry (which is the one that shifts $\a$). In the path integral, this soft breaking corresponds to the fact that the inter-site interaction gives a different action for different values of $\a$.

For $\eta>0$, the dominant saddle-point is the first saddle corresponding to $X_1$. The effect of $\a$ can be accounted by replacing $\eta$ with  $\eta/\sqrt{|\r{cos}(\tilde\a/2)|}$ starting from the $\tilde\a=0$ configuration. We can see that the action is strongly localized around $\tilde\a=\pi$ as we have
\be
e^{-S_1} \sim \r{exp}\le( {3 \b N \eta^{4/3}\/4 |\tilde\a-\pi|^{2/3}}\ri)\qq (\eta>0)~.
\ee
The divergence at $\tilde\a=\pi$ implies that we cannot use a saddle-point approximation here. Rather, the factor $e^{-S_1}$ inserts a delta function in the path integral which localizes it on the value $\tilde\a = \pi$. This value implies that $G_\r{LL}$ and $G_\r{LR}$ should be in-phase.    For $\eta<0$, the first and second saddle-point are exchanged, it is then $S_2$ that dominates and inserts a similar delta function in the path integral.

This localization mechanism explains that the Green's functions are in-phase for small $\eta$. For $\eta>\eta_c$, the restoration of  $\r{SL}(2,\R)$ symmetry described below imposes $\tau_L=\tau_R$. As a result, we must have $\tilde\a=0$ so that $G_\r{LL}$ and $G_\r{LR}$ must be out-of-phase.

\begin{figure}
	\begin{center}\begin{tabular}{cc}
\hspace{-0.5cm}		\subf{\includegraphics[width=7cm]{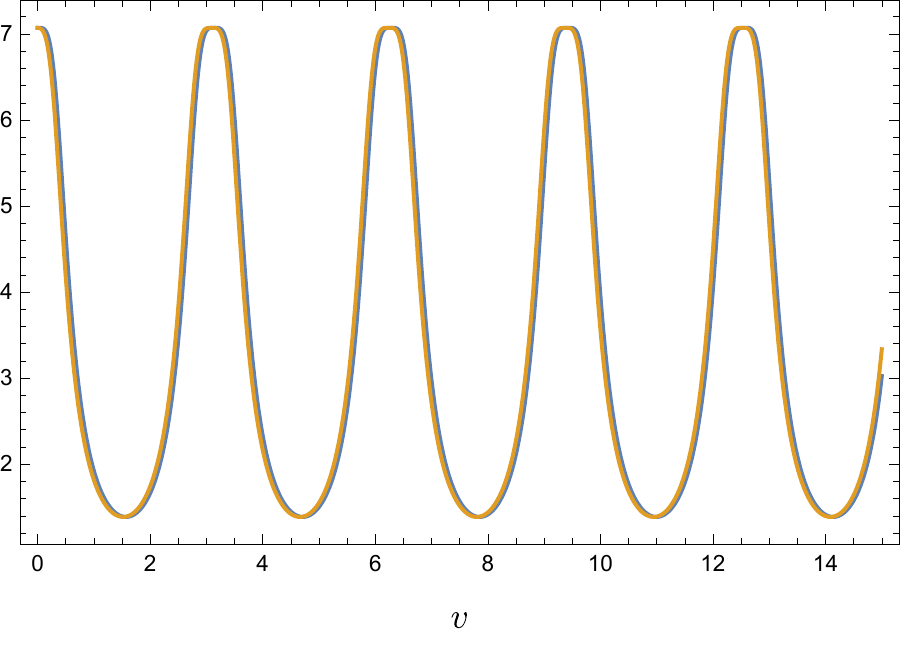}}{a. $\tilde\a=0 $} &
		\subf{\includegraphics[width=8.8cm]{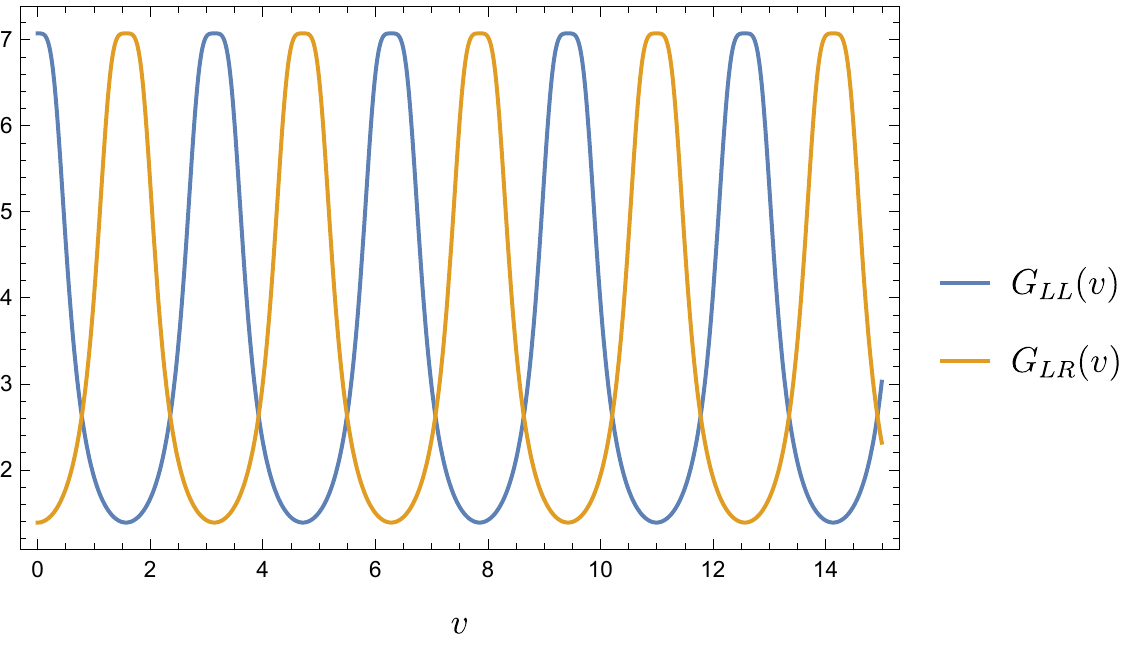}}{\hspace{-2cm}b. $\tilde\a=\pi$}
	\end{tabular}
\end{center}
\caption{Dephasing of $G_{LL}$ and $G_{LR}$ under the $\r{U}(1)\ra \r{SL}(2,\R)$ symmetry restoration. The plots are performed by using an $i\e$ regularization for the light-cone singularities.}\label{fig:GLRJT}
\end{figure}

 These two regimes for the Green's functions are plotted in Fig.~\ref{fig:GLRJT}.  The dephasing pattern matches precisely what is observed in the SYK model, see Figs.~\ref{fig:Gr_rholl_ka_p5_beta_1e4_lx_5e6_eta_2e-4_dm_25}, \ref{fig:Gr_rholl_Eg_T_ka_p4_ld_p12}, \ref{fig:Gr_rholl_ld_p12_ka_p4}.

\subsubsection{Order parameter}

The deformation by boundary sources $\pm i\tk$ corresponds to adding to the Euclidean action
\be
\d S = i \cO_{LR}~,
\ee
where we have defined the operator
\be
\cO_{LR} \equiv  \tk\int_{S^1}d\tau \,(\cO_L(\tau)-\cO_R(\tau))~.
\ee
Writing $Z$ as a path integral shows that its expectation value can be obtained by taking a derivative with respect to $k$:
\be
\ln \cO_{LR}\rn_\b =i\tk  {\p\/\p \tk}  \log Z_\b=i k  {\p\/\p k}  \log Z_\b~.
\ee
where we used that $k = \sqrt{2\/\pi N} \tk$.

We will see that the expectation value of this operator is an order parameter for the $\r{SL}(2,\R)$ symmetry breaking.  To study the $\r{SL}(2,\R)$ transformation, we formally continue to Lorentzian signature using $\tau = i t$ and use that under an infinitesimal $\r{SL}(2,\R)$ transformation
\bea
\d t_L \= \ve_0 +\ve_+ e^{\tau}+\ve_- e^{-\tau},\qq \d t_R = \ve_0 -\ve_+ e^{\tau}-\ve_- e^{-\tau}~.
\eea
This leads to
\be
\d \cO_{LR}= i\int d t \,(\ve_+e^{it}+\ve_- e^{-it})(\cO_L'(t)+ \cO_R'(t))~,
\ee
so this operator is invariant under $\r{U}(1)$ but transforms non-trivially under the other generators of $\r{SL}(2,\R)$.

The path integral computation of $\ln \cO_{LR}\rn$ requires a suitable choice of contour for the integral over $b$. The general procedure to selects the contour is not well understood, see for example \cite{Halliwell:1989dy,Bousso:1998na, Sorkin:2009ka, Witten:2021nzp}. In principle, the contour could be determined by the appropriate analytic continuation from Lorentzian signature if the system has a Lorentzian interpretation. In the theory defined by the Euclidean path integral, Picard-Lefschetz theory could be used to understand the correct contour prescription. Here, we will give a natural contour prescription that reproduces the SYK results, and leave a better understanding of the choice of contour for future work.

\begin{figure}
	\begin{center}\begin{tabular}{cc}
\hspace{-0.5cm}		\subf{\includegraphics[width=7.6cm]{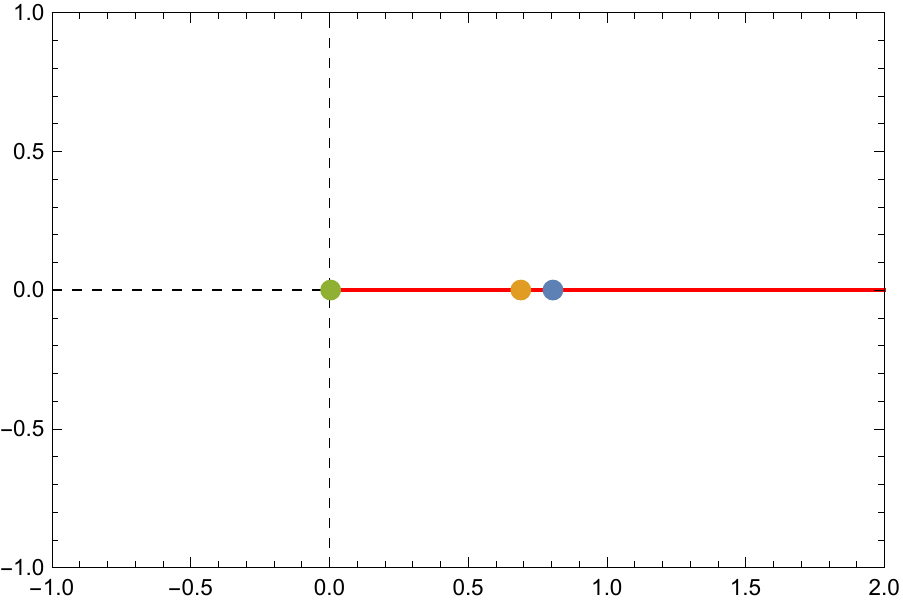}}{a. $\eta<\eta_c $} &
		\subf{\includegraphics[width=7.6cm]{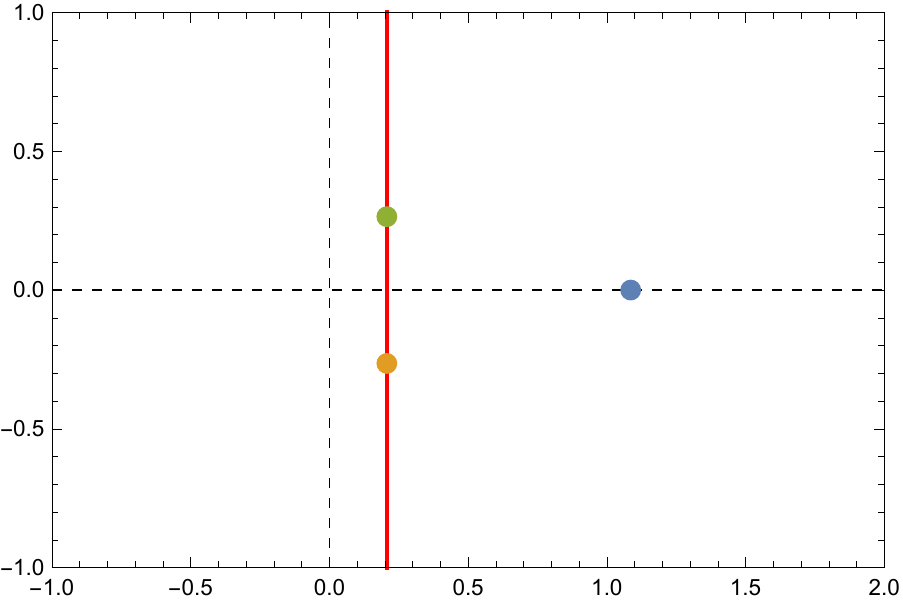}}{b. $\eta>\eta_c$}
	\end{tabular}
\end{center}
\caption{Contour of integration in the complex $b$-plane to compute $\ln O_{LR}\rn$. The dots represent the three saddle-points which are controlled by the cubic equation \eqref{cubicEq}. For $\eta>\eta_c$, two of the saddle-points acquire an imaginary part and become complex conjugate. The restoration of $\r{SL}(2,\R)$ symmetry can be explained by the transition to a vertical integration contour on which the saddle-point with lowest free energy stops contributing. }\label{Fig:bcontours}
\end{figure}

For $|\eta|<\eta_c$, we integrate over $b\in [0,+\infty)$ which is the contour in Fig.~\ref{Fig:bcontours}a. This gives a non-zero expectation value
\be
\ln \cO_{LR} \rn = 3 i N k^2 b_\ast ,\qq |\eta|<\eta_c
\ee
where $b_\ast$ is the size of the dominant wormhole. This non-zero expectation value spontaneously breaks the $\r{SL}(2,\R)$ symmetry to $\r{U}(1)$.

 For $\eta>\eta_c$, we propose that we should choose the vertical contour in Fig.~\ref{Fig:bcontours}b. This leads to
\be
\ln \cO_{LR} \rn  = 0~,\qq \eta>\eta_c~,
\ee
and the $\r{SL}(2,\R)$ symmetry is restored. For $\eta>\eta_c$, two-saddle points leave the real line, and we choose the contour that connects the two complex saddle points. The important point here is that the new contour should not include the real saddle-point. This ensures that  $\ln \cO_{LR} \rn =0$ in agreement with the SYK result.

\begin{figure}
	\centering
	\hspace{-2cm}\includegraphics[width=10cm]{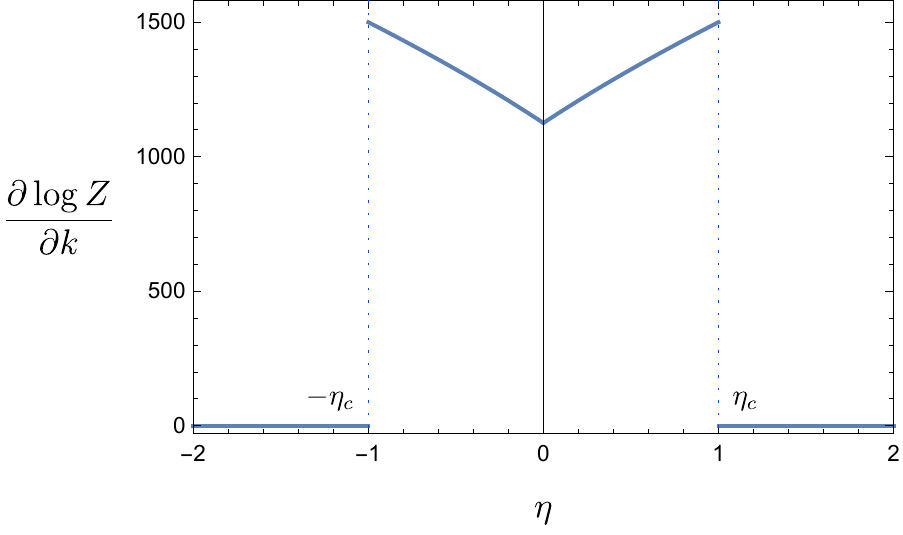}
	\caption{Expectation value $\ln O_{LR}\rn$ as a function of $\eta$. For $|\eta|>\eta_c = k^3$, we observe a transition where $\ln O_{LR}\rn=0$ and the $\r{SL}(2,\R)$ symmetry is restored. This is in good agreement with the SYK result Fig.~\ref{fig:orderpOLR}. Note that the additional peak in the latter is due to spin symmetry which is a non-universal featured related to the chosen minimal left-right coupling,}\label{Fig:orderparam}
\end{figure}

The result is depicted in Fig.~\ref{Fig:orderparam}. We see that at $\eta>\eta_c$, the order parameter vanishes. The transition also happens in the negative $\eta$ region, with the first (blue) and second (orange) saddle-points exchanged. We can compare this to Fig.~\ref{fig:orderpOLR} in SYK and we see a good qualitative match. This shows that the change of contour appears to be the right gravity mechanism to account for the complex-to-real transition observed in SYK. Note that the different choices of contour should reflect the ambiguity in the derivative of the partition function with respect to $k$ due to the branch cuts appearing in \eqref{C1C2}.

Note that there is a larger value of $\eta_\r{max} = 3\sqrt{3\/2} k^3$ above which the classical solution for the small wormholes cannot be trusted. This is the value at which the complex conjugated wormholes have $\r{Re}\,b = 0$. When $\r{Re}\,b \ll \b$,  one-loop effect cannot be ignored and will modify the analysis. Such one-loop effects were studied in \cite{garcia2021} for $\eta=0$. In this paper, we stay in the classical limit which is sufficient for the comparison with SYK results in the large $N$ limit.

This mechanism explains the surprising  observation in SYK that the Hamiltonian becomes real for $|\eta|>|\eta_c|$. Indeed, we have from \eqref{ImHexpct},
\be
\ln \cO_{LR}\rn_\b = \b \ln \r{Im}\,H\rn_\b
\ee
which shows that this is, up to the factor $\b$, the same order parameter that was previously identified in SYK. The higher moments of $\r{Im}\,H$ can be obtained by taking more derivatives with respect to $k$. The same mechanism suggests that these higher moments vanish above the transition, and the vanishing of all the moments implies that the energy spectrum must be real. This shows that the preservation of $\r{SL}(2,\R)$ symmetry implies that the energy spectrum must be real.

To summarize, we have proposed a mechanism for the restoration of $\r{SL}(2,\R)$ gauge symmetry in terms of a change of integration contour. We have shown that the restoration of $\r{SL}(2,\R)$ symmetry is equivalent to the energy spectrum becoming real by identifying an order parameter for the transition. At the moment, we cannot fully justify the change of contour in gravity, as the precise rules governing the path integral are not well understood.  We leave a better understanding of this mechanism for future work. In terms of the variable $X = \sqrt{b/\b}$, the path integral takes the form
\be
Z= \int dX \,\r{exp}\le( -\b N\le(X^4 -{3\/2}k^2 X^2 -\eta X \ri)\ri)~.
\ee
The Picard-Lefschetz theory of similar integrals was considered in \cite{Witten:2010zr} and one might hope that this could shed some light on this transition.

\subsection{Euclidean-to-Lorentzian transition}

In this section, we explain in what sense the complex-to-real transition can be understood as a Euclidean-to-Lorentzian transition.

The Lorentzian wormhole (eternal traversable wormhole) corresponds to the global AdS$_2$ geometry. The  $\r{SL}(2,\R)$ isometry group of the background translates into an $\r{SL}(2,\R)$ gauge constraint $Q_\xi = 0$ for any Killing vector $\xi$. In general, we can write
\be
Q_\xi = \int_\S T_\mn \xi^\mu {\ve}^\nu
\ee
where $\S$ is a Cauchy slice with volume form $\ve^\nu$. Here $T_\mn$ contains a gravity and matter part
\be
T_\mn = T_\mn^\r{grav} +T_\mn^\r{matter}
\ee
obtained by varying the action with respect to the metric. In JT gravity, $T_\mn^\r{grav} $ is the stress-tensor of the JT dilaton (it is proportional to the Einstein tensor in higher dimensional gravity). The equations of motion imply that $T_\mn=0$. This implies that the $\r{SL}(2,\R)$ charges have to vanish:
\be
Q_0 = 0,\qq Q_+ =0,\qq Q_-=0~.
\ee
The Euclidean wormhole is obtained by doing the Wick rotation from global AdS$_2$ and identifying periodically the time coordinate. This identification breaks the $\r{SL}(2,\R)$ symmetry to $\r{U}(1)$ so we should only impose the vanishing of $\r{U}(1)$ charge:
\be
Q_0 = 0~.
\ee
 In Euclidean signature, we must view this as a constraint on the configurations entering in the path integral. We see that the purely Euclidean system, where we only impose $Q_0=0$, has more configurations than the Lorentzian system. In the Schwarzian theory, this corresponds to configurations with $\tau_L\neq \tau_R$ while we must have $\tau_L=\tau_R$ in Lorentzian signature.

The system we study is viewed as a purely Euclidean system dual to a non-Hermitian Hamiltonian. The Euclidean-to-Lorentzian transition is the restoration of $\r{SL}(2,\R)$ symmetry of the Lorentzian geometry. Using the order parameter, we have seen that this implies that the energy spectrum is real.  Conversely, a real spectrum implies the existence of a Lorentzian continuation with unitary evolution, so the $\r{SL}(2,\R)$ symmetry must be restored. This shows that the complex-to-real transition in SYK corresponds to a Euclidean-to-Lorentzian transition in JT gravity.

The $\r{SL}(2,\R)$ symmetry discussed above is a  gauge symmetry in JT gravity that is part of the gravitational constraints of the Lorentzian wormhole. There is also a global $\r{SL}(2,\R)$ symmetry which corresponds to the isometry group of AdS$_2$ acting on the matter sector, appropriately dressed to commute with the $\r{SL}(2,\R)$ gauge symmetry. This global symmetry exists because the isometries are large diffeomorphisms at the asymptotic boundaries, see \cite{Lin:2019qwu, Harlow:2021dfp}. In the Euclidean wormhole, we similarly have a $\r{U}(1)$ gauge symmetry accompanied by a $\r{U}(1)$ global symmetry. The $\r{SL}(2,\R)\ra\r{U}(1)$ breaking/restoration involves both the gauge and global symmetry. These symmetries are gravitational in origin become they come from the isometries of the semi-classical background.

Although the system can always be viewed as a Euclidean wormhole, the restoration of $\r{SL}(2,\R)$ symmetry implies that above the transition, it can be continued to Lorentzian signature and is dual to an eternal traversable wormhole.\footnote{Note that for negative $\eta$, the gravity dual is really the time-reversed of the eternal traversable wormhole.}
The Hamiltonian is non-Hermitian but its eigenvalues are real so it can be used to define unitary evolution with a suitably modified inner product \cite{Mostafazadeh:2008pw}. In other words, the eigenstates of the Hamiltonian are not orthogonal with respect to the usual scalar product but they become orthogonal with a new inner product. Intriguingly, this indicates that pseudo-Hermitian Hamiltonians can appear in holography.

A final comment is that this Euclidean-to-Lorentzian transition is only interesting in the wormhole regime. For $T>T_c$, the system is dual to two black holes. In this case, the Euclidean and Lorentzian symmetries are both equal to $\r{U}(1)\times \r{U}(1)$ so there can be no symmetry breaking/restoration.

 \section{Level statistics and quantum chaos}
   We end the paper by investigating the nature of the quantum dynamics for long time scales of the order of the Heisenberg time.

  The real level statistics in the $\kappa = 0$ limit, was addressed in Ref.~\cite{garcia2019}. Sufficiently far from the wormhole ground state, the long time dynamics is quantum chaotic as level statistics agrees well with the random matrix prediction \cite{bohigas1984}. More specifically, it agrees with the Gaussian Orthogonal Ensemble which corresponds to systems with time reversal invariance. Unlike single-site SYK models whose global symmetries depend in general on the number of Majoranas, for an SYK model with two identical sites, time reversal invariance is always present. However, the fact its low energy excitations deviate strongly from this universal result suggests that, the low temperature phase transition between the wormhole and the two black holes  is accompanied by a qualitative change in the quantum dynamics.

  Level statistics in the limit $\lambda = 0$ has also been investigated recently \cite{garcia2021d}. The level statistics of the combined system is trivially Poisson because both SYK are not explicitly correlated. However, the spectral correlations of each SYK separately agrees well with the predictions of non-Hermitian random matrix theory. This SYK model, depending on the number of Majoranas $N$ and the $q-$body interacting Hamiltonian, reproduces many of the different universality classes predicted \cite{ueda2019} in non-Hermitian quantum chaotic systems.

  We now study the combined effect of a finite $\lambda$ and $\kappa$ in the spectral correlations.
 More specifically, we aim to clarify whether a weak explicit coupling $\lambda \ll 1$ is enough to make the dynamics of the combined non-Hermitian system quantum chaotic, at least for sufficiently high energies. Therefore the level statistics will be well described by random matrix theory. This is important as a further confirmation that even in a non-Hermitian setting, quantum black holes are related to quantum chaotic motion \cite{maldacena2015}. Likewise, we would like to explore whether the observed deviations from random matrix theory that characterize the quantum motion in the real case are also present in our model. It is also of interest to investigate whether in the region $\lambda >\lambda_c$, where the spectrum is real, a finite $\kappa$ is of any relevance in the description of the level statistics.

\begin{figure}
	\centering
	\subfigure[]{\includegraphics[width=8cm]{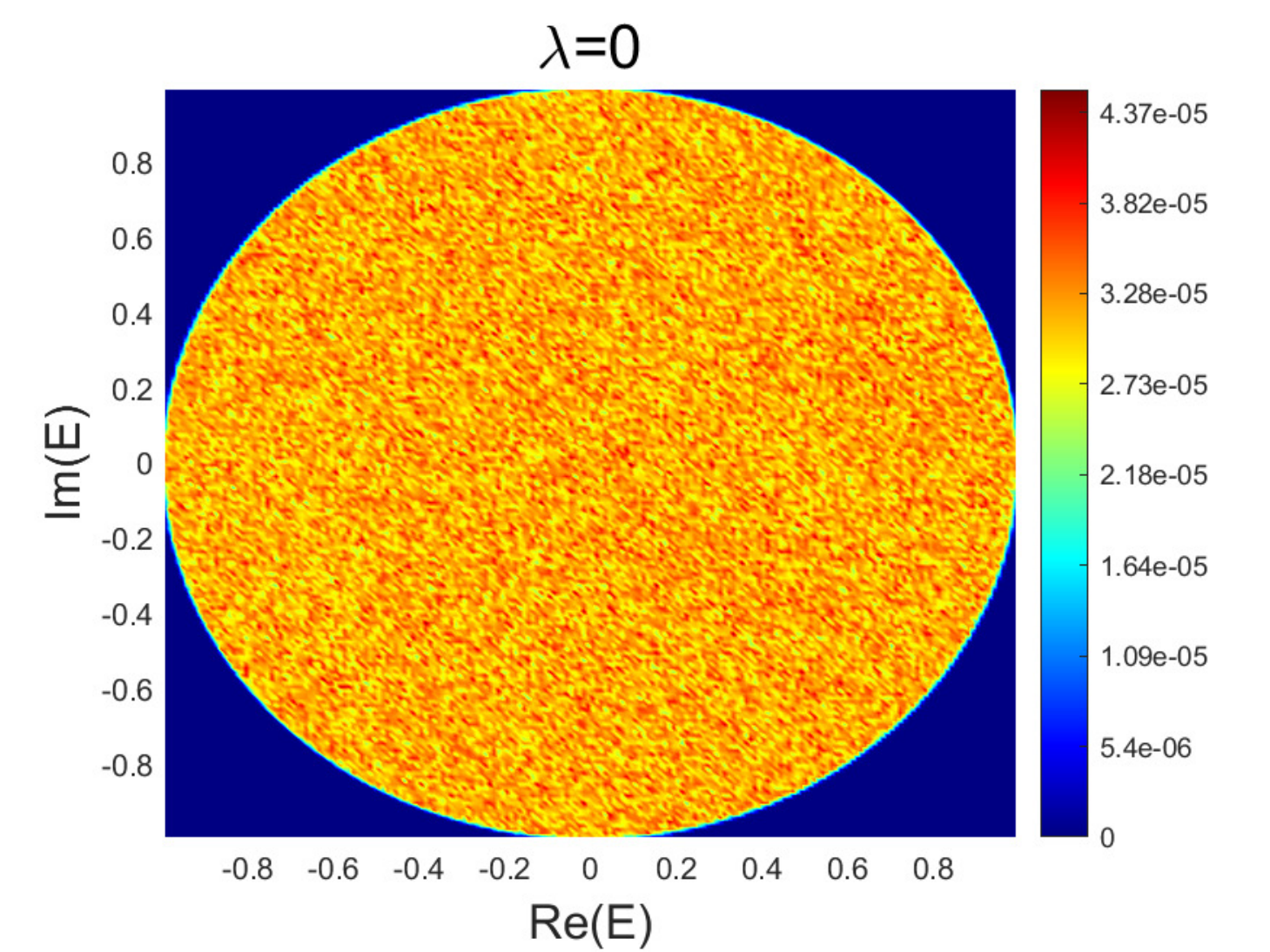}}\\
	\subfigure[]{\includegraphics[width=13cm]{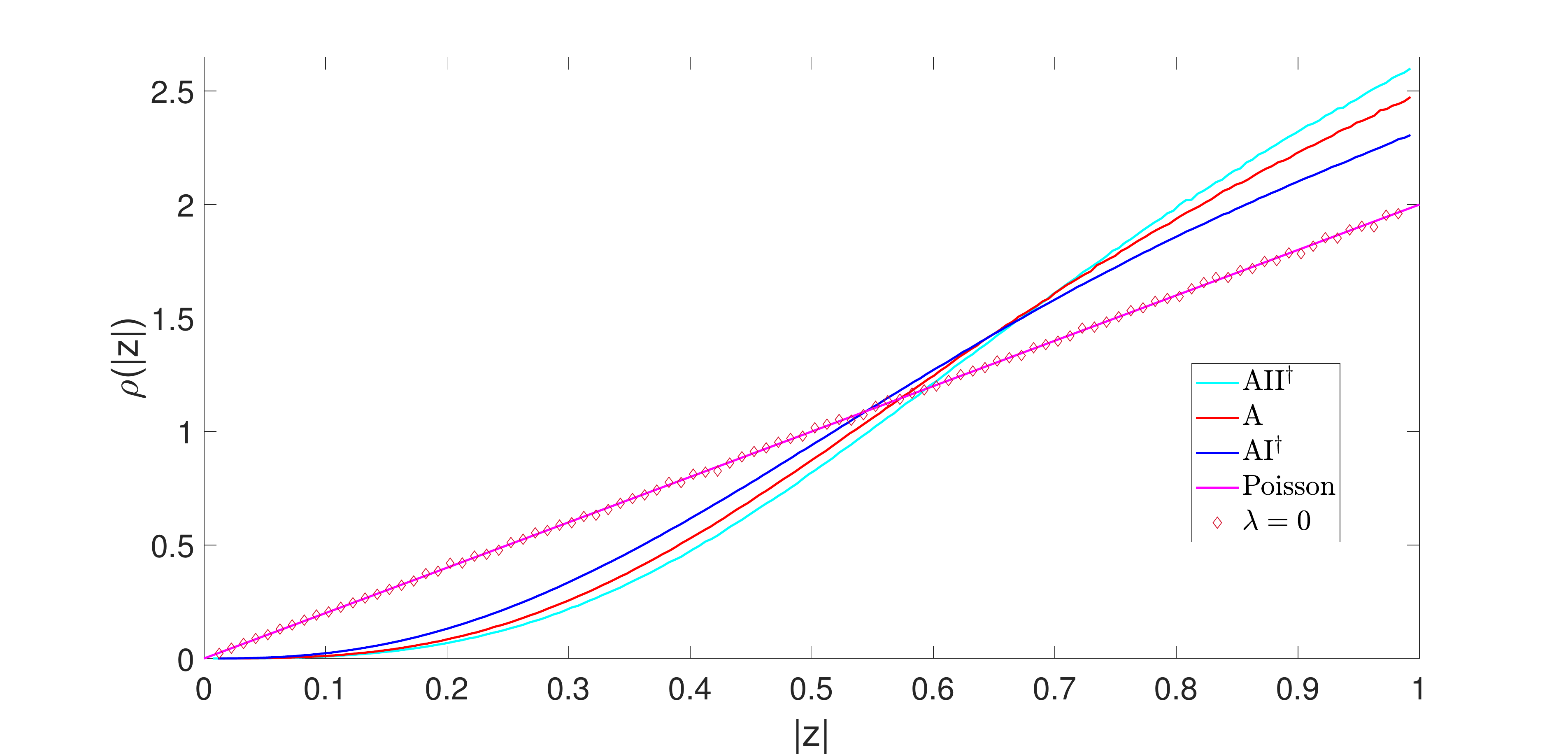}}\\
	\subfigure[]{\includegraphics[width=13cm]{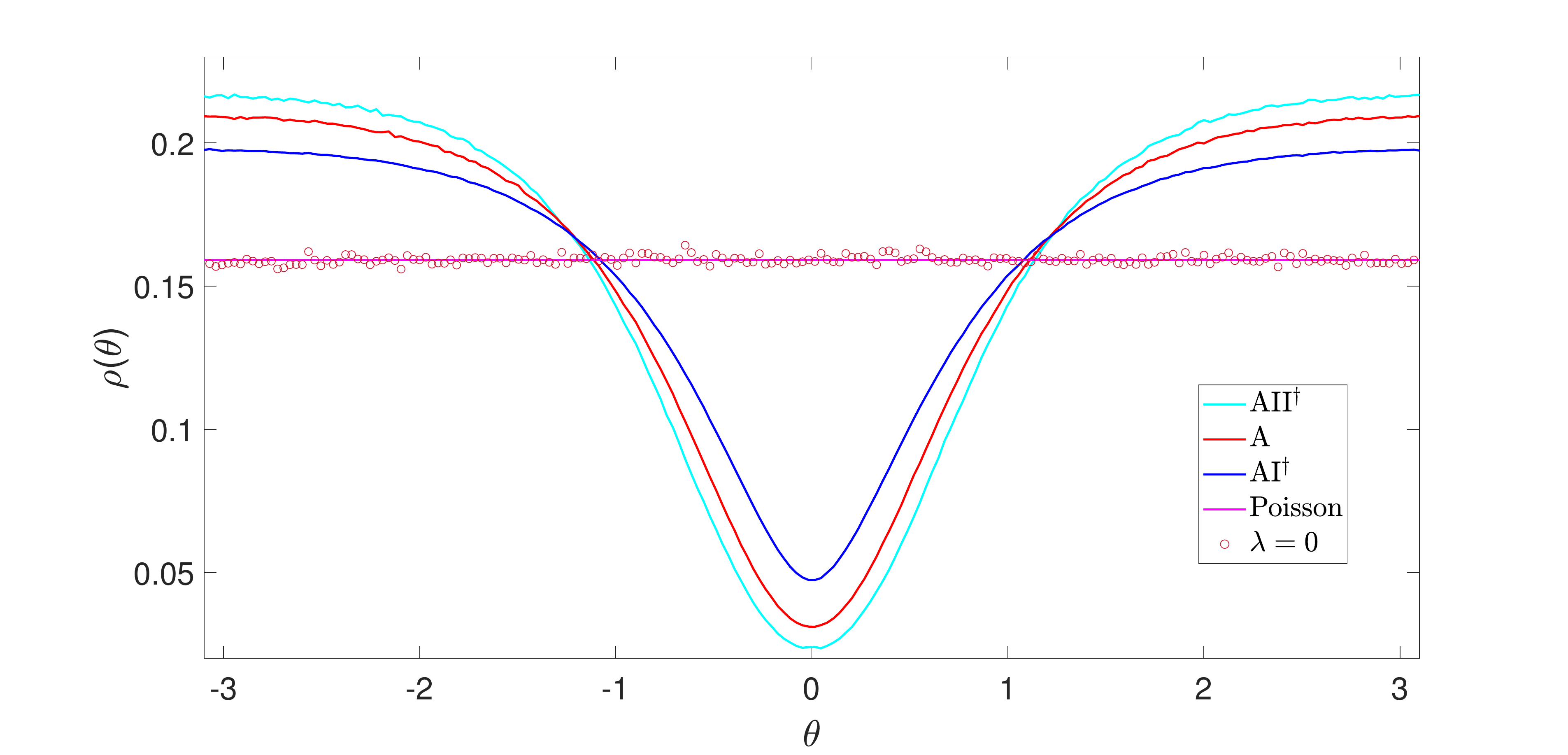}}
	\caption{Complex gap ratio for $N=16$, $\lambda=0$ and $\kappa=1$.  (a): full complex gap ratio distribution. The half eaten doughnut typical of random matrix correlations in not observed; (b): marginal radial distribution; (c): marginal angular distribution. After reorganizing the Hamiltonian into four blocks due to parity symmetry and taking out the complex-conjugate degeneracy for complex spectrum, the level statistics agrees well with Poisson level statistics. We choose a relatively large $N$ ($N=16$) to suppress finite size effects.  Note that level statistics for $\lambda = 0$ is qualitatively different from that at small $\lambda$ because the Hamiltonian for $\lambda = 0$ is the tensor product of two decoupled Hamiltonians.}\label{fig:ratiopoi}
\end{figure}

In order to avoid the spin-symmetry mentioned previously, we will perform the similarity transformation mentioned above and diagonalize numerically each parity block separately. We employ the complex spacing ratio as a spectral observable\cite{sa2020}, that does not require the unfolding of the spectrum.
This is especially important for a two-dimensional spectra where the unfolding procedure suffers in some cases from ambiguities. The complex spacing ratio is a short-range spectral observable that probes the quantum dynamics for times scales longer than the Heisenberg time, which is originally introduced to study correlations of real spectra~\cite{oganesyan2007,atas2016,brody1981}. In the complex case it is	defined as
\begin{equation}\label{eq:cspacing}
	z_k=\frac{E_k^\mathrm{NN}-E_k}{E_k^\mathrm{NNN}-E_k}.
\end{equation}
where $E_k$ is the complex spectrum for a given disorder realization, $E_k^\mathrm{NN}$ is the nearest eigenvalue to $E_k$ and $E_k^\mathrm{NNN}$ is the next to nearest eigenvalue to $E_k$. In order to eliminate statistical fluctuations, we carried out ensemble-averaging until we obtain at least $10^6$ eigenvalues for each choice of parameters $N, \lambda, \kappa$.
\begin{figure}
	\centering
	\includegraphics[scale=0.55]{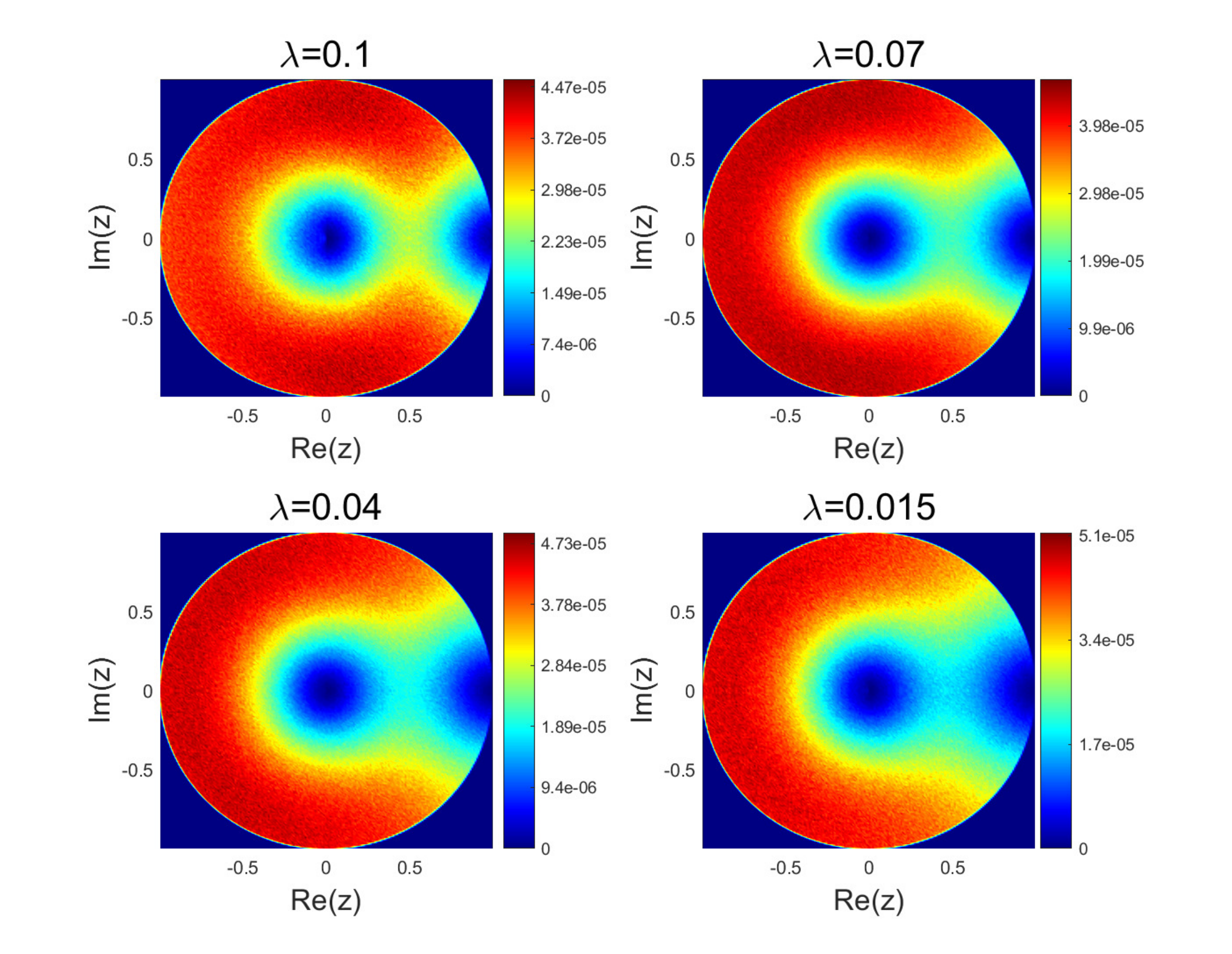}
	\caption{Complex gap ratios for different values of $\lambda$, with $\kappa = 1$ and $N = 12$. We do observe the half-eaten doughnut shape typical of random matrix behavior \cite{sa2020}. However, sizable deviations are observed as $\lambda$ increases because part of the spectrum becomes real and the complex spacing ratio is no longer applicable.}\label{fig:ratiodis_l}
\end{figure}
\begin{figure}
	\centering
	\includegraphics[scale=0.55]{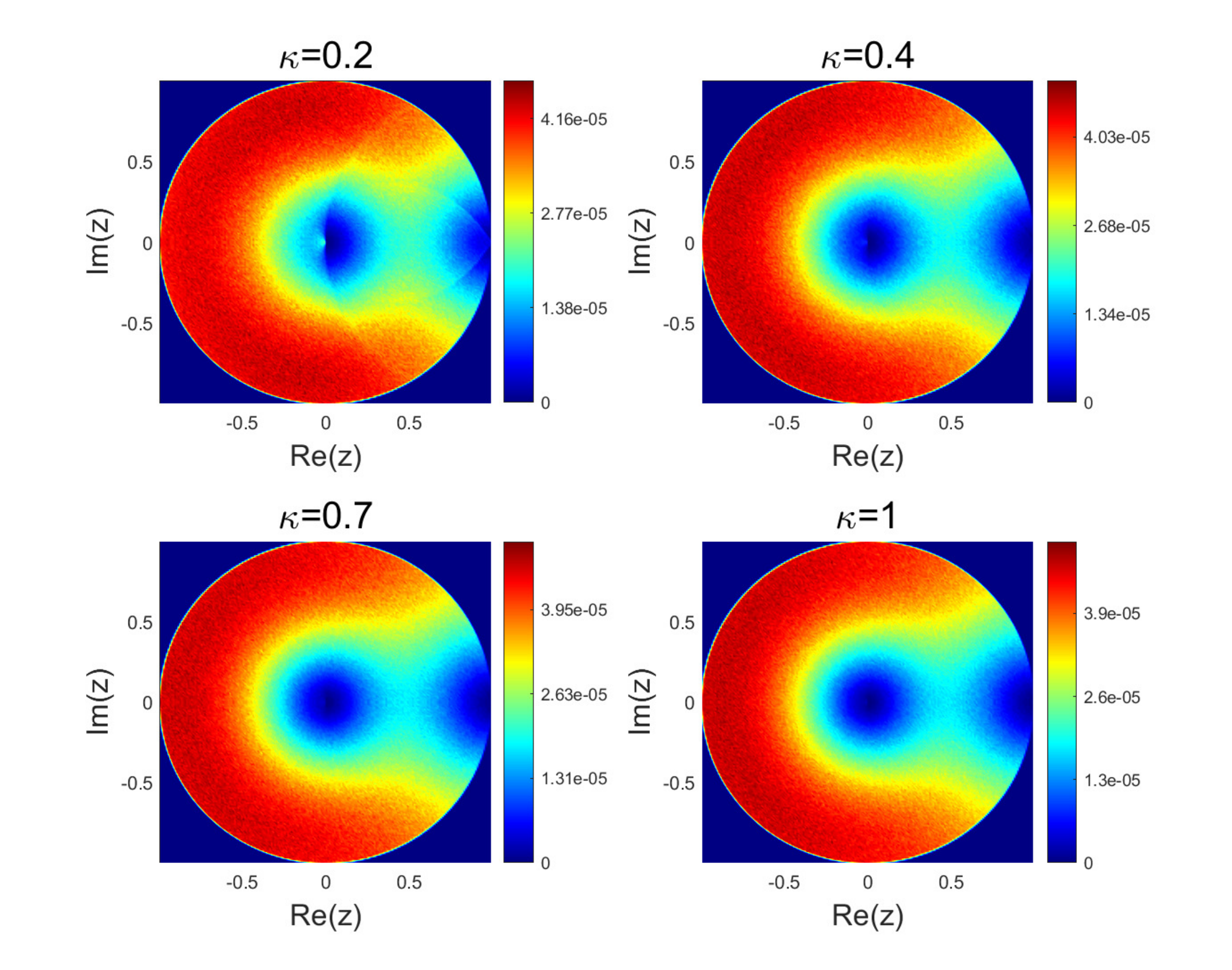}
	\caption{Complex gap ratios for different values of $\kappa$ with $\lambda= 0.02$ and $N = 12$. It fits with RMT well only for $\kappa$ not too small ($\kappa \geq 0.4$), since there exists a transition from non-RMT to near-RMT when $\kappa$ is small.}\label{fig:ratiodis_k}
\end{figure}
In Fig.~\ref{fig:ratiodis_l} and Fig.~\ref{fig:ratiopoi}(a), we depict the distribution of the complex spacing ratio $z_k$ for $\kappa = 1$ and different values of $\lambda$. For larger values of $\lambda$, it is necessary to use a better ensemble averaging to obtain similar results, since from Fig.~\ref{fig:null_kl}, the real eigenvalues become less with $\lambda$ larger. As was expected, for $\lambda =0$, we do not see any trace of revel repulsion for small distances. However, even for very small values of $\lambda =0.015$, the characteristic \cite{sa2020} half-eaten donuts shape that indicates level repulsion and potential quantum chaotic behavior is clearly visible.
\footnote{For $\lambda =0$, the symmetry operator is $E\otimes \gamma_c + a \gamma_c\otimes E$, E is the identity matrix, a is an arbitrary factor not equal to $\pm 1$. This operator will give rise to the four-block structure of the Hamiltonian.}

In Fig.~\ref{fig:ratiodis_k}, similar results are obtained for different values of $N$ and $\kappa > 0.2$. There exists obvious deviations from RMT for sufficiently small $\kappa$. This is expected because the RMT results assume a spectral density more or less locally symmetric in the complex plane. However, in the $\kappa \to 0$ limit, the spectral correlations are greatly enhanced along the real line so they  cannot be described by complex gap ratio.
As a result, it is necessary to use a minimum $\kappa_\r{min}$ so that for $\kappa > \kappa_\r{min}$, the complex part of the eigenvalues is much larger than the mean level spacing and non-Hermitian RMT results apply.

The distribution of the spacings does not allow a quantitative comparison with random matrix predictions. For that purpose, it is more convenient to use the angular $\rho(\theta)$ and radial $\rho(r)$ distributions \cite{sa2020,garcia2021d} of the complex spacing ratios
($\theta$ and $r$ corresponding to the angular and radial variable in polar coordinates).

\begin{figure}
	\centering
	\subfigure[]{\includegraphics[width=13cm]{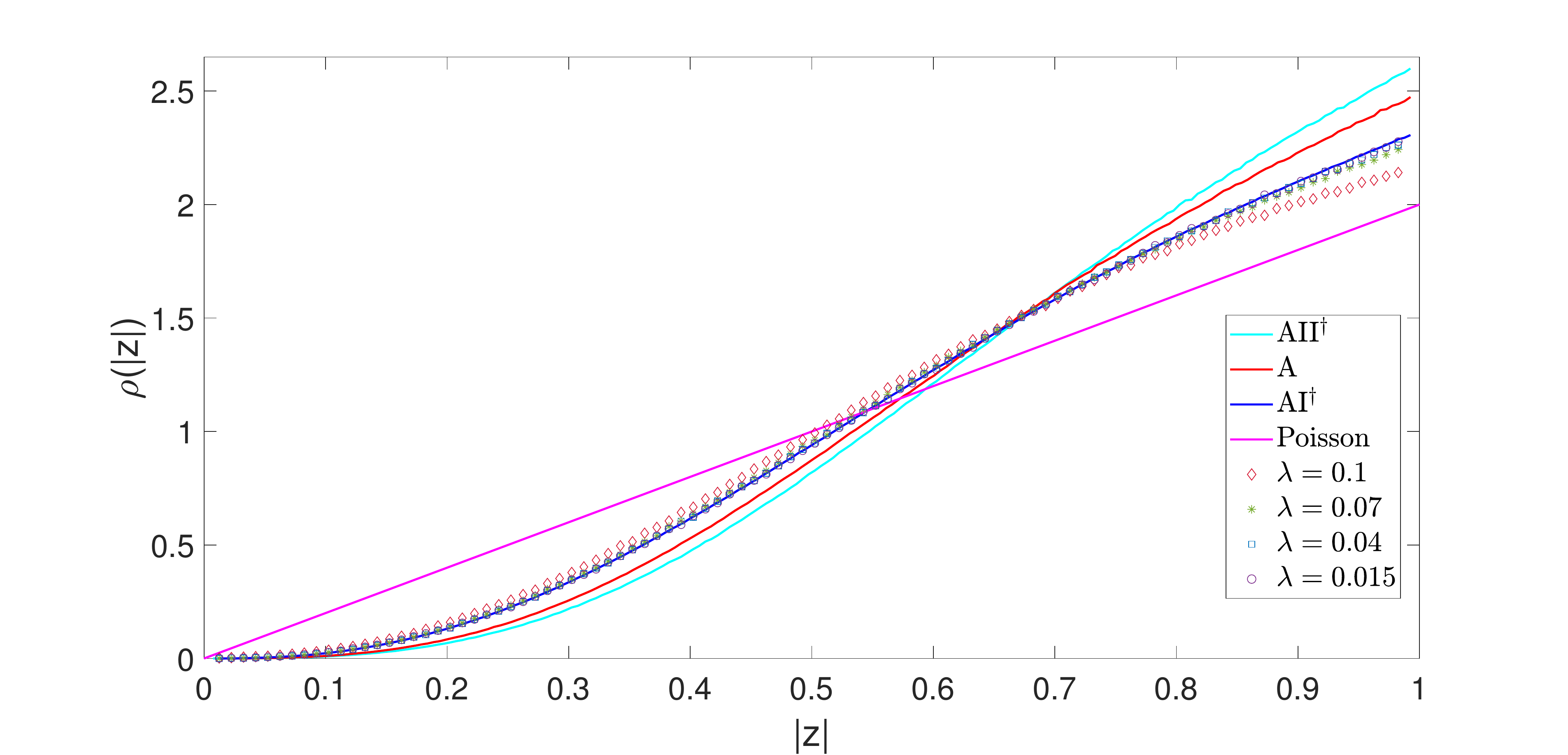}}\\
	\subfigure[]{\includegraphics[width=13cm]{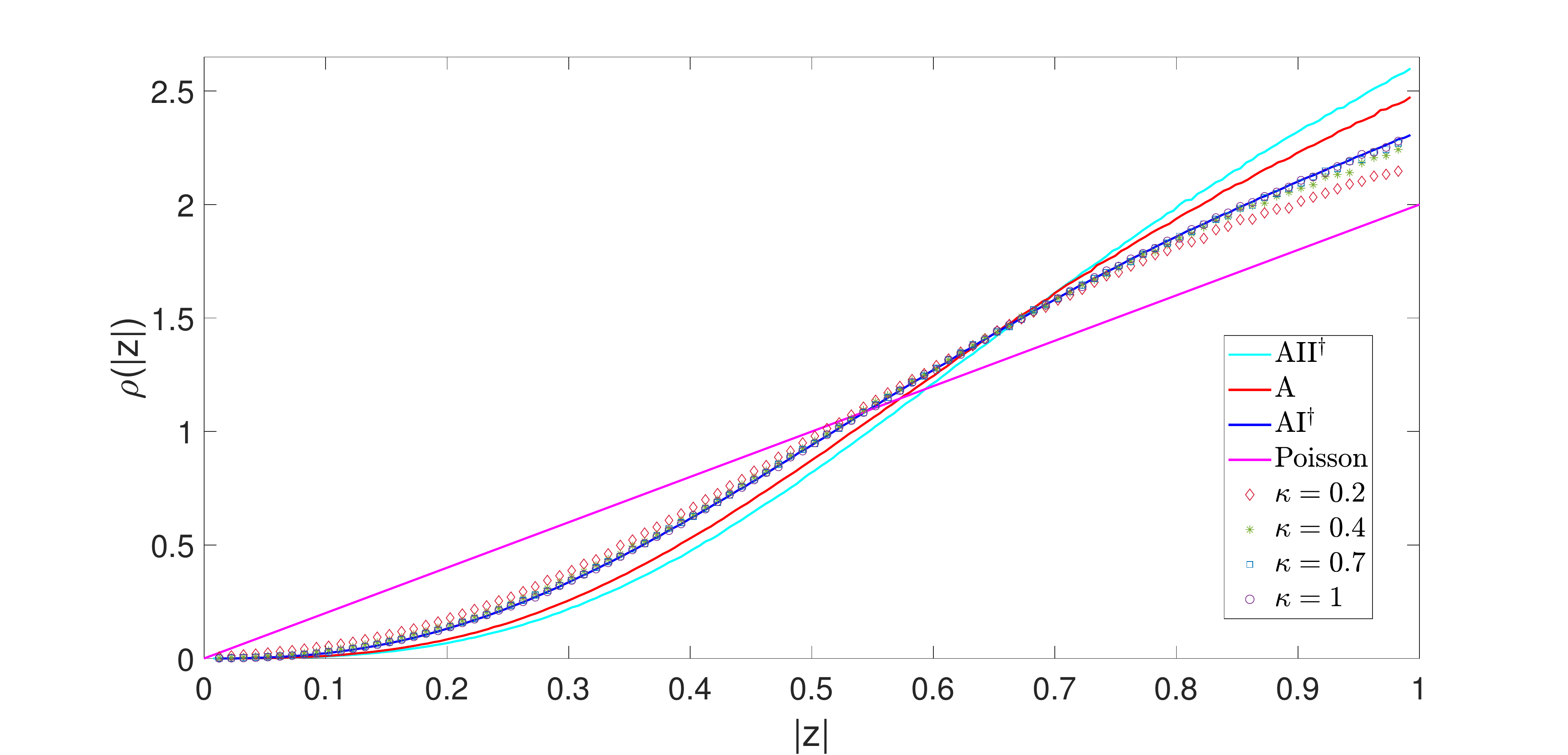}}
	\caption{Marginal radial distribution of the complex gap ratio for $N=12$. (a) $\kappa = 1$ and different values of $\lambda$; (b) $\lambda = 0.02$ and different values of $\kappa$. We compare the non-Hermitian RMT results for universality classes $A$, $AI^{\dagger}$ and $AII^{\dagger}$ \cite{ueda2019,garcia2021d} corresponding to examples with Orthogonal, Unitary and Symplectic symmetry respectively. The matrix size is $2048\times 2048$ in all cases, the same size as the two-site Sachdev-Ye-Kitaev model for $N=12$. For $\lambda=0.015$, we find excellent agreement with RMT for both radial and angular distributions.}\label{fig:ratiorho}
\end{figure}

\begin{figure}
	\centering
	\subfigure[]{\includegraphics[width=13cm]{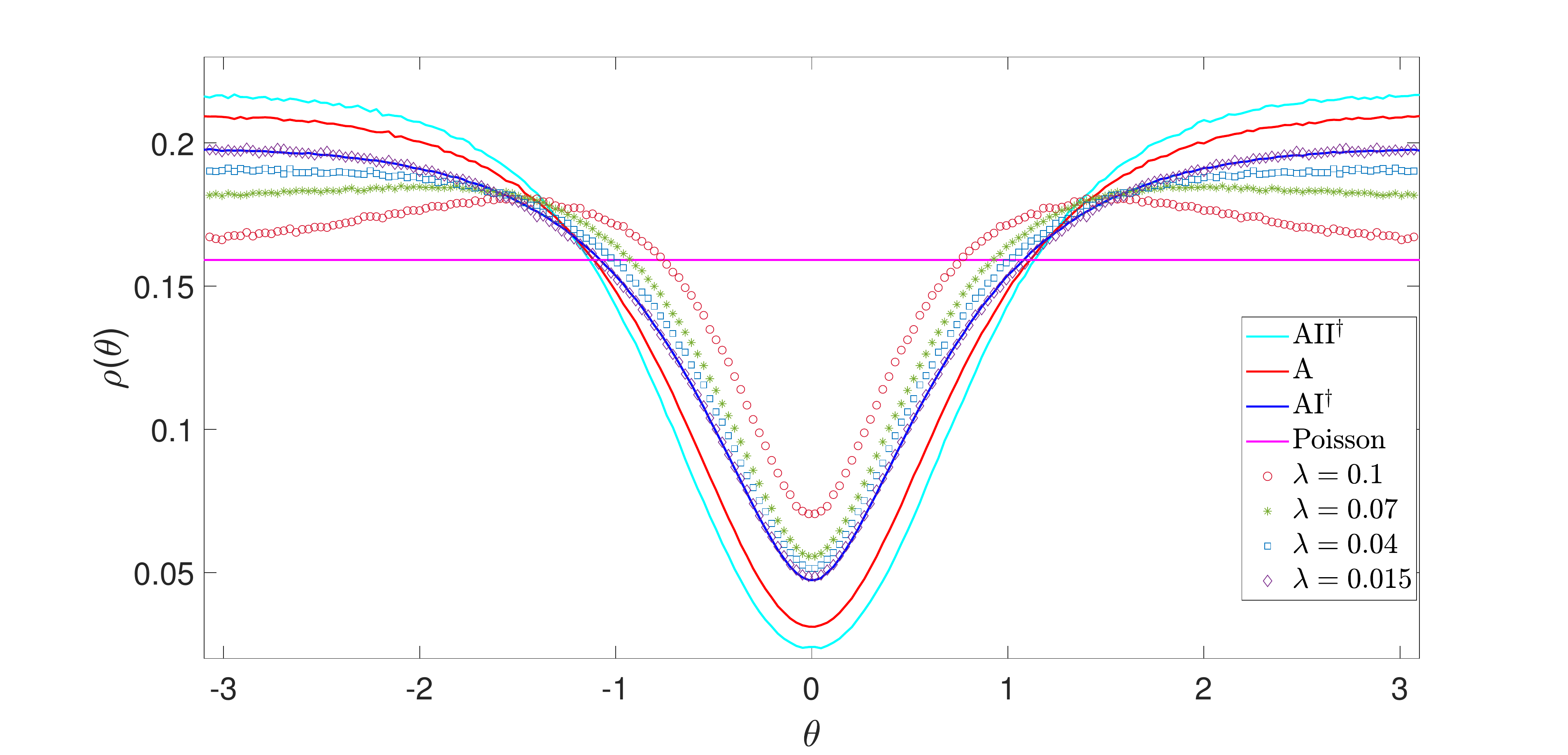}}\\
	\subfigure[]{\includegraphics[width=13cm]{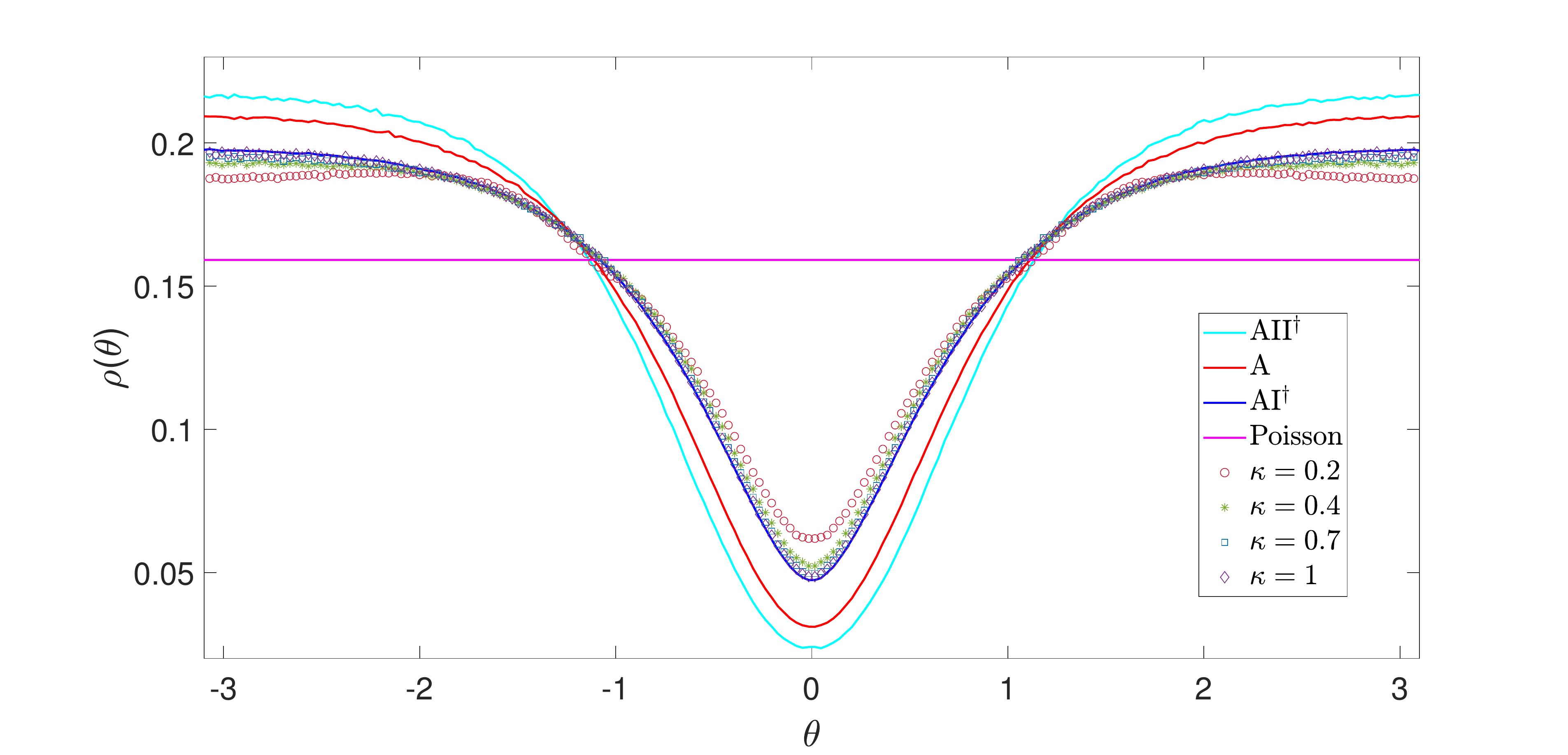}}
	\caption{Marginal angular distribution of the complex gap ratio for $N=12$. (a): $\kappa = 1$ and different values of $\lambda$; (b): $\lambda = 0.02$ and different values of $\kappa$.  Although we find a fair agreement with the $AI^\dagger$ class, results are rather sensitive to the value of $\kappa$, $\lambda$ and the matrix size.}\label{fig:ratiotheta}
\end{figure}

We observe in Fig.~\ref{fig:ratiopoi}(b,c) that, for $\lambda = 0$ and $\kappa =1$, both distributions are very close to the prediction of Poisson statistics typical of an integrable or localized systems. By contrast, see Fig.~\ref{fig:ratiorho} and Fig.~\ref{fig:ratiotheta}, even for small values of $\lambda =0.015$, the agreement with the random matrix prediction is excellent for all $\kappa$ and $N$ considered. Unlike other SYK models, the universality class is always that of systems with time reversal invariance corresponding to the universality class $AI^\dagger$ \cite{ueda2019}, related to transposition symmetry and not to the Ginibre Orthogonal Ensemble \cite{ginibre1965} related to complex conjugation symmetry. Although, strictly speaking, we do not have the equivalent of a Bohigas-Giannoni-Schmit conjecture \cite{bohigas1984} for non-Hermitian systems, we believe that this agreement with random matrix theory still provides evidence of quantum chaotic motion triggered by a weak explicit coupling $\lambda$. For larger values of $\lambda \geq 0.04$,  we start to observe growing deviations from the random matrix results, which is likely due to the fact that a growing number of eigenvalues become strictly real and the chosen spectral observables are intended for the analysis of complex eigenvalues. For instance, for $\lambda =0.1$ in Fig.~\ref{fig:null_kl}, about $85\%$ percent of the spectrum is real. In any case, this intermediate region is not universal and therefore is of less interest.

\begin{figure}
	\centering
	\subfigure[]{\includegraphics[width=7cm]{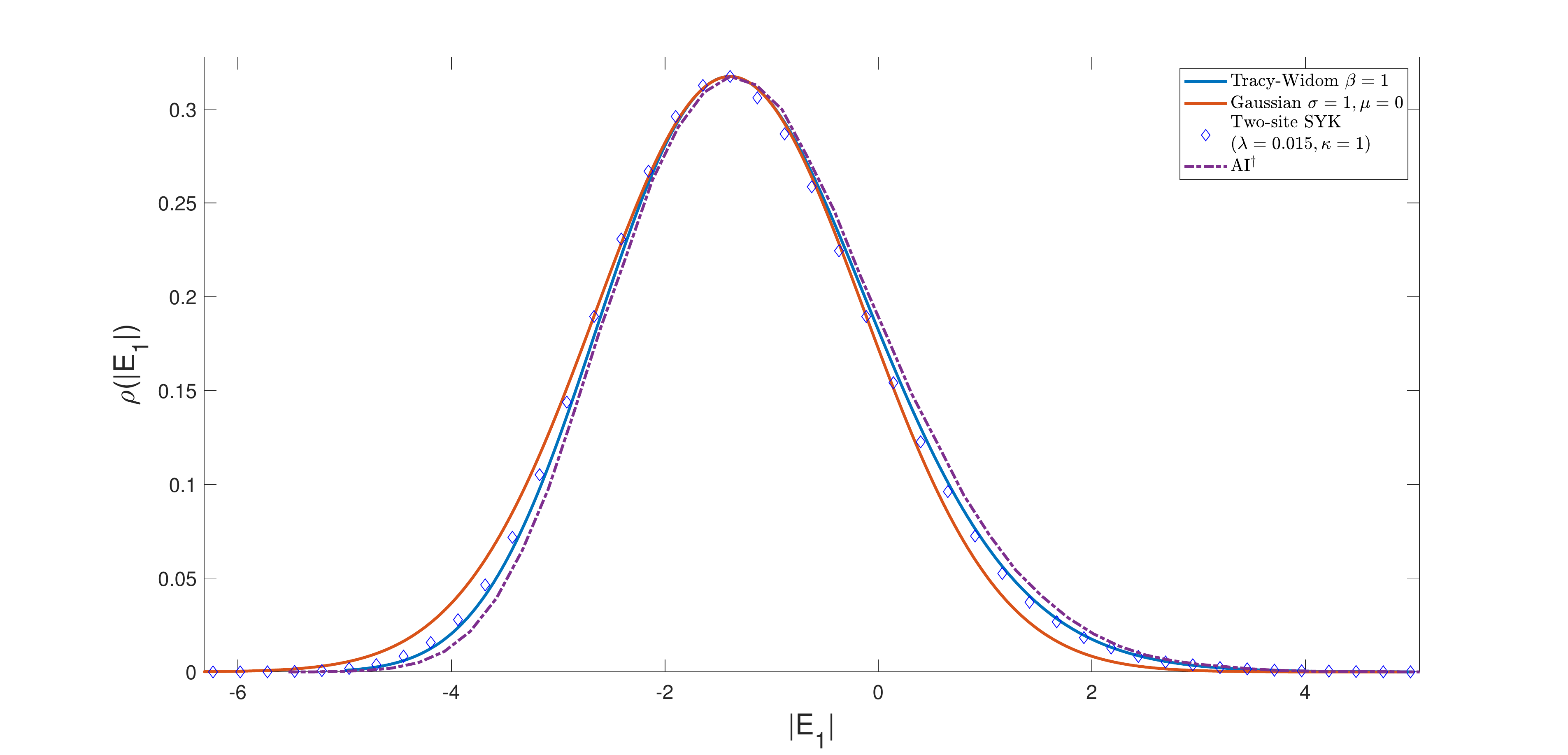}}
	\subfigure[]{\includegraphics[width=7cm]{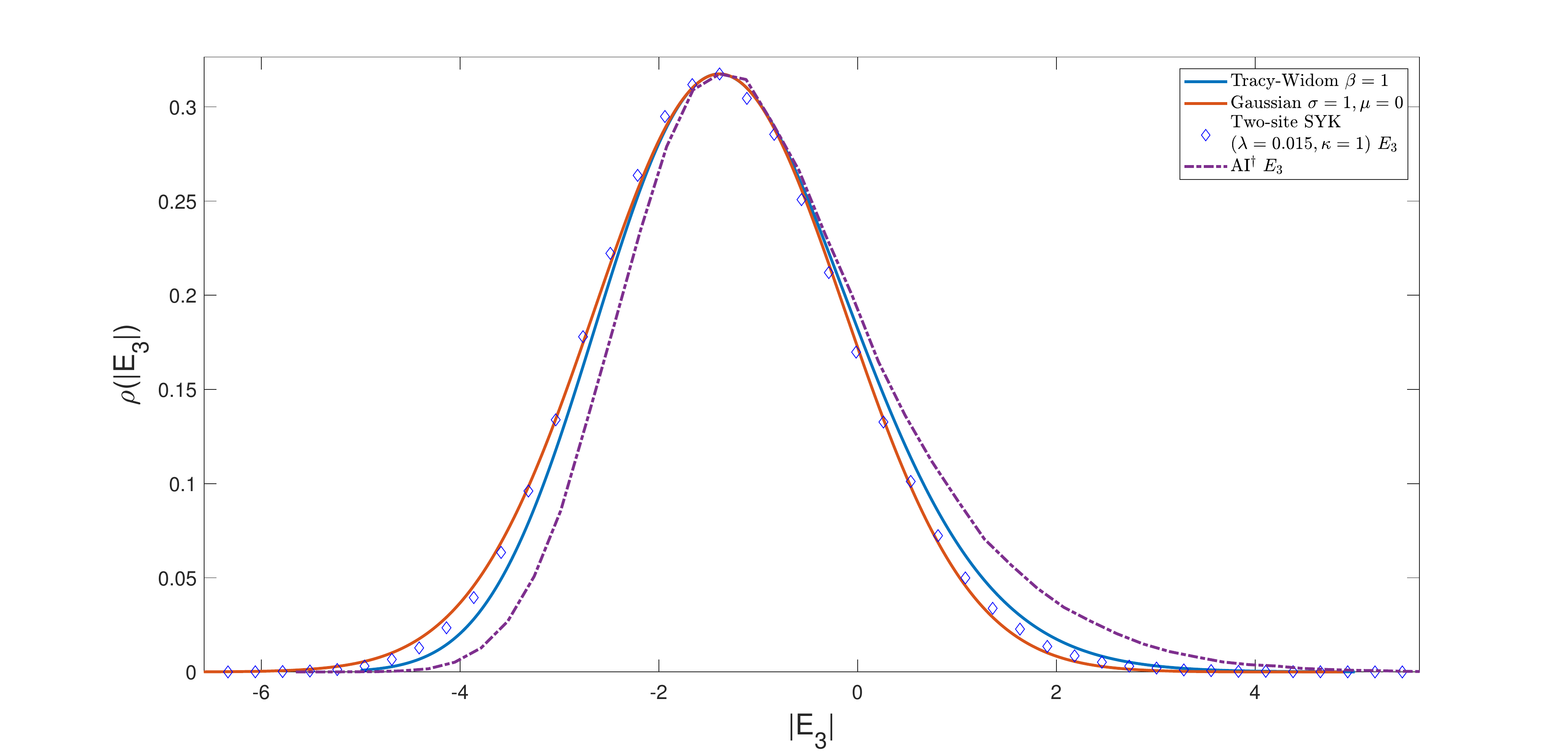}}
	\caption{(a) Probability distribution of $|E_1|$, the eigenvalue with the largest negative real part. (b) Probability distribution of $|E_3|$, the eigenvalue with the third largest negative real part. After a shifting and rescaling, only the distribution of $|E_1|$ agrees with the random matrix prediction, the Tracy-Widom \cite{tracy1994} distribution.}\label{fig:rhoE_1}
\end{figure}

We now investigate whether anomalies, even for small $\lambda$, are observed in the infrared part of the spectrum corresponding with the eigenvalues with the largest negative real part. This is precisely the region related to the Euclidean wormhole phase where the spectrum of the ensemble average system has a gap. An immediate problem is that it is not yet clear how to order the complex spectrum and therefore to define precisely the part of the spectrum related to the Euclidean wormhole.
Strictly speaking, the Euclidean wormhole is associated with the eigenvalue with the largest negative real part. This eigenvalue $E_1$ is always fully real for all range of parameters considered. We thus computed the variance $R_1$ of the probability distribution of $|E_1|$, and  normalize it by ensemble averaging. The results show its excellent agreement with the RMT prediction. For instance, for $\lambda = 0.015$  and $\kappa = 1$, $R_1$ equals to $ 1.00332$ while the random matrix prediction is $1.00315$. This agreement extends to other values of $\lambda$ where the spectrum is still complex. What's more, the agreement with the random matrix prediction goes beyond $R_1$. In Fig.~\ref{fig:rhoE_1}, we compare the full distribution of $|E_1|$ with the the random matrix prediction, namely, the Tracy-Widom distribution \cite{tracy1994} for systems with time reversal symmetry $\beta = 1$. After the preceptive rescaling and shifting, we obtain a good agreement with the Tracy-Widom distribution for the distribution of the eigenvalue with the largest real negative part. However, substantial differences are observed even for the distribution of the eigenvalue with the third largest real negative part. We note that the distribution of the eigenvalue with the largest real negative part corresponding to a random matrix belonging to the $AI^\dagger$ universality class is qualitatively different. As was expected, the agreement is worse if we fit it to a Gaussian distribution.
To some extent, the complex level statistics results are expected as there is no a visible gap in the spectrum for $\lambda = 0$ before any ensemble average. This is another indication that the wormhole phase, which is still characterized by a gap, requires dominance of off-diagonal replica configurations \cite{garcia2021a} and therefore ensemble average.

We now turn to the analysis of spectral correlation for $\lambda>\lambda_c$. The spectrum becomes real even if the Hamiltonian is non-Hermitian when $\kappa > 0$. We again employ the adjacent gap ratio which for a real spectrum \cite{luitz2015,oganesyan2007,bertrand2016,atas2016,brody1981,numasawa2019,Kourkoulou:2017zaj} is given by
\begin{equation}
	r_i = \frac{\min(\delta_i, \delta_{i+1})}{\max(\delta_i, \delta_{i+1})}
	\label{eq:agr}
\end{equation}
where $\delta_i = E_i - E_{i-1}$ and the spectrum is assumed to be ordered.

For random matrices, and for uncorrelated eigenvalues, it is possible \cite{atas2016} to find explicit analytic expressions for both its average and the full distribution function.
For instance, for a quantum chaotic system with no translational symmetry , the averaged gap ratio $\langle r \rangle \approx 0.530$ while $\left\langle r \right\rangle_\mathrm{P} \approx 0.386$ for Poisson distribution corresponding to uncorrelated eigenvalues. We shall also  study the
level spacing distribution $P(s)$, namely, the probability to find two consecutive eigenvalues
$E_{i}, E_{i+1}$ at a distance $s = (E_{i+1}-E_{i})/\Delta$, where $\Delta$ is the mean level spacing in that region of the spectrum.
For a fully quantum chaotic system, $P(s)$ is given by the random matrix theory results which depends on the global symmetries of the system. In the case of time reversal invariance, it is well approximated by the so-called Wigner surmise 
$
P_\mathrm{W,GOE}(s) \approx \frac{\pi}{2}s\exp(-\pi s^2/4)
$
for the Gaussian Orthogonal Ensemble (GOE), while for uncorrelated eigenvalues, corresponding to integrable non-degenerate or Anderson localized systems, it is given by Poisson statistics ($P_\mathrm{P}(s) = e^{-s}$). Technically, the calculation of $P(s)$ requires unfolding the spectrum so the average local level spacing is one. We carried it out by employing a low order, in most cases six order, polynomial to fit the average spectral density. The level spacing distribution, which is complementary to the adjacent gap ratio, provides information, specially its tail, about the dynamic of the system for time scales of the order of the Heisenberg time. The adjacent gap ratio, in the other hand, probes the dynamics to even longer scales.\\
\begin{figure}
	\centering
	\includegraphics[scale=0.4]{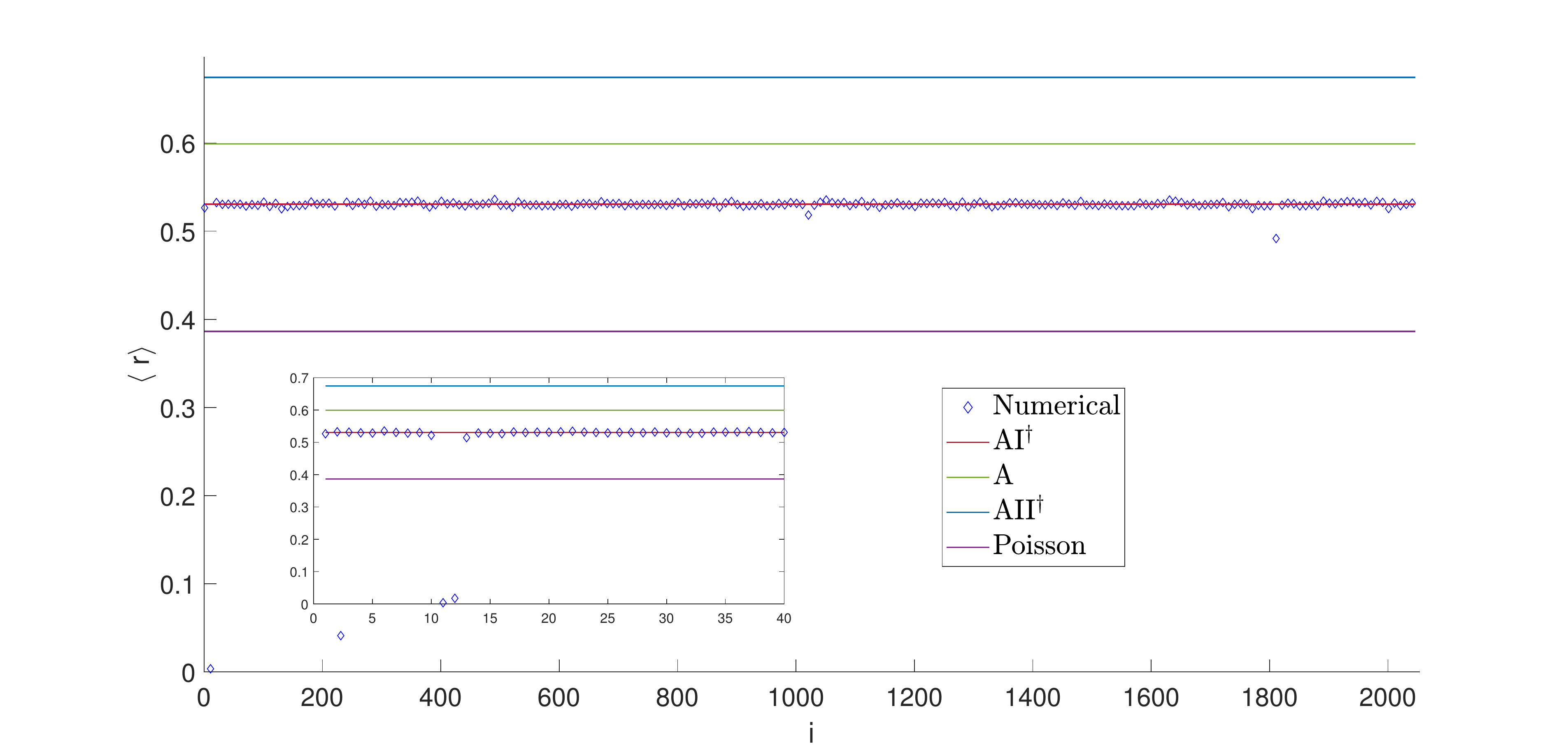}
	\caption{Average gap ratio for $\kappa =1$, $\lambda=0.15$ for the full spectrum of $N=12$, the deviation from RMT happens at the interval of nearby spectrum sectors clustering around eigenvalues of $\hat{S}$. Unlike the case of a SYK model dual to a traversable-wormhole \cite{maldacena2018}, our model does not have any clear deviation from the RMT result even for the most negative eigenvalues.}\label{fig:averreal}
\end{figure}
\begin{figure}
	\centering
	\includegraphics[scale=0.4]{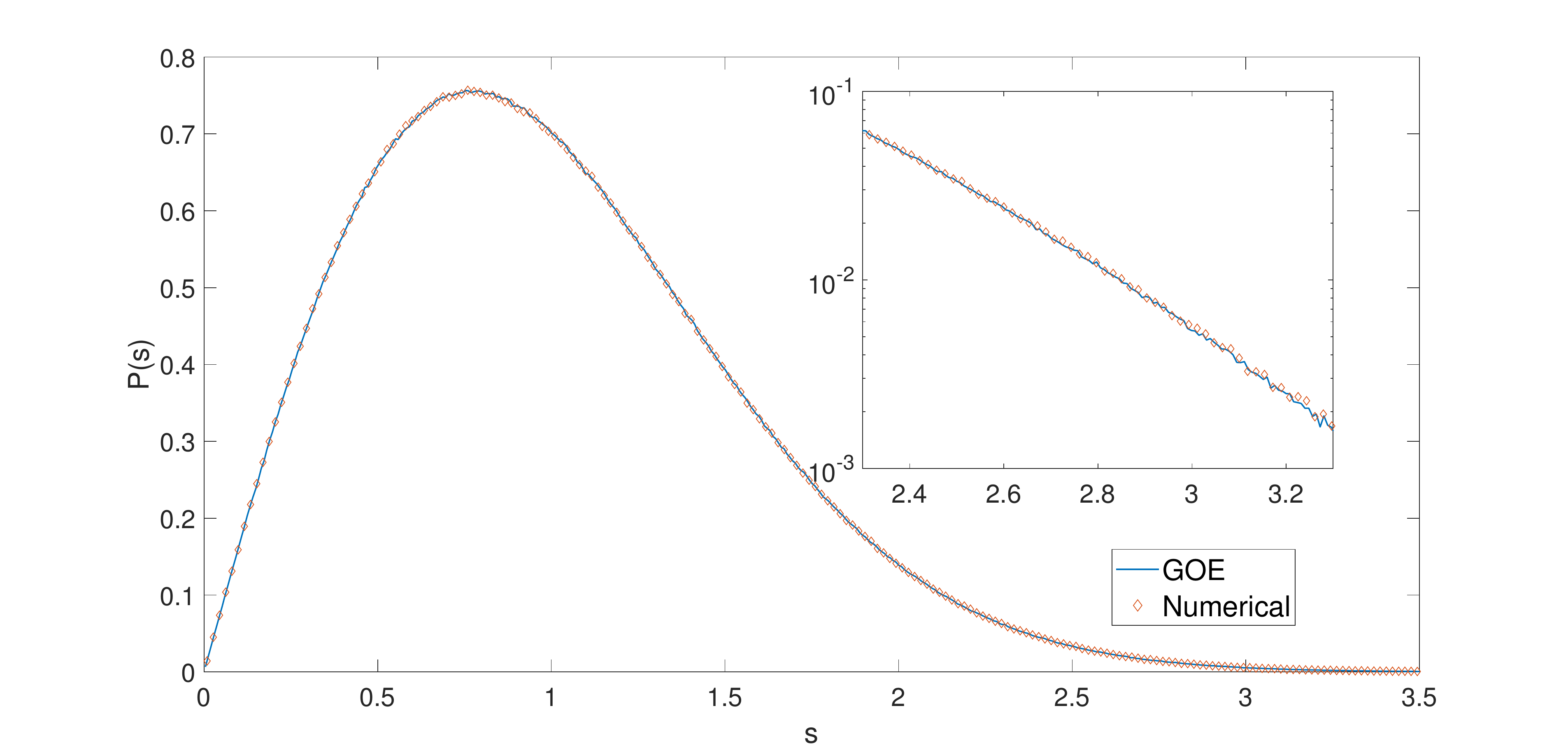}
	\caption{Level spacing distribution $P(s)$ for $\lambda > \lambda_c$ where the spectrum is real. We find excellent agreement with GOE level statistics which is obtained by exact diagonalization of random matrices of size $1000 \times 1000$, even at the tail of the spectrum. Unlike the non-hermitian case, almost no size dependence is observed.}\label{fig:ps}
\end{figure}

\begin{figure}
	\centering
	\includegraphics[scale=0.4]{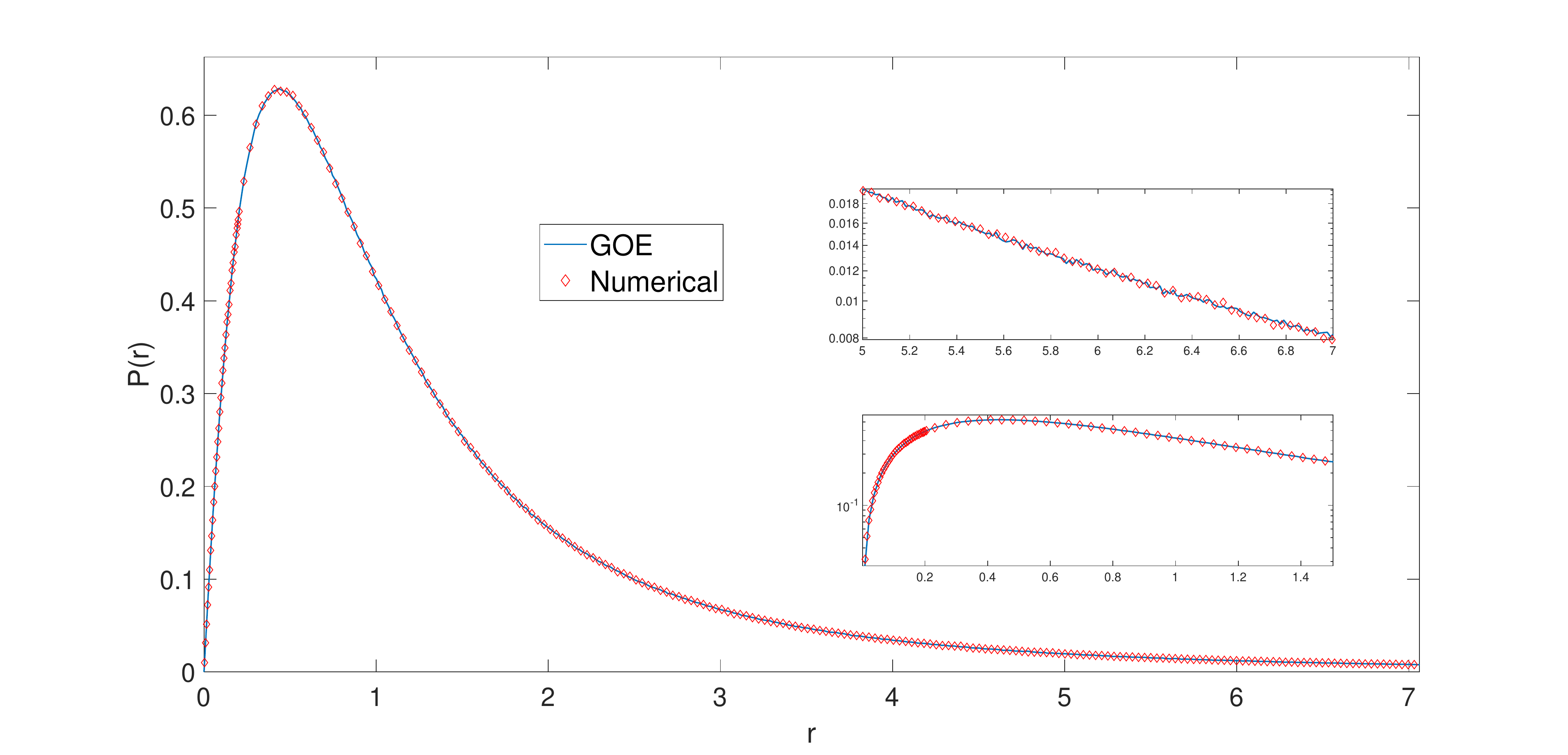}
	\caption{Distribution of the gap ratios $P(r)$ for $\lambda > \lambda_c$ where the spectrum is real. We observe an excellent agreement with the GOE prediction even in the tail of the $P(r)$, see  insets in log scale.}\label{fig:pr}
\end{figure}

As it can be seen in Fig.~\ref{fig:denreal}, the approximate spin-symmetry has an important effect: the spectrum is concentrated around the eigenvalues of $\hat S$. The results are indeed very similar to that of the $\kappa = 0$ case and therefore quite insensitive to $\kappa$. The same conclusion largely applies to level statistics. In Fig.~\ref{fig:averreal}, we present results for the averaged gap ratio $\langle r_i \rangle$ where $i = 1, \ldots, 2^{\frac{N}{2}}$ labels the eigenvalues with $i=1$ the ground state and $\langle \ldots \rangle$ stands for ensemble average. The values of $\kappa$  and $\lambda$ are such that the spectrum is always real. We find agreement for most parts of the spectrum with the random matrix prediction for systems with time reversal symmetry $\langle r \rangle_{RMT} \approx 0.529$ \cite{atas2016}. The observed deviations occur for gap ratios corresponding to eigenvalues of the Hamiltonian located at the edges of sectors belonging to different eigenvalues of $\hat{S}$. The gap ratio $\langle r\rangle$ is very small since $\max \{\delta_i,\delta_{i+1}\}$ is very large in this case. Therefore, these deviations with respect to the RMT prediction do not have a dynamical significance.
We note that the infrared part of the spectrum, related to the wormhole phase, also fits well with the RMT prediction.
This is in contrast with the $\kappa = 0$ case \cite{garcia2019} where the wormhole phase is characterized by strong deviations from the RMT result.

Similarly,  in Figs.~\ref{fig:ps} and ~\ref{fig:pr}, both the $P(s)$ and the distribution function of the gap ratio $P(r)$ of our model, agree well with the random matrix prediction. We note the agreement extends to even the tail of $P(s)$ which probes the dynamics at time scales of not only the order, but longer, than the Heisenberg time. For sufficiently large $\lambda$, the eigenvalues of the Hamiltonian cluster around the eigenvalues of the spin-like operator $\hat S$. Thus it is required that each of these clusters is considered separately for level statistics analysis. \\

\section{Conclusion and outlook}

We have investigated a two-site non-Hermitian SYK model with a weak inter-site coupling and its dual in JT gravity.  In the SYK model we have employed exact diagonalization techniques and the numerical solution of the Schwinger-Dyson equations describing the large $N$ saddle-points of the action. On the gravity side, we have derived an effective Schwarzian action and used it to compute the gravity path integral in the saddle-point approximation.

In both SYK and JT gravity, we have studied the thermodynamic properties and observed a thermal phase transition between the wormhole and two black holes. We have obtained an excellent match for the thermodynamic observables such as the free energy, the energy gap and the critical temperature, see Fig.~\ref{fig:Eg_ew_kappa_lambda} and Fig.~\ref{fig:Tc_ew_kappa_lambda}.

By tuning the inter-site coupling in the SYK model, we have found a dynamical transition where the energy spectrum becomes real despite the fact that the Hamiltonian is non-Hermitian. The existence of the transition has been demonstrated by an explicit exact diagonalization of the SYK Hamiltonian. In JT gravity, we have shown that this transition corresponds to a restoration of the gravitational $\r{SL}(2,\R)$ symmetry of the Lorentzian wormhole broken to $\r{U}(1)$ in the Euclidean wormhole, and can be viewed as a Euclidean-to-Lorentzian transition. This was shown by identifying an order parameter in both SYK and JT and showing a similar oscillating pattern of the real time Green's functions where the transition is characterized by the value of a phase shift.

 One of the motivation to introduce imaginary sources was to be able to study the gravitational path integral using saddle-points. We view here JT gravity as a low-energy approximation of the exact gravity dual of SYK, in a spirit similar to the higher-dimensional version of AdS/CFT. The transition observed here could also happen in higher dimensions. The Euclidean version of a Lorentzian geometry can have smaller isometry group because the periodic Euclidean time identification may only preserve a subset of the isometries. It is interesting to note that this requires the geometry to be similar to a wormhole as black hole spacetimes don't have this property. In the Euclidean system, defined by the gravity path integral, we impose a smaller number of gauge constraints than in Lorentzian, which defines a purely Euclidean system without sensible Lorentzian continuation. It would be interesting to see  whether a similar Euclidean-to-Lorentzian transition, \ie the dynamical restoration of the Lorentzian symmetries in a purely Euclidean system, can be observed in higher dimensional AdS/CFT. Higher-dimensional Euclidean wormholes supported by boundary sources were described in \cite{marolf2021} and have similar thermodynamical properties as our wormhole. Eternal traversable wormholes in higher dimensions are harder to construct \cite{Freivogel:2019lej, VanRaamsdonk:2020tlr} but can be obtained using the appropriate setup \cite{Maldacena:2018gjk, Bintanja:2021xfs}.

 The Euclidean wormhole studied in this paper can also be used to address the factorization puzzle \cite{Witten:1999xp, maldacena2004}, which comes from  the fact that the gravity path integral appears to compute an ensemble average, as recently discussed in \cite{Saad:2021rcu, Saad:2021uzi, Mukhametzhanov:2021nea,Iliesiu:2021are, garcia2022, Blommaert:2021gha, Mukhametzhanov:2021hdi,Blommaert:2021fob,  Schlenker:2022dyo, Collier:2022emf, Chandra:2022bqq}. For $\eta=0$, there is no interaction between the boundary and the Euclidean wormhole is purely a result of the average. It  was proposed in \cite{Saad:2021rcu} that  factorization could be restored by half-wormhole saddle-points. In our setup without inter-site coupling ($\eta=0$), half-wormhole saddle-points were constructed in \cite{garcia2022} and their behavior was matched with the SYK model at a single realization of the couplings. These half-wormhole solutions  should be generalizable to non-zero inter-site coupling $\eta$. This would correspond to a system of two coupled half-wormholes which, as suggested by our results, could possibly transition to an eternal traversable wormhole. It would be interesting to make this more precise and obtain a gravity picture of the complex-to-real transition for a single realization of the SYK couplings.

The long time dynamics has been explored by the study of level statistics in the SYK model. Both, for a real and a complex spectrum, spectral correlations are consistent with quantum chaotic motion as we find good agreement with the random matrix prediction in each case. Importantly, we have also found that the distribution of the energy level with the largest negative real part fit well the Tracy-Widom distribution which is an indication of quantum chaotic behavior in the wormhole phase at large $N$. In all cases, the universality class was that of systems with time reversal invariance. It would be interesting to investigate whether some aspects of the level statistics can be understood using the gravity path integral, which would require finer gravity observables than studied in this paper.

\acknowledgments{
 We were partially supported by the National Natural Science Foundation of China (NSFC) (Grant number 11874259), by the National
	Key R$\&$D Program of China (Project ID: 2019YFA0308603).  AMGG also acknowledges financial support from a Shanghai talent program. VG acknowledges the postdoctoral program at ICTS for funding support through the Department of Atomic Energy, Government of India, under project no. RTI4001. VG acknowledges useful  discussions with Raghu Mahajan and Suvrat Raju.  AMG acknowledges illuminating correspondence with Zhenbin Yang, Dario Rosa, Jac Verbaarschot and Yiyang Jia. JPZ and CY thank Pengfei Zhang and Stephen Plugge for help with the real-time calculation of Green's functions.}

\appendix

\section{Energy gap in JT gravity and SYK}\label{app:Eg}

We provide additional details about the method we have employed to compare the gap $E_g$ in JT gravity (\ref{eq:gapgravity}) with that in the SYK model.

The JT and SYK parameters should be proportional $k=A\kappa$, $\eta=B\lambda$ where $A, B$ are some constants.  In order to determine these constants, we compare the numerical results for $E_g$ in SYK with the analytical expression $E_g$ in JT gravity (\ref{eq:gapgravity}) using $A, B$ as fitting parameter. More specifically, we note that $X=C_1+C_2$, related to $E_g$ by (\ref{eq:EgX}), is the only real-valued solution of (\ref{cubicEq}). For the sake of convenience, we define,
\begin{equation}\begin{aligned}
z &=X_1^3=\left(\frac{E_g}{\Delta}\right)^{3/2}= 8 E_g^{3/2}\\
x &=\frac{3\kappa^2}{4}\left(\frac{E_g}{\Delta}\right)^{1/2} =1.5\kappa^2 E_g^{1/2} \\
y &=\frac{\lambda}{4}
\end{aligned}\end{equation}
we can then rewrite  (\ref{cubicEq}) as $z(\kappa,\lambda)=A^2 x(\kappa,\lambda)+By(\kappa,\lambda)$. The functions $x(\kappa,\lambda)$, $y(\kappa,\lambda)$, $z(\kappa,\lambda)$ can be calculated from the above expressions where $E_g$ is taken to be the numerical SYK result. We see that $A$, $B$ are just the coefficients of the linear equation $z(x,y)=A^2 x+y$ which are easily calculated by numerical fitting. We find that for $\kappa=0,0.1,\cdots,0.7$, $\lambda=0,0.01,\cdots,0.09$ the best fit corresponds to $A=1.58497$ and $B=27.05755$. In Fig.~\ref{fig:Eg_ew_kappa_lambda} of the main text, we compare explicitly the numerical SYK results with those from the fitting above calculation and find that they are fully consistent especially when $\kappa$ is small. We also compare $T_c$ in Fig.~\ref{fig:Tc_ew_kappa_lambda} using the same values of $A$ and $B$ and obtain an excellent match.

\section{Real time calculation}

We consider the Hamiltonian \eqref{hamie}. In the large $N$ limit, we obtain the effective action
\begin{equation}\begin{aligned}
I_\r{eff}= -\frac{1}{2}\log\det(\delta_{ab}\partial-\Sigma_{ab}) +\frac{1}{2}\sum_{ab}\int\!\!\!\int\left(\Sigma_{ab}G_{ab}-\frac{(1-t_{ab}\kappa^2)J^2}{4}G_{ab}^4 \right) +\frac{i\lambda}{2}\int G_{LR}(\tau,\tau)-G_{RL}(\tau,\tau).
\end{aligned}\end{equation}
from the use of the replica trick after ensemble average. Here $t_{ab}$ is defined as 
\be
t_{LL}=t_{RR}=1,\qq t_{LR}=t_{LR} = -1~.
\ee
A saddle-point analysis leads to the Schwinger-Dyson (SD) equations,
\begin{equation}\left\{\begin{aligned}
& -i\omega G_{LL}-\Sigma_{LL}G_{LL}-\Sigma_{LR}G_{RL}=1, \quad -i\omega G_{LR}-\Sigma_{LL}G_{LR}-\Sigma_{LR}G_{RR}=0 \\
& \Sigma_{LL}(\tau)=(1-\kappa^2)J^2 G_{LL}^3(\tau), \qquad \Sigma_{LR}(\tau)=(1+\kappa^2)J^2 G_{LR}^3(\tau) -i\lambda\delta(\tau)
\end{aligned}\right.\end{equation}
where the first two equations are expressed in the frequency domain, while the last two are in the imaginary time one.

The real time dynamics is studied after performing a Wick rotation $-i\omega\to \omega+i\epsilon$ to the Schwinger-Dyson equations for the imaginary time Green's function. We closely follow the method of Ref.~\cite{sahoo2020,plugge2020a}.

We first introduce \cite{plugge2020} $G_+=G_{LL}+iG_{LR}$ and $\Sigma_+=\Sigma_{LL}+i\hat{\Sigma}_{LR}$, where $\Sigma_{LR}=\hat{\Sigma}_{LR}-i\lambda\delta(\tau)$. With these definitions, we have
\begin{equation}\begin{aligned}
G_+=\frac{1}{-i\omega-\Sigma_+ -\lambda}
\end{aligned}\end{equation}
and the relations $G_{LL}=G_{RR}, G_{LR}=-G_{RL}$. The retarded Green's function, after the Wick rotation, takes the form
\begin{equation}\begin{aligned}
G^r_+=\frac{1}{\omega+i\epsilon-\Sigma^r_+-\lambda}.
\end{aligned}\end{equation}
One of the main technical difficulties is the calculation of $\Sigma^r_+$. Following Refs.~\cite{plugge2020a,plugge2020}, we first calculate $\Sigma_{LL}(\omega_n)$ and $\hat{\Sigma}_{LR}(\omega_n)$, which are simply the self-energy $\Sigma_{LL}(\tau)=J^2 G_{LL}^3(\tau)$ and $\hat{\Sigma}_{LR}(\tau)=J^2 G_{LR}^3(\tau)$ but in the frequency domain. We then apply the Wick rotation to obtain, $\Sigma_+(\omega)=\Sigma_{LL}(\omega)+i\Sigma_{LR}(\omega)$, with
\begin{equation}\begin{aligned}
\Sigma^r_{LL}(\omega) &= -2i(1-\kappa^2)J^2\int_0^{\infty}e^{i(\omega+i\epsilon)t}\text{Re}[n_{LL}^3(t)]dt \\
\Sigma^r_{LR}(\omega) &= -2(1+\kappa^2)J^2\int_0^{\infty}e^{i(\omega+i\epsilon)t}\text{Im}[n_{LR}^3]dt \\
\end{aligned}\end{equation}
where
\begin{equation}\begin{aligned}
n_{LL/LR}(t) &= \int_{-\infty}^{\infty}d\omega \rho_{LL/LR}(\omega)n_F(\omega)e^{-i\omega t}, \qquad n_F(\omega)=\frac{1}{e^{\beta\omega}+1}. \\
\end{aligned}\end{equation}
More details of the calculation can be found in Appendix F of \cite{plugge2020a} and Appendix D of \cite{sahoo2020}.

In order to find $n_{LL/LR}$, we still need to know $\rho_{LL/LR}$. A direct way to calculate $\rho_{LL/LR}(\omega)$ is from the retarded Green's function: $\rho_{LL}= -\frac{1}{\pi}\text{Im}[G^r_{LL}(\omega)]$ and $\rho_{LR}= -\frac{1}{\pi}\text{Re}[iG^r_{LR}(\omega)]$, by noticing that
\begin{equation}\begin{aligned}
G^r_{LL}(\omega) &=\int\frac{d\omega'}{2\pi}\frac{\rho_{LL}(\omega')}{\omega-\omega'+i\epsilon} \\
G^r_{LR}(\omega) &=\int\frac{d\omega'}{2\pi}\frac{-i\rho_{LR}(\omega')}{\omega-\omega'+i\epsilon} \\
\end{aligned}\end{equation}
and using the relation
\be
\frac{1}{x+i\epsilon}=\mathcal{P}\frac{1}{x}-i\pi \delta(x)~.
\ee
An alternative way to calculate $\rho_{LL/LR}$ is from $\rho_{+}=-\frac{1}{\pi}\text{Im}G^r_+(\omega)$, by
\begin{equation}\begin{aligned}
\rho_{LL/LR}(\omega) &= \frac{1}{2}(\rho_+(\omega) \pm \rho_+(-\omega)) \\
\end{aligned}\end{equation}
after imposing $\rho_{LL}(\omega)=\rho_{LL}(-\omega)$ and $\rho_{LR}(\omega)=-\rho_{LR}(-\omega)$. Mathematically, these two methods are related by $G^r_+=G^r_{LL}+iG^r_{LR}$. For simplicity, the latter is chosen in our calculation. We have already obtained all the Schwinger-Dyson real time equations,
\begin{equation}\begin{aligned}
		\rho_+(\omega) &= -\frac{1}{\pi}\text{Im}G^r_+(\omega) \\
		\rho_{LL/LR}(\omega) &= \frac{1}{2}(\rho_+(\omega) \pm \rho_+(-\omega)) \\
		n_{LL/LR}(t) &= \int_{-\infty}^{\infty}d\omega \rho_{LL/LR}(\omega)n_F(\omega)e^{-i\omega t}, \qquad n_F(\omega)=\frac{1}{e^{\beta\omega}+1} \\
		\Sigma^r_+(\omega) &= -2iJ^2\int_0^{\infty}dt e^{i(\omega+i\epsilon)t}[(1-\kappa^2)\text{Re}[n_{LL}^3(t)]-i(1+\kappa^2)\text{Im}[n_{LR}^3]]\\
		G^r_{+}(\omega) &= \frac{1}{\omega + i\epsilon -\Sigma^{r}_{+} -\lambda}  \\
\end{aligned}\end{equation}
We employ the following Green's function $G^>_{LL/LR}(t)$,
\begin{equation}\begin{aligned}
		G^>_{ab}(t) &= -i\frac{1}{N}\sum_i\langle\psi_{i,a}(t)\psi_{i,b}(0)\rangle \qq
		\Rightarrow \qquad G^>_{ab}(\omega) &= -i(1-n_F(\omega))\rho_{ab}(\omega)~,
\end{aligned}\end{equation}
with $a,b = L,R$, to study the time evolution of the Hamiltonian. In the main text, we provide a detailed analysis of $|G^>_{ab}(t)|$.

Here we focus on the technical details required to solve numerically these real time saddle-point equations. The main difficulty lies in the choice of a proper cutoff $L$ in $t$, and also a small but finite $\epsilon$ and the number of discrete points $N$ in the sums, for given parameters $T$, $\lambda$, $\kappa$, $J$. 

 It can be seen from $G(\omega)=\int\frac{\rho(\omega')d\omega'}{\omega-\omega'+i\epsilon}$ that the first relation that must hold is $d\omega=\frac{2\pi}{L}\ll\epsilon$. To make sure that $\frac{\epsilon}{(\omega-\omega')+\epsilon^2}\approx \pi\delta(\omega-\omega')$, $\epsilon$ should be small enough so that we obtain a sharp peak, mimicking a delta function. We also need $d\omega\ll\epsilon$ so that the profile is still smooth. For a given set of parameters, it is important to find the right balance between the necessary suppression of discretization artifacts that may obscure real physical effects and the optimization of computational resources in terms for instance to computation times are RAM usage.

The wormhole phase \cite{maldacena2018} requires low temperature, i.e. $\beta J\gg 1$ and small coupling $\lambda \ll 1$. Moreover, the choices of physical parameters and numerical parameters are not independent. From a rewriting of $G^r_{+}(\omega)$,
\begin{equation}\begin{aligned}
G^r_{+}(\omega) &= \frac{1}{\omega + i\epsilon -\Sigma^{r}_{+} -\lambda} \\
&= \frac{\omega-\lambda -\text{Re}[\Sigma^{r}_{+}]  }{(\omega-\lambda -\text{Re}[\Sigma^{r}_{+}])^2 +(\epsilon -\text{Im}[\Sigma^{r}_{+}])^2} -i\frac{\epsilon -\text{Im}[\Sigma^{r}_{+}] }{(\omega-\lambda -\text{Re}[\Sigma^{r}_{+}])^2 +(\epsilon -\text{Im}[\Sigma^{r}_{+}])^2} \\
\end{aligned}\end{equation}
we observe that we must impose that $\epsilon \ll \lambda$ so that the effect of $\lambda$ is not obscured by a too large $\epsilon$.
Finally, the choice of the parameter $N$ is driven by the required accuracy of the numerical integral over $t$. Since $dt=\frac{L}{N}$, $N$ should be small enough so that the accuracy of the integral meets a certain minimum. In addition, the range of $\omega$ is approximately $\sim \frac{2\pi}{L}N$. Therefore a small enough $L/N$ will also guarantee the irrelevance of the truncation for large $\omega$. Taking all this into consideration we set $L=5\times 10^6$ and $N=2^{25}\approx 3\times 10^7$.

\bibliography{libgenWH}
\end{document}

%% file: victorsmacros.tex
\usepackage{float}

\definecolor{Mblue}{rgb}{0.37,0.51,0.71}
\definecolor{Morange}{rgb}{0.88,0.61,0.14}
\definecolor{Mgreen}{rgb}{0.56,0.69,0.19}

\def\ie{\emph{i.e.} }
\def\eg{\emph{e.g.} }

\def\nt{\notag}

\def\wt{\widetilde}

\def\R{\mathbb{R}}

\def\cJ{\mathcal{J}}

\def\cO{\mathcal{O}}

\def\p{\partial}

\def\/{\over}

\def\rn{\rangle}
\def\ln{\langle}

\def\e{\epsilon}
\def\ve{\varepsilon}
\def\vphi{\varphi}
\def\a{\alpha}
\def\b{\beta}
\def\d{\delta}
\def\k{\kappa}

\def\la {\lambda}
\def\w {\omega}

\def\l{\ell}
\def\mn{{\mu\nu}}

\def\D{\Delta}

\def\S{\Sigma}

\def\ra{\rightarrow}

\def\r{\mathrm}

\def\_{\hspace{2cm}}
\def\-{\\\notag}
\def\={&=&}

\newcommand{\bea}{\begin{eqnarray}}
\newcommand{\eea}{\end{eqnarray}}

\newcommand{\bpm}{\begin{pmatrix}}
\newcommand{\epm}{\end{pmatrix}}

\newcommand{\bit}{\begin{itemize}}
\newcommand{\eit}{\end{itemize}}

\newcommand{\ben}{\begin{enumerate}}
\newcommand{\een}{\end{enumerate}}

\newcommand\bsp{\begin{split}}
\newcommand\esp{\end{split}}

\def\le{\left}
\def\ri{\right}

\def\l{\ell}

\def\qq{\qquad}

\def\cos{\r{cos}}
\def\sin{\r{sin}}

%% file: main_published.bbl
\begin{thebibliography}{115}%
\makeatletter
\providecommand \@ifxundefined [1]{%
 \@ifx{#1\undefined}
}%
\providecommand \@ifnum [1]{%
 \ifnum #1\expandafter \@firstoftwo
 \else \expandafter \@secondoftwo
 \fi
}%
\providecommand \@ifx [1]{%
 \ifx #1\expandafter \@firstoftwo
 \else \expandafter \@secondoftwo
 \fi
}%
\providecommand \natexlab [1]{#1}%
\providecommand \enquote  [1]{``#1''}%
\providecommand \bibnamefont  [1]{#1}%
\providecommand \bibfnamefont [1]{#1}%
\providecommand \citenamefont [1]{#1}%
\providecommand \href@noop [0]{\@secondoftwo}%
\providecommand \href [0]{\begingroup \@sanitize@url \@href}%
\providecommand \@href[1]{\@@startlink{#1}\@@href}%
\providecommand \@@href[1]{\endgroup#1\@@endlink}%
\providecommand \@sanitize@url [0]{\catcode `\\12\catcode `\$12\catcode
  `\&12\catcode `\#12\catcode `\^12\catcode `\_12\catcode `\%12\relax}%
\providecommand \@@startlink[1]{}%
\providecommand \@@endlink[0]{}%
\providecommand \url  [0]{\begingroup\@sanitize@url \@url }%
\providecommand \@url [1]{\endgroup\@href {#1}{\urlprefix }}%
\providecommand \urlprefix  [0]{URL }%
\providecommand \Eprint [0]{\href }%
\providecommand \doibase [0]{https://doi.org/}%
\providecommand \selectlanguage [0]{\@gobble}%
\providecommand \bibinfo  [0]{\@secondoftwo}%
\providecommand \bibfield  [0]{\@secondoftwo}%
\providecommand \translation [1]{[#1]}%
\providecommand \BibitemOpen [0]{}%
\providecommand \bibitemStop [0]{}%
\providecommand \bibitemNoStop [0]{.\EOS\space}%
\providecommand \EOS [0]{\spacefactor3000\relax}%
\providecommand \BibitemShut  [1]{\csname bibitem#1\endcsname}%
\let\auto@bib@innerbib\@empty
\bibitem [{\citenamefont {Sachdev}\ and\ \citenamefont
  {Ye}(1993)}]{sachdev1993}%
  \BibitemOpen
  \bibfield  {author} {\bibinfo {author} {\bibfnamefont {S.}~\bibnamefont
  {Sachdev}}\ and\ \bibinfo {author} {\bibfnamefont {J.}~\bibnamefont {Ye}},\
  }\bibfield  {title} {\bibinfo {title} {Gapless spin-fluid ground state in a
  random quantum heisenberg magnet},\ }\href
  {https://doi.org/10.1103/PhysRevLett.70.3339} {\bibfield  {journal} {\bibinfo
   {journal} {Phys. Rev. Lett.}\ }\textbf {\bibinfo {volume} {70}},\ \bibinfo
  {pages} {3339} (\bibinfo {year} {1993})},\ \Eprint
  {https://arxiv.org/abs/cond-mat/9212030} {arXiv:cond-mat/9212030 [cond-mat]}
  \BibitemShut {NoStop}%
\bibitem [{\citenamefont {Kitaev}()}]{kitaev2015}%
  \BibitemOpen
  \bibfield  {author} {\bibinfo {author} {\bibfnamefont {A.}~\bibnamefont
  {Kitaev}},\ }\href {http://online.kitp.ucsb.edu/online/entangled15/}
  {\bibinfo {title} {A simple model of quantum holography}},\ \bibinfo {note}
  {string seminar at KITP and Entanglement 2015 program, 12 February, 7 April
  and 27 May 2015, http://online.kitp.ucsb.edu/online/entangled15/}\BibitemShut
  {NoStop}%
\bibitem [{\citenamefont {French}\ and\ \citenamefont
  {Wong}(1970)}]{french1970}%
  \BibitemOpen
  \bibfield  {author} {\bibinfo {author} {\bibfnamefont {J.}~\bibnamefont
  {French}}\ and\ \bibinfo {author} {\bibfnamefont {S.}~\bibnamefont {Wong}},\
  }\bibfield  {title} {\bibinfo {title} {Validity of random matrix theories for
  many-particle systems},\ }\href
  {https://doi.org/http://dx.doi.org/10.1016/0370-2693(70)90213-3} {\bibfield
  {journal} {\bibinfo  {journal} {Physics Letters B}\ }\textbf {\bibinfo
  {volume} {33}},\ \bibinfo {pages} {449 } (\bibinfo {year}
  {1970})}\BibitemShut {NoStop}%
\bibitem [{\citenamefont {Bohigas}\ and\ \citenamefont
  {Flores}(1971{\natexlab{a}})}]{bohigas1971}%
  \BibitemOpen
  \bibfield  {author} {\bibinfo {author} {\bibfnamefont {O.}~\bibnamefont
  {Bohigas}}\ and\ \bibinfo {author} {\bibfnamefont {J.}~\bibnamefont
  {Flores}},\ }\bibfield  {title} {\bibinfo {title} {Two-body random
  hamiltonian and level density},\ }\href
  {https://doi.org/http://dx.doi.org/10.1016/0370-2693(71)90598-3} {\bibfield
  {journal} {\bibinfo  {journal} {Physics Letters B}\ }\textbf {\bibinfo
  {volume} {34}},\ \bibinfo {pages} {261 } (\bibinfo {year}
  {1971}{\natexlab{a}})}\BibitemShut {NoStop}%
\bibitem [{\citenamefont {Bohigas}\ and\ \citenamefont
  {Flores}(1971{\natexlab{b}})}]{bohigas1971a}%
  \BibitemOpen
  \bibfield  {author} {\bibinfo {author} {\bibfnamefont {O.}~\bibnamefont
  {Bohigas}}\ and\ \bibinfo {author} {\bibfnamefont {J.}~\bibnamefont
  {Flores}},\ }\bibfield  {title} {\bibinfo {title} {Spacing and individual
  eigenvalue distributions of two-body random hamiltonians},\ }\href
  {https://doi.org/http://dx.doi.org/10.1016/0370-2693(71)90399-6} {\bibfield
  {journal} {\bibinfo  {journal} {Physics Letters B}\ }\textbf {\bibinfo
  {volume} {35}},\ \bibinfo {pages} {383 } (\bibinfo {year}
  {1971}{\natexlab{b}})}\BibitemShut {NoStop}%
\bibitem [{\citenamefont {French}\ and\ \citenamefont
  {Wong}(1971)}]{french1971}%
  \BibitemOpen
  \bibfield  {author} {\bibinfo {author} {\bibfnamefont {J.}~\bibnamefont
  {French}}\ and\ \bibinfo {author} {\bibfnamefont {S.}~\bibnamefont {Wong}},\
  }\bibfield  {title} {\bibinfo {title} {Some random-matrix level and spacing
  distributions for fixed-particle-rank interactions},\ }\href
  {https://doi.org/http://dx.doi.org/10.1016/0370-2693(71)90424-2} {\bibfield
  {journal} {\bibinfo  {journal} {Physics Letters B}\ }\textbf {\bibinfo
  {volume} {35}},\ \bibinfo {pages} {5 } (\bibinfo {year} {1971})}\BibitemShut
  {NoStop}%
\bibitem [{\citenamefont {Mon}\ and\ \citenamefont {French}(1975)}]{mon1975}%
  \BibitemOpen
  \bibfield  {author} {\bibinfo {author} {\bibfnamefont {K.}~\bibnamefont
  {Mon}}\ and\ \bibinfo {author} {\bibfnamefont {J.}~\bibnamefont {French}},\
  }\bibfield  {title} {\bibinfo {title} {Statistical properties of
  many-particle spectra},\ }\href
  {https://doi.org/http://dx.doi.org/10.1016/0003-4916(75)90045-7} {\bibfield
  {journal} {\bibinfo  {journal} {Annals of Physics}\ }\textbf {\bibinfo
  {volume} {95}},\ \bibinfo {pages} {90 } (\bibinfo {year} {1975})}\BibitemShut
  {NoStop}%
\bibitem [{\citenamefont {Benet}\ \emph {et~al.}(2001)\citenamefont {Benet},
  \citenamefont {Rupp},\ and\ \citenamefont {Weidenm\"uller}}]{benet2001}%
  \BibitemOpen
  \bibfield  {author} {\bibinfo {author} {\bibfnamefont {L.}~\bibnamefont
  {Benet}}, \bibinfo {author} {\bibfnamefont {T.}~\bibnamefont {Rupp}},\ and\
  \bibinfo {author} {\bibfnamefont {H.~A.}\ \bibnamefont {Weidenm\"uller}},\
  }\bibfield  {title} {\bibinfo {title} {Nonuniversal behavior of the
  $\mathit{k}$-body embedded gaussian unitary ensemble of random matrices},\
  }\href {https://doi.org/10.1103/PhysRevLett.87.010601} {\bibfield  {journal}
  {\bibinfo  {journal} {Phys. Rev. Lett.}\ }\textbf {\bibinfo {volume} {87}},\
  \bibinfo {pages} {010601} (\bibinfo {year} {2001})},\ \Eprint
  {https://arxiv.org/abs/cond-mat/0010425} {arXiv:cond-mat/0010425 [cond-mat]}
  \BibitemShut {NoStop}%
\bibitem [{\citenamefont {Jackiw}(1985)}]{jackiw1985}%
  \BibitemOpen
  \bibfield  {author} {\bibinfo {author} {\bibfnamefont {R.}~\bibnamefont
  {Jackiw}},\ }\bibfield  {title} {\bibinfo {title} {Lower dimensional
  gravity},\ }\href
  {https://doi.org/https://doi.org/10.1016/0550-3213(85)90448-1} {\bibfield
  {journal} {\bibinfo  {journal} {Nuclear Physics B}\ }\textbf {\bibinfo
  {volume} {252}},\ \bibinfo {pages} {343 } (\bibinfo {year}
  {1985})}\BibitemShut {NoStop}%
\bibitem [{\citenamefont {Teitelboim}(1983)}]{teitelboim1983}%
  \BibitemOpen
  \bibfield  {author} {\bibinfo {author} {\bibfnamefont {C.}~\bibnamefont
  {Teitelboim}},\ }\bibfield  {title} {\bibinfo {title} {Gravitation and
  hamiltonian structure in two spacetime dimensions},\ }\href
  {https://doi.org/https://doi.org/10.1016/0370-2693(83)90012-6} {\bibfield
  {journal} {\bibinfo  {journal} {Physics Letters B}\ }\textbf {\bibinfo
  {volume} {126}},\ \bibinfo {pages} {41 } (\bibinfo {year}
  {1983})}\BibitemShut {NoStop}%
\bibitem [{\citenamefont {Almheiri}\ and\ \citenamefont
  {Polchinski}(2015)}]{Almheiri:2014cka}%
  \BibitemOpen
  \bibfield  {author} {\bibinfo {author} {\bibfnamefont {A.}~\bibnamefont
  {Almheiri}}\ and\ \bibinfo {author} {\bibfnamefont {J.}~\bibnamefont
  {Polchinski}},\ }\bibfield  {title} {\bibinfo {title} {{Models of AdS$_{2}$
  backreaction and holography}},\ }\href
  {https://doi.org/10.1007/JHEP11(2015)014} {\bibfield  {journal} {\bibinfo
  {journal} {Journal of High Energy Physics}\ }\textbf {\bibinfo {volume}
  {11}},\ \bibinfo {pages} {014} (\bibinfo {year} {2015})},\ \Eprint
  {https://arxiv.org/abs/1402.6334} {arXiv:1402.6334 [hep-th]} \BibitemShut
  {NoStop}%
\bibitem [{\citenamefont {Jensen}(2016)}]{jensen2016}%
  \BibitemOpen
  \bibfield  {author} {\bibinfo {author} {\bibfnamefont {K.}~\bibnamefont
  {Jensen}},\ }\bibfield  {title} {\bibinfo {title} {Chaos in
  ${\mathrm{ads}}_{2}$ holography},\ }\href
  {https://doi.org/10.1103/PhysRevLett.117.111601} {\bibfield  {journal}
  {\bibinfo  {journal} {Phys. Rev. Lett.}\ }\textbf {\bibinfo {volume} {117}},\
  \bibinfo {pages} {111601} (\bibinfo {year} {2016})},\ \Eprint
  {https://arxiv.org/abs/1605.06098} {arXiv:1605.06098 [hep-th]} \BibitemShut
  {NoStop}%
\bibitem [{\citenamefont {Engels{\"o}y}\ \emph {et~al.}(2016)\citenamefont
  {Engels{\"o}y}, \citenamefont {Mertens},\ and\ \citenamefont
  {Verlinde}}]{engels2016}%
  \BibitemOpen
  \bibfield  {author} {\bibinfo {author} {\bibfnamefont {J.}~\bibnamefont
  {Engels{\"o}y}}, \bibinfo {author} {\bibfnamefont {T.~G.}\ \bibnamefont
  {Mertens}},\ and\ \bibinfo {author} {\bibfnamefont {H.}~\bibnamefont
  {Verlinde}},\ }\bibfield  {title} {\bibinfo {title} {An investigation of ads2
  backreaction and holography},\ }\href
  {https://doi.org/10.1007/JHEP07(2016)139} {\bibfield  {journal} {\bibinfo
  {journal} {Journal of High Energy Physics}\ }\textbf {\bibinfo {volume}
  {07}},\ \bibinfo {pages} {1} (\bibinfo {year} {2016})},\ \Eprint
  {https://arxiv.org/abs/1606.03438} {arXiv:1606.03438 [hep-th]} \BibitemShut
  {NoStop}%
\bibitem [{\citenamefont {Maldacena}\ \emph
  {et~al.}(2016{\natexlab{a}})\citenamefont {Maldacena}, \citenamefont
  {Stanford},\ and\ \citenamefont {Yang}}]{maldacena2016a}%
  \BibitemOpen
  \bibfield  {author} {\bibinfo {author} {\bibfnamefont {J.}~\bibnamefont
  {Maldacena}}, \bibinfo {author} {\bibfnamefont {D.}~\bibnamefont
  {Stanford}},\ and\ \bibinfo {author} {\bibfnamefont {Z.}~\bibnamefont
  {Yang}},\ }\bibfield  {title} {\bibinfo {title} {Conformal symmetry and its
  breaking in two-dimensional nearly anti-de sitter space},\ }\href
  {https://doi.org/10.1093/ptep/ptw124} {\bibfield  {journal} {\bibinfo
  {journal} {Progress of Theoretical and Experimental Physics}\ }\textbf
  {\bibinfo {volume} {2016}},\ \bibinfo {pages} {12C104} (\bibinfo {year}
  {2016}{\natexlab{a}})},\ \Eprint {https://arxiv.org/abs/1606.01857}
  {arXiv:1606.01857 [hep-th]} \BibitemShut {NoStop}%
\bibitem [{\citenamefont {Nayak}\ \emph {et~al.}(2018)\citenamefont {Nayak},
  \citenamefont {Shukla}, \citenamefont {Soni}, \citenamefont {Trivedi},\ and\
  \citenamefont {Vishal}}]{Nayak:2018qej}%
  \BibitemOpen
  \bibfield  {author} {\bibinfo {author} {\bibfnamefont {P.}~\bibnamefont
  {Nayak}}, \bibinfo {author} {\bibfnamefont {A.}~\bibnamefont {Shukla}},
  \bibinfo {author} {\bibfnamefont {R.~M.}\ \bibnamefont {Soni}}, \bibinfo
  {author} {\bibfnamefont {S.~P.}\ \bibnamefont {Trivedi}},\ and\ \bibinfo
  {author} {\bibfnamefont {V.}~\bibnamefont {Vishal}},\ }\bibfield  {title}
  {\bibinfo {title} {{On the Dynamics of Near-Extremal Black Holes}},\ }\href
  {https://doi.org/10.1007/JHEP09(2018)048} {\bibfield  {journal} {\bibinfo
  {journal} {JHEP}\ }\textbf {\bibinfo {volume} {09}},\ \bibinfo {pages}
  {048}},\ \Eprint {https://arxiv.org/abs/1802.09547} {arXiv:1802.09547
  [hep-th]} \BibitemShut {NoStop}%
\bibitem [{\citenamefont {Moitra}\ \emph {et~al.}(2019)\citenamefont {Moitra},
  \citenamefont {Sake}, \citenamefont {Trivedi},\ and\ \citenamefont
  {Vishal}}]{Moitra:2019bub}%
  \BibitemOpen
  \bibfield  {author} {\bibinfo {author} {\bibfnamefont {U.}~\bibnamefont
  {Moitra}}, \bibinfo {author} {\bibfnamefont {S.~K.}\ \bibnamefont {Sake}},
  \bibinfo {author} {\bibfnamefont {S.~P.}\ \bibnamefont {Trivedi}},\ and\
  \bibinfo {author} {\bibfnamefont {V.}~\bibnamefont {Vishal}},\ }\bibfield
  {title} {\bibinfo {title} {{Jackiw-Teitelboim Gravity and Rotating Black
  Holes}},\ }\href {https://doi.org/10.1007/JHEP11(2019)047} {\bibfield
  {journal} {\bibinfo  {journal} {JHEP}\ }\textbf {\bibinfo {volume} {11}},\
  \bibinfo {pages} {047}},\ \Eprint {https://arxiv.org/abs/1905.10378}
  {arXiv:1905.10378 [hep-th]} \BibitemShut {NoStop}%
\bibitem [{\citenamefont {Castro}\ and\ \citenamefont
  {Godet}(2020)}]{Castro:2019crn}%
  \BibitemOpen
  \bibfield  {author} {\bibinfo {author} {\bibfnamefont {A.}~\bibnamefont
  {Castro}}\ and\ \bibinfo {author} {\bibfnamefont {V.}~\bibnamefont {Godet}},\
  }\bibfield  {title} {\bibinfo {title} {{Breaking away from the near horizon
  of extreme Kerr}},\ }\href {https://doi.org/10.21468/SciPostPhys.8.6.089}
  {\bibfield  {journal} {\bibinfo  {journal} {SciPost Phys.}\ }\textbf
  {\bibinfo {volume} {8}},\ \bibinfo {pages} {089} (\bibinfo {year} {2020})},\
  \Eprint {https://arxiv.org/abs/1906.09083} {arXiv:1906.09083 [hep-th]}
  \BibitemShut {NoStop}%
\bibitem [{\citenamefont {Iliesiu}\ and\ \citenamefont
  {Turiaci}(2021)}]{Iliesiu:2020qvm}%
  \BibitemOpen
  \bibfield  {author} {\bibinfo {author} {\bibfnamefont {L.~V.}\ \bibnamefont
  {Iliesiu}}\ and\ \bibinfo {author} {\bibfnamefont {G.~J.}\ \bibnamefont
  {Turiaci}},\ }\bibfield  {title} {\bibinfo {title} {{The statistical
  mechanics of near-extremal black holes}},\ }\href
  {https://doi.org/10.1007/JHEP05(2021)145} {\bibfield  {journal} {\bibinfo
  {journal} {JHEP}\ }\textbf {\bibinfo {volume} {05}},\ \bibinfo {pages}
  {145}},\ \Eprint {https://arxiv.org/abs/2003.02860} {arXiv:2003.02860
  [hep-th]} \BibitemShut {NoStop}%
\bibitem [{\citenamefont {Cotler}\ \emph {et~al.}(2017)\citenamefont {Cotler},
  \citenamefont {Gur-Ari}, \citenamefont {Hanada}, \citenamefont {Polchinski},
  \citenamefont {Saad}, \citenamefont {Shenker}, \citenamefont {Stanford},
  \citenamefont {Streicher},\ and\ \citenamefont {Tezuka}}]{Cotler:2016fpe}%
  \BibitemOpen
  \bibfield  {author} {\bibinfo {author} {\bibfnamefont {J.~S.}\ \bibnamefont
  {Cotler}}, \bibinfo {author} {\bibfnamefont {G.}~\bibnamefont {Gur-Ari}},
  \bibinfo {author} {\bibfnamefont {M.}~\bibnamefont {Hanada}}, \bibinfo
  {author} {\bibfnamefont {J.}~\bibnamefont {Polchinski}}, \bibinfo {author}
  {\bibfnamefont {P.}~\bibnamefont {Saad}}, \bibinfo {author} {\bibfnamefont
  {S.~H.}\ \bibnamefont {Shenker}}, \bibinfo {author} {\bibfnamefont
  {D.}~\bibnamefont {Stanford}}, \bibinfo {author} {\bibfnamefont
  {A.}~\bibnamefont {Streicher}},\ and\ \bibinfo {author} {\bibfnamefont
  {M.}~\bibnamefont {Tezuka}},\ }\bibfield  {title} {\bibinfo {title} {{Black
  Holes and Random Matrices}},\ }\href
  {https://doi.org/10.1007/JHEP05(2017)118} {\bibfield  {journal} {\bibinfo
  {journal} {Journal of High Energy Physics}\ }\textbf {\bibinfo {volume}
  {05}},\ \bibinfo {pages} {118} (\bibinfo {year} {2017})},\ \Eprint
  {https://arxiv.org/abs/1611.04650} {arXiv:1611.04650 [hep-th]} \BibitemShut
  {NoStop}%
\bibitem [{\citenamefont {Garc\'{\i}a-Garc\'{\i}a}\ and\ \citenamefont
  {Verbaarschot}(2016)}]{garcia2016}%
  \BibitemOpen
  \bibfield  {author} {\bibinfo {author} {\bibfnamefont {A.~M.}\ \bibnamefont
  {Garc\'{\i}a-Garc\'{\i}a}}\ and\ \bibinfo {author} {\bibfnamefont {J.~J.~M.}\
  \bibnamefont {Verbaarschot}},\ }\bibfield  {title} {\bibinfo {title}
  {Spectral and thermodynamic properties of the sachdev-ye-kitaev model},\
  }\href {https://doi.org/10.1103/PhysRevD.94.126010} {\bibfield  {journal}
  {\bibinfo  {journal} {Phys. Rev. D}\ }\textbf {\bibinfo {volume} {94}},\
  \bibinfo {pages} {126010} (\bibinfo {year} {2016})},\ \Eprint
  {https://arxiv.org/abs/1610.03816} {arXiv:1610.03816 [hep-th]} \BibitemShut
  {NoStop}%
\bibitem [{\citenamefont {Garc\'{\i}a-Garc\'{\i}a}\ and\ \citenamefont
  {Verbaarschot}(2017)}]{garcia2017}%
  \BibitemOpen
  \bibfield  {author} {\bibinfo {author} {\bibfnamefont {A.~M.}\ \bibnamefont
  {Garc\'{\i}a-Garc\'{\i}a}}\ and\ \bibinfo {author} {\bibfnamefont {J.~J.~M.}\
  \bibnamefont {Verbaarschot}},\ }\bibfield  {title} {\bibinfo {title}
  {Analytical spectral density of the sachdev-ye-kitaev model at finite $n$},\
  }\href {https://doi.org/10.1103/PhysRevD.96.066012} {\bibfield  {journal}
  {\bibinfo  {journal} {Phys. Rev. D}\ }\textbf {\bibinfo {volume} {96}},\
  \bibinfo {pages} {066012} (\bibinfo {year} {2017})},\ \Eprint
  {https://arxiv.org/abs/1701.06593} {arXiv:1701.06593 [hep-th]} \BibitemShut
  {NoStop}%
\bibitem [{\citenamefont {Saad}\ \emph {et~al.}(2018)\citenamefont {Saad},
  \citenamefont {Shenker},\ and\ \citenamefont {Stanford}}]{Saad:2018bqo}%
  \BibitemOpen
  \bibfield  {author} {\bibinfo {author} {\bibfnamefont {P.}~\bibnamefont
  {Saad}}, \bibinfo {author} {\bibfnamefont {S.~H.}\ \bibnamefont {Shenker}},\
  and\ \bibinfo {author} {\bibfnamefont {D.}~\bibnamefont {Stanford}},\
  }\bibfield  {title} {\bibinfo {title} {{A semiclassical ramp in SYK and in
  gravity}},\ }\href@noop {} {\bibfield  {journal} {\bibinfo  {journal}
  {eprint}\ } (\bibinfo {year} {2018})},\ \Eprint
  {https://arxiv.org/abs/1806.06840} {arXiv:1806.06840 [hep-th]} \BibitemShut
  {NoStop}%
\bibitem [{\citenamefont {Saad}\ \emph {et~al.}(2019)\citenamefont {Saad},
  \citenamefont {Shenker},\ and\ \citenamefont {Stanford}}]{Saad:2019lba}%
  \BibitemOpen
  \bibfield  {author} {\bibinfo {author} {\bibfnamefont {P.}~\bibnamefont
  {Saad}}, \bibinfo {author} {\bibfnamefont {S.~H.}\ \bibnamefont {Shenker}},\
  and\ \bibinfo {author} {\bibfnamefont {D.}~\bibnamefont {Stanford}},\
  }\bibfield  {title} {\bibinfo {title} {{JT gravity as a matrix integral}},\
  }\href@noop {} {\bibfield  {journal} {\bibinfo  {journal} {eprint}\ }
  (\bibinfo {year} {2019})},\ \Eprint {https://arxiv.org/abs/1903.11115}
  {arXiv:1903.11115 [hep-th]} \BibitemShut {NoStop}%
\bibitem [{\citenamefont {Saad}(2019)}]{Saad:2019pqd}%
  \BibitemOpen
  \bibfield  {author} {\bibinfo {author} {\bibfnamefont {P.}~\bibnamefont
  {Saad}},\ }\bibfield  {title} {\bibinfo {title} {{Late Time Correlation
  Functions, Baby Universes, and ETH in JT Gravity}},\ }\href@noop {} {\
  (\bibinfo {year} {2019})},\ \Eprint {https://arxiv.org/abs/1910.10311}
  {arXiv:1910.10311 [hep-th]} \BibitemShut {NoStop}%
\bibitem [{\citenamefont {Maldacena}\ \emph {et~al.}(2017)\citenamefont
  {Maldacena}, \citenamefont {Stanford},\ and\ \citenamefont
  {Yang}}]{maldacena2017}%
  \BibitemOpen
  \bibfield  {author} {\bibinfo {author} {\bibfnamefont {J.}~\bibnamefont
  {Maldacena}}, \bibinfo {author} {\bibfnamefont {D.}~\bibnamefont
  {Stanford}},\ and\ \bibinfo {author} {\bibfnamefont {Z.}~\bibnamefont
  {Yang}},\ }\bibfield  {title} {\bibinfo {title} {{Diving into traversable
  wormholes}},\ }\href {https://doi.org/10.1002/prop.201700034} {\bibfield
  {journal} {\bibinfo  {journal} {Fortsch. Phys.}\ }\textbf {\bibinfo {volume}
  {65}},\ \bibinfo {pages} {1700034} (\bibinfo {year} {2017})},\ \Eprint
  {https://arxiv.org/abs/1704.05333} {arXiv:1704.05333 [hep-th]} \BibitemShut
  {NoStop}%
\bibitem [{\citenamefont {Maldacena}\ and\ \citenamefont
  {Qi}(2018)}]{maldacena2018}%
  \BibitemOpen
  \bibfield  {author} {\bibinfo {author} {\bibfnamefont {J.}~\bibnamefont
  {Maldacena}}\ and\ \bibinfo {author} {\bibfnamefont {X.-L.}\ \bibnamefont
  {Qi}},\ }\bibfield  {title} {\bibinfo {title} {{Eternal traversable
  wormhole}},\ }\href@noop {} {\bibfield  {journal} {\bibinfo  {journal}
  {eprint}\ } (\bibinfo {year} {2018})},\ \Eprint
  {https://arxiv.org/abs/1804.00491} {arXiv:1804.00491 [hep-th]} \BibitemShut
  {NoStop}%
\bibitem [{\citenamefont {Maldacena}\ and\ \citenamefont
  {Milekhin}(2021)}]{milekhin2019}%
  \BibitemOpen
  \bibfield  {author} {\bibinfo {author} {\bibfnamefont {J.}~\bibnamefont
  {Maldacena}}\ and\ \bibinfo {author} {\bibfnamefont {A.}~\bibnamefont
  {Milekhin}},\ }\bibfield  {title} {\bibinfo {title} {Syk wormhole formation
  in real time},\ }\href {https://doi.org/10.1007/JHEP04(2021)258} {\bibfield
  {journal} {\bibinfo  {journal} {Journal of High Energy Physics}\ }\textbf
  {\bibinfo {volume} {04}},\ \bibinfo {pages} {258} (\bibinfo {year} {2021})},\
  \Eprint {https://arxiv.org/abs/1912.03276} {arXiv:1912.03276 [hep-th]}
  \BibitemShut {NoStop}%
\bibitem [{\citenamefont {Maldacena}\ \emph {et~al.}(2021)\citenamefont
  {Maldacena}, \citenamefont {Turiaci},\ and\ \citenamefont
  {Yang}}]{Maldacena:2019cbz}%
  \BibitemOpen
  \bibfield  {author} {\bibinfo {author} {\bibfnamefont {J.}~\bibnamefont
  {Maldacena}}, \bibinfo {author} {\bibfnamefont {G.~J.}\ \bibnamefont
  {Turiaci}},\ and\ \bibinfo {author} {\bibfnamefont {Z.}~\bibnamefont
  {Yang}},\ }\bibfield  {title} {\bibinfo {title} {{Two dimensional Nearly de
  Sitter gravity}},\ }\href {https://doi.org/10.1007/JHEP01(2021)139}
  {\bibfield  {journal} {\bibinfo  {journal} {Journal of High Energy Physics}\
  }\textbf {\bibinfo {volume} {01}},\ \bibinfo {pages} {139} (\bibinfo {year}
  {2021})},\ \Eprint {https://arxiv.org/abs/1904.01911} {arXiv:1904.01911
  [hep-th]} \BibitemShut {NoStop}%
\bibitem [{\citenamefont {Anous}\ \emph {et~al.}(2020)\citenamefont {Anous},
  \citenamefont {Kruthoff},\ and\ \citenamefont {Mahajan}}]{Anous:2020lka}%
  \BibitemOpen
  \bibfield  {author} {\bibinfo {author} {\bibfnamefont {T.}~\bibnamefont
  {Anous}}, \bibinfo {author} {\bibfnamefont {J.}~\bibnamefont {Kruthoff}},\
  and\ \bibinfo {author} {\bibfnamefont {R.}~\bibnamefont {Mahajan}},\
  }\bibfield  {title} {\bibinfo {title} {{Density matrices in quantum
  gravity}},\ }\href {https://doi.org/10.21468/SciPostPhys.9.4.045} {\bibfield
  {journal} {\bibinfo  {journal} {SciPost Phys.}\ }\textbf {\bibinfo {volume}
  {9}},\ \bibinfo {pages} {045} (\bibinfo {year} {2020})},\ \Eprint
  {https://arxiv.org/abs/2006.17000} {arXiv:2006.17000 [hep-th]} \BibitemShut
  {NoStop}%
\bibitem [{\citenamefont {Chen}\ \emph {et~al.}(2021)\citenamefont {Chen},
  \citenamefont {Gorbenko},\ and\ \citenamefont {Maldacena}}]{Chen:2020tes}%
  \BibitemOpen
  \bibfield  {author} {\bibinfo {author} {\bibfnamefont {Y.}~\bibnamefont
  {Chen}}, \bibinfo {author} {\bibfnamefont {V.}~\bibnamefont {Gorbenko}},\
  and\ \bibinfo {author} {\bibfnamefont {J.}~\bibnamefont {Maldacena}},\
  }\bibfield  {title} {\bibinfo {title} {{Bra-ket wormholes in gravitationally
  prepared states}},\ }\href
  {https://doi.org/https://doi.org/10.1007/JHEP02(2021)009} {\bibfield
  {journal} {\bibinfo  {journal} {Journal of High Energy Physics}\ }\textbf
  {\bibinfo {volume} {2021}},\ \bibinfo {pages} {9} (\bibinfo {year} {2021})},\
  \Eprint {https://arxiv.org/abs/2007.16091} {arXiv:2007.16091 [hep-th]}
  \BibitemShut {NoStop}%
\bibitem [{\citenamefont {Hartman}\ \emph {et~al.}(2020)\citenamefont
  {Hartman}, \citenamefont {Jiang},\ and\ \citenamefont
  {Shaghoulian}}]{Hartman:2020khs}%
  \BibitemOpen
  \bibfield  {author} {\bibinfo {author} {\bibfnamefont {T.}~\bibnamefont
  {Hartman}}, \bibinfo {author} {\bibfnamefont {Y.}~\bibnamefont {Jiang}},\
  and\ \bibinfo {author} {\bibfnamefont {E.}~\bibnamefont {Shaghoulian}},\
  }\bibfield  {title} {\bibinfo {title} {{Islands in cosmology}},\ }\href
  {https://doi.org/10.1007/JHEP11(2020)111} {\bibfield  {journal} {\bibinfo
  {journal} {JHEP}\ }\textbf {\bibinfo {volume} {11}},\ \bibinfo {pages}
  {111}},\ \Eprint {https://arxiv.org/abs/2008.01022} {arXiv:2008.01022
  [hep-th]} \BibitemShut {NoStop}%
\bibitem [{\citenamefont {Aalsma}\ and\ \citenamefont
  {Sybesma}(2021)}]{Aalsma:2021bit}%
  \BibitemOpen
  \bibfield  {author} {\bibinfo {author} {\bibfnamefont {L.}~\bibnamefont
  {Aalsma}}\ and\ \bibinfo {author} {\bibfnamefont {W.}~\bibnamefont
  {Sybesma}},\ }\bibfield  {title} {\bibinfo {title} {{The Price of Curiosity:
  Information Recovery in de Sitter Space}},\ }\href
  {https://doi.org/10.1007/JHEP05(2021)291} {\bibfield  {journal} {\bibinfo
  {journal} {Journal of High Energy Physics}\ }\textbf {\bibinfo {volume}
  {05}},\ \bibinfo {pages} {291} (\bibinfo {year} {2021})},\ \Eprint
  {https://arxiv.org/abs/2104.00006} {arXiv:2104.00006 [hep-th]} \BibitemShut
  {NoStop}%
\bibitem [{\citenamefont {Kames-King}\ \emph {et~al.}(2022)\citenamefont
  {Kames-King}, \citenamefont {Verheijden},\ and\ \citenamefont
  {Verlinde}}]{Kames-King:2021etp}%
  \BibitemOpen
  \bibfield  {author} {\bibinfo {author} {\bibfnamefont {J.}~\bibnamefont
  {Kames-King}}, \bibinfo {author} {\bibfnamefont {E.~M.~H.}\ \bibnamefont
  {Verheijden}},\ and\ \bibinfo {author} {\bibfnamefont {E.~P.}\ \bibnamefont
  {Verlinde}},\ }\bibfield  {title} {\bibinfo {title} {{No Page curves for the
  de Sitter horizon}},\ }\href {https://doi.org/10.1007/JHEP03(2022)040}
  {\bibfield  {journal} {\bibinfo  {journal} {Journal of High Energy Physics}\
  }\textbf {\bibinfo {volume} {03}},\ \bibinfo {pages} {040} (\bibinfo {year}
  {2022})},\ \Eprint {https://arxiv.org/abs/2108.09318} {arXiv:2108.09318
  [hep-th]} \BibitemShut {NoStop}%
\bibitem [{\citenamefont {Dubovsky}\ \emph {et~al.}(2017)\citenamefont
  {Dubovsky}, \citenamefont {Gorbenko},\ and\ \citenamefont
  {Mirbabayi}}]{Dubovsky:2017cnj}%
  \BibitemOpen
  \bibfield  {author} {\bibinfo {author} {\bibfnamefont {S.}~\bibnamefont
  {Dubovsky}}, \bibinfo {author} {\bibfnamefont {V.}~\bibnamefont {Gorbenko}},\
  and\ \bibinfo {author} {\bibfnamefont {M.}~\bibnamefont {Mirbabayi}},\
  }\bibfield  {title} {\bibinfo {title} {{Asymptotic fragility, near AdS$_{2}$
  holography and $ T\overline{T} $}},\ }\href
  {https://doi.org/10.1007/JHEP09(2017)136} {\bibfield  {journal} {\bibinfo
  {journal} {JHEP}\ }\textbf {\bibinfo {volume} {09}},\ \bibinfo {pages}
  {136}},\ \Eprint {https://arxiv.org/abs/1706.06604} {arXiv:1706.06604
  [hep-th]} \BibitemShut {NoStop}%
\bibitem [{\citenamefont {Afshar}\ \emph {et~al.}(2020)\citenamefont {Afshar},
  \citenamefont {Gonz\'alez}, \citenamefont {Grumiller},\ and\ \citenamefont
  {Vassilevich}}]{Afshar:2019axx}%
  \BibitemOpen
  \bibfield  {author} {\bibinfo {author} {\bibfnamefont {H.}~\bibnamefont
  {Afshar}}, \bibinfo {author} {\bibfnamefont {H.~A.}\ \bibnamefont
  {Gonz\'alez}}, \bibinfo {author} {\bibfnamefont {D.}~\bibnamefont
  {Grumiller}},\ and\ \bibinfo {author} {\bibfnamefont {D.}~\bibnamefont
  {Vassilevich}},\ }\bibfield  {title} {\bibinfo {title} {{Flat space
  holography and the complex Sachdev-Ye-Kitaev model}},\ }\href
  {https://doi.org/10.1103/PhysRevD.101.086024} {\bibfield  {journal} {\bibinfo
   {journal} {Phys. Rev. D}\ }\textbf {\bibinfo {volume} {101}},\ \bibinfo
  {pages} {086024} (\bibinfo {year} {2020})},\ \Eprint
  {https://arxiv.org/abs/1911.05739} {arXiv:1911.05739 [hep-th]} \BibitemShut
  {NoStop}%
\bibitem [{\citenamefont {Gautason}\ \emph {et~al.}(2020)\citenamefont
  {Gautason}, \citenamefont {Schneiderbauer}, \citenamefont {Sybesma},\ and\
  \citenamefont {Thorlacius}}]{Gautason:2020tmk}%
  \BibitemOpen
  \bibfield  {author} {\bibinfo {author} {\bibfnamefont {F.~F.}\ \bibnamefont
  {Gautason}}, \bibinfo {author} {\bibfnamefont {L.}~\bibnamefont
  {Schneiderbauer}}, \bibinfo {author} {\bibfnamefont {W.}~\bibnamefont
  {Sybesma}},\ and\ \bibinfo {author} {\bibfnamefont {L.}~\bibnamefont
  {Thorlacius}},\ }\bibfield  {title} {\bibinfo {title} {{Page Curve for an
  Evaporating Black Hole}},\ }\href {https://doi.org/10.1007/JHEP05(2020)091}
  {\bibfield  {journal} {\bibinfo  {journal} {JHEP}\ }\textbf {\bibinfo
  {volume} {05}},\ \bibinfo {pages} {091}},\ \Eprint
  {https://arxiv.org/abs/2004.00598} {arXiv:2004.00598 [hep-th]} \BibitemShut
  {NoStop}%
\bibitem [{\citenamefont {Godet}\ and\ \citenamefont
  {Marteau}(2021)}]{Godet:2021cdl}%
  \BibitemOpen
  \bibfield  {author} {\bibinfo {author} {\bibfnamefont {V.}~\bibnamefont
  {Godet}}\ and\ \bibinfo {author} {\bibfnamefont {C.}~\bibnamefont
  {Marteau}},\ }\bibfield  {title} {\bibinfo {title} {{From black holes to baby
  universes in CGHS gravity}},\ }\href
  {https://doi.org/10.1007/JHEP07(2021)138} {\bibfield  {journal} {\bibinfo
  {journal} {JHEP}\ }\textbf {\bibinfo {volume} {07}},\ \bibinfo {pages}
  {138}},\ \Eprint {https://arxiv.org/abs/2103.13422} {arXiv:2103.13422
  [hep-th]} \BibitemShut {NoStop}%
\bibitem [{\citenamefont {Lin}\ and\ \citenamefont
  {Susskind}(2020)}]{Lin:2019kpf}%
  \BibitemOpen
  \bibfield  {author} {\bibinfo {author} {\bibfnamefont {H.~W.}\ \bibnamefont
  {Lin}}\ and\ \bibinfo {author} {\bibfnamefont {L.}~\bibnamefont {Susskind}},\
  }\bibfield  {title} {\bibinfo {title} {{Complexity Geometry and Schwarzian
  Dynamics}},\ }\href {https://doi.org/10.1007/JHEP01(2020)087} {\bibfield
  {journal} {\bibinfo  {journal} {Journal of High Energy Physics}\ }\textbf
  {\bibinfo {volume} {01}},\ \bibinfo {pages} {087} (\bibinfo {year} {2020})},\
  \Eprint {https://arxiv.org/abs/1911.02603} {arXiv:1911.02603 [hep-th]}
  \BibitemShut {NoStop}%
\bibitem [{\citenamefont {Iliesiu}\ \emph
  {et~al.}(2021{\natexlab{a}})\citenamefont {Iliesiu}, \citenamefont {Mezei},\
  and\ \citenamefont {S\'arosi}}]{Iliesiu:2021ari}%
  \BibitemOpen
  \bibfield  {author} {\bibinfo {author} {\bibfnamefont {L.~V.}\ \bibnamefont
  {Iliesiu}}, \bibinfo {author} {\bibfnamefont {M.}~\bibnamefont {Mezei}},\
  and\ \bibinfo {author} {\bibfnamefont {G.}~\bibnamefont {S\'arosi}},\
  }\bibfield  {title} {\bibinfo {title} {{The volume of the black hole interior
  at late times}},\ }\href@noop {} {\bibfield  {journal} {\bibinfo  {journal}
  {eprint}\ } (\bibinfo {year} {2021}{\natexlab{a}})},\ \Eprint
  {https://arxiv.org/abs/2107.06286} {arXiv:2107.06286 [hep-th]} \BibitemShut
  {NoStop}%
\bibitem [{\citenamefont {Almheiri}\ \emph {et~al.}(2018)\citenamefont
  {Almheiri}, \citenamefont {Mousatov},\ and\ \citenamefont
  {Shyani}}]{Almheiri:2018ijj}%
  \BibitemOpen
  \bibfield  {author} {\bibinfo {author} {\bibfnamefont {A.}~\bibnamefont
  {Almheiri}}, \bibinfo {author} {\bibfnamefont {A.}~\bibnamefont {Mousatov}},\
  and\ \bibinfo {author} {\bibfnamefont {M.}~\bibnamefont {Shyani}},\
  }\bibfield  {title} {\bibinfo {title} {{Escaping the Interiors of Pure
  Boundary-State Black Holes}},\ }\href@noop {} {\  (\bibinfo {year} {2018})},\
  \Eprint {https://arxiv.org/abs/1803.04434} {arXiv:1803.04434 [hep-th]}
  \BibitemShut {NoStop}%
\bibitem [{\citenamefont {Almheiri}\ \emph {et~al.}(2019)\citenamefont
  {Almheiri}, \citenamefont {Engelhardt}, \citenamefont {Marolf},\ and\
  \citenamefont {Maxfield}}]{Almheiri:2019psf}%
  \BibitemOpen
  \bibfield  {author} {\bibinfo {author} {\bibfnamefont {A.}~\bibnamefont
  {Almheiri}}, \bibinfo {author} {\bibfnamefont {N.}~\bibnamefont
  {Engelhardt}}, \bibinfo {author} {\bibfnamefont {D.}~\bibnamefont {Marolf}},\
  and\ \bibinfo {author} {\bibfnamefont {H.}~\bibnamefont {Maxfield}},\
  }\bibfield  {title} {\bibinfo {title} {{The entropy of bulk quantum fields
  and the entanglement wedge of an evaporating black hole}},\ }\href
  {https://doi.org/10.1007/JHEP12(2019)063} {\bibfield  {journal} {\bibinfo
  {journal} {Journal of High Energy Physics}\ }\textbf {\bibinfo {volume}
  {12}},\ \bibinfo {pages} {063} (\bibinfo {year} {2019})},\ \Eprint
  {https://arxiv.org/abs/1905.08762} {arXiv:1905.08762 [hep-th]} \BibitemShut
  {NoStop}%
\bibitem [{\citenamefont {Almheiri}\ \emph
  {et~al.}(2020{\natexlab{a}})\citenamefont {Almheiri}, \citenamefont
  {Mahajan}, \citenamefont {Maldacena},\ and\ \citenamefont
  {Zhao}}]{Almheiri:2019hni}%
  \BibitemOpen
  \bibfield  {author} {\bibinfo {author} {\bibfnamefont {A.}~\bibnamefont
  {Almheiri}}, \bibinfo {author} {\bibfnamefont {R.}~\bibnamefont {Mahajan}},
  \bibinfo {author} {\bibfnamefont {J.}~\bibnamefont {Maldacena}},\ and\
  \bibinfo {author} {\bibfnamefont {Y.}~\bibnamefont {Zhao}},\ }\bibfield
  {title} {\bibinfo {title} {{The Page curve of Hawking radiation from
  semiclassical geometry}},\ }\href {https://doi.org/10.1007/JHEP03(2020)149}
  {\bibfield  {journal} {\bibinfo  {journal} {Journal of High Energy Physics}\
  }\textbf {\bibinfo {volume} {03}},\ \bibinfo {pages} {149} (\bibinfo {year}
  {2020}{\natexlab{a}})},\ \Eprint {https://arxiv.org/abs/1908.10996}
  {arXiv:1908.10996 [hep-th]} \BibitemShut {NoStop}%
\bibitem [{\citenamefont {Almheiri}\ \emph
  {et~al.}(2020{\natexlab{b}})\citenamefont {Almheiri}, \citenamefont
  {Hartman}, \citenamefont {Maldacena}, \citenamefont {Shaghoulian},\ and\
  \citenamefont {Tajdini}}]{almheiri2020}%
  \BibitemOpen
  \bibfield  {author} {\bibinfo {author} {\bibfnamefont {A.}~\bibnamefont
  {Almheiri}}, \bibinfo {author} {\bibfnamefont {T.}~\bibnamefont {Hartman}},
  \bibinfo {author} {\bibfnamefont {J.}~\bibnamefont {Maldacena}}, \bibinfo
  {author} {\bibfnamefont {E.}~\bibnamefont {Shaghoulian}},\ and\ \bibinfo
  {author} {\bibfnamefont {A.}~\bibnamefont {Tajdini}},\ }\bibfield  {title}
  {\bibinfo {title} {Replica wormholes and the entropy of hawking radiation},\
  }\href {https://doi.org/10.1007/jhep05(2020)013} {\bibfield  {journal}
  {\bibinfo  {journal} {Journal of High Energy Physics}\ }\textbf {\bibinfo
  {volume} {2020}},\ \bibinfo {pages} {13} (\bibinfo {year}
  {2020}{\natexlab{b}})},\ \Eprint {https://arxiv.org/abs/1911.12333}
  {arXiv:1911.12333 [hep-th]} \BibitemShut {NoStop}%
\bibitem [{\citenamefont {Penington}\ \emph {et~al.}(2022)\citenamefont
  {Penington}, \citenamefont {Shenker}, \citenamefont {Stanford},\ and\
  \citenamefont {Yang}}]{penington2020}%
  \BibitemOpen
  \bibfield  {author} {\bibinfo {author} {\bibfnamefont {G.}~\bibnamefont
  {Penington}}, \bibinfo {author} {\bibfnamefont {S.~H.}\ \bibnamefont
  {Shenker}}, \bibinfo {author} {\bibfnamefont {D.}~\bibnamefont {Stanford}},\
  and\ \bibinfo {author} {\bibfnamefont {Z.}~\bibnamefont {Yang}},\ }\bibfield
  {title} {\bibinfo {title} {Replica wormholes and the black hole interior},\
  }\href {https://doi.org/10.1007/JHEP03(2022)205} {\bibfield  {journal}
  {\bibinfo  {journal} {Journal of High Energy Physics}\ }\textbf {\bibinfo
  {volume} {2022}},\ \bibinfo {pages} {205} (\bibinfo {year} {2022})},\ \Eprint
  {https://arxiv.org/abs/1911.11977} {arXiv:1911.11977 [hep-th]} \BibitemShut
  {NoStop}%
\bibitem [{\citenamefont {Almheiri}\ \emph {et~al.}(2021)\citenamefont
  {Almheiri}, \citenamefont {Hartman}, \citenamefont {Maldacena}, \citenamefont
  {Shaghoulian},\ and\ \citenamefont {Tajdini}}]{almheiri2020a}%
  \BibitemOpen
  \bibfield  {author} {\bibinfo {author} {\bibfnamefont {A.}~\bibnamefont
  {Almheiri}}, \bibinfo {author} {\bibfnamefont {T.}~\bibnamefont {Hartman}},
  \bibinfo {author} {\bibfnamefont {J.}~\bibnamefont {Maldacena}}, \bibinfo
  {author} {\bibfnamefont {E.}~\bibnamefont {Shaghoulian}},\ and\ \bibinfo
  {author} {\bibfnamefont {A.}~\bibnamefont {Tajdini}},\ }\bibfield  {title}
  {\bibinfo {title} {The entropy of hawking radiation},\ }\href
  {https://doi.org/10.1103/RevModPhys.93.035002} {\bibfield  {journal}
  {\bibinfo  {journal} {Rev. Mod. Phys.}\ }\textbf {\bibinfo {volume} {93}},\
  \bibinfo {pages} {035002} (\bibinfo {year} {2021})},\ \Eprint
  {https://arxiv.org/abs/2006.06872} {arXiv:2006.06872 [hep-th]} \BibitemShut
  {NoStop}%
\bibitem [{\citenamefont {Stanford}(2020)}]{Stanford:2020wkf}%
  \BibitemOpen
  \bibfield  {author} {\bibinfo {author} {\bibfnamefont {D.}~\bibnamefont
  {Stanford}},\ }\bibfield  {title} {\bibinfo {title} {{More quantum noise from
  wormholes}},\ }\href@noop {} {\  (\bibinfo {year} {2020})},\ \Eprint
  {https://arxiv.org/abs/2008.08570} {arXiv:2008.08570 [hep-th]} \BibitemShut
  {NoStop}%
\bibitem [{\citenamefont {Gao}\ and\ \citenamefont
  {Lamprou}(2021)}]{Gao:2021tzr}%
  \BibitemOpen
  \bibfield  {author} {\bibinfo {author} {\bibfnamefont {P.}~\bibnamefont
  {Gao}}\ and\ \bibinfo {author} {\bibfnamefont {L.}~\bibnamefont {Lamprou}},\
  }\bibfield  {title} {\bibinfo {title} {{Seeing behind black hole horizons in
  SYK}},\ }\href@noop {} {\bibfield  {journal} {\bibinfo  {journal} {eprint}\ }
  (\bibinfo {year} {2021})},\ \Eprint {https://arxiv.org/abs/2111.14010}
  {arXiv:2111.14010 [hep-th]} \BibitemShut {NoStop}%
\bibitem [{\citenamefont {Witten}\ and\ \citenamefont
  {Yau}(1999)}]{Witten:1999xp}%
  \BibitemOpen
  \bibfield  {author} {\bibinfo {author} {\bibfnamefont {E.}~\bibnamefont
  {Witten}}\ and\ \bibinfo {author} {\bibfnamefont {S.-T.}\ \bibnamefont
  {Yau}},\ }\bibfield  {title} {\bibinfo {title} {{Connectedness of the
  boundary in the AdS / CFT correspondence}},\ }\href
  {https://doi.org/10.4310/ATMP.1999.v3.n6.a1} {\bibfield  {journal} {\bibinfo
  {journal} {Adv. Theor. Math. Phys.}\ }\textbf {\bibinfo {volume} {3}},\
  \bibinfo {pages} {1635} (\bibinfo {year} {1999})},\ \Eprint
  {https://arxiv.org/abs/hep-th/9910245} {arXiv:hep-th/9910245} \BibitemShut
  {NoStop}%
\bibitem [{\citenamefont {Maldacena}\ and\ \citenamefont
  {Maoz}(2004)}]{maldacena2004}%
  \BibitemOpen
  \bibfield  {author} {\bibinfo {author} {\bibfnamefont {J.}~\bibnamefont
  {Maldacena}}\ and\ \bibinfo {author} {\bibfnamefont {L.}~\bibnamefont
  {Maoz}},\ }\bibfield  {title} {\bibinfo {title} {Wormholes in {AdS}},\ }\href
  {https://doi.org/10.1088/1126-6708/2004/02/053} {\bibfield  {journal}
  {\bibinfo  {journal} {Journal of High Energy Physics}\ }\textbf {\bibinfo
  {volume} {2004}},\ \bibinfo {pages} {053} (\bibinfo {year} {2004})},\ \Eprint
  {https://arxiv.org/abs/hep-th/0401024} {arXiv:hep-th/0401024 [hep-th]}
  \BibitemShut {NoStop}%
\bibitem [{\citenamefont {Saad}\ \emph
  {et~al.}(2021{\natexlab{a}})\citenamefont {Saad}, \citenamefont {Shenker},
  \citenamefont {Stanford},\ and\ \citenamefont {Yao}}]{Saad:2021rcu}%
  \BibitemOpen
  \bibfield  {author} {\bibinfo {author} {\bibfnamefont {P.}~\bibnamefont
  {Saad}}, \bibinfo {author} {\bibfnamefont {S.~H.}\ \bibnamefont {Shenker}},
  \bibinfo {author} {\bibfnamefont {D.}~\bibnamefont {Stanford}},\ and\
  \bibinfo {author} {\bibfnamefont {S.}~\bibnamefont {Yao}},\ }\bibfield
  {title} {\bibinfo {title} {{Wormholes without averaging}},\ }\href@noop {} {\
   (\bibinfo {year} {2021}{\natexlab{a}})},\ \Eprint
  {https://arxiv.org/abs/2103.16754} {arXiv:2103.16754 [hep-th]} \BibitemShut
  {NoStop}%
\bibitem [{\citenamefont {Saad}\ \emph
  {et~al.}(2021{\natexlab{b}})\citenamefont {Saad}, \citenamefont {Shenker},\
  and\ \citenamefont {Yao}}]{Saad:2021uzi}%
  \BibitemOpen
  \bibfield  {author} {\bibinfo {author} {\bibfnamefont {P.}~\bibnamefont
  {Saad}}, \bibinfo {author} {\bibfnamefont {S.}~\bibnamefont {Shenker}},\ and\
  \bibinfo {author} {\bibfnamefont {S.}~\bibnamefont {Yao}},\ }\bibfield
  {title} {\bibinfo {title} {{Comments on wormholes and factorization}},\
  }\href@noop {} {\  (\bibinfo {year} {2021}{\natexlab{b}})},\ \Eprint
  {https://arxiv.org/abs/2107.13130} {arXiv:2107.13130 [hep-th]} \BibitemShut
  {NoStop}%
\bibitem [{\citenamefont
  {Mukhametzhanov}(2021{\natexlab{a}})}]{Mukhametzhanov:2021nea}%
  \BibitemOpen
  \bibfield  {author} {\bibinfo {author} {\bibfnamefont {B.}~\bibnamefont
  {Mukhametzhanov}},\ }\bibfield  {title} {\bibinfo {title} {{Half-wormholes in
  SYK with one time point}},\ }\href@noop {} {\  (\bibinfo {year}
  {2021}{\natexlab{a}})},\ \Eprint {https://arxiv.org/abs/2105.08207}
  {arXiv:2105.08207 [hep-th]} \BibitemShut {NoStop}%
\bibitem [{\citenamefont {Iliesiu}\ \emph
  {et~al.}(2021{\natexlab{b}})\citenamefont {Iliesiu}, \citenamefont
  {Kologlu},\ and\ \citenamefont {Turiaci}}]{Iliesiu:2021are}%
  \BibitemOpen
  \bibfield  {author} {\bibinfo {author} {\bibfnamefont {L.~V.}\ \bibnamefont
  {Iliesiu}}, \bibinfo {author} {\bibfnamefont {M.}~\bibnamefont {Kologlu}},\
  and\ \bibinfo {author} {\bibfnamefont {G.~J.}\ \bibnamefont {Turiaci}},\
  }\bibfield  {title} {\bibinfo {title} {{Supersymmetric indices factorize}},\
  }\href@noop {} {\  (\bibinfo {year} {2021}{\natexlab{b}})},\ \Eprint
  {https://arxiv.org/abs/2107.09062} {arXiv:2107.09062 [hep-th]} \BibitemShut
  {NoStop}%
\bibitem [{\citenamefont {Garc\'\i{}a-Garc\'\i{}a}\ and\ \citenamefont
  {Godet}(2021)}]{garcia2022}%
  \BibitemOpen
  \bibfield  {author} {\bibinfo {author} {\bibfnamefont {A.~M.}\ \bibnamefont
  {Garc\'\i{}a-Garc\'\i{}a}}\ and\ \bibinfo {author} {\bibfnamefont
  {V.}~\bibnamefont {Godet}},\ }\bibfield  {title} {\bibinfo {title}
  {Half-wormholes in nearly ads$_2$ holography},\ }\href@noop {} {\bibfield
  {journal} {\bibinfo  {journal} {eprint}\ } (\bibinfo {year} {2021})},\
  \Eprint {https://arxiv.org/abs/2107.07720} {arXiv:2107.07720 [hep-th]}
  \BibitemShut {NoStop}%
\bibitem [{\citenamefont {Blommaert}\ and\ \citenamefont
  {Kruthoff}(2022)}]{Blommaert:2021gha}%
  \BibitemOpen
  \bibfield  {author} {\bibinfo {author} {\bibfnamefont {A.}~\bibnamefont
  {Blommaert}}\ and\ \bibinfo {author} {\bibfnamefont {J.}~\bibnamefont
  {Kruthoff}},\ }\bibfield  {title} {\bibinfo {title} {{Gravity without
  averaging}},\ }\href {https://doi.org/10.21468/SciPostPhys.12.2.073}
  {\bibfield  {journal} {\bibinfo  {journal} {SciPost Phys.}\ }\textbf
  {\bibinfo {volume} {12}},\ \bibinfo {pages} {073} (\bibinfo {year} {2022})},\
  \Eprint {https://arxiv.org/abs/2107.02178} {arXiv:2107.02178 [hep-th]}
  \BibitemShut {NoStop}%
\bibitem [{\citenamefont
  {Mukhametzhanov}(2021{\natexlab{b}})}]{Mukhametzhanov:2021hdi}%
  \BibitemOpen
  \bibfield  {author} {\bibinfo {author} {\bibfnamefont {B.}~\bibnamefont
  {Mukhametzhanov}},\ }\bibfield  {title} {\bibinfo {title} {{Factorization and
  complex couplings in SYK and in Matrix Models}},\ }\href@noop {} {\
  (\bibinfo {year} {2021}{\natexlab{b}})},\ \Eprint
  {https://arxiv.org/abs/2110.06221} {arXiv:2110.06221 [hep-th]} \BibitemShut
  {NoStop}%
\bibitem [{\citenamefont {Blommaert}\ \emph {et~al.}(2021)\citenamefont
  {Blommaert}, \citenamefont {Iliesiu},\ and\ \citenamefont
  {Kruthoff}}]{Blommaert:2021fob}%
  \BibitemOpen
  \bibfield  {author} {\bibinfo {author} {\bibfnamefont {A.}~\bibnamefont
  {Blommaert}}, \bibinfo {author} {\bibfnamefont {L.~V.}\ \bibnamefont
  {Iliesiu}},\ and\ \bibinfo {author} {\bibfnamefont {J.}~\bibnamefont
  {Kruthoff}},\ }\bibfield  {title} {\bibinfo {title} {{Gravity factorized}},\
  }\href@noop {} {\  (\bibinfo {year} {2021})},\ \Eprint
  {https://arxiv.org/abs/2111.07863} {arXiv:2111.07863 [hep-th]} \BibitemShut
  {NoStop}%
\bibitem [{\citenamefont {Heckman}\ \emph {et~al.}(2022)\citenamefont
  {Heckman}, \citenamefont {Turner},\ and\ \citenamefont
  {Yu}}]{Heckman:2021vzx}%
  \BibitemOpen
  \bibfield  {author} {\bibinfo {author} {\bibfnamefont {J.~J.}\ \bibnamefont
  {Heckman}}, \bibinfo {author} {\bibfnamefont {A.~P.}\ \bibnamefont
  {Turner}},\ and\ \bibinfo {author} {\bibfnamefont {X.}~\bibnamefont {Yu}},\
  }\bibfield  {title} {\bibinfo {title} {{Disorder averaging and its UV
  discontents}},\ }\href {https://doi.org/10.1103/PhysRevD.105.086021}
  {\bibfield  {journal} {\bibinfo  {journal} {Phys. Rev. D}\ }\textbf {\bibinfo
  {volume} {105}},\ \bibinfo {pages} {086021} (\bibinfo {year} {2022})},\
  \Eprint {https://arxiv.org/abs/2111.06404} {arXiv:2111.06404 [hep-th]}
  \BibitemShut {NoStop}%
\bibitem [{\citenamefont {Schlenker}\ and\ \citenamefont
  {Witten}(2022)}]{Schlenker:2022dyo}%
  \BibitemOpen
  \bibfield  {author} {\bibinfo {author} {\bibfnamefont {J.-M.}\ \bibnamefont
  {Schlenker}}\ and\ \bibinfo {author} {\bibfnamefont {E.}~\bibnamefont
  {Witten}},\ }\bibfield  {title} {\bibinfo {title} {{No Ensemble Averaging
  Below the Black Hole Threshold}},\ }\href@noop {} {\  (\bibinfo {year}
  {2022})},\ \Eprint {https://arxiv.org/abs/2202.01372} {arXiv:2202.01372
  [hep-th]} \BibitemShut {NoStop}%
\bibitem [{\citenamefont {Collier}\ and\ \citenamefont
  {Perlmutter}(2022)}]{Collier:2022emf}%
  \BibitemOpen
  \bibfield  {author} {\bibinfo {author} {\bibfnamefont {S.}~\bibnamefont
  {Collier}}\ and\ \bibinfo {author} {\bibfnamefont {E.}~\bibnamefont
  {Perlmutter}},\ }\bibfield  {title} {\bibinfo {title} {{Harnessing S-Duality
  in $\mathcal{N}=4$ SYM \& Supergravity as $SL(2,\mathbb{Z})$-Averaged
  Strings}},\ }\href@noop {} {\  (\bibinfo {year} {2022})},\ \Eprint
  {https://arxiv.org/abs/2201.05093} {arXiv:2201.05093 [hep-th]} \BibitemShut
  {NoStop}%
\bibitem [{\citenamefont {Chandra}\ \emph {et~al.}(2022)\citenamefont
  {Chandra}, \citenamefont {Collier}, \citenamefont {Hartman},\ and\
  \citenamefont {Maloney}}]{Chandra:2022bqq}%
  \BibitemOpen
  \bibfield  {author} {\bibinfo {author} {\bibfnamefont {J.}~\bibnamefont
  {Chandra}}, \bibinfo {author} {\bibfnamefont {S.}~\bibnamefont {Collier}},
  \bibinfo {author} {\bibfnamefont {T.}~\bibnamefont {Hartman}},\ and\ \bibinfo
  {author} {\bibfnamefont {A.}~\bibnamefont {Maloney}},\ }\bibfield  {title}
  {\bibinfo {title} {{Semiclassical 3D gravity as an average of large-c
  CFTs}},\ }\href@noop {} {\  (\bibinfo {year} {2022})},\ \Eprint
  {https://arxiv.org/abs/2203.06511} {arXiv:2203.06511 [hep-th]} \BibitemShut
  {NoStop}%
\bibitem [{\citenamefont {Halliwell}\ and\ \citenamefont
  {Hartle}(1990)}]{Halliwell:1989dy}%
  \BibitemOpen
  \bibfield  {author} {\bibinfo {author} {\bibfnamefont {J.~J.}\ \bibnamefont
  {Halliwell}}\ and\ \bibinfo {author} {\bibfnamefont {J.~B.}\ \bibnamefont
  {Hartle}},\ }\bibfield  {title} {\bibinfo {title} {{Integration Contours for
  the No Boundary Wave Function of the Universe}},\ }\href
  {https://doi.org/10.1103/PhysRevD.41.1815} {\bibfield  {journal} {\bibinfo
  {journal} {Phys. Rev. D}\ }\textbf {\bibinfo {volume} {41}},\ \bibinfo
  {pages} {1815} (\bibinfo {year} {1990})}\BibitemShut {NoStop}%
\bibitem [{\citenamefont {Bousso}\ and\ \citenamefont
  {Hawking}(1999)}]{Bousso:1998na}%
  \BibitemOpen
  \bibfield  {author} {\bibinfo {author} {\bibfnamefont {R.}~\bibnamefont
  {Bousso}}\ and\ \bibinfo {author} {\bibfnamefont {S.~W.}\ \bibnamefont
  {Hawking}},\ }\bibfield  {title} {\bibinfo {title} {{Lorentzian condition in
  quantum gravity}},\ }\href {https://doi.org/10.1103/PhysRevD.59.103501}
  {\bibfield  {journal} {\bibinfo  {journal} {Phys. Rev. D}\ }\textbf {\bibinfo
  {volume} {59}},\ \bibinfo {pages} {103501} (\bibinfo {year} {1999})},\
  \Eprint {https://arxiv.org/abs/hep-th/9807148} {arXiv:hep-th/9807148}
  \BibitemShut {NoStop}%
\bibitem [{\citenamefont {Sorkin}(2009)}]{Sorkin:2009ka}%
  \BibitemOpen
  \bibfield  {author} {\bibinfo {author} {\bibfnamefont {R.~D.}\ \bibnamefont
  {Sorkin}},\ }\bibfield  {title} {\bibinfo {title} {{Is the spacetime metric
  Euclidean rather than Lorentzian?}},\ }\href@noop {} {\  (\bibinfo {year}
  {2009})},\ \Eprint {https://arxiv.org/abs/0911.1479} {arXiv:0911.1479
  [gr-qc]} \BibitemShut {NoStop}%
\bibitem [{\citenamefont {Witten}(2021)}]{Witten:2021nzp}%
  \BibitemOpen
  \bibfield  {author} {\bibinfo {author} {\bibfnamefont {E.}~\bibnamefont
  {Witten}},\ }\bibfield  {title} {\bibinfo {title} {{A Note On Complex
  Spacetime Metrics}},\ }\href@noop {} {\  (\bibinfo {year} {2021})},\ \Eprint
  {https://arxiv.org/abs/2111.06514} {arXiv:2111.06514 [hep-th]} \BibitemShut
  {NoStop}%
\bibitem [{\citenamefont {Plugge}\ \emph {et~al.}(2020)\citenamefont {Plugge},
  \citenamefont {Lantagne-Hurtubise},\ and\ \citenamefont
  {Franz}}]{plugge2020}%
  \BibitemOpen
  \bibfield  {author} {\bibinfo {author} {\bibfnamefont {S.}~\bibnamefont
  {Plugge}}, \bibinfo {author} {\bibfnamefont {E.}~\bibnamefont
  {Lantagne-Hurtubise}},\ and\ \bibinfo {author} {\bibfnamefont
  {M.}~\bibnamefont {Franz}},\ }\bibfield  {title} {\bibinfo {title} {Revival
  dynamics in a traversable wormhole},\ }\href
  {https://doi.org/10.1103/PhysRevLett.124.221601} {\bibfield  {journal}
  {\bibinfo  {journal} {Phys. Rev. Lett.}\ }\textbf {\bibinfo {volume} {124}},\
  \bibinfo {pages} {221601} (\bibinfo {year} {2020})},\ \Eprint
  {https://arxiv.org/abs/2003.03914} {arXiv:2003.03914 [cond-mat.str-el]}
  \BibitemShut {NoStop}%
\bibitem [{\citenamefont {Garc\'{\i}a-Garc\'{\i}a}\ \emph
  {et~al.}(2019)\citenamefont {Garc\'{\i}a-Garc\'{\i}a}, \citenamefont
  {Nosaka}, \citenamefont {Rosa},\ and\ \citenamefont
  {Verbaarschot}}]{garcia2019}%
  \BibitemOpen
  \bibfield  {author} {\bibinfo {author} {\bibfnamefont {A.~M.}\ \bibnamefont
  {Garc\'{\i}a-Garc\'{\i}a}}, \bibinfo {author} {\bibfnamefont
  {T.}~\bibnamefont {Nosaka}}, \bibinfo {author} {\bibfnamefont
  {D.}~\bibnamefont {Rosa}},\ and\ \bibinfo {author} {\bibfnamefont {J.~J.~M.}\
  \bibnamefont {Verbaarschot}},\ }\bibfield  {title} {\bibinfo {title} {Quantum
  chaos transition in a two-site sachdev-ye-kitaev model dual to an eternal
  traversable wormhole},\ }\href {https://doi.org/10.1103/PhysRevD.100.026002}
  {\bibfield  {journal} {\bibinfo  {journal} {Phys. Rev. D}\ }\textbf {\bibinfo
  {volume} {100}},\ \bibinfo {pages} {026002} (\bibinfo {year} {2019})},\
  \Eprint {https://arxiv.org/abs/1901.06031} {arXiv:1901.06031 [hep-th]}
  \BibitemShut {NoStop}%
\bibitem [{\citenamefont {Zhang}\ \emph
  {et~al.}(2021{\natexlab{a}})\citenamefont {Zhang}, \citenamefont {Jian},
  \citenamefont {Liu},\ and\ \citenamefont {Chen}}]{pengfei2021}%
  \BibitemOpen
  \bibfield  {author} {\bibinfo {author} {\bibfnamefont {P.}~\bibnamefont
  {Zhang}}, \bibinfo {author} {\bibfnamefont {S.-K.}\ \bibnamefont {Jian}},
  \bibinfo {author} {\bibfnamefont {C.}~\bibnamefont {Liu}},\ and\ \bibinfo
  {author} {\bibfnamefont {X.}~\bibnamefont {Chen}},\ }\bibfield  {title}
  {\bibinfo {title} {{SYK Meets Non-Hermiticity I: Emergent Replica Conformal
  Symmetry}},\ }\href@noop {} {\bibfield  {journal} {\bibinfo  {journal}
  {eprint}\ } (\bibinfo {year} {2021}{\natexlab{a}})},\ \Eprint
  {https://arxiv.org/abs/2104.04088} {arXiv:2104.04088 [cond-mat.str-el]}
  \BibitemShut {NoStop}%
\bibitem [{\citenamefont {Sahoo}\ \emph {et~al.}(2020)\citenamefont {Sahoo},
  \citenamefont {Lantagne-Hurtubise}, \citenamefont {Plugge},\ and\
  \citenamefont {Franz}}]{sahoo2020}%
  \BibitemOpen
  \bibfield  {author} {\bibinfo {author} {\bibfnamefont {S.}~\bibnamefont
  {Sahoo}}, \bibinfo {author} {\bibfnamefont {E.}~\bibnamefont
  {Lantagne-Hurtubise}}, \bibinfo {author} {\bibfnamefont {S.}~\bibnamefont
  {Plugge}},\ and\ \bibinfo {author} {\bibfnamefont {M.}~\bibnamefont
  {Franz}},\ }\bibfield  {title} {\bibinfo {title} {Traversable wormhole and
  hawking-page transition in coupled complex syk models},\ }\href
  {https://doi.org/10.1103/PhysRevResearch.2.043049} {\bibfield  {journal}
  {\bibinfo  {journal} {Phys. Rev. Research}\ }\textbf {\bibinfo {volume}
  {2}},\ \bibinfo {pages} {043049} (\bibinfo {year} {2020})},\ \Eprint
  {https://arxiv.org/abs/2006.06019} {arXiv:2006.06019 [cond-mat.str-el]}
  \BibitemShut {NoStop}%
\bibitem [{\citenamefont {Zhang}(2021)}]{zhang2020}%
  \BibitemOpen
  \bibfield  {author} {\bibinfo {author} {\bibfnamefont {P.}~\bibnamefont
  {Zhang}},\ }\bibfield  {title} {\bibinfo {title} {{More on Complex
  Sachdev-Ye-Kitaev Eternal Wormholes}},\ }\href
  {https://doi.org/10.1007/JHEP03(2021)087} {\bibfield  {journal} {\bibinfo
  {journal} {Journal of High Energy Physics}\ }\textbf {\bibinfo {volume}
  {03}},\ \bibinfo {pages} {087} (\bibinfo {year} {2021})},\ \Eprint
  {https://arxiv.org/abs/2011.10360} {arXiv:2011.10360 [hep-th]} \BibitemShut
  {NoStop}%
\bibitem [{\citenamefont {Garc\'{\i}a-Garc\'{\i}a}\ \emph
  {et~al.}(2021{\natexlab{a}})\citenamefont {Garc\'{\i}a-Garc\'{\i}a},
  \citenamefont {Zheng},\ and\ \citenamefont {Ziogas}}]{garcia2021e}%
  \BibitemOpen
  \bibfield  {author} {\bibinfo {author} {\bibfnamefont {A.~M.}\ \bibnamefont
  {Garc\'{\i}a-Garc\'{\i}a}}, \bibinfo {author} {\bibfnamefont {J.~P.}\
  \bibnamefont {Zheng}},\ and\ \bibinfo {author} {\bibfnamefont
  {V.}~\bibnamefont {Ziogas}},\ }\bibfield  {title} {\bibinfo {title} {Phase
  diagram of a two-site coupled complex syk model},\ }\href
  {https://doi.org/10.1103/PhysRevD.103.106023} {\bibfield  {journal} {\bibinfo
   {journal} {Phys. Rev. D}\ }\textbf {\bibinfo {volume} {103}},\ \bibinfo
  {pages} {106023} (\bibinfo {year} {2021}{\natexlab{a}})},\ \Eprint
  {https://arxiv.org/abs/2008.00039} {arXiv:2008.00039 [hep-th]} \BibitemShut
  {NoStop}%
\bibitem [{\citenamefont {Garc\'{\i}a-Garc\'{\i}a}\ \emph
  {et~al.}(2021{\natexlab{b}})\citenamefont {Garc\'{\i}a-Garc\'{\i}a},
  \citenamefont {Jia}, \citenamefont {Rosa},\ and\ \citenamefont
  {Verbaarschot}}]{garcia2021c}%
  \BibitemOpen
  \bibfield  {author} {\bibinfo {author} {\bibfnamefont {A.~M.}\ \bibnamefont
  {Garc\'{\i}a-Garc\'{\i}a}}, \bibinfo {author} {\bibfnamefont
  {Y.}~\bibnamefont {Jia}}, \bibinfo {author} {\bibfnamefont {D.}~\bibnamefont
  {Rosa}},\ and\ \bibinfo {author} {\bibfnamefont {J.~J.~M.}\ \bibnamefont
  {Verbaarschot}},\ }\bibfield  {title} {\bibinfo {title} {Sparse
  sachdev-ye-kitaev model, quantum chaos, and gravity duals},\ }\href
  {https://doi.org/10.1103/PhysRevD.103.106002} {\bibfield  {journal} {\bibinfo
   {journal} {Phys. Rev. D}\ }\textbf {\bibinfo {volume} {103}},\ \bibinfo
  {pages} {106002} (\bibinfo {year} {2021}{\natexlab{b}})},\ \Eprint
  {https://arxiv.org/abs/2007.13837} {arXiv:2007.13837 [hep-th]} \BibitemShut
  {NoStop}%
\bibitem [{\citenamefont {Xu}\ \emph {et~al.}(2020)\citenamefont {Xu},
  \citenamefont {Susskind}, \citenamefont {Su},\ and\ \citenamefont
  {Swingle}}]{swingle2020}%
  \BibitemOpen
  \bibfield  {author} {\bibinfo {author} {\bibfnamefont {S.}~\bibnamefont
  {Xu}}, \bibinfo {author} {\bibfnamefont {L.}~\bibnamefont {Susskind}},
  \bibinfo {author} {\bibfnamefont {Y.}~\bibnamefont {Su}},\ and\ \bibinfo
  {author} {\bibfnamefont {B.}~\bibnamefont {Swingle}},\ }\bibfield  {title}
  {\bibinfo {title} {A sparse model of quantum holography},\ }\href@noop {}
  {\bibfield  {journal} {\bibinfo  {journal} {eprint}\ } (\bibinfo {year}
  {2020})},\ \Eprint {https://arxiv.org/abs/2008.02303} {arXiv:2008.02303
  [cond-mat.str-el]} \BibitemShut {NoStop}%
\bibitem [{\citenamefont {C{\'{a}}ceres}\ \emph {et~al.}(2021)\citenamefont
  {C{\'{a}}ceres}, \citenamefont {Misobuchi},\ and\ \citenamefont
  {Pimentel}}]{caceres2021}%
  \BibitemOpen
  \bibfield  {author} {\bibinfo {author} {\bibfnamefont {E.}~\bibnamefont
  {C{\'{a}}ceres}}, \bibinfo {author} {\bibfnamefont {A.}~\bibnamefont
  {Misobuchi}},\ and\ \bibinfo {author} {\bibfnamefont {R.}~\bibnamefont
  {Pimentel}},\ }\bibfield  {title} {\bibinfo {title} {Sparse {SYK} and
  traversable wormholes},\ }\href {https://doi.org/10.1007/JHEP11(2021)015}
  {\bibfield  {journal} {\bibinfo  {journal} {Journal of High Energy Physics}\
  }\textbf {\bibinfo {volume} {2021}},\ \bibinfo {pages} {11} (\bibinfo {year}
  {2021})},\ \Eprint {https://arxiv.org/abs/2108.08808} {arXiv:2108.08808
  [hep-th]} \BibitemShut {NoStop}%
\bibitem [{\citenamefont {Zhou}\ and\ \citenamefont
  {Zhang}(2020)}]{pengfei2021a}%
  \BibitemOpen
  \bibfield  {author} {\bibinfo {author} {\bibfnamefont {T.-G.}\ \bibnamefont
  {Zhou}}\ and\ \bibinfo {author} {\bibfnamefont {P.}~\bibnamefont {Zhang}},\
  }\bibfield  {title} {\bibinfo {title} {Tunneling through an eternal
  traversable wormhole},\ }\href {https://doi.org/10.1103/PhysRevB.102.224305}
  {\bibfield  {journal} {\bibinfo  {journal} {Phys. Rev. B}\ }\textbf {\bibinfo
  {volume} {102}},\ \bibinfo {pages} {224305} (\bibinfo {year} {2020})},\
  \Eprint {https://arxiv.org/abs/2009.02641} {arXiv:2009.02641
  [cond-mat.str-el]} \BibitemShut {NoStop}%
\bibitem [{\citenamefont {Garc\'{\i}a-Garc\'{\i}a}\ and\ \citenamefont
  {Godet}(2021)}]{garcia2021}%
  \BibitemOpen
  \bibfield  {author} {\bibinfo {author} {\bibfnamefont {A.~M.}\ \bibnamefont
  {Garc\'{\i}a-Garc\'{\i}a}}\ and\ \bibinfo {author} {\bibfnamefont
  {V.}~\bibnamefont {Godet}},\ }\bibfield  {title} {\bibinfo {title} {Euclidean
  wormhole in the sachdev-ye-kitaev model},\ }\href
  {https://doi.org/10.1103/PhysRevD.103.046014} {\bibfield  {journal} {\bibinfo
   {journal} {Phys. Rev. D}\ }\textbf {\bibinfo {volume} {103}},\ \bibinfo
  {pages} {046014} (\bibinfo {year} {2021})},\ \Eprint
  {https://arxiv.org/abs/2010.11633} {arXiv:2010.11633 [hep-th]} \BibitemShut
  {NoStop}%
\bibitem [{\citenamefont {Garc\'{\i}a-Garc\'{\i}a}\ \emph
  {et~al.}(2022)\citenamefont {Garc\'{\i}a-Garc\'{\i}a}, \citenamefont {Jia},
  \citenamefont {Rosa},\ and\ \citenamefont {Verbaarschot}}]{garcia2021a}%
  \BibitemOpen
  \bibfield  {author} {\bibinfo {author} {\bibfnamefont {A.~M.}\ \bibnamefont
  {Garc\'{\i}a-Garc\'{\i}a}}, \bibinfo {author} {\bibfnamefont
  {Y.}~\bibnamefont {Jia}}, \bibinfo {author} {\bibfnamefont {D.}~\bibnamefont
  {Rosa}},\ and\ \bibinfo {author} {\bibfnamefont {J.~J.~M.}\ \bibnamefont
  {Verbaarschot}},\ }\bibfield  {title} {\bibinfo {title} {Dominance of replica
  off-diagonal configurations and phase transitions in a $pt$ symmetric
  sachdev-ye-kitaev model},\ }\href
  {https://doi.org/10.1103/PhysRevLett.128.081601} {\bibfield  {journal}
  {\bibinfo  {journal} {Phys. Rev. Lett.}\ }\textbf {\bibinfo {volume} {128}},\
  \bibinfo {pages} {081601} (\bibinfo {year} {2022})},\ \Eprint
  {https://arxiv.org/abs/2102.06630} {arXiv:2102.06630 [hep-th]} \BibitemShut
  {NoStop}%
\bibitem [{\citenamefont {García-García}\ \emph {et~al.}(2022)\citenamefont
  {García-García}, \citenamefont {Jia}, \citenamefont {Rosa},\ and\
  \citenamefont {Verbaarschot}}]{garcia2022a}%
  \BibitemOpen
  \bibfield  {author} {\bibinfo {author} {\bibfnamefont {A.~M.}\ \bibnamefont
  {García-García}}, \bibinfo {author} {\bibfnamefont {Y.}~\bibnamefont
  {Jia}}, \bibinfo {author} {\bibfnamefont {D.}~\bibnamefont {Rosa}},\ and\
  \bibinfo {author} {\bibfnamefont {J.~J.~M.}\ \bibnamefont {Verbaarschot}},\
  }\bibfield  {title} {\bibinfo {title} {Replica symmetry breaking in random
  non-hermitian systems},\ }\href@noop {} {\bibfield  {journal} {\bibinfo
  {journal} {eprint}\ } (\bibinfo {year} {2022})},\ \Eprint
  {https://arxiv.org/abs/2203.13080} {arXiv:2203.13080 [hep-th]} \BibitemShut
  {NoStop}%
\bibitem [{\citenamefont {García-García}\ \emph {et~al.}(2021)\citenamefont
  {García-García}, \citenamefont {Sá},\ and\ \citenamefont
  {Verbaarschot}}]{garcia2021d}%
  \BibitemOpen
  \bibfield  {author} {\bibinfo {author} {\bibfnamefont {A.~M.}\ \bibnamefont
  {García-García}}, \bibinfo {author} {\bibfnamefont {L.}~\bibnamefont
  {Sá}},\ and\ \bibinfo {author} {\bibfnamefont {J.~J.~M.}\ \bibnamefont
  {Verbaarschot}},\ }\bibfield  {title} {\bibinfo {title} {Symmetry
  classification and universality in non-hermitian many-body quantum chaos by
  the sachdev-ye-kitaev model},\ }\href@noop {} {\bibfield  {journal} {\bibinfo
   {journal} {eprint}\ } (\bibinfo {year} {2021})},\ \Eprint
  {https://arxiv.org/abs/2110.03444} {arXiv:2110.03444 [hep-th]} \BibitemShut
  {NoStop}%
\bibitem [{\citenamefont {Rathi}\ and\ \citenamefont
  {Roychowdhury}(2021)}]{Rathi:2021mla}%
  \BibitemOpen
  \bibfield  {author} {\bibinfo {author} {\bibfnamefont {H.}~\bibnamefont
  {Rathi}}\ and\ \bibinfo {author} {\bibfnamefont {D.}~\bibnamefont
  {Roychowdhury}},\ }\bibfield  {title} {\bibinfo {title} {{Phases of complex
  SYK from Euclidean wormholes}},\ }\href@noop {} {\  (\bibinfo {year}
  {2021})},\ \Eprint {https://arxiv.org/abs/2111.11279} {arXiv:2111.11279
  [hep-th]} \BibitemShut {NoStop}%
\bibitem [{\citenamefont {Zhang}\ \emph
  {et~al.}(2021{\natexlab{b}})\citenamefont {Zhang}, \citenamefont {Jian},
  \citenamefont {Liu},\ and\ \citenamefont {Chen}}]{zhang2021}%
  \BibitemOpen
  \bibfield  {author} {\bibinfo {author} {\bibfnamefont {P.}~\bibnamefont
  {Zhang}}, \bibinfo {author} {\bibfnamefont {S.-K.}\ \bibnamefont {Jian}},
  \bibinfo {author} {\bibfnamefont {C.}~\bibnamefont {Liu}},\ and\ \bibinfo
  {author} {\bibfnamefont {X.}~\bibnamefont {Chen}},\ }\bibfield  {title}
  {\bibinfo {title} {Emergent replica conformal symmetry in non-hermitian
  {SYK}2 chains},\ }\href {https://doi.org/10.22331/q-2021-11-16-579}
  {\bibfield  {journal} {\bibinfo  {journal} {Quantum}\ }\textbf {\bibinfo
  {volume} {5}},\ \bibinfo {pages} {579} (\bibinfo {year}
  {2021}{\natexlab{b}})},\ \Eprint {https://arxiv.org/abs/2104.04088}
  {arXiv:2104.04088 [cond-mat.str-el]} \BibitemShut {NoStop}%
\bibitem [{\citenamefont {Jian}\ \emph {et~al.}(2021)\citenamefont {Jian},
  \citenamefont {Liu}, \citenamefont {Chen}, \citenamefont {Swingle},\ and\
  \citenamefont {Zhang}}]{pengfei2021c}%
  \BibitemOpen
  \bibfield  {author} {\bibinfo {author} {\bibfnamefont {S.-K.}\ \bibnamefont
  {Jian}}, \bibinfo {author} {\bibfnamefont {C.}~\bibnamefont {Liu}}, \bibinfo
  {author} {\bibfnamefont {X.}~\bibnamefont {Chen}}, \bibinfo {author}
  {\bibfnamefont {B.}~\bibnamefont {Swingle}},\ and\ \bibinfo {author}
  {\bibfnamefont {P.}~\bibnamefont {Zhang}},\ }\bibfield  {title} {\bibinfo
  {title} {Measurement-induced phase transition in the monitored
  sachdev-ye-kitaev model},\ }\href
  {https://doi.org/10.1103/PhysRevLett.127.140601} {\bibfield  {journal}
  {\bibinfo  {journal} {Phys. Rev. Lett.}\ }\textbf {\bibinfo {volume} {127}},\
  \bibinfo {pages} {140601} (\bibinfo {year} {2021})},\ \Eprint
  {https://arxiv.org/abs/2106.09635} {arXiv:2106.09635 [quant-ph]} \BibitemShut
  {NoStop}%
\bibitem [{\citenamefont {Sá}\ \emph {et~al.}(2021)\citenamefont {Sá},
  \citenamefont {Ribeiro},\ and\ \citenamefont {Prosen}}]{sa2022}%
  \BibitemOpen
  \bibfield  {author} {\bibinfo {author} {\bibfnamefont {L.}~\bibnamefont
  {Sá}}, \bibinfo {author} {\bibfnamefont {P.}~\bibnamefont {Ribeiro}},\ and\
  \bibinfo {author} {\bibfnamefont {T.}~\bibnamefont {Prosen}},\ }\bibfield
  {title} {\bibinfo {title} {Lindbladian dissipation of strongly-correlated
  quantum matter},\ }\href@noop {} {\bibfield  {journal} {\bibinfo  {journal}
  {eprint}\ } (\bibinfo {year} {2021})},\ \Eprint
  {https://arxiv.org/abs/2112.12109} {arXiv:2112.12109 [cond-mat.stat-mech]}
  \BibitemShut {NoStop}%
\bibitem [{\citenamefont {Kulkarni}\ \emph {et~al.}()\citenamefont {Kulkarni},
  \citenamefont {Numasawa},\ and\ \citenamefont {Ryu}}]{kulkarni2022}%
  \BibitemOpen
  \bibfield  {author} {\bibinfo {author} {\bibfnamefont {A.}~\bibnamefont
  {Kulkarni}}, \bibinfo {author} {\bibfnamefont {T.}~\bibnamefont {Numasawa}},\
  and\ \bibinfo {author} {\bibfnamefont {S.}~\bibnamefont {Ryu}},\ }\bibfield
  {title} {\bibinfo {title} {Syk lindbladian},\ }\href@noop {} {\bibfield
  {journal} {\bibinfo  {journal} {eprint}\ }}\Eprint
  {https://arxiv.org/abs/2112.13489} {arXiv:2112.13489 [cond-mat.stat-mech]}
  \BibitemShut {NoStop}%
\bibitem [{\citenamefont {Lin}\ \emph {et~al.}(2019)\citenamefont {Lin},
  \citenamefont {Maldacena},\ and\ \citenamefont {Zhao}}]{Lin:2019qwu}%
  \BibitemOpen
  \bibfield  {author} {\bibinfo {author} {\bibfnamefont {H.~W.}\ \bibnamefont
  {Lin}}, \bibinfo {author} {\bibfnamefont {J.}~\bibnamefont {Maldacena}},\
  and\ \bibinfo {author} {\bibfnamefont {Y.}~\bibnamefont {Zhao}},\ }\bibfield
  {title} {\bibinfo {title} {Symmetries near the horizon},\ }\href
  {https://doi.org/10.1007/JHEP08(2019)049} {\bibfield  {journal} {\bibinfo
  {journal} {Journal of High Energy Physics}\ }\textbf {\bibinfo {volume}
  {08}},\ \bibinfo {pages} {049} (\bibinfo {year} {2019})},\ \Eprint
  {https://arxiv.org/abs/1904.12820} {arXiv:1904.12820 [hep-th]} \BibitemShut
  {NoStop}%
\bibitem [{\citenamefont {Harlow}\ and\ \citenamefont
  {Wu}(2021)}]{Harlow:2021dfp}%
  \BibitemOpen
  \bibfield  {author} {\bibinfo {author} {\bibfnamefont {D.}~\bibnamefont
  {Harlow}}\ and\ \bibinfo {author} {\bibfnamefont {J.-q.}\ \bibnamefont
  {Wu}},\ }\bibfield  {title} {\bibinfo {title} {{Algebra of
  diffeomorphism-invariant observables in Jackiw-Teitelboim Gravity}},\
  }\href@noop {} {\bibfield  {journal} {\bibinfo  {journal} {eprint}\ }
  (\bibinfo {year} {2021})},\ \Eprint {https://arxiv.org/abs/2108.04841}
  {arXiv:2108.04841 [hep-th]} \BibitemShut {NoStop}%
\bibitem [{\citenamefont {Mostafazadeh}(2010)}]{Mostafazadeh:2008pw}%
  \BibitemOpen
  \bibfield  {author} {\bibinfo {author} {\bibfnamefont {A.}~\bibnamefont
  {Mostafazadeh}},\ }\bibfield  {title} {\bibinfo {title} {{Pseudo-Hermitian
  Representation of Quantum Mechanics}},\ }\href
  {https://doi.org/10.1142/S0219887810004816} {\bibfield  {journal} {\bibinfo
  {journal} {Int. J. Geom. Meth. Mod. Phys.}\ }\textbf {\bibinfo {volume}
  {7}},\ \bibinfo {pages} {1191} (\bibinfo {year} {2010})},\ \Eprint
  {https://arxiv.org/abs/0810.5643} {arXiv:0810.5643 [quant-ph]} \BibitemShut
  {NoStop}%
\bibitem [{\citenamefont {Goel}\ and\ \citenamefont
  {Verlinde}(2021)}]{Goel:2021wim}%
  \BibitemOpen
  \bibfield  {author} {\bibinfo {author} {\bibfnamefont {A.}~\bibnamefont
  {Goel}}\ and\ \bibinfo {author} {\bibfnamefont {H.}~\bibnamefont
  {Verlinde}},\ }\bibfield  {title} {\bibinfo {title} {{Towards a String Dual
  of SYK}},\ }\href@noop {} {\  (\bibinfo {year} {2021})},\ \Eprint
  {https://arxiv.org/abs/2103.03187} {arXiv:2103.03187 [hep-th]} \BibitemShut
  {NoStop}%
\bibitem [{\citenamefont {Marolf}\ and\ \citenamefont
  {Santos}(2021)}]{marolf2021}%
  \BibitemOpen
  \bibfield  {author} {\bibinfo {author} {\bibfnamefont {D.}~\bibnamefont
  {Marolf}}\ and\ \bibinfo {author} {\bibfnamefont {J.~E.}\ \bibnamefont
  {Santos}},\ }\bibfield  {title} {\bibinfo {title} {Ads euclidean wormholes},\
  }\href@noop {} {\bibfield  {journal} {\bibinfo  {journal} {eprint}\ }
  (\bibinfo {year} {2021})},\ \Eprint {https://arxiv.org/abs/2101.08875}
  {arXiv:2101.08875 [hep-th]} \BibitemShut {NoStop}%
\bibitem [{\citenamefont {Maldacena}\ \emph {et~al.}(2018)\citenamefont
  {Maldacena}, \citenamefont {Milekhin},\ and\ \citenamefont
  {Popov}}]{Maldacena:2018gjk}%
  \BibitemOpen
  \bibfield  {author} {\bibinfo {author} {\bibfnamefont {J.}~\bibnamefont
  {Maldacena}}, \bibinfo {author} {\bibfnamefont {A.}~\bibnamefont
  {Milekhin}},\ and\ \bibinfo {author} {\bibfnamefont {F.}~\bibnamefont
  {Popov}},\ }\bibfield  {title} {\bibinfo {title} {{Traversable wormholes in
  four dimensions}},\ }\href@noop {} {\  (\bibinfo {year} {2018})},\ \Eprint
  {https://arxiv.org/abs/1807.04726} {arXiv:1807.04726 [hep-th]} \BibitemShut
  {NoStop}%
\bibitem [{\citenamefont {Bintanja}\ \emph {et~al.}(2021)\citenamefont
  {Bintanja}, \citenamefont {Esp\'\i{}ndola}, \citenamefont {Freivogel},\ and\
  \citenamefont {Nikolakopoulou}}]{Bintanja:2021xfs}%
  \BibitemOpen
  \bibfield  {author} {\bibinfo {author} {\bibfnamefont {S.}~\bibnamefont
  {Bintanja}}, \bibinfo {author} {\bibfnamefont {R.}~\bibnamefont
  {Esp\'\i{}ndola}}, \bibinfo {author} {\bibfnamefont {B.}~\bibnamefont
  {Freivogel}},\ and\ \bibinfo {author} {\bibfnamefont {D.}~\bibnamefont
  {Nikolakopoulou}},\ }\bibfield  {title} {\bibinfo {title} {{How to make
  traversable wormholes: eternal AdS$_{4}$ wormholes from coupled
  CFT\textquoteright{}s}},\ }\href {https://doi.org/10.1007/JHEP10(2021)173}
  {\bibfield  {journal} {\bibinfo  {journal} {Journal of High Energy Physics}\
  }\textbf {\bibinfo {volume} {10}},\ \bibinfo {pages} {173} (\bibinfo {year}
  {2021})},\ \Eprint {https://arxiv.org/abs/2102.06628} {arXiv:2102.06628
  [hep-th]} \BibitemShut {NoStop}%
\bibitem [{\citenamefont {Qi}\ and\ \citenamefont {Zhang}(2020)}]{qi2020}%
  \BibitemOpen
  \bibfield  {author} {\bibinfo {author} {\bibfnamefont {X.-L.}\ \bibnamefont
  {Qi}}\ and\ \bibinfo {author} {\bibfnamefont {P.}~\bibnamefont {Zhang}},\
  }\bibfield  {title} {\bibinfo {title} {The coupled syk model at finite
  temperature},\ }\href {https://doi.org/10.1007/jhep05(2020)129} {\bibfield
  {journal} {\bibinfo  {journal} {Journal of High Energy Physics}\ }\textbf
  {\bibinfo {volume} {2020}},\ \bibinfo {pages} {129} (\bibinfo {year}
  {2020})},\ \Eprint {https://arxiv.org/abs/2003.03916} {arXiv:2003.03916
  [hep-th]} \BibitemShut {NoStop}%
\bibitem [{\citenamefont {Roberge}\ and\ \citenamefont
  {Weiss}(1986)}]{Roberge:1986mm}%
  \BibitemOpen
  \bibfield  {author} {\bibinfo {author} {\bibfnamefont {A.}~\bibnamefont
  {Roberge}}\ and\ \bibinfo {author} {\bibfnamefont {N.}~\bibnamefont
  {Weiss}},\ }\bibfield  {title} {\bibinfo {title} {{Gauge Theories With
  Imaginary Chemical Potential and the Phases of \{QCD\}}},\ }\href
  {https://doi.org/10.1016/0550-3213(86)90582-1} {\bibfield  {journal}
  {\bibinfo  {journal} {Nucl. Phys. B}\ }\textbf {\bibinfo {volume} {275}},\
  \bibinfo {pages} {734} (\bibinfo {year} {1986})}\BibitemShut {NoStop}%
\bibitem [{\citenamefont {de~Forcrand}\ and\ \citenamefont
  {Philipsen}(2002)}]{deForcrand:2002hgr}%
  \BibitemOpen
  \bibfield  {author} {\bibinfo {author} {\bibfnamefont {P.}~\bibnamefont
  {de~Forcrand}}\ and\ \bibinfo {author} {\bibfnamefont {O.}~\bibnamefont
  {Philipsen}},\ }\bibfield  {title} {\bibinfo {title} {{The QCD phase diagram
  for small densities from imaginary chemical potential}},\ }\href
  {https://doi.org/10.1016/S0550-3213(02)00626-0} {\bibfield  {journal}
  {\bibinfo  {journal} {Nucl. Phys. B}\ }\textbf {\bibinfo {volume} {642}},\
  \bibinfo {pages} {290} (\bibinfo {year} {2002})},\ \Eprint
  {https://arxiv.org/abs/hep-lat/0205016} {arXiv:hep-lat/0205016} \BibitemShut
  {NoStop}%
\bibitem [{\citenamefont {Gao}\ \emph {et~al.}(2017)\citenamefont {Gao},
  \citenamefont {Jafferis},\ and\ \citenamefont {Wall}}]{gao2016}%
  \BibitemOpen
  \bibfield  {author} {\bibinfo {author} {\bibfnamefont {P.}~\bibnamefont
  {Gao}}, \bibinfo {author} {\bibfnamefont {D.~L.}\ \bibnamefont {Jafferis}},\
  and\ \bibinfo {author} {\bibfnamefont {A.}~\bibnamefont {Wall}},\ }\bibfield
  {title} {\bibinfo {title} {{Traversable Wormholes via a Double Trace
  Deformation}},\ }\href {https://doi.org/10.1007/JHEP12(2017)151} {\bibfield
  {journal} {\bibinfo  {journal} {Journal of High Energy Physics}\ }\textbf
  {\bibinfo {volume} {12}},\ \bibinfo {pages} {151} (\bibinfo {year} {2017})},\
  \Eprint {https://arxiv.org/abs/1608.05687} {arXiv:1608.05687 [hep-th]}
  \BibitemShut {NoStop}%
\bibitem [{\citenamefont {Chowdhury}\ \emph {et~al.}(2022)\citenamefont
  {Chowdhury}, \citenamefont {Godet}, \citenamefont {Papadoulaki},\ and\
  \citenamefont {Raju}}]{Chowdhury:2021nxw}%
  \BibitemOpen
  \bibfield  {author} {\bibinfo {author} {\bibfnamefont {C.}~\bibnamefont
  {Chowdhury}}, \bibinfo {author} {\bibfnamefont {V.}~\bibnamefont {Godet}},
  \bibinfo {author} {\bibfnamefont {O.}~\bibnamefont {Papadoulaki}},\ and\
  \bibinfo {author} {\bibfnamefont {S.}~\bibnamefont {Raju}},\ }\bibfield
  {title} {\bibinfo {title} {{Holography from the Wheeler-DeWitt equation}},\
  }\href {https://doi.org/10.1007/JHEP03(2022)019} {\bibfield  {journal}
  {\bibinfo  {journal} {JHEP}\ }\textbf {\bibinfo {volume} {03}},\ \bibinfo
  {pages} {019}},\ \Eprint {https://arxiv.org/abs/2107.14802} {arXiv:2107.14802
  [hep-th]} \BibitemShut {NoStop}%
\bibitem [{\citenamefont {Raju}(2022)}]{Raju:2020smc}%
  \BibitemOpen
  \bibfield  {author} {\bibinfo {author} {\bibfnamefont {S.}~\bibnamefont
  {Raju}},\ }\bibfield  {title} {\bibinfo {title} {{Lessons from the
  information paradox}},\ }\href
  {https://doi.org/10.1016/j.physrep.2021.10.001} {\bibfield  {journal}
  {\bibinfo  {journal} {Phys. Rept.}\ }\textbf {\bibinfo {volume} {943}},\
  \bibinfo {pages} {1} (\bibinfo {year} {2022})},\ \Eprint
  {https://arxiv.org/abs/2012.05770} {arXiv:2012.05770 [hep-th]} \BibitemShut
  {NoStop}%
\bibitem [{\citenamefont {Cardano}\ and\ \citenamefont
  {Spon}(1968)}]{cardano1968ars}%
  \BibitemOpen
  \bibfield  {author} {\bibinfo {author} {\bibfnamefont {G.}~\bibnamefont
  {Cardano}}\ and\ \bibinfo {author} {\bibfnamefont {C.}~\bibnamefont {Spon}},\
  }\bibfield  {title} {\bibinfo {title} {Ars magna (1545)},\ }\href@noop {}
  {\bibfield  {journal} {\bibinfo  {journal} {Opera Omnia}\ }\textbf {\bibinfo
  {volume} {4}},\ \bibinfo {pages} {221} (\bibinfo {year} {1968})}\BibitemShut
  {NoStop}%
\bibitem [{\citenamefont {Witten}(2010)}]{Witten:2010zr}%
  \BibitemOpen
  \bibfield  {author} {\bibinfo {author} {\bibfnamefont {E.}~\bibnamefont
  {Witten}},\ }\bibfield  {title} {\bibinfo {title} {A new look at the path
  integral of quantum mechanics},\ }\href@noop {} {\bibfield  {journal}
  {\bibinfo  {journal} {eprint}\ } (\bibinfo {year} {2010})},\ \Eprint
  {https://arxiv.org/abs/1009.6032} {arXiv:1009.6032 [hep-th]} \BibitemShut
  {NoStop}%
\bibitem [{\citenamefont {Bohigas}\ \emph {et~al.}(1984)\citenamefont
  {Bohigas}, \citenamefont {Giannoni},\ and\ \citenamefont
  {Schmit}}]{bohigas1984}%
  \BibitemOpen
  \bibfield  {author} {\bibinfo {author} {\bibfnamefont {O.}~\bibnamefont
  {Bohigas}}, \bibinfo {author} {\bibfnamefont {M.~J.}\ \bibnamefont
  {Giannoni}},\ and\ \bibinfo {author} {\bibfnamefont {C.}~\bibnamefont
  {Schmit}},\ }\bibfield  {title} {\bibinfo {title} {Characterization of
  chaotic quantum spectra and universality of level fluctuation laws},\ }\href
  {https://doi.org/10.1103/PhysRevLett.52.1} {\bibfield  {journal} {\bibinfo
  {journal} {Phys. Rev. Lett.}\ }\textbf {\bibinfo {volume} {52}},\ \bibinfo
  {pages} {1} (\bibinfo {year} {1984})}\BibitemShut {NoStop}%
\bibitem [{\citenamefont {Kawabata}\ \emph {et~al.}(2019)\citenamefont
  {Kawabata}, \citenamefont {Shiozaki}, \citenamefont {Ueda},\ and\
  \citenamefont {Sato}}]{ueda2019}%
  \BibitemOpen
  \bibfield  {author} {\bibinfo {author} {\bibfnamefont {K.}~\bibnamefont
  {Kawabata}}, \bibinfo {author} {\bibfnamefont {K.}~\bibnamefont {Shiozaki}},
  \bibinfo {author} {\bibfnamefont {M.}~\bibnamefont {Ueda}},\ and\ \bibinfo
  {author} {\bibfnamefont {M.}~\bibnamefont {Sato}},\ }\bibfield  {title}
  {\bibinfo {title} {Symmetry and topology in non-hermitian physics},\ }\href
  {https://doi.org/10.1103/PhysRevX.9.041015} {\bibfield  {journal} {\bibinfo
  {journal} {Phys. Rev. X}\ }\textbf {\bibinfo {volume} {9}},\ \bibinfo {pages}
  {041015} (\bibinfo {year} {2019})},\ \Eprint
  {https://arxiv.org/abs/1812.09133} {arXiv:1812.09133 [cond-mat.mes-hall]}
  \BibitemShut {NoStop}%
\bibitem [{\citenamefont {Maldacena}\ \emph
  {et~al.}(2016{\natexlab{b}})\citenamefont {Maldacena}, \citenamefont
  {Shenker},\ and\ \citenamefont {Stanford}}]{maldacena2015}%
  \BibitemOpen
  \bibfield  {author} {\bibinfo {author} {\bibfnamefont {J.}~\bibnamefont
  {Maldacena}}, \bibinfo {author} {\bibfnamefont {S.~H.}\ \bibnamefont
  {Shenker}},\ and\ \bibinfo {author} {\bibfnamefont {D.}~\bibnamefont
  {Stanford}},\ }\bibfield  {title} {\bibinfo {title} {A bound on chaos},\
  }\href {https://doi.org/10.1007/JHEP08(2016)106} {\bibfield  {journal}
  {\bibinfo  {journal} {Journal of High Energy Physics}\ }\textbf {\bibinfo
  {volume} {08}},\ \bibinfo {pages} {106} (\bibinfo {year}
  {2016}{\natexlab{b}})},\ \Eprint {https://arxiv.org/abs/1503.01409}
  {arXiv:1503.01409 [hep-th]} \BibitemShut {NoStop}%
\bibitem [{\citenamefont {S\'a}\ \emph {et~al.}(2020)\citenamefont {S\'a},
  \citenamefont {Ribeiro},\ and\ \citenamefont {Prosen}}]{sa2020}%
  \BibitemOpen
  \bibfield  {author} {\bibinfo {author} {\bibfnamefont {L.}~\bibnamefont
  {S\'a}}, \bibinfo {author} {\bibfnamefont {P.}~\bibnamefont {Ribeiro}},\ and\
  \bibinfo {author} {\bibfnamefont {T.}~\bibnamefont {Prosen}},\ }\bibfield
  {title} {\bibinfo {title} {Complex spacing ratios: A signature of dissipative
  quantum chaos},\ }\href {https://doi.org/10.1103/PhysRevX.10.021019}
  {\bibfield  {journal} {\bibinfo  {journal} {Phys. Rev. X}\ }\textbf {\bibinfo
  {volume} {10}},\ \bibinfo {pages} {021019} (\bibinfo {year} {2020})},\
  \Eprint {https://arxiv.org/abs/1910.12784} {arXiv:1910.12784
  [cond-mat.stat-mech]} \BibitemShut {NoStop}%
\bibitem [{\citenamefont {Oganesyan}\ and\ \citenamefont
  {Huse}(2007)}]{oganesyan2007}%
  \BibitemOpen
  \bibfield  {author} {\bibinfo {author} {\bibfnamefont {V.}~\bibnamefont
  {Oganesyan}}\ and\ \bibinfo {author} {\bibfnamefont {D.~A.}\ \bibnamefont
  {Huse}},\ }\bibfield  {title} {\bibinfo {title} {Localization of interacting
  fermions at high temperature},\ }\href
  {https://doi.org/10.1103/PhysRevB.75.155111} {\bibfield  {journal} {\bibinfo
  {journal} {Phys. Rev. B}\ }\textbf {\bibinfo {volume} {75}},\ \bibinfo
  {pages} {155111} (\bibinfo {year} {2007})},\ \Eprint
  {https://arxiv.org/abs/cond-mat/0610854} {arXiv:cond-mat/0610854
  [cond-mat.str-el]} \BibitemShut {NoStop}%
\bibitem [{\citenamefont {Atas}\ \emph {et~al.}(2013)\citenamefont {Atas},
  \citenamefont {Bogomolny}, \citenamefont {Giraud},\ and\ \citenamefont
  {Roux}}]{atas2016}%
  \BibitemOpen
  \bibfield  {author} {\bibinfo {author} {\bibfnamefont {Y.~Y.}\ \bibnamefont
  {Atas}}, \bibinfo {author} {\bibfnamefont {E.}~\bibnamefont {Bogomolny}},
  \bibinfo {author} {\bibfnamefont {O.}~\bibnamefont {Giraud}},\ and\ \bibinfo
  {author} {\bibfnamefont {G.}~\bibnamefont {Roux}},\ }\bibfield  {title}
  {\bibinfo {title} {Distribution of the ratio of consecutive level spacings in
  random matrix ensembles},\ }\href
  {https://doi.org/10.1103/PhysRevLett.110.084101} {\bibfield  {journal}
  {\bibinfo  {journal} {Phys. Rev. Lett.}\ }\textbf {\bibinfo {volume} {110}},\
  \bibinfo {pages} {084101} (\bibinfo {year} {2013})},\ \Eprint
  {https://arxiv.org/abs/1212.5611} {arXiv:1212.5611 [math-ph]} \BibitemShut
  {NoStop}%
\bibitem [{\citenamefont {Brody}\ \emph {et~al.}(1981)\citenamefont {Brody},
  \citenamefont {Flores}, \citenamefont {French}, \citenamefont {Mello},
  \citenamefont {Pandey},\ and\ \citenamefont {Wong}}]{brody1981}%
  \BibitemOpen
  \bibfield  {author} {\bibinfo {author} {\bibfnamefont {T.~A.}\ \bibnamefont
  {Brody}}, \bibinfo {author} {\bibfnamefont {J.}~\bibnamefont {Flores}},
  \bibinfo {author} {\bibfnamefont {J.~B.}\ \bibnamefont {French}}, \bibinfo
  {author} {\bibfnamefont {P.~A.}\ \bibnamefont {Mello}}, \bibinfo {author}
  {\bibfnamefont {A.}~\bibnamefont {Pandey}},\ and\ \bibinfo {author}
  {\bibfnamefont {S.~S.~M.}\ \bibnamefont {Wong}},\ }\bibfield  {title}
  {\bibinfo {title} {Random-matrix physics: spectrum and strength
  fluctuations},\ }\href {https://doi.org/10.1103/RevModPhys.53.385} {\bibfield
   {journal} {\bibinfo  {journal} {Rev. Mod. Phys.}\ }\textbf {\bibinfo
  {volume} {53}},\ \bibinfo {pages} {385} (\bibinfo {year} {1981})}\BibitemShut
  {NoStop}%
\bibitem [{\citenamefont {Ginibre}(1965)}]{ginibre1965}%
  \BibitemOpen
  \bibfield  {author} {\bibinfo {author} {\bibfnamefont {J.}~\bibnamefont
  {Ginibre}},\ }\bibfield  {title} {\bibinfo {title} {Statistical ensembles of
  complex, quaternion, and real matrices},\ }\href
  {https://doi.org/10.1063/1.1704292} {\bibfield  {journal} {\bibinfo
  {journal} {Journal of Mathematical Physics}\ }\textbf {\bibinfo {volume}
  {6}},\ \bibinfo {pages} {440} (\bibinfo {year} {1965})}\BibitemShut {NoStop}%
\bibitem [{\citenamefont {Tracy}\ and\ \citenamefont
  {Widom}(1994)}]{tracy1994}%
  \BibitemOpen
  \bibfield  {author} {\bibinfo {author} {\bibfnamefont {C.~A.}\ \bibnamefont
  {Tracy}}\ and\ \bibinfo {author} {\bibfnamefont {H.}~\bibnamefont {Widom}},\
  }\bibfield  {title} {\bibinfo {title} {Level-spacing distributions and the
  airy kernel},\ }\href {https://doi.org/10.1007/BF02100489} {\bibfield
  {journal} {\bibinfo  {journal} {Communications in Mathematical Physics}\
  }\textbf {\bibinfo {volume} {159}},\ \bibinfo {pages} {151} (\bibinfo {year}
  {1994})},\ \Eprint {https://arxiv.org/abs/hep-th/9211141}
  {arXiv:hep-th/9211141 [hep-th]} \BibitemShut {NoStop}%
\bibitem [{\citenamefont {Luitz}\ \emph {et~al.}(2015)\citenamefont {Luitz},
  \citenamefont {Laflorencie},\ and\ \citenamefont {Alet}}]{luitz2015}%
  \BibitemOpen
  \bibfield  {author} {\bibinfo {author} {\bibfnamefont {D.~J.}\ \bibnamefont
  {Luitz}}, \bibinfo {author} {\bibfnamefont {N.}~\bibnamefont {Laflorencie}},\
  and\ \bibinfo {author} {\bibfnamefont {F.}~\bibnamefont {Alet}},\ }\bibfield
  {title} {\bibinfo {title} {Many-body localization edge in the random-field
  heisenberg chain},\ }\href {https://doi.org/10.1103/PhysRevB.91.081103}
  {\bibfield  {journal} {\bibinfo  {journal} {Phys. Rev. B}\ }\textbf {\bibinfo
  {volume} {91}},\ \bibinfo {pages} {081103} (\bibinfo {year} {2015})},\
  \Eprint {https://arxiv.org/abs/1411.0660} {arXiv:1411.0660 [cond-mat.dis-nn]}
  \BibitemShut {NoStop}%
\bibitem [{\citenamefont {Bertrand}\ and\ \citenamefont
  {Garc\'{\i}a-Garc\'{\i}a}(2016)}]{bertrand2016}%
  \BibitemOpen
  \bibfield  {author} {\bibinfo {author} {\bibfnamefont {C.~L.}\ \bibnamefont
  {Bertrand}}\ and\ \bibinfo {author} {\bibfnamefont {A.~M.}\ \bibnamefont
  {Garc\'{\i}a-Garc\'{\i}a}},\ }\bibfield  {title} {\bibinfo {title} {Anomalous
  thouless energy and critical statistics on the metallic side of the many-body
  localization transition},\ }\href
  {https://doi.org/10.1103/PhysRevB.94.144201} {\bibfield  {journal} {\bibinfo
  {journal} {Phys. Rev. B}\ }\textbf {\bibinfo {volume} {94}},\ \bibinfo
  {pages} {144201} (\bibinfo {year} {2016})},\ \Eprint
  {https://arxiv.org/abs/1606.08419} {arXiv:1606.08419 [cond-mat.dis-nn]}
  \BibitemShut {NoStop}%
\bibitem [{\citenamefont {Numasawa}(2019)}]{numasawa2019}%
  \BibitemOpen
  \bibfield  {author} {\bibinfo {author} {\bibfnamefont {T.}~\bibnamefont
  {Numasawa}},\ }\bibfield  {title} {\bibinfo {title} {{Late Time Quantum Chaos
  of pure states in the SYK model}},\ }\href
  {https://doi.org/10.1103/PhysRevD.100.126017} {\bibfield  {journal} {\bibinfo
   {journal} {Phys. Rev. D}\ }\textbf {\bibinfo {volume} {100}},\ \bibinfo
  {pages} {126017} (\bibinfo {year} {2019})},\ \Eprint
  {https://arxiv.org/abs/1901.02025} {arXiv:1901.02025 [hep-th]} \BibitemShut
  {NoStop}%
\bibitem [{\citenamefont {Kourkoulou}\ and\ \citenamefont
  {Maldacena}(2017)}]{Kourkoulou:2017zaj}%
  \BibitemOpen
  \bibfield  {author} {\bibinfo {author} {\bibfnamefont {I.}~\bibnamefont
  {Kourkoulou}}\ and\ \bibinfo {author} {\bibfnamefont {J.}~\bibnamefont
  {Maldacena}},\ }\bibfield  {title} {\bibinfo {title} {{Pure states in the SYK
  model and nearly-$AdS_2$ gravity}},\ }\href@noop {} {\bibfield  {journal}
  {\bibinfo  {journal} {eprint}\ } (\bibinfo {year} {2017})},\ \Eprint
  {https://arxiv.org/abs/1707.02325} {arXiv:1707.02325 [hep-th]} \BibitemShut
  {NoStop}%
\bibitem [{\citenamefont {Freivogel}\ \emph {et~al.}(2019)\citenamefont
  {Freivogel}, \citenamefont {Godet}, \citenamefont {Morvan}, \citenamefont
  {Pedraza},\ and\ \citenamefont {Rotundo}}]{Freivogel:2019lej}%
  \BibitemOpen
  \bibfield  {author} {\bibinfo {author} {\bibfnamefont {B.}~\bibnamefont
  {Freivogel}}, \bibinfo {author} {\bibfnamefont {V.}~\bibnamefont {Godet}},
  \bibinfo {author} {\bibfnamefont {E.}~\bibnamefont {Morvan}}, \bibinfo
  {author} {\bibfnamefont {J.~F.}\ \bibnamefont {Pedraza}},\ and\ \bibinfo
  {author} {\bibfnamefont {A.}~\bibnamefont {Rotundo}},\ }\bibfield  {title}
  {\bibinfo {title} {{Lessons on Eternal Traversable Wormholes in AdS}},\
  }\href {https://doi.org/10.1007/JHEP07(2019)122} {\bibfield  {journal}
  {\bibinfo  {journal} {Journal of High Energy Physics}\ }\textbf {\bibinfo
  {volume} {07}},\ \bibinfo {pages} {122} (\bibinfo {year} {2019})},\ \Eprint
  {https://arxiv.org/abs/1903.05732} {arXiv:1903.05732 [hep-th]} \BibitemShut
  {NoStop}%
\bibitem [{\citenamefont {Van~Raamsdonk}(2021)}]{VanRaamsdonk:2020tlr}%
  \BibitemOpen
  \bibfield  {author} {\bibinfo {author} {\bibfnamefont {M.}~\bibnamefont
  {Van~Raamsdonk}},\ }\bibfield  {title} {\bibinfo {title} {{Comments on
  wormholes, ensembles, and cosmology}},\ }\href
  {https://doi.org/10.1007/JHEP12(2021)156} {\bibfield  {journal} {\bibinfo
  {journal} {Journal of High Energy Physics}\ }\textbf {\bibinfo {volume}
  {12}},\ \bibinfo {pages} {156} (\bibinfo {year} {2021})},\ \Eprint
  {https://arxiv.org/abs/2008.02259} {arXiv:2008.02259 [hep-th]} \BibitemShut
  {NoStop}%
\bibitem [{\citenamefont {Lantagne-Hurtubise}\ \emph
  {et~al.}(2020)\citenamefont {Lantagne-Hurtubise}, \citenamefont {Plugge},
  \citenamefont {Can},\ and\ \citenamefont {Franz}}]{plugge2020a}%
  \BibitemOpen
  \bibfield  {author} {\bibinfo {author} {\bibfnamefont {E.}~\bibnamefont
  {Lantagne-Hurtubise}}, \bibinfo {author} {\bibfnamefont {S.}~\bibnamefont
  {Plugge}}, \bibinfo {author} {\bibfnamefont {O.}~\bibnamefont {Can}},\ and\
  \bibinfo {author} {\bibfnamefont {M.}~\bibnamefont {Franz}},\ }\bibfield
  {title} {\bibinfo {title} {Diagnosing quantum chaos in many-body systems
  using entanglement as a resource},\ }\href
  {https://doi.org/10.1103/PhysRevResearch.2.013254} {\bibfield  {journal}
  {\bibinfo  {journal} {Phys. Rev. Research}\ }\textbf {\bibinfo {volume}
  {2}},\ \bibinfo {pages} {013254} (\bibinfo {year} {2020})},\ \Eprint
  {https://arxiv.org/abs/1907.01628} {arXiv:1907.01628 [cond-mat.str-el]}
  \BibitemShut {NoStop}%
\end{thebibliography}%
